\documentclass[fleqn,usenatbib]{mnras}

\usepackage{newtxtext,newtxmath}
\usepackage[T1]{fontenc}

\DeclareRobustCommand{\VAN}[3]{#2}
\let\VANthebibliography\thebibliography
\def\thebibliography{\DeclareRobustCommand{\VAN}[3]{##3}\VANthebibliography}

\usepackage{amsmath,bm}
\usepackage{graphicx}
\usepackage{subcaption}
\usepackage{comment}
\usepackage{booktabs}
\usepackage{float}
\usepackage{fontawesome}
\usepackage{url}
\usepackage{hyperref}
\usepackage[nameinlink,capitalise]{cleveref} % must be loaded after hyperref

\newcommand{\fNL}{f_\mathrm{NL}}

\newcommand{\aMpch}{h/\mathrm{Mpc}}
\newcommand{\dd}{\mathrm{d}}

\usepackage{xcolor}
\usepackage[normalem]{ulem}

\defcitealias{Addis_2024}{Paper I}
\defcitealias{Addis_2025}{Paper II}

\title[Light-cone effects on $P$+$B$ forecasts]{Constraints and biases: impact of light-cone effects on power spectrum and bispectrum forecasts}

\author[C. Addis et al.]{
Chris Addis$^{1}$\thanks{E-mail: c.l.j.addis@qmul.ac.uk}, Chris Clarkson$^{1,3}$,
Dionysios Karagiannis$^{2,3}$
\\
$^{1}$Centre for Theoretical Physics and Astronomy, School of Physical \& Chemical Sciences, Queen Mary University of London, London E1 4NS, UK\\
$^{2}$Van Swinderen Institute for Particle Physics and Gravity, University of Groningen, 9747 AG Groningen, The Netherlands\\
$^{3}$Department of Physics \& Astronomy, University of Western Cape, Cape Town 7535, South Africa}

\date{Accepted XXX. Received YYY; in original form ZZZ}

\pubyear{2026}

\begin{document}
\label{firstpage}
\pagerange{\pageref{firstpage}--\pageref{lastpage}}
\maketitle

\begin{abstract}

Next-generation galaxy surveys will map large-scale structure over unprecedented volumes, bringing two goals within reach: constraining primordial non-Gaussianity (PNG) and detecting relativistic effects on the largest scales. On these scales, relativistic and wide-separation (light-cone) effects share the scale dependence of the local-type PNG signal, and thus they must be modelled both to avoid biased inferences and to recover the true parameter uncertainties. We present joint power spectrum and bispectrum forecasts that consider the impact of these effects, with a multi-tracer bispectrum covariance including the odd-parity multipoles and leading non-Gaussian contributions. Neglecting light-cone effects leads to $\sim 2\sigma$ and $> 10\, \sigma$ biases on local-type $\fNL$ for a \textit{Euclid}-like H$\alpha$ and a MegaMapper-like LBG survey, respectively, mostly through lensing; wide-separation and local relativistic effects contribute less than $1\sigma$. Additionally, when we include these corrections, marginalising over the evolution and magnification biases without informative priors leads to a $\sim 60\%$ increase in the uncertainty on local-type $\fNL$ for a \textit{Euclid}-like power spectrum analysis and a $\sim 50\%$ increase for a joint power spectrum and bispectrum analysis of the even multipoles. However, we can break most of this degeneracy by considering the odd-parity multipoles, which are sensitive to leading-order relativistic corrections, both in the single-tracer bispectrum and the multi-tracer power spectrum. Non-local PNG shapes are less affected; for these, a bright-faint split of the MegaMapper-like survey improves the joint power spectrum and bispectrum constraints on the equilateral and orthogonal shapes by a factor of $\sim2$. Finally, local relativistic effects should be detectable at SNR $\sim13$, with a bright-faint split of a DESI-like BGS survey, provided the magnification bias is known to a few per cent, but only marginally, SNR $\lesssim2.5$, for the higher-redshift surveys, and not with the single-tracer bispectrum dipole. The code, \textsc{CosmoWAP}, is publicly available.
\end{abstract}

% Select between one and six entries from the list of approved MNRAS keywords.

%%%%%%%%%%%%%%%%% BODY OF PAPER %%%%%%%%%%%%%%%%%%

\section{Introduction}\label{sec:intro}

A wealth of cosmological information is encoded in the manner in which matter is clustered across the universe. On the very largest observable scales, this can offer a unique window to constrain the primordial universe and test fundamental models of gravity. While the complexities of non-linear clustering are largely mitigated at these scales, mapping the universe over such large distances on our past light-cone introduces a different set of observational and theoretical challenges. Specifically, as we probe larger and larger scales, phenomena such as dynamical and light-cone relativistic effects \citep{Macdonald_2009,Yoo_2010,Challinor_2011,Bonvin_2011,Tram_2016, 2016VillaRampf}, survey window effects \cite[e.g.][]{Feldman_1994,Wilson_2017}, and wide-separation corrections \citep{Yoo_2015,Reimberg_2016,Beutler_2020,Elkhashab_2022,Noorikuhani_2022,Addis_2024} to our statistics become increasingly relevant. 

Ongoing Stage IV galaxy surveys, such as the Dark Energy Spectroscopic Instrument (DESI) \citep{desi}, \textit{Euclid} \citep{euclid_2}, \textit{SPHEREx} \citep{Dore_SPHEREx} and the Nancy Grace Roman Space Telescope (formerly WFIRST) \citep{wfirst}, along with future planned missions like SIRMOS \citep{Wirth_2026}, MegaMapper \citep{MegaMapper} and WST \citep{WST_2024}, will not only allow us to access these larger scales but will also do so with increased statistical power. It is therefore important that we are confident about the systematics and uncertainties in our Large-Scale Structure (LSS) analyses.

A major motivation for these surveys is to constrain primordial non-Gaussianity (PNG), a key probe of the physics of the early universe \citep{inflation_2022}. Local-type \citep{Salopek_1990,Komatsu_2001} PNG is considered a key indicator of multi-field inflation, while the equilateral \citep{Creminelli_2006} and orthogonal \citep{Senatore_2010} shapes largely constrain the nature of single-field inflation models; together, these three shapes can therefore distinguish between different inflationary scenarios. Leading constraints, for all three shapes, derive from the \textit{Planck} cosmic microwave background (CMB) data \citep{Akrami_2020}, but LSS constraints can, in theory, surpass those of the CMB with access to a far greater number of Fourier modes. However, one has to isolate this PNG signal from the proverbial haystack of late-time non-Gaussianity generated by the non-linear evolution of matter, and as such it is critical that we understand this late-time background noise if we are to measure PNG in LSS data.

The PNG signal enters into higher-order correlators, which are sensitive to non-Gaussian signals, the lowest order of which is the bispectrum (or 3-point function, its configuration-space counterpart). However, it also leaves an imprint in the galaxy power spectrum through its modification to halo bias. In the case of local-type PNG this leads to a scale-dependent correction which is a key target of power spectrum analyses on large scales \citep{Dalal_2008,Slosar_2008, Matarrese_2008,Afshordi_2008}, but for non-local shapes of PNG, higher-order statistics are necessary \cite[e.g.][]{Cabass_2022a}. Next-generation galaxy surveys should provide constraints competitive with those from \textit{Planck} \citep{Akrami_2020}, with $\mathcal{O}(1)$ uncertainties for local-type non-Gaussianity \citep{Giannantonio_2012,FontRibera_2014,Amendola_2018,Mueller_2019,Karagiannis_2020,Karagiannis_2021,Bull_2024}.

Additionally, access to these surveys should open up novel observables; while detections of local projection effects have so far been limited to observations of galaxy clusters \citep{Wotjak_2011,Iftach_2015}, previous studies have forecasted that local relativistic projection effects should be detectable in the odd-parity multipoles of the multi-tracer 2-point function \citep{Bonvin_2016, Gaztanaga_2017, Montano_2024a,Montano_2024b,Novara_2025,Montano_2026} and single-tracer 3-point function \citep{Maartens_2020,Noorikuhani_2022,Addis_2024,Rossiter_2024,Montano_2026,Rossiter_2026} for Stage IV and V galaxy surveys. Due to their sensitivity to the metric potentials, $\Phi$ and $\Psi$, these corrections can be used to construct novel tests of gravity on the largest scales \citep{Lombriser_2013,Bonvin_2018,Bonvin_2020,Castello_2022,Blanco_2023,Castello_2025,Camera_2026}.

These relativistic effects generate signals that share the same scale-dependence as that sourced by local-type PNG \citep{Bruni_2012,Camera_2015,Maartens_2021}. Consequently, neglecting these effects in an analysis can bias the measurement of local-type $\fNL$, or produce a false detection, both in the 2-point \citep{Wang_2020,Guedezounme_2025,Addis_2025} and 3-point functions \citep{Umeh_2017,Dio_2017,Rossiter_2024,Addis_2025}. Ignoring these contributions can lead to other biases on our inference, such as on cosmological parameters \citep{Addis_2025} and the power spectrum turnover scale \citep{Dube_2026}. More recent work has also considered the uncertainty in the relativistic biases, the magnification bias, $\mathcal{Q}$, and the evolution bias, $b_e$, which, through their degeneracy with local-type $\fNL$, increases its uncertainty \citep{Wang_2020,Addis_2025}.

Multi-tracer methods, whereby one correlates different samples, have been proposed to enable one to measure the leading-order $(\mathcal{H}/k)$ odd-parity relativistic effects in the cross-power spectrum \citep{Macdonald_2009,Bonvin_2014_b}. Bright-faint splits, in which a single galaxy population is split by luminosity into two non-overlapping samples, allow such a measurement within a single survey \citep{Bonvin_2016,Gaztanaga_2017,Montano_2024a,Montano_2024b,Novara_2025,Montano_2026}. Furthermore, as these approaches enhance our knowledge of the luminosity function \cite[e.g.][]{Blanco_2024}, they can also strengthen our constraining power on $\fNL$ \citep{Addis_2025}. Additionally, by cross-correlating highly biased samples one can boost the PNG signal \citep{Seljak_2009,Barreira_2023,Karagiannis_2024}. Therefore, multi-tracer approaches can improve LSS constraints on both relativistic effects and PNG.

The 3D Fourier power spectrum and bispectrum (or their configuration-space counterparts) are common bases for this type of analysis. These statistics retain full 3D information and allow for the isolation of leading-order relativistic effects in the odd-parity multipoles. However, one also has to consider the impact of wide-separation effects; these arise when correlating large scales on our past light-cone and exhibit the same scale dependence as the signals of interest. Consequently, a degeneracy exists between these effects, PNG, and relativistic corrections \citep{Noorikuhani_2022,Pardede_2023,Addis_2024,Guedezounme_2025,Addis_2025,Wen_2026}.

On large scales, clustering statistics therefore present a complex picture, with several degeneracies between effects that are not yet part of standard LSS analysis pipelines. Robustly accounting for these effects both in the power spectrum and bispectrum is critical not only for an unbiased analysis but also for recovering the true uncertainties on our parameters of interest. Therefore, understanding when these effects are important or when they can be neglected is useful for future work in this space.

This work builds directly on \citet{Addis_2024} and \citet{Addis_2025}, hereafter \citetalias{Addis_2024} and \citetalias{Addis_2025}, and presents comprehensive joint power spectrum and bispectrum forecasts that focus on relativistic and wide-separation effects, their impact on local, equilateral, and orthogonal PNG, and the detectability of the local relativistic effects themselves, for ongoing and future surveys. We consider the uncertainty in our relativistic biases and the impact it has on our measurement of PNG and relativistic effects; in particular, we show that the odd-parity multipoles break most of the resulting degeneracy with local-type $\fNL$, leaving the magnification bias as the limiting nuisance parameter, and we quantify how precisely it must be determined independently of the clustering. We include integrated effects in the power spectrum using the results from \citetalias{Addis_2025} but leave their computation in the bispectrum to future work.

Alternative bases to the 3D Fourier summary statistics are potentially beneficial on ultra-large scales, such as the spherical Fourier--Bessel basis \citep{Fisher_1995a,Heavens_1995} or the total angular momentum basis \citep{Dai_2012}, which retain both the full small-scale redshift information and the complete wide-separation information without the need for a series expansion. Indeed, several works have investigated local-type PNG and relativistic corrections in the spherical Fourier--Bessel basis due to these properties \citep{Yoo_2013,Wang_2020a,Wen_2025,Wen_2026,Lukens_2026}; however, challenges remain due to its computational complexity, and its extension to higher-order statistics is non-trivial \cite[e.g.][]{Benabou_2024}.

In this work we use the term relativistic effects broadly, to encompass integrated (e.g.\ lensing and ISW), local relativistic projection (e.g.\ relativistic Doppler and gravitational redshift) and `dynamical' relativistic effects that arise in second-order perturbation theory from non-linear evolution (these are also non-integrated and therefore local). We use the term light-cone effects to include both relativistic effects and wide-separation (wide-angle and radial evolution) corrections. 

The paper is laid out as follows: \cref{sec:theory} defines our estimators and focuses on the PNG contributions to them, including how we model the PNG biases. \cref{sec:forecast_details} outlines the key forecast method and details, along with the specific choices and assumptions that we have made. \cref{sec:cov} extends the covariance calculation of \citetalias{Addis_2025} to the bispectrum and defines our covariance matrix. Results focused on PNG and relativistic effects are presented in \cref{sec:results1PNG} and \cref{sec:results2GR}, respectively. Further details about the survey and bias modelling choices we make are included in the appendices.

All of our results are produced and implemented using the public \textit{Python} code \textsc{CosmoWAP} \href{https://github.com/craddis1/CosmoWAP}{\faGithub} \citep{Addis_2024,Addis_2025,cosmowap_paper} with companion \textit{Mathematica} notebooks included in \textsc{MathWAP} \href{https://github.com/craddis1/MathWAP}{\faGithub} \citep{mathwap}.

\section{Theory and Observables}\label{sec:theory}

This section briefly describes the theoretical modelling we use in our forecasts. First, we define our choice of estimators and their connection to what we compute theoretically. \Cref{sec:PNG_theory} outlines the PNG contributions to the power spectrum and bispectrum, including through the modification to the halo bias (\cref{sec:png_bias_params}), and \cref{sec:GR_theory} the corrections to the same kernels from local relativistic projection effects. Finally, \cref{sec:galaxy_bias} describes how we connect our halo bias quantities to the galaxy biases we use in our forecasts.

In this work we consider the observable summary statistics that are measured using the standard FFT estimators: the ``Yamamoto'' estimator \citep{Feldman_1994,beutler2014,Bianchi_2015,Scoccimarro_2015} for the power spectrum,
\begin{multline}\label{eq:yamamoto}
    \hat{P}^{ab}_{\ell}(k) = \frac{2\, \ell+1}{N_{k}}\int_{\mathcal{S}_1} \frac{\dd^3\bm{q}_1}{(2 \pi)^3} \int_{\mathcal{S}_2} \dd^3\bm{q}_2 \, \delta_D(\bm{q}_{1}+\bm{q}_{2}) \\
    \times \int_{\bm{x}_1,\bm{x}_2} \mathrm{e}^{-i\,(\bm{q}_1\cdot\bm{x}_1+\bm{q}_2\cdot\bm{x}_{2})}\,\Delta^{a}_W(\bm{x}_1)\,\Delta^{b}_W(\bm{x}_2)\,\mathcal{L}_{\ell}(\hat{\bm{q}}_1 \cdot \hat{\bm{d}}),
\end{multline}
and the ``Scoccimarro'' estimator\footnote{We restrict this analysis to the Legendre polynomial decomposed bispectrum and ignore the $m\neq0$ multipoles of the full spherical harmonic decomposition as these are no longer computable with separable FFTs. One could, however, instead work with alternative decompositions of the anisotropic bispectrum, such as tri-polar spherical harmonics (TripoSH) \citep{Sugiyama_2018}.} for the bispectrum \citep{Scoccimarro_2015},
\begin{multline}
    \hat{B}^{abc}_{\ell}(k_1,k_2,k_3) = \frac{2 \ell +1}{N^{T}_{123}\,k_{\rm f}^{3}}\int_{\mathcal{S}_1} \frac{\dd^3 \bm{q}_1}{(2 \pi)^3} \int_{\mathcal{S}_2} \dd^3 \bm{q}_2\int_{\mathcal{S}_3} \dd^3 \bm{q}_3 \\
    \times \, \delta_D(\bm{q}_{123}) \int_{\bm{x}_1,\bm{x}_2,\bm{x}_3} \mathrm{e}^{-i(\bm{q}_1 \cdot \bm{x}_{1} + \bm{q}_2 \cdot \bm{x}_{2}+ \bm{q}_3 \cdot \bm{x}_{3})} \\
    \times \, \Delta^a_W(\bm{x}_1)\,\Delta^b_W(\bm{x}_2)\,\Delta^c_W(\bm{x}_3) \, \mathcal{L}_{\ell}(\hat{\bm{q}}_1 \cdot \hat{\bm{d}}),
\end{multline}
where $\Delta_W(\bm{x})=\Delta(\bm{x})\,W(\bm{x})$ represents the windowed field for a given survey window function, $W(\bm{x})$. We normalise such that $\langle \Delta(\boldsymbol{k})\Delta(\boldsymbol{k}')\rangle = (2\pi)^3\delta_D(\boldsymbol{k}+\boldsymbol{k}')P(k)$, with $\delta_D(\boldsymbol{0}) = V/(2\pi)^3 = k_{\rm f}^{-3}$ for survey volume $V$, and $N_k$ and $N^T_{123}$ denote the number of fundamental modes and triangles in each bin. In practice, the integrals are evaluated as discrete sums over Fourier modes, $\int_{\mathcal{S}_i}{\rm d}^3\boldsymbol{q}_i\,F(\boldsymbol{q}_i) \to k_{\rm f}^3\sum_{\boldsymbol{q}_i\in\mathcal{S}_i}F(\boldsymbol{q}_i)$, over thin $k$-space shells, $\mathcal{S}_i$, of width $\Delta k$, centred at $k_i$, such that $\mathcal{S}_i \equiv \mathcal{S}(k_i|\Delta k)$ is the region of $k$-magnitudes contained by a given $k_i$-bin, $k_i - \Delta k/2 \leq k < k_i + \Delta k/2$.

\subsection{Local statistics}

In order to compute what these estimators measure in perturbation theory, it is helpful to introduce the `local' power spectrum and bispectrum \citep{hamilton1998,Zaroubi_1993, Scoccimarro_2015, castorina2018}, defined for a region of space described by a single line of sight (LOS), $\bm{d}$. For a given LOS, the local power spectrum is
\begin{equation}\label{eq:local_pk}
P^{ab}_{\text{loc}}(k,\mu ;\bm{d})=\int_{\bm{x}_{12}} \mathrm{e}^{-i\,\bm{k} \cdot \bm{x}_{12}} \, \langle\Delta^a(\bm{x}_1)\,\Delta^b(\bm{x}_2)\rangle,
\end{equation}
where the LOS orientation is defined as $\mu=\hat{\bm{k}}\cdot\hat{\bm{d}}$ and $\bm{x}_{12}=\bm{x}_1-\bm{x}_2$. Similarly, for the local bispectrum,
\begin{multline}\label{eq:local_bispectrum}
    B^{abc}_\mathrm{loc}(k_1,k_2,k_3, \mu_1, \phi_{12};\bm{d}) = \int \dd^3 \bm{x}_{13}\,\dd^3 \bm{x}_{23} \\
    \times \,\mathrm{e}^{-i(\bm{k}_1 \cdot \bm{x}_{13} + \bm{k}_2 \cdot \bm{x}_{23})}\,\langle\Delta^a(\bm{x}_1)\Delta^b(\bm{x}_2)\Delta^c(\bm{x}_3)\rangle,
\end{multline}
where the LOS orientation is defined by $\mu_1=\hat{\bm{k}}_1\cdot\hat{\bm{d}}$ and $\phi_{12}$ is the azimuthal angle such that $\mu_2 = \mu_1\cos{\theta} + \sqrt{1-\mu_1^2}\sin{\theta}\cos{\phi_{12}}$, where $\cos{\theta}= \hat{\bm{k}}_1 \cdot \hat{\bm{k}}_2$ (see \citealt{scoccimarro1999} for details on bispectrum geometry).

These configuration-space density fields can then be expressed as the Fourier transforms of the Fourier-space kernels and linear density field, and we can contract the Dirac deltas after computing our wide-separation expansions. By computing these contributions directly in Cartesian Fourier space, this framework enables the consistent inclusion of both wide-separation corrections and integrated relativistic effects \citep{Noorikuhani_2022, Addis_2024,Guedezounme_2025, Addis_2025}.

If we consider the `true' unwindowed theory\footnote{Window effects will be relevant for the large scales considered here, but we do not include them in these forecasts. For the power spectrum, the window convolution is implemented on the signal side in \textsc{CosmoWAP} (\href{https://github.com/craddis1/CosmoWAP/blob/v0.10.0/src/cosmo_wap/lib/survey_window.py}{\texttt{survey\_window.py}}): following \citet{Wilson_2017}, the window multiplies the correlation function multipoles, and the transforms to and from configuration space combine into a single matrix acting on the model multipoles \citep{Beutler_2021}, in which each wide-separation order can be given its own window \citep{beutler2019}. A consistent forecast would also need the window in the covariance, and for the bispectrum the convolution is a more involved computation \citep[see][]{Wang_2024}; we leave both to future work.} then we can straightforwardly relate these local quantities to the survey-averaged statistics one measures with the FFT estimators \citep{Scoccimarro_2015}. The LOS orientation can be projected onto a Legendre basis such that we obtain the local multipoles, which are then related to the survey-averaged statistics by averaging over the survey volume. Thus, the survey-averaged unwindowed power spectrum can be written as
\begin{equation}\label{eq:d_integral}
\hat{P}^{ab}_{\ell}(k) = \frac{(2\,\ell+1)}{2}\int_{V} \frac{\dd^3 \bm{d}}{ V} \int^{1}_{-1} \dd \mu \, P^{ab}_\mathrm{loc}(k,\mu;\bm{d})\,\mathcal{L}_{\ell}(\mu),
\end{equation}
and similarly for the `Scoccimarro' bispectrum multipole estimator,
\begin{multline}\label{eq:d_integral_bk}
\hat{B}^{abc}_{\ell}(k_1,k_2,k_3) = \frac{(2\,\ell+1)}{4 \pi}\int_{V} \frac{\dd^3 \bm{d}}{ V} \int^{1}_{-1} \dd \mu_1 \int^{2 \pi}_{0} \dd \phi_{12} \\
\times \, B^{abc}_\mathrm{loc}(k_1,k_2,k_3,\mu_1,\phi_{12};\bm{d})\,\mathcal{L}_{\ell}(\mu_1).
\end{multline}

Below, we give the tree-level kernels at zeroth order in the wide-separation expansion: the PNG contributions in \cref{sec:PNG_theory,sec:png_bias_params} and the local relativistic projection effects in \cref{sec:GR_theory}. We do not reproduce the derivations for the wide-separation corrections, or the integrated (lensing, ISW and time-delay) contributions that we include in the power spectrum; for these we adopt the formalism of \citetalias{Addis_2025} for the power spectrum and \citetalias{Addis_2024} for the bispectrum, both of which are implemented in the \textit{Python} package \textsc{CosmoWAP}.

\subsubsection{PNG contributions}\label{sec:PNG_theory}

PNG generates non-zero higher-order correlators in the primordial curvature perturbations; the primordial bispectrum $B_{\Phi}(k_1,k_2,k_3)$, which can be related to the primordial curvature bispectrum, can be linearly extrapolated to source the PNG contribution to the late-time matter density bispectrum in real space,
\begin{equation}
    B_\mathrm{PNG}(k_1,k_2,k_3, z)=M(k_1, z)M(k_2, z)M(k_3, z)\,B_{\Phi}(k_1,k_2,k_3),
\end{equation}
using the transfer function
\begin{equation}
    M(k,z) = \frac{2 \, c^2 \, k^2 \, D(z)}{3 \, \Omega_m \, H_0^2} T(k),
\end{equation}
which relates the Bardeen power spectrum to the late-time matter power spectrum
\begin{equation}
    P(k,z) = M^2(k,z)P_{\Phi}(k).
\end{equation}
Here, $D(z)$ and $T(k)$ are the linear growth factor and the matter transfer function, respectively.

The primordial bispectrum templates, for the local, equilateral, and orthogonal shapes considered here, are defined as:
\begin{subequations}
\begin{equation}
B^{\rm loc}_{\Phi}(k_1, k_2, k_3) = 2 \fNL^{\rm loc} \left[ P_{\Phi}(k_1)P_{\Phi}(k_2) + \text{2 perms} \right],
\end{equation}
\begin{multline}
B^{\rm eq}_{\Phi}(k_1, k_2, k_3) = 6 \fNL^{\rm eq} \Big\{ - \left[ P_{\Phi}(k_1)P_{\Phi}(k_2) + \text{2 perms} \right] \\
- 2 \left[ P_{\Phi}(k_1)P_{\Phi}(k_2)P_{\Phi}(k_3) \right]^{2/3} \\
+ \left[ P_{\Phi}^{1/3}(k_1)P_{\Phi}^{2/3}(k_2)P_{\Phi}(k_3) + \text{5 perms} \right] \Big\},
\end{multline}
\begin{multline}
B^{\rm orth}_{\Phi}(k_1, k_2, k_3) = 6 \fNL^{\rm orth} \left[ P_{\Phi}(k_1)P_{\Phi}(k_2)P_{\Phi}(k_3) \right]^{2/3} \\
\times \left[ (1+q)\,\frac{\Delta}{k_1k_2k_3} - q\,\frac{\Gamma^3}{(k_1k_2k_3)^2} \right],
\end{multline}
\end{subequations}
where $k_T = k_1 + k_2 + k_3$, $\Delta = (k_T - 2k_1)(k_T - 2k_2)(k_T - 2k_3)$, $\Gamma = \tfrac{2}{3}(k_1k_2 + k_2k_3 + k_3k_1) - \tfrac{1}{3}(k_1^2 + k_2^2 + k_3^2)$ and $q = 27/[-21 + 743/(7(20\pi^2 - 193))] \approx 8.53$. This is the orthogonal-LSS approximation of \citet[appendix~B]{Senatore_2010}, which, unlike the standard (CMB) orthogonal template, has the squeezed limit $\propto P_\Phi(k_1)P_\Phi(k_3)\,(k_3/k_1)^2$, as the equilateral template does, rather than as $k_3/k_1$. The two templates agree on equilateral triangles, and the orthogonal-LSS one is used in LSS analyses \citep[e.g.][]{Cabass_2022a}, since the squeezed limit sets the scale-dependent bias (\cref{sec:png_bias_params}). Local-type PNG peaks in squeezed configurations ($k_3 \ll k_2 \approx k_1$), the equilateral and orthogonal types in equilateral ones ($k_3 \approx k_2 \approx k_1$), and the orthogonal type also in folded ones ($k_1 \approx 2\,k_2 \approx 2\,k_3$).

The matter bispectrum is also sourced by non-linear gravitational evolution, and as such one can write the local multi-tracer galaxy bispectrum as the standard tree-level contribution plus the intrinsic PNG contribution\footnote{Note that here we drop the explicit LOS dependence and ignore wide-separation corrections, taking the zeroth-order term in the expansion. For details of the wide-separation corrections in the bispectrum see \citetalias{Addis_2024}.}
\begin{multline}\label{eq:bk_tree}
    B^{abc}_\mathrm{loc}(k_1,k_2,k_3,\mu_1,\phi_{12}) = \mathcal{K}^a_1(\bm{k}_1,\bm{d})\mathcal{K}^b_1(\bm{k}_2,\bm{d})\mathcal{K}^c_1(\bm{k}_3,\bm{d}) \\
    \times \, D(d)^3 B_\mathrm{PNG}(k_1,k_2,k_3) \\
    + 2\big[\mathcal{K}^a_1(\bm{k}_1,\bm{d})\mathcal{K}^b_1(\bm{k}_2,\bm{d})\mathcal{K}^c_2(-\bm{k}_1, -\bm{k}_2, \bm{d}) \\
    \times \, D(d)^4 P(k_1)P(k_2) + \text{perms}\big]
\end{multline}
where $D(d)$ is the linear growth factor expressed as a function of comoving distance. Additionally, the local power spectrum in linear perturbation theory is 
\begin{equation}\label{eq:pk_tree}
P^{ab}_\mathrm{loc}(k,\mu) = \mathcal{K}^a_1(\bm{k},\bm{d})\,\mathcal{K}^b_1(-\bm{k},\bm{d})\,D(d)^2P(k).
\end{equation}
The second kernel is evaluated at $-\bm{k}$, since $\langle\delta(\bm{k})\delta(\bm{k}')\rangle\propto\delta_D(\bm{k}+\bm{k}')$, and similarly the second-order kernel in \cref{eq:bk_tree}, which is contracted with the linear fields at $\bm{k}_1$ and $\bm{k}_2$, is evaluated at $(-\bm{k}_1,-\bm{k}_2)$. The redshift-space kernels $\mathcal{K}_1$ and $\mathcal{K}_2$, which also contain PNG contributions through the modification to the dark matter halo bias \citep{Baldauf_2011,tellarini2016}, can be expressed, up to linear order in $\fNL$ and dropping the explicit redshift dependence, as
\begin{subequations}
\begin{equation}\label{eq:first_order}
\mathcal{K}^a_1(\bm{q}_i,\bm{d}) = b^a_1+f\mu_i^2 + \frac{\fNL b^a_{\Psi}q^{\alpha}}{M(q)}
\end{equation}
\begin{multline}
\mathcal{K}^a_2(\bm{q}_i, \bm{q}_j, \bm{d}) = b^a_1 F_2(\bm{q}_i, \bm{q}_j) + f (\hat{\bm{q}}_{ij}\cdot\hat{\bm{d}})^2 G_2(\bm{q}_i, \bm{q}_j) \\
+ \frac{b^a_2}{2}+ b_{\Gamma_2}S_2(\bm{q}_i, \bm{q}_j) \\
+ \frac{f}{2} (\bm{q}_{ij}\cdot\hat{\bm{d}}) \left[\frac{\mu_i}{q_i} \mathcal{K}^a_1(\bm{q}_j,\bm{d}) + \frac{\mu_j}{q_j} \mathcal{K}^a_1(\bm{q}_i,\bm{d}) \right] \\
+ \fNL \bigg\{ \left[b^a_{\Psi\delta} + b^a_{\Psi} N_2(\bm{q}_j, \bm{q}_i)\right] \frac{q_i^\alpha}{2M(q_i)} \\
+ \left[b^a_{\Psi\delta} + b^a_{\Psi} N_2(\bm{q}_i, \bm{q}_j)\right] \frac{q_j^\alpha}{ 2M(q_j)} \bigg\}.
\end{multline}
\end{subequations}
where $\mu_i = \hat{\bm{q}}_{i}\cdot\hat{\bm{d}}$, $\alpha = 0$ for the local PNG case, and $\alpha = 2$ for both the equilateral and orthogonal shapes. We also define the shorthand notation $\bm{q}_{ij} \equiv \bm{q}_i + \bm{q}_j$. Further, $f$ is the linear growth rate, $N_2 = (\bm{q}_1\cdot\bm{q}_2)/q_1^2 $, $F_2$ and $G_2$ are the second-order symmetric standard perturbation theory kernels (see \citealt{Bernardeau_2001} for a review), while $b_{\Gamma_2}$ is the tidal bias and $S_2$ is the tidal kernel $S_2(\bm{q}_1,\bm{q}_2)= (\hat{\bm{q}}_1 \cdot \hat{\bm{q}}_2)^2-1$ \citep{Mcdonald_2009}.

\subsubsection{PNG biases}\label{sec:png_bias_params}
We model the PNG biases, $b_{\Psi}$ and $b^a_{\Psi\delta}$, using results from the peak-background split (PBS) formalism \citep{Mo_1997,Slosar_2008,Schmidt_2013}. The bias, $b_{\Psi}$, which results from PNG modulating the amplitude of small-scale fluctuations, can be defined by the logarithmic derivative of the halo number density, $n_h$, with respect to the amplitude of the smoothed density field, $\sigma_8$,
\begin{equation}
    b_{\Psi}\equiv 2\frac{\partial \ln n_h}{\partial \ln \sigma_8}.
\end{equation}
For a universal mass function we can derive analytic expressions for the non-Gaussian bias parameters \citep{Schmidt_2013,Desjacques_2018},
\begin{equation}
    b_{\Psi}(M,z) = A\left[2 \, \delta_c \,(b_{1}-1)+ 4 \left(\frac{\dd \ln{\sigma^2_{R,-\alpha}}}{\dd \ln{\sigma^2_{R}}}-1\right)\right]\frac{\sigma^2_{R,-\alpha}}{\sigma^2_{R}}
\end{equation}
and \citep{Giannantonio_2010,Karagiannis_2018}
\begin{multline}
    b_{\Psi \delta}(M,z) = 2A\bigg[\delta_c\left(b_2-\frac{8}{21}(b_1-1)\right) \\
    + (b_1-1)\left(2\frac{\dd \ln{\sigma^2_{R,-\alpha}}}{\dd \ln{\sigma^2_{R}}}-3\right)\bigg]\frac{\sigma^2_{R,-\alpha}}{\sigma^2_{R}}+b_{\Psi}
\end{multline}
where $M$ is halo mass, $\delta_c=1.686$ is the critical overdensity for spherical collapse,
\begin{equation*}
\sigma^2_{R,n} = \frac{1}{2\pi^2}\int \dd k \, k^{2+n} W_R(k)^2 P^L_m(k),
\end{equation*}
and $W_R$ is a spherical top hat filter. For local PNG, $\alpha=0$, $A=1$, and thus the biases reduce to their well-known results. For equilateral PNG, $A=3$ \citep{Schmidt_2010, Giannantonio_2012}, and for the orthogonal-LSS template $A = 3 - 5q/9 \approx -1.74$, from its squeezed limit at $k_1 = k_2$, taken as for the equilateral template \citep[section~V]{Schmidt_2010}\footnote{For $\alpha = 2$ the squeezed limit also depends on the angle between the long and short wavevectors. Following the literature, we take $k_1 = k_2$, where the long mode is perpendicular to the short ones; averaging over the angle instead, as the response of the small-scale variance to a long mode does, gives $A = 2$ and $2 - 14q/9 \approx -11.3$. Since $b_\Psi \propto A$, the non-local PNG biases are uncertain at this level.}. For the non-local cases, $\sigma_{R,-\alpha}$ can be computed for a given halo mass function (HMF). How we connect these halo biases to the galaxy biases we use in our forecasts is described in \cref{sec:galaxy_bias}.

\subsubsection{Local relativistic effects}\label{sec:GR_theory}
The local relativistic projection effects add terms suppressed by powers of $\mathcal{H}/k$, where $\mathcal{H}$ is the conformal Hubble rate, so that $\mathcal{K}_n\to\mathcal{K}_n+\mathcal{K}_{n,\mathrm{GR}}$. At first order,
\begin{equation}\label{eq:K1_GR}
\mathcal{K}^a_{1,\mathrm{GR}}(\bm{q}_i,\bm{d}) = \mathrm{i}\,\mu_i\,f\gamma^a_1\,\frac{\mathcal{H}}{q_i} + \gamma^a_2\,\frac{\mathcal{H}^2}{q_i^2},
\end{equation}%
where the first term is the Doppler contribution and the second collects those of the velocity potential and the gravitational potentials. In terms of the evolution and magnification biases of the sample, $\gamma_1=b_e-2\mathcal{Q}-2(1-\mathcal{Q})/(x\mathcal{H})-\mathcal{H}'/\mathcal{H}^2$ and $\gamma_2=f(3-b_e)+\tfrac{3}{2}\Omega_m(2-f-2\mathcal{Q}+\gamma_1)$ \citep[e.g.][]{Maartens_2021}, where $x$ is the comoving distance, the prime denotes a conformal-time derivative and $\Omega_m$ is evaluated at the redshift of the sample. Substituting \cref{eq:be_nevol} expresses them in terms of the count slope, $n_{\rm evol}$, and $\mathcal{Q}$, the relativistic nuisance parameters of our forecasts (\cref{sec:forecast_details}): for a sample selected at a fixed flux,
\begin{subequations}\label{eq:gamma}
\begin{align}
\gamma_1 &= -\left(1+n_{\rm evol}+4\mathcal{Q}\right), \label{eq:gamma1}\\
\gamma_2 &= f\left[4+n_{\rm evol}+2\mathcal{Q}-\frac{2(1-\mathcal{Q})}{x\mathcal{H}}-\frac{\mathcal{H}'}{\mathcal{H}^2}\right] \nonumber\\
&\quad+\frac{3}{2}\,\Omega_m\left(1-f-n_{\rm evol}-6\mathcal{Q}\right). \label{eq:gamma2}
\end{align}
\end{subequations}
A K-correction, or a bright-faint split at a fixed bright fraction, changes only the relation between $n_{\rm evol}$ and $b_e$ (\cref{sec:magnification_evolution,sec:bf_split}), and so adds terms proportional to $\mathcal{Q}$ to \cref{eq:gamma} but leaves the expressions in terms of $b_e$ unchanged.

The second-order kernel is longer and contains relativistic terms up to fourth order in $\mathcal{H}/k$, with coefficients that depend on $b_1$, $b_e$, $\mathcal{Q}$ and their derivatives with respect to time and luminosity (\citealt{Jolicoeur_2018}; we use the form of \citealt{Maartens_2021}). Since the fields are real in configuration space, every kernel satisfies $\mathcal{K}_n(-\bm{q}_1,\ldots,-\bm{q}_n,\bm{d})=\mathcal{K}^*_n(\bm{q}_1,\ldots,\bm{q}_n,\bm{d})$: the Newtonian terms and the relativistic terms of even order in $\mathcal{H}/k$ are real and even in the wavevectors, while those of odd order, such as the Doppler term in \cref{eq:K1_GR}, are imaginary and odd.

\subsubsection{Galaxy bias}\label{sec:galaxy_bias}
To connect halo-based quantities to galaxy-based quantities one needs a prescription for how galaxies populate haloes. One way of doing this is to use a Halo Occupation Distribution (HOD) \citep{Cooray_2002} to describe the mean number of galaxies per halo, $\langle N(M,z)\rangle$. Thus, the galaxy bias values can be obtained by a weighted integral of the halo bias, $b_i^h(M,z)$, above a minimum halo mass, $M_\mathrm{min}$,
\begin{equation}
    b_i(z)= \frac{\int^{\infty}_{M_\mathrm{min}(z)} \dd \ln{M} \, n_h(M,z)b_i^h(M,z)\langle N(M,z)\rangle}{\bar{n}_g(z)},
\end{equation}
where $n_h(M,z)$ is the mean number density of haloes for a given mass and $\bar{n}_g(z)$ is the mean galaxy density. The mean galaxy density can be computed from the integral
\begin{equation}\label{eq:n_bar}
    \bar{n}_g(z) = \int_{M_{\text{min}}(z)}^{\infty} \dd \ln M \, n_h(M, z) \langle N(M, z) \rangle.
\end{equation}
The HOD models we use in our forecasts are outlined in \cref{app:survey_details}.

\section{Forecast Details}\label{sec:forecast_details}

This section outlines the details of our data vector, forecast set-up and the specific fiducial parameters that we choose. The full covariance matrix is detailed in \cref{sec:cov}.

Throughout this study, we assume a fiducial cosmology from the \textit{Planck} results \citep{Planck_2018}, and in all our forecasts, for both PNG and the local relativistic effects, we marginalise over the cosmological parameters with \textit{Planck} constraints as priors; in the Fisher forecasts these are $\Omega_m$, $h$, $n_s$ and $\ln(10^{10}A_s)$. We assume a fiducial value of $\fNL=0$ for all PNG shapes and forecast for each PNG shape independently. For multipoles beyond the monopole we assume an endpoint LOS, as enforced by the FFT estimators, specifically $\bm{d}=\bm{x}_1$ for the power spectrum and $\bm{d}=\bm{x}_3$ for the bispectrum. This choice is relevant for wide-separation type corrections including the integrated effects beyond the Limber approximation (see \citetalias{Addis_2025}). In the monopole, where we are free to choose our LOS as it only denotes the point about which we do our wide-separation expansion, we adopt a midpoint LOS for the power spectrum and a centre-of-mass LOS for the bispectrum. This choice has minimal impact on our forecasts; see \citetalias{Addis_2024} and \citetalias{Addis_2025} for more details.

\subsection{Bias nuisance parameters}\label{sec:nuisance}

We marginalise over the bias parameters in each redshift bin, namely the linear bias, $b_1$, and, where we include relativistic effects, the magnification bias, $\mathcal{Q}$, and the count slope, $n_{\rm evol}$, which we use in place of the evolution bias, $b_e$\footnote{The evolution bias is defined at fixed luminosity (rather than at the fixed flux of the survey) and so is not directly observable. $n_{\rm evol}\equiv\mathrm{d}\ln N/\mathrm{d}\ln(1+z)$ is the slope of the observed redshift distribution of the sample, $N(z)$, and so, like $\mathcal{Q}$, can be measured from the catalogue itself (\cref{sec:magnification_evolution}). Evolution bias then follows from $n_{\rm evol}$ and $\mathcal{Q}$ through \cref{eq:be_nevol}, so that varying $\mathcal{Q}$, or the cosmology, at fixed $n_{\rm evol}$ also changes $b_e$.}. Whenever the bispectrum is included, we also marginalise over the quadratic and tidal biases, $b_2$ and $b_{\Gamma_2}$, in each redshift bin (and for each sample of a bright-faint split) with no prior; as they enter only the bispectrum, they are not varied in the power spectrum analyses. The fiducial values of $b_2$ come from the HOD (\cref{sec:galaxy_bias}) and we use the local Lagrangian relation, $b_{\Gamma_2}=-\frac{2}{7}(b_1-1)$, for the tidal bias. The fiducial bias model is described in \cref{app:survey_details}. We consider two limiting treatments of the relativistic biases, either holding them at their fiducial values (`fixed'), which assumes that they are known exactly, or marginalising over them in each redshift bin with no prior (`free'), giving $3$ and $5$ nuisance parameters per bin, respectively, for a single tracer when the bispectrum is included. The two use the same signal and covariance.

\subsubsection{Priors on the bias parameters}\label{sec:bias_priors}

For the \textit{Euclid}-like H$\alpha$ survey we also consider Gaussian priors on $n_{\rm evol}$ and $\mathcal{Q}$ forward-modelled from the uncertainty in its luminosity function, following \citet{Wang_2020}, as an example of a sample-motivated prior; the method is described in \cref{sec:lumfunc_prior}. We discuss how else these biases can be determined, including from the survey catalogue itself, in \cref{sec:beQ_constraints}.

We do not, however, account for uncertainty in the PNG biases, $b_{\Psi}$ and $b_{\Psi \delta}$, and hold them fixed in our forecasts. Under the assumption of universality of the mass function we can derive expressions for the PNG biases for a given HOD and linear bias for each shape using a peak-background split approach \citep{Slosar_2008,Matarrese_2008,Dalal_2008,Giannantonio_2010,Baldauf_2011}. While extensive results from $N$-body simulations have shown PNG biases to depart from their universal mass function expectations \citep{Grossi_2009,Barreira_2020,Lazeryas_2023,PNGUNIT_2024,Fondi_2024}, and one could account for this in our forecasts by introducing modifications to the PNG biases along with further nuisance parameters, we leave a discussion of non-universal estimates of the bias parameters for the different PNG shapes to future work. We expect this to mainly affect the uncertainty on the determination of $\fNL$, but to have less impact on the core focus of this work -- the degeneracy of $\fNL$ with light-cone effects (since it does not change the $k$-dependence). Breaking the degeneracy between the PNG biases and $\fNL$ is non-trivial without strict priors on $b_{\Psi}$ \citep{Barreira_2020,Barreira_2022a,Dizgah_2021,Linde_euclid_2026}, such as from simulations calibrated to the selected sample \citep{Fondi_2024} or from the redshift evolution of its number counts \citep{Nascimento_2026}, which is related to the evolution bias (\cref{sec:be_bpsi}); this motivates the inclusion not only of the bispectrum but also of multi-tracer approaches \cite[e.g.][]{Barreira_2023}.

\subsection{Data vector}\label{sec:data_vector} 

Our full data vector, $D$, is a sum over redshift bins\footnote{We assume each redshift bin is independent -- one would expect this to be a reasonable assumption due to the thick redshift bins considered in our analysis. The cross-bin correlation will be dominated by the lensing contribution, as is typical in angular analyses.}. The data vector in each bin, $D_z$, then contains a given selection of power spectrum and bispectrum multipoles:
\begin{equation}
D_z = \{P_{\ell_i}, ... , P_{\ell_n},B_{\ell_i}, ... , B_{\ell_m}\}.
\end{equation}
For our single-tracer analysis, we include even multipoles up to the hexadecapole, $\ell=0,2,4$, for the power spectrum and up to the quadrupole for the bispectrum, $\ell=0,2$, and can also include the non-zero bispectrum dipole, $\ell=1$. In the full multi-tracer analysis, the data vector for each power spectrum multipole consists of three combinations, $P_{\ell_i=\text{even}} = \{P_{\ell_i}^{BB}, P_{\ell_i}^{BF}, P_{\ell_i}^{FF}\}$ from the bright, $B$, and faint, $F$, samples. In the case of the bright-faint split for the bispectrum, each multipole consists of four independent combinations, $B_{\ell_i=\text{even}} = \{B_{\ell_i}^{BBB}, B_{\ell_i}^{BBF}, B_{\ell_i}^{BFF}, B_{\ell_i}^{FFF}\}$. Additionally, in the multi-tracer case one can now also consider the odd multipoles of the power spectrum, $(\ell=1,3)$; however, the auto-correlations are still zero and therefore we only include the cross-correlations for the odd multipoles, $P_{\ell_i= \text{odd}} = P_{\ell_i}^{BF}$.

For each multipole, we use $k$-bins of width $\Delta k = 4 k_\mathrm{f}$, with the largest scale set by the fundamental frequency of the survey, $k_\mathrm{f} = 2 \pi/V^{1/3}$, and impose a constant cut-off at $k_\mathrm{max} =0.15\,[\aMpch]$ for the power spectrum and $k_\mathrm{max} =0.1\,[\aMpch]$ for the bispectrum. We also impose an effective $k_\mathrm{min}$ by excluding every $k$-bin that contains Fourier modes with $k < 2 \pi/x(z_\mathrm{min})$, where $x$ is the comoving distance and $z_\mathrm{min}$ is the minimum redshift of the redshift bin, as above these scales we expect the perturbative wide-separation expansion to have broken down, which is particularly relevant for low-redshift surveys.

Each estimator averages over the Fourier modes in its bin, so we average the model in the same way rather than evaluating it at the bin centre. For the power spectrum, the multipoles, their derivatives and $\textsf{C}_{PP}$ are averaged over the modes of each $k$-shell. For the bispectrum we follow \citet{Oddo_2020} and use the effective triangle of each bin, whose sides are the mean longest, middle and shortest sides of its closed triangles, weighted by their number; as the estimator sums over every labelling of the sides in the same shell, isosceles and equilateral bins are averaged over the corresponding relabellings. The bispectrum, its derivatives and $C^{PPP}$ are evaluated on this triangle, with the triangle count of the full bin (\cref{eq:beta}), while the non-Gaussian covariance terms use the bispectrum averaged over the triangles of each bin (\cref{sec:cov_ng}). Averaging the bispectrum and its derivatives over the triangles of each bin instead changes our single-tracer \textit{Euclid}-like constraints on $\fNL$ and on the amplitude of the local relativistic effects by $\lesssim1\%$.

For simplicity, in the bispectrum we restrict the analysis to triangular configurations satisfying $k_1 \geq k_2 \geq k_3$, although beyond the monopole the bispectrum is no longer symmetric under cyclic permutations of $\bm{k}_1, \bm{k}_2, \bm{k}_3$.

\subsection{Methods} 

To explore the constraints and biases on the parameters of interest, we either sample the posteriors with a Markov Chain Monte Carlo (MCMC) pipeline or use Fisher forecasts\footnote{We use MCMC for the bias plots as Fisher forecasts are evaluated only at the fiducial values, and so one would expect the approximation to break down for large biases; we use Fisher forecasts elsewhere as the MCMCs become more computationally expensive when we move to multi-tracer forecasts, marginalise over several parameters per bin and include integrated effects.}. These methods are described in Section 3 of \citetalias{Addis_2025} and in \citet{cosmowap_paper}; we summarise the key points below. The MCMCs are implemented in \textsc{CosmoWAP} using \textsc{Cobaya} \citep{Neal_2005,Lewis_2002,Lewis_2013,Torrado_2021} where we speed up the sampling over cosmological parameters by using \textsc{CosmoPower} \citep{Mancini_2022} emulators for the linear power spectrum. For all chains the tolerance for convergence is set by the Gelman--Rubin parameter $R-1 = 0.001$.

\section{Multi-tracer Multipole Covariance}\label{sec:cov}

While computing covariances from mock catalogues offers many advantages, it presents several key challenges on ultra-large scales. Most notably, the need for large-volume mocks significantly increases computational costs, and incorporating the relativistic effects that become relevant at these scales remains non-trivial. As a result, analytical covariance computations are essential for probing the information content in this large, linear regime. Here, we introduce a novel calculation of the multi-tracer complex bispectrum multipole covariance that includes the odd-parity multipoles, together with its leading non-Gaussian contributions and the power spectrum--bispectrum cross-covariance.

One can define the full covariance for a combined power spectrum and bispectrum analysis as
\begin{equation}
    C=
    \begin{pmatrix}
    {\textsf{C}}_{PP} & {\textsf{C}}_{PB} \\
    {\textsf{C}}_{BP} & {\textsf{C}}_{BB}
    \end{pmatrix}.
\end{equation}
This is defined for a single redshift bin and we assume each redshift bin is independent.

The multi-tracer power spectrum covariance including odd-parity multipoles is detailed for the first time in \citetalias{Addis_2025} with the inclusion of integrated effects and wide-separation corrections (see also \citealt{Montano_2026_cov} for a similar derivation of just the local relativistic terms); here we extend this computation to the bispectrum. As in that work, we take the power spectrum covariance, $\textsf{C}_{PP}$, to be Gaussian. For the bispectrum, we compute the Gaussian covariance and model the leading non-Gaussian contributions, which also correlate the power spectrum and bispectrum through $\textsf{C}_{PB}$, in a squeezed-limit approximation (\cref{sec:cov_ng}). We neglect wide-separation corrections here, though they are computed for the power spectrum covariance in \citetalias{Addis_2025}. Unlike \citetalias{Addis_2025}, we also neglect integrated relativistic contributions to the covariance, as these first require a treatment of the radial window (\cref{sec:cov_element}). We also neglect super-sample covariance from modes larger than the survey \citep{Chan_2018}.

\subsection{Multi-tracer bispectrum multipole covariance}

For a given redshift bin, we can express the bispectrum-bispectrum covariance matrix as a block matrix for each multipole pair
\begin{equation}
\textsf{C}_{BB} =
\begin{pmatrix}
{\textsf{C}}[B_{\ell_i}, B_{\ell_i}] & ... & {\textsf{C}}[B_{\ell_i}, B_{\ell_n}] \\
\vdots & \ddots & \vdots \\
{\textsf{C}}[B_{\ell_n}, B_{\ell_i}] & ... & {\textsf{C}}[B_{\ell_n}, B_{\ell_n}]
\end{pmatrix},
\end{equation}
where each cross/auto-multipole covariance block is a $4 \times 4$ sub-matrix. Writing $C^{abc,def} \equiv C[B^{abc}_{\ell_i},B^{def}_{\ell_j}]$ for brevity, for two distinct populations, $X$ and $Y$, the sub-matrix can be written as
\begingroup
\setlength{\arraycolsep}{2.5pt}
\begin{equation}\label{eq:multi_tracer_cov}
{\textsf{C}}(B_i, B_j) =
\begin{pmatrix}
C^{XXX,XXX} & C^{XXX,XXY} & C^{XXX,XYY} & C^{XXX,YYY}\\
C^{XXY,XXX} & C^{XXY,XXY} & C^{XXY,XYY} & C^{XXY,YYY}\\
C^{XYY,XXX} & C^{XYY,XXY} & C^{XYY,XYY} & C^{XYY,YYY}\\
C^{YYY,XXX} & C^{YYY,XXY} & C^{YYY,XYY} & C^{YYY,YYY}
\end{pmatrix},
\end{equation}
\endgroup
where the tracers are listed in the order of the sides, $k_1\geq k_2\geq k_3$, so that the second tracer occupies the shortest sides.

This two-tracer expression is equivalent to the covariance matrices in \citet{Karagiannis_2024}. The details of the calculation for a given covariance matrix element are given in the next section.

\subsubsection{Covariance matrix element}\label{sec:cov_element}

In this section we compute the covariance of the multi-tracer bispectrum multipoles for the Scoccimarro estimator. The covariance matrix is defined for complex variables as
\begin{multline}
C[\hat{B}^{abc}_{\ell_i},\hat{B}^{def}_{\ell_j}](\bm{k}, \bm{q}) = \\
\langle \hat{B}^{abc}_{\ell_i}(\bm{k}_1,\bm{k}_2,\bm{k}_3) \, \hat{B}^{def}_{\ell_j}(\bm{q}_1,\bm{q}_2,\bm{q}_3)^* \rangle \\
- \langle \hat{B}^{abc}_{\ell_i}(\bm{k}_1,\bm{k}_2,\bm{k}_3) \rangle \, \langle \hat{B}^{def}_{\ell_j}(\bm{q}_1,\bm{q}_2,\bm{q}_3)^* \rangle.
\end{multline}
For convenience, we define the function,
\begin{equation}\label{eq:F_est}
F_{\ell_i}(\bm{k}) = \int_{\bm{x}} \, \mathrm{e}^{-i\,\bm{k}\cdot\bm{x}}\mathcal{L}_{\ell_i}(\hat{\bm{k}} \cdot \hat{\bm{x}}) \Delta(\bm{x}),
\end{equation}
as well as the shortened integral notation 
\begin{equation}
\int_{\hat{q}_1}=\int_{\mathcal{S}_1} \dd^3 \bm{q}_1
\end{equation}
such that, using $k_{\rm f}^3 = (2\pi)^3/V$ to write the estimator normalisation as $1/(V N^T_{123} k_{\rm f}^6)$, the bispectrum-bispectrum multipole matrix element can be expressed, ignoring weights and windows, in terms of a 6-point function
\begin{multline}\label{eq:cov_6pt}
C[\hat{B}^{abc}_{\ell_i},\hat{B}^{def}_{\ell_j}](\bm{k}, \bm{q}) = \frac{(2\ell_i+1)(2\ell_j+1)}{V^2 k_{\rm f}^{12}\,N^{T}_{\bm{k}}\,N^{T}_{\bm{q}}} \\
\times \int_{\hat{k}_1,\hat{k}_2,\hat{k}_3,\hat{q}_1,\hat{q}_2,\hat{q}_3} \delta_D(\bm{k}_{123})\delta_D(\bm{q}_{123}) \\
\times \big[ \langle F^a_{\ell_i}(\bm{k}_1) F^b_{0}(\bm{k}_2) F^c_{0}(\bm{k}_3) F^d_{\ell_j}(\bm{q}_1)^* F^e_{0}(\bm{q}_2)^* F^f_{0}(\bm{q}_3)^* \rangle \\
- \langle F^a_{\ell_i}(\bm{k}_1) F^b_{0}(\bm{k}_2) F^c_{0}(\bm{k}_3) \rangle \langle F^d_{\ell_j}(\bm{q}_1)^* F^e_{0}(\bm{q}_2)^* F^f_{0}(\bm{q}_3)^* \rangle\big],
\end{multline}
where $N^T_{\bm{k}}$ and $N^T_{\bm{q}}$ are the numbers of triangles in the two bins. In the continuum limit, $\Delta k \ll k$, we can re-express the integrals over spherical shells as angular integrals
\begin{equation}\label{eq:continuum_limit}
\frac{1}{N^{T}_{123}k_{\rm f}^6}\int_{\hat q_1,\hat q_2,\hat q_3}\delta_D(\boldsymbol q_{123}) = \int\frac{{\rm d}\Omega}{4\pi},
\end{equation}
where $N^{T}_{123} = 8\pi^2 k_1 k_2 k_3\,(\Delta k)^3\,\beta/k_{\rm f}^6$ is the number of fundamental triangles in the bin. Here $\beta$ is the fraction of the bin $\mathcal{S}_1 \times \mathcal{S}_2 \times \mathcal{S}_3$ satisfying the closure condition $q_1 \leq q_2 + q_3$. Writing $q_i = k_i + \Delta k\, \delta_i$, so that $\delta_i \in [-1/2,1/2]$ spans the $i$-th shell, this is
\begin{multline}\label{eq:beta}
\beta = \frac{1}{k_1 k_2 k_3} \int_{-1/2}^{1/2} \dd \delta_1 \dd \delta_2 \dd \delta_3 \; \prod_{i=1}^{3} \left( k_i + \Delta k\, \delta_i \right) \\
\times \Theta\left( u - \delta_1 + \delta_2 + \delta_3 \right),
\end{multline}
where $\Theta$ is the Heaviside step function and $u \equiv (k_2 + k_3 - k_1)/\Delta k$ measures the distance from the closure boundary in units of the bin width. The radial integrals contribute $\int_{\mathcal{S}_i} q \dd q = k_i \Delta k$ exactly, so the prefactor in $N^{T}_{123}$ receives no further correction in $\Delta k$ and the entire bin-width dependence resides in $\beta$. Since $|\delta_1 - \delta_2 - \delta_3| \leq 3/2$, the step function is unity across the whole bin for $u \geq 3/2$ and vanishes across it for $u \leq -3/2$, giving $\beta = 1$ and $\beta = 0$ respectively; only bins straddling the closure boundary need be evaluated. In the thin-bin limit, $\Delta k /k \to 0$, \cref{eq:beta} recovers the familiar $\beta = 1/2$ for folded triangles, $k_1 = k_2 + k_3$. At the bin width adopted here, $\Delta k = 4 k_\mathrm{f}$, these configurations instead have $0.53 \lesssim \beta \lesssim 0.60$, and we therefore use \cref{eq:beta} throughout.

The 6-point function in \cref{eq:cov_6pt} decomposes into products of connected correlators: three power spectra ($PPP$), two bispectra ($BB$), a power spectrum and a trispectrum ($PT$), and the connected 6-point function. Of the ten ways of splitting the six fields into two triplets, the one that separates the two estimators cancels the disconnected product in \cref{eq:cov_6pt}, and any power spectrum of two sides of the same triangle vanishes, as $\bm{k}_1+\bm{k}_2=\bm{0}$ would require $\bm{k}_3=\bm{0}$. Every remaining term therefore contracts fields of both triangles, and neglecting the connected 6-point function, which is subleading on the scales of interest \citep{Biagetti_2022,Salvalaggio_2024}, the covariance is
\begin{equation}\label{eq:cov_split}
C = C^{PPP} + C^{BB} + C^{PT},
\end{equation}
with six, nine and nine terms, respectively. The Gaussian term, $C^{PPP}$, pairs every side of one triangle with a side of the other, so it is non-zero only when the two bins coincide, $N^T_{\bm{k}}=N^T_{\bm{q}}\equiv N^T_{123}$, whereas $C^{BB}$ and $C^{PT}$ link the two triangles through a single shared mode; we compute the former here and return to the latter in \cref{sec:cov_ng}. With the assumption that we have non-overlapping samples \citep{Smith_2009}, the Gaussian term can be written as
\begin{equation}\label{eq:full_cov}
\begin{aligned}
&C^{PPP}[\hat{B}^{abc}_{\ell_i},\hat{B}^{def}_{\ell_j}](\bm{k}, \bm{q}) = \frac{(2\ell_i+1)(2\ell_j+1)}{\left(V\,N^{T}_{123}\,k_{\rm f}^6\right)^2} \int_{\hat{k}_1,\hat{k}_2,\hat{k}_3,\hat{q}_1,\hat{q}_2,\hat{q}_3} \\
&\quad \times \delta_D(\bm{k}_{123})\delta_D(\bm{q}_{123}) \\
&\quad \times \bigg\{ \langle F^a_{\ell_i}(\bm{k}_1) F^d_{\ell_j}(\bm{q}_1)^* \rangle \Big[ \langle F^b_{0}(\bm{k}_2) F^e_{0}(\bm{q}_2)^* \rangle \langle F^c_{0}(\bm{k}_3) F^f_{0}(\bm{q}_3)^* \rangle \\
&\qquad\qquad\quad + \langle F^b_{0}(\bm{k}_2) F^f_{0}(\bm{q}_3)^* \rangle \langle F^c_{0}(\bm{k}_3) F^e_{0}(\bm{q}_2)^* \rangle \Big] \\
&\qquad + \langle F^a_{\ell_i}(\bm{k}_1) F^e_{0}(\bm{q}_2)^* \rangle \Big[ \langle F^b_{0}(\bm{k}_2) F^d_{\ell_j}(\bm{q}_1)^* \rangle \langle F^c_{0}(\bm{k}_3) F^f_{0}(\bm{q}_3)^* \rangle \\
&\qquad\qquad\quad + \langle F^b_{0}(\bm{k}_2) F^f_{0}(\bm{q}_3)^* \rangle \langle F^c_{0}(\bm{k}_3) F^d_{\ell_j}(\bm{q}_1)^* \rangle \Big] \\
&\qquad + \langle F^a_{\ell_i}(\bm{k}_1) F^f_{0}(\bm{q}_3)^* \rangle \Big[ \langle F^b_{0}(\bm{k}_2) F^e_{0}(\bm{q}_2)^* \rangle \langle F^c_{0}(\bm{k}_3) F^d_{\ell_j}(\bm{q}_1)^* \rangle \\
&\qquad\qquad\quad + \langle F^b_{0}(\bm{k}_2) F^d_{\ell_j}(\bm{q}_1)^* \rangle \langle F^c_{0}(\bm{k}_3) F^e_{0}(\bm{q}_2)^* \rangle \Big] \bigg\}.
\end{aligned}
\end{equation}

In the next section, we restrict the calculation to the zeroth-order term in the wide-separation expansion (equivalent to assuming the plane-parallel constant-redshift limit) for simplicity, as one would need to define wide-separation corrections on our local 6-point function (see \citetalias{Addis_2025} for details of this calculation in the power spectrum multipole covariance). We also neglect integrated effects in the covariance. In principle they can be included in the Gaussian term, provided they are defined with respect to the LOS of the full local 6-point function (see \citetalias{Addis_2025} for relevant discussion of the 4-point function in the power spectrum covariance). However, without a radial window function the lensing contribution stays correlated along the LOS out to the observer, so in $P(\bm{k})$ it is concentrated in a narrow spike at $\mu\simeq0$. This spike integrates to a finite contribution to the multipoles of the signal, but the covariance, as a product of power spectra, strongly overweights it. In practice, the redshift bin limits the LOS separations that contribute, which spreads this spike over a width set by the inverse radial depth of the bin. We therefore leave integrated contributions to the covariance to future work, alongside the window convolution.

\subsubsection{Gaussian plane-parallel constant-redshift covariance}

At zeroth order in the wide-separation expansion, as in the power spectrum case, we can remove the Legendre polynomial from the integral
\begin{equation}
F^a_{\ell_i}(\bm{k}) = \mathcal{L}_{\ell_i} (\hat{\bm{k}} \cdot \hat{\bm{x}}) F^a_0 (\bm{k}),
\end{equation}
such that the covariance matrix element can be expressed in terms of power spectrum contributions,
\begin{equation}\label{eq:pp_cov}
\begin{aligned}
&C^{PPP}[\hat{B}^{abc}_{\ell_i},\hat{B}^{def}_{\ell_j}](\bm{k}, \bm{q}) = \frac{(2\pi)^9\,(2\ell_i+1)(2\ell_j+1)}{\left(V\,N^{T}_{123}\,k_{\rm f}^6\right)^2} \\
&\quad \times \int_{\hat{k}_1,\hat{k}_2,\hat{k}_3,\hat{q}_1,\hat{q}_2,\hat{q}_3} \delta_D(\bm{k}_{123})\delta_D(\bm{q}_{123}) \\
&\quad \times \mathcal{L}_{\ell_i} (\hat{\bm{k}}_1 \cdot \hat{\bm{d}})\mathcal{L}_{\ell_j}(\hat{\bm{q}}_1 \cdot \hat{\bm{d}}) \bigg\{ \hat{P}^{ad}(k_1,\mu_1)\delta_D(\bm{k}_1-\bm{q}_1) \\
&\qquad \times \Big[ \hat{P}^{be}(k_2,\mu_2) \hat{P}^{cf}(k_3,\mu_3)\delta_D(\bm{k}_2-\bm{q}_2)\delta_D(\bm{k}_3-\bm{q}_3) \\
&\qquad\quad + \hat{P}^{bf}(k_2,\mu_2) \hat{P}^{ce}(k_3,\mu_3)\delta_D(\bm{k}_2-\bm{q}_3)\delta_D(\bm{k}_3-\bm{q}_2) \Big] \\
&\qquad + \hat{P}^{ae}(k_1,\mu_1)\delta_D(\bm{k}_1-\bm{q}_2) \\
&\qquad\quad \times \Big[ \hat{P}^{bd}(k_2,\mu_2) \hat{P}^{cf}(k_3,\mu_3)\delta_D(\bm{k}_2-\bm{q}_1)\delta_D(\bm{k}_3-\bm{q}_3) \\
&\qquad\qquad + \hat{P}^{bf}(k_2,\mu_2) \hat{P}^{cd}(k_3,\mu_3)\delta_D(\bm{k}_2-\bm{q}_3)\delta_D(\bm{k}_3-\bm{q}_1) \Big] \\
&\qquad + \hat{P}^{af}(k_1,\mu_1)\delta_D(\bm{k}_1-\bm{q}_3) \\
&\qquad\quad \times \Big[ \hat{P}^{be}(k_2,\mu_2) \hat{P}^{cd}(k_3,\mu_3) \delta_D(\bm{k}_2-\bm{q}_2)\delta_D(\bm{k}_3-\bm{q}_1) \\
&\qquad\qquad + \hat{P}^{bd}(k_2,\mu_2) \hat{P}^{ce}(k_3,\mu_3)\delta_D(\bm{k}_2-\bm{q}_1)\delta_D(\bm{k}_3-\bm{q}_2) \Big] \bigg\},
\end{aligned}
\end{equation}
where each Wick contraction gives $\langle F^a(\boldsymbol{k}_i)F^b(\boldsymbol{q}_j)^*\rangle = (2\pi)^3\delta_D(\boldsymbol{k}_i-\boldsymbol{q}_j)\hat{P}^{ab}(\boldsymbol{k}_i)$, and the power spectrum includes the shot-noise contribution in the auto-correlation case, assuming we have non-overlapping samples,
\begin{equation}\label{eq:pk_shot}
\hat{P}^{ab}(\bm{k},\bm{d}) = P^{ab}_\mathrm{loc}(\bm{k},\bm{d}) + \frac{\delta^K_{a,b}}{n_a},
\end{equation}
where $n_a$ is the number density of our given sample. $P^{ab}(\bm{k},\bm{d})$ can in principle contain all relevant contributions to the power spectrum, but here we include only the local relativistic corrections (\cref{sec:cov_element}). Also, note the relative sign change compared to the standard bispectrum covariance literature \citep{Ivanov_2023,Rizzo_2023,Salvalaggio_2024}, which arises as this is the covariance of a complex variable.

For the Gaussian part of the covariance distinct triangle bins are uncorrelated. Since the two triangles lie in the same bin, a term of \cref{eq:pp_cov} survives only if it pairs each side $\bm{q}_i$ with a side of $\bm{k}$ in the same shell, so each surviving term carries the tracers of the second triangle, $(d,e,f)$, in some order $(d',e',f')$ on the sides $\bm{k}_1$, $\bm{k}_2$ and $\bm{k}_3$. Integrating over the $\bm{q}_i$ then sets $\bm{q}_1$ to the side carrying $d$, which moves the Legendre polynomial of the second triangle onto that side, and leaves $\delta_D(\bm{k}_{123})^2 = \delta_D(\bm{0})\,\delta_D(\bm{k}_{123})$, with $\delta_D(\bm{0}) = V/(2\pi)^3$. Using the continuum limit, \cref{eq:continuum_limit}, and $k_{\rm f}^3 = (2\pi)^3/V$, the covariance then reduces to an angular integral,
\begin{multline}\label{eq:ppp_final}
C^{PPP}[\hat{B}^{abc}_{\ell_i},\hat{B}^{def}_{\ell_j}](\bm{k}) = \frac{(2\ell_i+1)(2\ell_j+1)\,V}{N^{T}_{123}} \\
\times \sum_{(d',e',f')} \int\frac{\dd\Omega}{4\pi}\, \mathcal{L}_{\ell_i}(\mu_1)\,\mathcal{L}_{\ell_j}(\mu_{d}) \\
\times \hat{P}^{ad'}(k_1,\mu_1)\,\hat{P}^{be'}(k_2,\mu_2)\,\hat{P}^{cf'}(k_3,\mu_3),
\end{multline}
where the sum runs over the orderings of $(d,e,f)$ that only exchange tracers between sides in the same shell, and $\mu_d$ is the cosine of the side carrying $d$. For scalene triangles only $(d,e,f)$ survives; $k_2=k_3$ adds $(d,f,e)$, $k_1=k_2$ adds $(e,d,f)$, and equilateral triangles include all six. For a single tracer the $(d,f,e)$ term equals the first, and for $\ell_i=\ell_j=0$ every term does, recovering the triangle counting factor $s_B=1,2,6$. The $(e,d,f)$ term, however, moves the Legendre weight to $\mu_2$, so the higher multipoles of equilateral and $k_1=k_2$ triangles are not simply rescaled by $s_B$ \citep{Rizzo_2023,Salvalaggio_2024}.

\begin{figure*}
\centering

\includegraphics[width=0.485\linewidth]{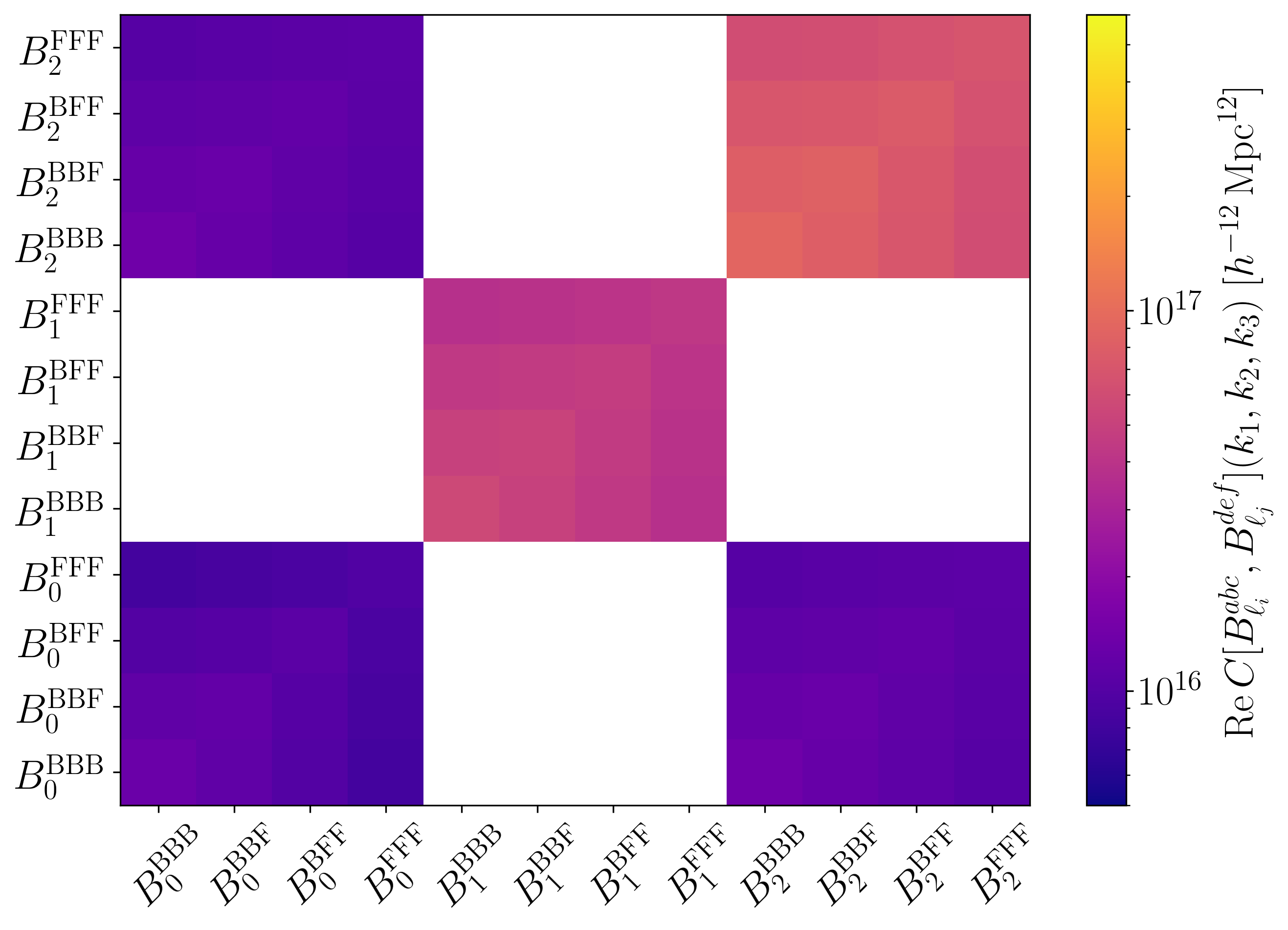}
\includegraphics[width=0.495\linewidth]{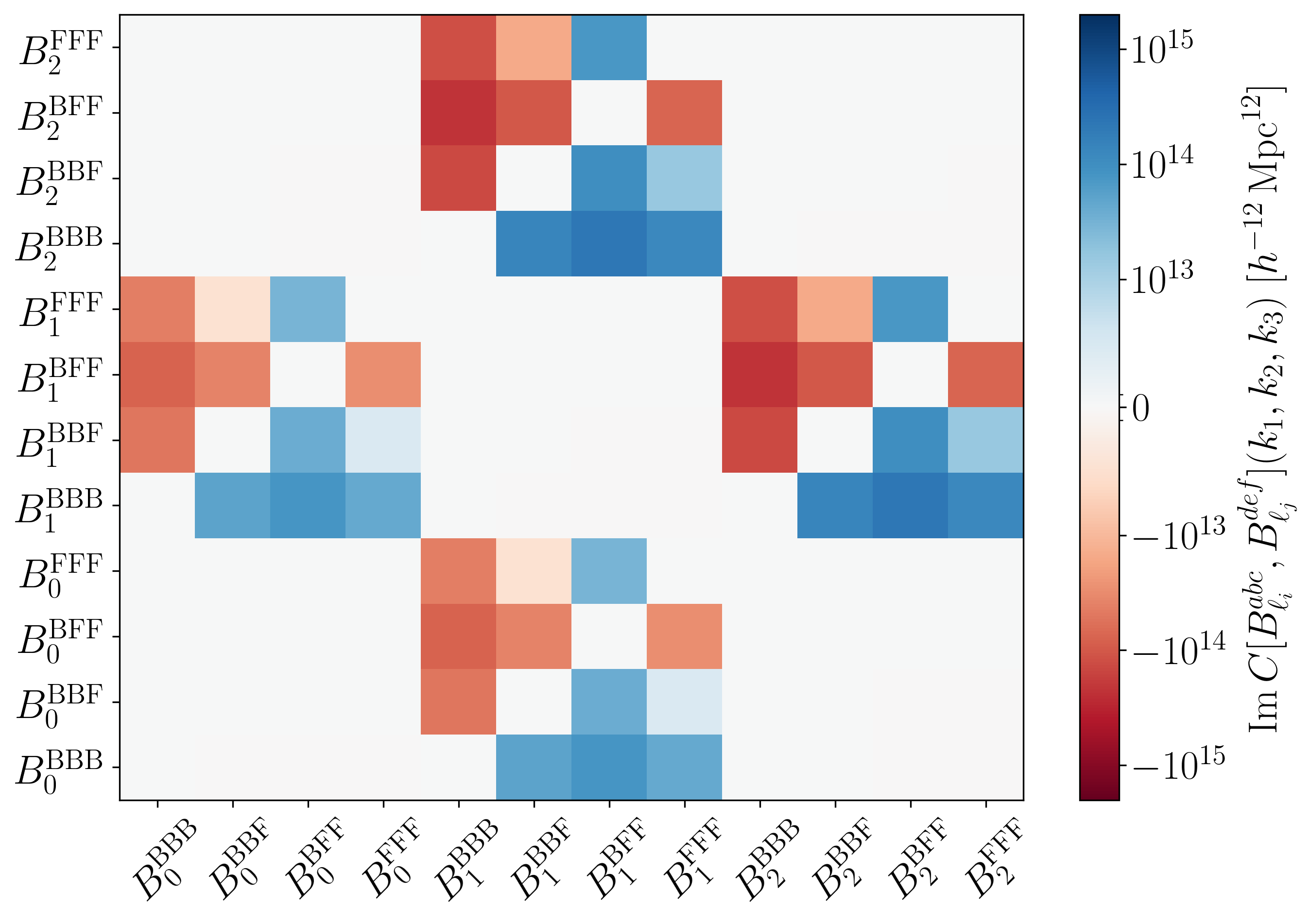}

\includegraphics[width=0.485\linewidth]{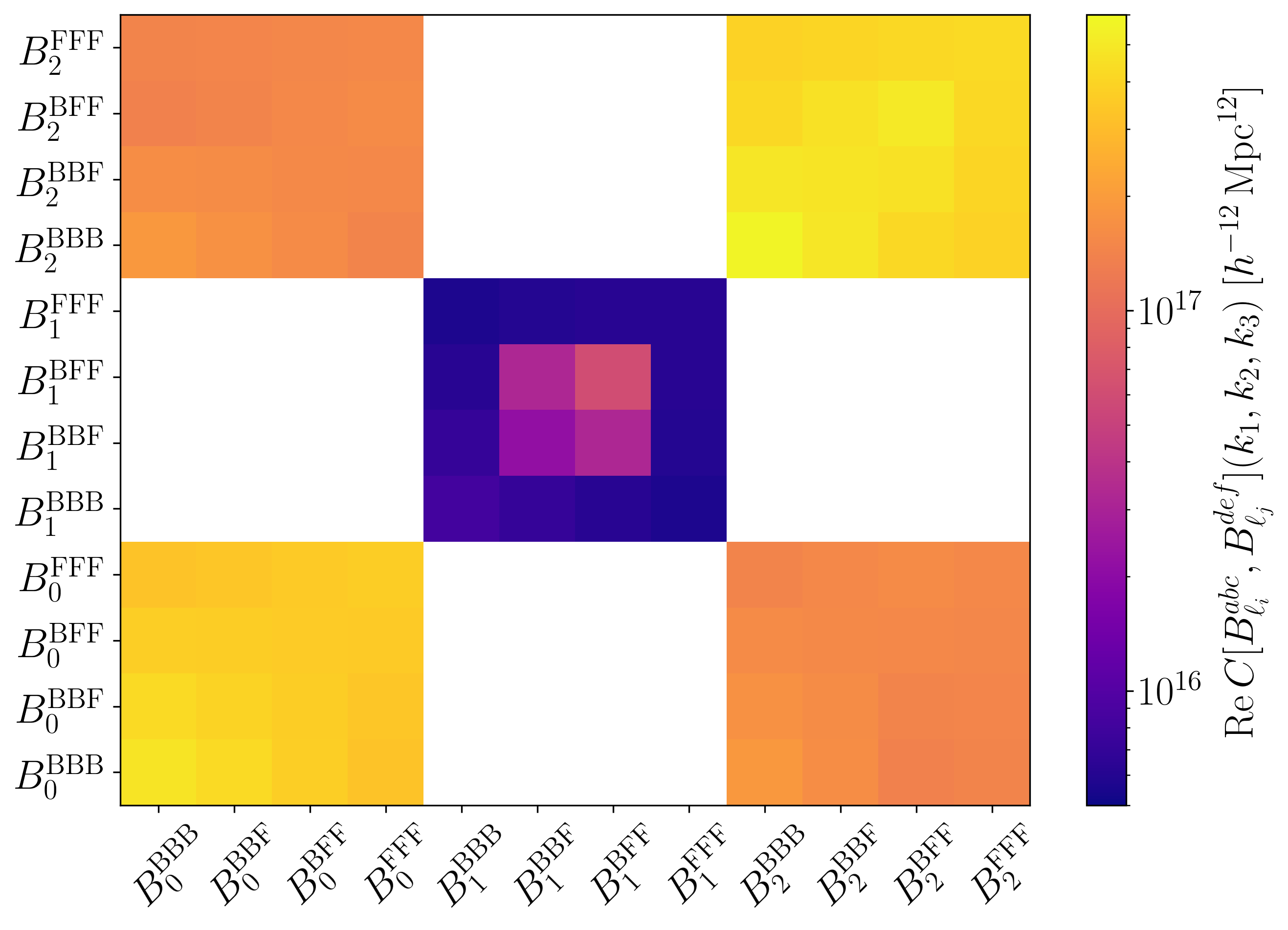}
\includegraphics[width=0.495\linewidth]{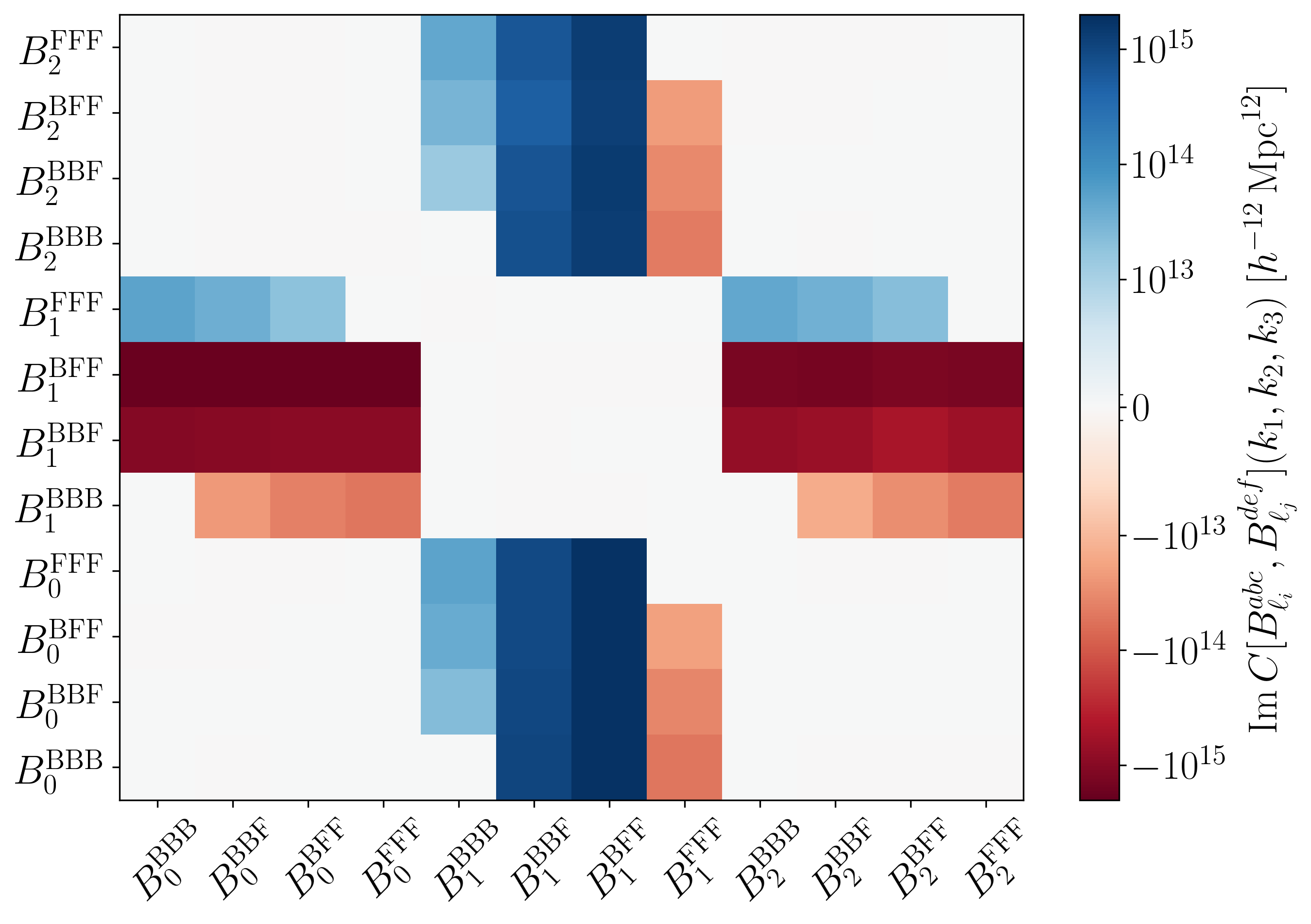}
\caption{Real (left) and imaginary (right) parts of the Gaussian multipole covariance, $\textsf{C}^{PPP}_{BB}$, of a single triangle bin for a multi-tracer analysis. The top row is for a scalene configuration with $k$-bins centred at $k_1=0.049, k_2=0.036, k_3 = 0.024 \, [\aMpch]$ and the bottom row represents an equilateral configuration $k_1=k_2=k_3 = 0.024 \, [\aMpch]$. We assume a bright-faint split for a \textit{Euclid}-like survey with $F_s=3\times10^{-16}\,[\mathrm{erg\,cm^{-2}\,s^{-1}}]$, for a redshift bin $(1<z<1.2)$, including the local relativistic contributions to the covariance. White entries vanish. For the equilateral configuration the single-tracer dipoles, $B^{\rm BBB}_1$ and $B^{\rm FFF}_1$, are strongly suppressed, as the estimator sums over every labelling of the sides and $\mu_1+\mu_2+\mu_3=0$ for an exactly equilateral triangle; they are non-zero only through the spread of the triangles across the bin (\cref{sec:data_vector}). For the equivalent power spectrum plot, see Figure 9 of \citetalias{Addis_2025}.\label{fig:cov}}
\end{figure*}

The Gaussian part of the multi-tracer bispectrum-bispectrum covariance, $\textsf{C}^{PPP}_{BB}$, is therefore block-diagonal in triangle bins, with each block constructed of $4 \times 4$ sub-matrices, see \cref{eq:multi_tracer_cov}, for each multipole pair. This is shown for $\ell=0,1,2$ in \cref{fig:cov} for equilateral and scalene configurations; the non-Gaussian terms of \cref{sec:cov_ng} then couple different triangle bins. The imaginary part of each $4 \times 4$ sub-matrix (right panels of \cref{fig:cov}) is non-zero only for odd--even cross-multipole pairs, and the even--odd multipole pairs behave such that the matrix is Hermitian and positive semi-definite.

\subsection{Non-Gaussian contributions}\label{sec:cov_ng}

For the bispectrum, the Gaussian covariance overestimates the information content \citep{Chan_2017,Byun_2017}, and in squeezed configurations the non-Gaussian contributions can exceed the Gaussian one \citep{Barreira_2019,Biagetti_2022}. Beyond the Gaussian term, the $BB$ and $PT$ terms of \cref{eq:cov_split} are the leading contributions to the bispectrum covariance, and for a joint analysis the power spectrum and bispectrum are correlated at the same order through a term with one power spectrum and one bispectrum ($PB$) \citep{Chan_2017,Sugiyama_2020,Biagetti_2022,Salvalaggio_2024}. Whereas $C^{PPP}$ pairs every side of one triangle with a side of the other, these terms link the two estimators through a single shared mode, so they correlate different configurations whenever these share a $k$-shell. As for $C^{PPP}$, we derive each term from \cref{eq:cov_6pt}, or its 5-point analogue for $PB$, and evaluate it in the plane-parallel constant-redshift limit. The power spectra and bispectra in these terms include the local relativistic contributions, but we neglect the integrated ones.

\subsubsection{$BB$ term}

Each of the nine $BB$ permutations connects two fields of one triangle with one field of the other, such that
\begin{equation}\label{eq:bb_full}
\begin{aligned}
&C^{BB}[\hat{B}^{abc}_{\ell_i},\hat{B}^{def}_{\ell_j}](\bm{k}, \bm{q}) = \frac{(2\ell_i+1)(2\ell_j+1)}{V^2 k_{\rm f}^{12}\,N^{T}_{\bm{k}}\,N^{T}_{\bm{q}}} \int_{\hat{k}_1,\hat{k}_2,\hat{k}_3,\hat{q}_1,\hat{q}_2,\hat{q}_3} \\
&\quad \times \delta_D(\bm{k}_{123})\delta_D(\bm{q}_{123}) \Big[ \langle F^a_{\ell_i}(\bm{k}_1) F^e_{0}(\bm{q}_2)^* F^f_{0}(\bm{q}_3)^*\rangle \\
&\qquad \times \langle F^b_{0}(\bm{k}_2) F^c_{0}(\bm{k}_3) F^d_{\ell_j}(\bm{q}_1)^* \rangle + 8\ \text{perms}\Big].
\end{aligned}
\end{equation}
In the plane-parallel limit each 3-point function gives a bispectrum, e.g.\ $\langle F^a(\bm{k}_1)F^e(\bm{q}_2)^*F^f(\bm{q}_3)^*\rangle = (2\pi)^3\delta_D(\bm{k}_1-\bm{q}_2-\bm{q}_3)\,\hat{B}^{aef}(\bm{k}_1,-\bm{q}_2,-\bm{q}_3)$, so that
\begin{equation}\label{eq:bb_pp}
\begin{aligned}
&C^{BB}[\hat{B}^{abc}_{\ell_i},\hat{B}^{def}_{\ell_j}](\bm{k}, \bm{q}) = \frac{(2\pi)^6\,(2\ell_i+1)(2\ell_j+1)}{V^2 k_{\rm f}^{12}\,N^{T}_{\bm{k}}\,N^{T}_{\bm{q}}} \\
&\quad \times \int_{\hat{k}_1,\hat{k}_2,\hat{k}_3,\hat{q}_1,\hat{q}_2,\hat{q}_3} \delta_D(\bm{k}_{123})\delta_D(\bm{q}_{123}) \\
&\quad \times \mathcal{L}_{\ell_i} (\hat{\bm{k}}_1 \cdot \hat{\bm{d}})\mathcal{L}_{\ell_j}(\hat{\bm{q}}_1 \cdot \hat{\bm{d}}) \Big[ \delta_D(\bm{k}_{1}-\bm{q}_{2}-\bm{q}_{3})\, \delta_D(\bm{k}_{2}+\bm{k}_{3}-\bm{q}_{1}) \\
&\qquad \times \hat{B}^{aef}(\bm{k}_1,-\bm{q}_2,-\bm{q}_3)\, \hat{B}^{dbc}(-\bm{q}_1,\bm{k}_2,\bm{k}_3) + 8\ \text{perms}\Big].
\end{aligned}
\end{equation}
where, as for $\hat{P}^{ab}$ in \cref{eq:pk_shot}, the bispectrum includes shot noise for non-overlapping samples. Since the shot noise of each estimator is subtracted, only galaxies shared between the two estimators contribute, so the single field from the other estimator, here $\bm{k}_1$ in $\hat{B}^{aef}(\bm{k}_1,-\bm{q}_2,-\bm{q}_3)$, must coincide with one of the other two \citep{Sugiyama_2020}. With this side first,
\begin{equation}\label{eq:bk_shot}
\hat{B}^{abc}(\bm{k}_1,\bm{k}_2,\bm{k}_3) = B^{abc} + \frac{\delta^K_{ab}}{n_a}P^{ca}(\bm{k}_3) + \frac{\delta^K_{ac}}{n_a}P^{ba}(\bm{k}_2),
\end{equation}
without the $\delta^K_{bc}P^{ab}(\bm{k}_1)/n_b$ and $\delta^K_{abc}/n_a^2$ terms of the shot noise of a single bispectrum estimator.
Once both triangles close, the two Dirac deltas in the bracket each set $\bm{q}_1=-\bm{k}_1$, so one gives $\delta_D(\bm{0})=V/(2\pi)^3$, and since $\Delta(\bm{x})$ is real, $\hat{B}^{aef}(\bm{k}_1,-\bm{q}_2,-\bm{q}_3)=\hat{B}^{aef}(\bm{q}_1,\bm{q}_2,\bm{q}_3)^*$. The term is therefore $\hat{B}^{dbc}(\bm{k}_1,\bm{k}_2,\bm{k}_3)\,\hat{B}^{aef}(\bm{q}_1,\bm{q}_2,\bm{q}_3)^*$, such that each triangle keeps its own shape but the tracers on the shared side are exchanged, and it is non-zero only if $k_1$ and $q_1$ lie in the same shell. For a fixed shared mode, the integral over the rest of each triangle counts the triangles of its bin that contain the mode, on average $N^T_{\bm{k}}/N_{k_1}$ and $N^T_{\bm{q}}/N_{k_1}$, and averages the bispectrum over their rotations about it; here $N_k = 4\pi k^2\Delta k\,[1+(\Delta k/k)^2/12]/k_{\rm f}^3$ is the number of modes in the shell, which reduces to the familiar $4\pi k^2\Delta k/k_{\rm f}^3$ for thin shells. These counts cancel the normalisation, and the plane-parallel element is
\begin{multline}\label{eq:cov_bb}
C^{BB}[\hat{B}^{abc}_{\ell_i},\hat{B}^{def}_{\ell_j}](\bm{k},\bm{q}) = (2\ell_i+1)(2\ell_j+1)\bigg[\frac{\delta^K_{k_1 q_1}}{N_{k_1}} \\
\times \int\frac{\dd\Omega}{4\pi}\int_0^{2\pi}\frac{\dd\phi'_{12}}{2\pi}\,\mathcal{L}_{\ell_i}(\mu_1)\,\mathcal{L}_{\ell_j}(-\mu_1) \\
\times \hat{B}^{dbc}(\bm{k}_1,\bm{k}_2,\bm{k}_3)\,\hat{B}^{aef}(-\bm{k}_1,\bm{q}_2,\bm{q}_3)^* + 8\ \text{perms}\bigg],
\end{multline}
where, as in \cref{eq:ppp_final}, $\int\dd\Omega/4\pi$ averages over the orientations of the $\bm{k}$ triangle, parametrised by $\mu_1$ and the azimuthal angle $\phi_{12}$ of $\bm{k}_2$ about $\bm{k}_1$ (\cref{eq:local_bispectrum}), and $\phi'_{12}$ is the corresponding angle of the $\bm{q}$ triangle about its first side, now fixed to $\bm{q}_1=-\bm{k}_1$. The permutations run over the nine pairings of a side $k_m$ with a side $q_n$ in the same shell, with the Legendre polynomials remaining on $\bm{k}_1$ and $\bm{q}_1$; the pairing shown places both on the shared mode, so that, as for $C^{PPP}$, they depend only on $\mu_1$.

Using the average counts assumes that the triangles of each bin are spread evenly over the modes of the shell. This holds for closed bins, but not for (nearly) flattened bins, where the closure condition cuts through the shell (\cref{eq:beta}), so that the number of triangles containing a mode varies across its width \citep{Biagetti_2022}. We therefore evaluate the shared mode at three Gauss--Legendre nodes across the width of its shell, weighting each triangle by its closure fraction there, which is sufficient for the mode counting of flattened bins to converge. For the same reason, at each node we average the bispectra over the triangles of the bin whose shared side lies there, rather than using the effective triangle of the whole bin (\cref{sec:data_vector}), and the same averaged bispectra enter $\textsf{C}_{PB}$ (\cref{eq:cov_pb}). Evaluated on the bin centres instead, these terms would be overestimated by a factor of a few for flattened, squeezed bins.

For $\ell_i=\ell_j=0$ in real space \cref{eq:cov_bb} reduces to the familiar $\hat{B}\hat{B}'/N_k$ per shared side \citep{Biagetti_2022}. However, since $\hat{B}(-\bm{k}_1,-\bm{k}_2,-\bm{k}_3)=\hat{B}(\bm{k}_1,\bm{k}_2,\bm{k}_3)^*$, the average over $\phi'_{12}$ of $\hat{B}^{aef}(-\bm{k}_1,\bm{q}_2,\bm{q}_3)$ is the complex conjugate of that with $\bm{q}_1=+\bm{k}_1$, while $\mathcal{L}_{\ell_j}(-\mu_1)=(-1)^{\ell_j}\mathcal{L}_{\ell_j}(\mu_1)$. The $BB$ term alone is therefore indefinite: for odd multipoles it contributes a negative variance and can spuriously increase their signal-to-noise.

\subsubsection{$PT$ term}

The nine $PT$ terms instead contract one field of each triangle into a power spectrum and the remaining four fields into a trispectrum. Taking the same pairing of sides as in \cref{eq:bb_full}, i.e.\ $\langle F^a_{\ell_i}(\bm{k}_1)F^d_{\ell_j}(\bm{q}_1)^*\rangle\langle F^b_{0}(\bm{k}_2)F^c_{0}(\bm{k}_3)F^e_{0}(\bm{q}_2)^*F^f_{0}(\bm{q}_3)^*\rangle$, the plane-parallel element is
\begin{equation}\label{eq:pt_pp}
\begin{aligned}
&C^{PT}[\hat{B}^{abc}_{\ell_i},\hat{B}^{def}_{\ell_j}](\bm{k}, \bm{q}) = \frac{(2\pi)^6\,(2\ell_i+1)(2\ell_j+1)}{V^2 k_{\rm f}^{12}\,N^{T}_{\bm{k}}\,N^{T}_{\bm{q}}} \\
&\quad \times \int_{\hat{k}_1,\hat{k}_2,\hat{k}_3,\hat{q}_1,\hat{q}_2,\hat{q}_3} \delta_D(\bm{k}_{123})\delta_D(\bm{q}_{123}) \\
&\quad \times \mathcal{L}_{\ell_i} (\hat{\bm{k}}_1 \cdot \hat{\bm{d}})\mathcal{L}_{\ell_j}(\hat{\bm{q}}_1 \cdot \hat{\bm{d}}) \Big[ \delta_D(\bm{k}_{1}-\bm{q}_{1})\, \delta_D(\bm{k}_{2}+\bm{k}_{3}-\bm{q}_{2}-\bm{q}_{3}) \\
&\qquad \times \hat{P}^{ad}(\bm{k}_1)\, T^{bcef}(\bm{k}_2,\bm{k}_3,-\bm{q}_2,-\bm{q}_3) + 8\ \text{perms}\Big].
\end{aligned}
\end{equation}
The shared side is now parallel, $\bm{q}_1=+\bm{k}_1$, and the trispectrum is in the collapsed configuration $\bm{k}_2+\bm{k}_3=-\bm{k}_1$, whose diagonal is the shared mode. In general this does not factorise over the two triangles, but it does when the shared mode is soft compared to the other sides, as for squeezed triangles that share their shortest side.

We therefore model the $BB$ and $PT$ terms together in this limit. On the shared mode, each triangle is the product of the field on its shared side, $\Delta^a(\bm{k}_1)$, and its other two fields, which we denote $Y^{bc}(-\bm{k}_1)$. The $BB$ term contracts the shared field of each triangle with the other two fields of the other, and the $PT$ term contracts the two shared fields with each other and the remaining fields with each other. In the squeezed limit, the other two fields respond linearly to the shared mode \citep{Barreira_2019}. Writing $\Delta^X=Z_1^X\delta_{\rm L}+\epsilon_X$, where $Z_1^X$ is the linear kernel of tracer $X$ and $\epsilon_X$ its Poisson noise, with $\langle\epsilon_X\epsilon_{X'}^*\rangle'=\delta^K_{XX'}/n_X$ (the prime denotes a correlator without its factor of $(2\pi)^3\delta_D$), we take
\begin{equation}\label{eq:Y_model}
Y^{bc}(-\bm{k}_1) = \frac{R^{bc}}{P_{\rm L}(k_1)}\,\delta_{\rm L}(-\bm{k}_1) + \sum_X n_X\, s^{bc}_X\, \epsilon_X(-\bm{k}_1),
\end{equation}
where $R^{bc}$ is the response of the pair to the linear density on the shared mode, multiplied by $P_{\rm L}(k_1)$, and $s^{bc}_X=\hat{B}^{Xbc}-B^{Xbc}$ is the shot noise of \cref{eq:bk_shot} with $X$ on the shared side. This is the minimal form that gives $\langle Y^{bc}(-\bm{k}_1)\Delta^X(\bm{k}_1)\rangle'=Z_1^XR^{bc}+s^{bc}_X$, i.e.\ the bispectrum with any tracer $X$ on the shared side, $\hat{B}^{Xbc}$, in the squeezed limit for its clustering part and exactly for its shot noise. The $PT$ contraction then gives
\begin{equation}\label{eq:pt_sep}
\hat{P}^{ad}(\bm{k}_1)\,T^{bcef} \to \hat{P}^{ad}(\bm{k}_1)\left[\frac{R^{bc}R^{ef*}}{P_{\rm L}(k_1)} + \sum_X n_X\, s^{bc}_X\, s^{ef*}_X\right],
\end{equation}
whose clustering part is the collapsed exchange trispectrum. The shot noise part is a model rather than a derivation, as the shot noise of the trispectrum has no separable collapsed limit; without the shot noise of the bispectra it reduces to $[1+\delta^K_{ad}/(n_aP^{aa})]\,B^{abc}B^{def*}$, as kept in $C^{PT}$ by \citet{Salvalaggio_2024}.

In the same model, the $BB$ term of \cref{eq:cov_bb} pairs $\hat{B}^{dbc}=Z_1^dR^{bc}+s^{bc}_d$, the $\bm{k}$ triangle with the other triangle's tracer on its shared side, with its counterpart for the $\bm{q}$ triangle. Its shot noise is therefore the exact exchange of \cref{eq:bk_shot}, while its clustering part takes the exchange in the squeezed limit, $B^{dbc}\to(Z_1^d/Z_1^a)\,B^{abc}$, whose ratios cancel in the product. The integrals for the $PT$ term reduce as for \cref{eq:cov_bb}, but with $\bm{q}_1=+\bm{k}_1$, so that both Legendre polynomials are evaluated at $+\mu_1$, and the two terms combine into
\begin{multline}\label{eq:cov_ng}
C^{BB}+C^{PT} \simeq (2\ell_i+1)(2\ell_j+1)\bigg[\frac{\delta^K_{k_1 q_1}}{N_{k_1}}\int\frac{\dd\Omega}{4\pi}\int_0^{2\pi}\frac{\dd\phi'_{12}}{2\pi} \\
\times \mathcal{L}_{\ell_i}(\mu_1)\,\mathcal{L}_{\ell_j}(\mu_1)\,\bigg\{\hat{P}^{ad}(\bm{k}_1) \\
\times \bigg[\frac{R^{bc}R^{ef*}}{P_{\rm L}(k_1)} + \sum_X n_X\, s^{bc}_X\, s^{ef*}_X\bigg] \\
+ (-1)^{\ell_j}\,\hat{B}^{dbc}(\bm{k}_1,\bm{k}_2,\bm{k}_3)\,\hat{B}^{aef}(-\bm{k}_1,\bm{q}_2,\bm{q}_3)^*\bigg\} \\
+ 8\ \text{perms}\bigg],
\end{multline}
where $R^{bc}$ and $s^{bc}_X$ are evaluated on the $\bm{k}$ triangle, $R^{ef}$ and $s^{ef}_X$ on the $\bm{q}$ triangle with $\bm{q}_1=+\bm{k}_1$, and in each term $\bm{q}_2$ and $\bm{q}_3$ follow from $\bm{q}_1$ and $\phi'_{12}$. At fixed $\bm{k}_1$ the integrals over $\phi_{12}$ and $\phi'_{12}$ factorise, so each triangle enters only through azimuthal averages about its shared side.

Since the $BB$ and $PT$ terms are the two contractions of products of the jointly Gaussian variables of \cref{eq:Y_model}, their sum is the covariance of these products and is therefore positive semi-definite for any number of tracers, even though the $BB$ term alone is not. We therefore use this separable form for all nine pairings, extending the squeezed-limit approximation to configurations where the shared mode is not soft. With the full tree-level trispectrum instead, the $PT$ term on its own need not be positive semi-definite, so we use it only to test the separable form (\cref{sec:cov_ng_impl}). Only the tracers of the two triangles enter, so the block of two single-tracer triangles, $XXX$, reduces to the single-tracer covariance of $X$.

\begin{figure*}
\centering

\includegraphics[width=\linewidth]{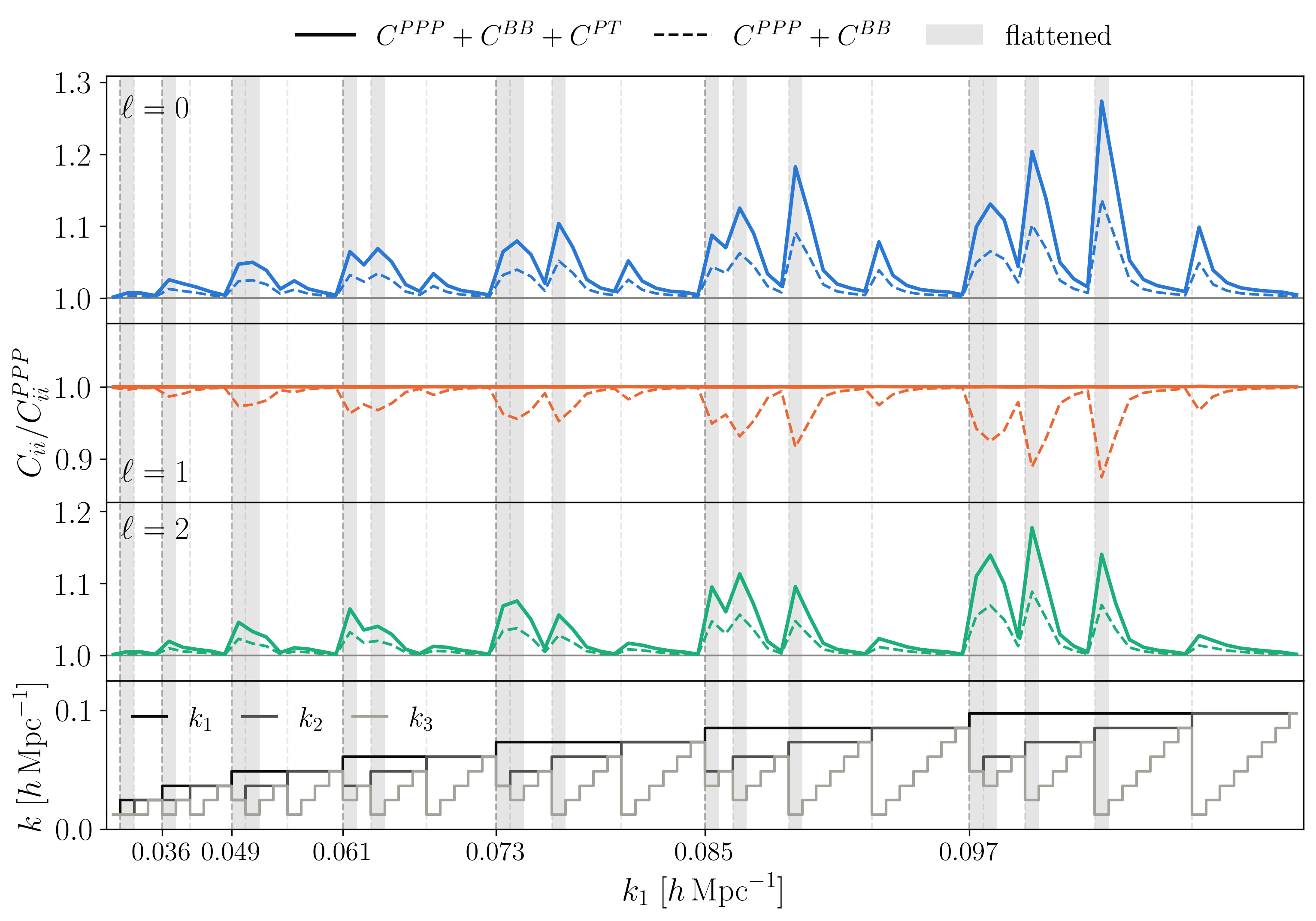}
\caption{Diagonal of the single-tracer bispectrum multipole covariance including the non-Gaussian terms of \cref{eq:cov_ng} (solid), and with the $BB$ term alone (dashed), relative to the Gaussian term, for the monopole (top), dipole and quadrupole of each triangle bin. The bottom panel shows the sides of each triangle: the triangles are ordered by $k_1$, then $k_2$, then $k_3$, with the dark (faint) dashed lines marking where $k_1$ ($k_2$) changes. The grey bands mark flattened configurations, $k_1=k_2+k_3$ at the bin centres. We assume a \textit{Euclid}-like survey with a redshift bin $(1<z<1.2)$ and bins of width $\Delta k=4k_{\rm f}$ up to $k_{\rm max}=0.1\,[\aMpch]$, including the local relativistic contributions to the covariance.}\label{fig:cov_ng}
\end{figure*}

For the real-space monopole without shot noise the two clustering parts are equal, recovering the $C^{PT}\simeq C^{BB}$ approximation used for squeezed configurations \citep{Biagetti_2022,Salvalaggio_2024}. For odd multipoles, however, the $PT$ term cancels the $BB$ term rather than doubling it: for a real bispectrum (the Newtonian limit) the clustering parts of \cref{eq:cov_ng} cancel, leaving only shot noise, so these multipoles, which carry the leading local relativistic signal, are largely insensitive to these terms (\cref{fig:cov_ng}).

\Cref{fig:cov_ng} shows the effect of these terms on the variance of the single-tracer bispectrum multipoles. For the monopole and quadrupole, they are concentrated in (nearly) flattened configurations, $k_1\simeq k_2+k_3$, where two sides are aligned and the bispectrum is largest relative to the power spectra, and are largest for those that are also squeezed, as their long-wavelength side lies in a shell with few modes, each shared by many triangles; there they reach a significant fraction of the Gaussian variance. Here the $PT$ term roughly doubles the $BB$ term, close to the $C^{PT}\simeq C^{BB}$ limit. For the dipole, the $BB$ term alone reduces the variance of the same configurations, whereas with the $PT$ term it remains very close to Gaussian.

\subsubsection{Power spectrum--bispectrum cross-covariance}

For a joint analysis we also require $\textsf{C}_{PB}$. Writing the normalisation of \cref{eq:yamamoto} as $1/(V N_k k_{\rm f}^3)$, the analogue of \cref{eq:cov_6pt} is a 5-point function,
\begin{multline}\label{eq:cov_5pt}
C[\hat{P}^{ab}_{\ell_i},\hat{B}^{def}_{\ell_j}](k, \bm{q}) = \frac{(2\ell_i+1)(2\ell_j+1)}{V^2 k_{\rm f}^{9}\,N_k\,N^{T}_{\bm{q}}} \\
\times \int_{\hat{k}_1,\hat{k}_2,\hat{q}_1,\hat{q}_2,\hat{q}_3} \delta_D(\bm{k}_{12})\delta_D(\bm{q}_{123}) \\
\times \big[ \langle F^a_{\ell_i}(\bm{k}_1) F^b_{0}(\bm{k}_2) F^d_{\ell_j}(\bm{q}_1)^* F^e_{0}(\bm{q}_2)^* F^f_{0}(\bm{q}_3)^* \rangle \\
- \langle F^a_{\ell_i}(\bm{k}_1) F^b_{0}(\bm{k}_2) \rangle \langle F^d_{\ell_j}(\bm{q}_1)^* F^e_{0}(\bm{q}_2)^* F^f_{0}(\bm{q}_3)^* \rangle\big],
\end{multline}
where $\bm{k}_1$ and $\bm{k}_2$ both run over the shell of $k$. Of the ten ways of splitting the five fields into a pair and a triplet, the one that separates the two estimators cancels the disconnected product, and a power spectrum of two sides of the triangle vanishes as before. Neglecting the connected (tetraspectrum) contribution, which is subleading on the scales of interest \citep{Biagetti_2022,Salvalaggio_2024}, each of the six remaining terms contracts one field of $\hat{P}$ with a side of the triangle and closes a bispectrum with the other field and the remaining two sides, e.g.\ $\langle F^a_{\ell_i}(\bm{k}_1)F^d_{\ell_j}(\bm{q}_1)^*\rangle\langle F^b_{0}(\bm{k}_2)F^e_{0}(\bm{q}_2)^*F^f_{0}(\bm{q}_3)^*\rangle$. In the plane-parallel limit this sets $\bm{q}_1=\bm{k}_1$ and gives $\hat{P}^{ad}(\bm{k}_1)\,\hat{B}^{bef}(\bm{q}_1,\bm{q}_2,\bm{q}_3)^*$, so, as in \cref{eq:cov_bb}, the other tracer of $\hat{P}$ takes the place of the shared side. Reducing the integrals as before,
\begin{multline}\label{eq:cov_pb}
C[\hat{P}^{ab}_{\ell_i},\hat{B}^{def}_{\ell_j}](k,\bm{q}) = (2\ell_i+1)(2\ell_j+1)\bigg[\frac{\delta^K_{k q_1}}{N_{k}}\int\frac{\dd\Omega}{4\pi} \\
\times\mathcal{L}_{\ell_i}(\mu_1)\,\mathcal{L}_{\ell_j}(\mu_1)\Big(\hat{P}^{ad}(\bm{k}_1)\,\hat{B}^{bef}(\bm{k}_1,\bm{q}_2,\bm{q}_3)^* \\
+ (-1)^{\ell_i}\,\hat{P}^{bd}(\bm{k}_1)\,\hat{B}^{aef}(\bm{k}_1,\bm{q}_2,\bm{q}_3)^*\Big) + 2\ \text{perms}\bigg],
\end{multline}
where $\int\dd\Omega/4\pi$ now averages over the orientations of the $\bm{q}$ triangle, with its first side along the power spectrum mode, $\bm{q}_1=\bm{k}_1$. The two terms correspond to the two fields of $\hat{P}$, with $\bm{k}_1\to-\bm{k}_1$ in the second giving the factor $(-1)^{\ell_i}$, and the permutations to the three sides of the triangle. For $\ell_i=\ell_j=0$ in real space this reduces to $2\hat{P}\hat{B}/N_k$ per shared side \citep{Biagetti_2022}. Here the mode counting is exact, as the triangles on each mode of a shell sum to $N^T_{\bm{q}}$, and $N_k$ is the exact number of modes in the shell rather than its thin-shell approximation.

For consistency with \cref{eq:cov_ng} we take the tracer exchange of the clustering part in the squeezed limit, $B^{bef}\to (Z_1^b/Z_1^d)\,B^{def}$, keeping the exact exchange in the shot noise, so that $\hat{B}^{bef}=Z_1^bR^{ef}+s^{ef}_b$ is the correlator of \cref{eq:Y_model} that also enters the $BB$ term. With this choice, \cref{eq:cov_pb} follows from the same model as \cref{eq:cov_ng} and the Gaussian $\textsf{C}_{PP}$, so the joint covariance of the power spectrum and bispectrum is consistent. This requires $\textsf{C}_{PP}$ to be averaged over the modes of each shell (\cref{sec:data_vector}) on the same nodes, with the same exact mode count. This matters for multi-tracer analyses: we find that using only the triangles' own tracers in the shot noise of the $PT$ term, or the exact tracer exchange for the clustering part of \cref{eq:cov_pb}, can make the joint covariance indefinite, as can the thin-shell mode count at the bin centre in $\textsf{C}_{PP}$ and $\textsf{C}_{PB}$.

Away from the diagonal, these terms correlate the triangle bins that share a $k$-shell, and $\textsf{C}_{PB}$ correlates each triangle with the power spectrum in the shells of its sides, as shown by the correlation matrix of the joint data vector in \cref{fig:cov_ng_corr}. As for the Gaussian term, the real part couples only multipoles of the same parity and the imaginary part only odd--even pairs, so the bispectrum dipole is correlated with the even multipoles and the power spectrum only through the imaginary part, which arises from the local relativistic contributions to the bispectrum and is small.

\subsubsection{Implementation and validation}\label{sec:cov_ng_impl}

After the azimuthal integrals, \cref{eq:cov_ng,eq:cov_pb} depend on the configurations only through $\mu_1$ of the shared mode, so on a Gauss--Legendre grid in $\mu_1$ the non-Gaussian covariance is a low-rank product, $\textsf{C}^{\rm NG}=W\Xi W^\dagger$. Here $W$ has a fixed number of columns per $k$-shell, built from the azimuthally averaged power spectra, bispectra and their shot noise, and $\Xi$ holds the quadrature weights and couples columns only within a shell. The rank is therefore independent of the number of triangles, and the inverse of the full covariance follows from the Woodbury identity in terms of the inverse of the Gaussian covariance, which is block-diagonal in $k$-bins and triangles, without forming the dense matrix.

We have verified \cref{eq:cov_bb,eq:cov_ng,eq:cov_pb} against direct sums over the orientations of the triangles and power spectrum modes, and the mode counting of \cref{eq:cov_bb} against a count of the triangle pairs that share a Fourier mode on a lattice. To test the separable form of \cref{eq:pt_sep}, we have also computed the $PT$ term with the full tree-level trispectrum, for which the $s$ channel, the exchange of the shared mode, is exactly separable, but the $t$ and $u$ channels and the $T_{3111}$ contribution depend on the relative orientation of the two triangles and must be integrated directly. The two give consistent signal-to-noise for both the Newtonian and the local relativistic bispectrum.\footnote{These terms are implemented in \textsc{CosmoWAP} in \href{https://github.com/craddis1/CosmoWAP/blob/v0.10.0/src/cosmo_wap/forecast/covariances.py}{\texttt{covariances.py}}, with the tree-level comparison in the class \texttt{PTTreeCovBk}, and the tests described here are in \href{https://github.com/craddis1/CosmoWAP/blob/v0.10.0/tests/test_bb_cov.py}{\texttt{test\_bb\_cov.py}}.}

\section{Results I: PNG}\label{sec:results1PNG}

This section presents the results of the forecasts on PNG, with \cref{sec:png_bias} covering the bias in the determination of $\fNL$ if one neglects relativistic and wide-separation effects. Including these effects introduces the relativistic biases, $n_{\rm evol}$ and $\mathcal{Q}$, as additional nuisance parameters, so in \cref{sec:beQ_constraints} we consider how well they are constrained by the clustering data themselves and how they can be determined independently. Then, in \cref{sec:png_constraints} we investigate the constraining power on $\fNL$ from single-tracer and multi-tracer forecasts and the impact that the uncertainty introduced by relativistic effects has on the inference. In particular, we focus on the underestimation of the true parameter covariance when one neglects this uncertainty. 

 We constrain each shape separately, holding the amplitudes of the other two at zero.  We give the joint equilateral and orthogonal constraints for comparison in \cref{sec:nonlocal_png}. We focus primarily on a \textit{Euclid}-like H$\alpha$ survey, but also consider a \textit{SPHEREx}-like survey and a MegaMapper-like Stage V Lyman-break galaxy (LBG) survey; details of the survey modelling are given in \cref{app:survey_details}. For the \textit{SPHEREx}-like survey we use only the sample with the most precise redshifts, which is partly why our constraints are weaker than published \textit{SPHEREx} forecasts \citep{Dore_SPHEREx,Heinrich_2024} that combine all five redshift-accuracy samples (see \cref{app:spherex}). We do not consider the DESI-like BGS survey here since, in our forecasts, it lacks the volume for competitive constraints on PNG.

In \cref{sec:png_bias} we sample the posterior on local-type $\fNL$ with MCMCs, and elsewhere use Fisher forecasts (\cref{sec:forecast_details}).

\subsection{Systematic bias}\label{sec:png_bias}

\begin{figure*}
\centering
\includegraphics[width=0.85\linewidth]{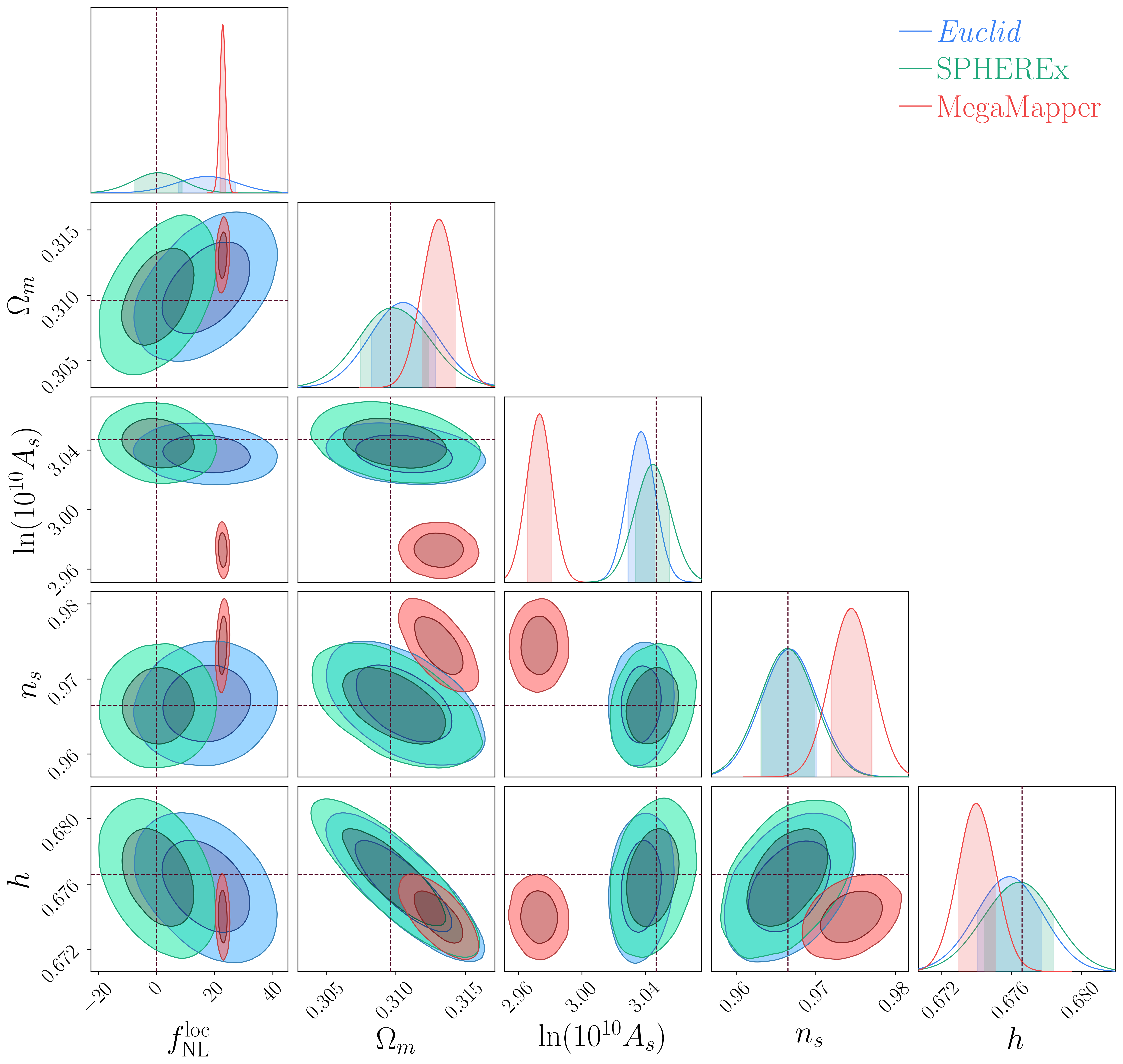}
\caption{Forecasted marginalised and joint $1$ and $2 \, \sigma$ parameter covariances for a \textit{Euclid}-like H$\alpha$ survey (blue), a \textit{SPHEREx}-like survey (green) and a MegaMapper-like LBG survey (red), if we neglect integrated, local relativistic and wide-separation effects, for a joint power spectrum and bispectrum analysis with priors on the cosmological parameters from the \textit{Planck} 2018 results. $b_1$, $b_2$ and $b_{\Gamma_2}$ are marginalised in each redshift bin.}\label{fig:fnl_bias}
\end{figure*}

The forecasted parameter constraints for a `naive' analysis that neglects relativistic and wide-separation effects are shown in the ellipses in \cref{fig:fnl_bias} for a joint power spectrum and bispectrum analysis. For the \textit{Euclid}-like survey the inferred value is shifted to $\fNL^{\rm loc}=17.2^{+9.2}_{-9.4}$, a $\sim2\sigma$ shift from the true fiducial value of $\fNL=0$ considered in this forecast.  

This bias is comparable to the \textit{Euclid} error bar but negligible for \textit{SPHEREx}, $\fNL^{\rm loc}=0.5^{+7.7}_{-7.7}$, whose lower redshift coverage and magnification bias suppress the lensing contribution. It dominates the error budget, however, for the MegaMapper-like LBG survey constraints shown in the red posteriors. This arises for two main reasons, firstly due to the improved constraining power, and secondly due to the greater dominance of the lensing effect at higher redshifts. Thus, we forecast that one would `detect' a positive value at high significance, $\fNL^{\rm loc}=22.8^{+0.9}_{-0.9}$, a $\sim25\sigma$ shift\footnote{The precise size of these shifts should be treated with caution. The lensing contribution peaks for modes transverse to the line of sight ($\mu=0$), which are the modes most suppressed on large scales by the survey window for all but very thick redshift bins. From tests run using \textsc{CosmoWAP}, including a radial window function reduces the shift by a factor of $\sim2$. The shift is also sensitive to the covariance modelling, since in the joint analysis the power spectrum–bispectrum cross-covariance partially offsets the separate power spectrum and bispectrum contributions to it. In all the cases we have considered, however, the shift remains large and positive.}, generated purely by neglecting relativistic effects despite there being no fiducial PNG signal.

As in \citetalias{Addis_2025}, local relativistic effects shift $\fNL^{\rm loc}$ towards negative values, but this is countered by a larger positive shift from integrated lensing: if we only neglect the local relativistic and wide-separation terms (\cref{fig:fnl_bias_noint}), the shift is small and negative for \textit{Euclid} and MegaMapper, $\approx0.1\sigma$ and $\approx0.4\sigma$, and negligible for \textit{SPHEREx}, while the shifts in the non-local shapes are negligible and those in the cosmological parameters are $\lesssim0.2\sigma$.

Additionally, for MegaMapper, the inference on the orthogonal shape is also biased, by $\sim3\sigma$, $\fNL^{\rm orth}=26\pm9$, while the equilateral shape by $\sim1\sigma$, $\fNL^{\rm eq}=-6\pm6$\footnote{These non-local results are from Fisher forecasts, which reproduce the MCMC shift in local-type $\fNL$ for \textit{Euclid} ($1.9\sigma$) but underestimate it for MegaMapper, where the shift is far outside the regime in which the Fisher approximation holds.}. For \textit{Euclid} these shifts are forecasted to be well within the survey error budget, $\fNL^{\rm eq}\approx-9\pm34$ and $\fNL^{\rm orth}\approx16\pm40$. Neglecting lensing also biases the inference of the cosmological parameters for the MegaMapper-like case, despite the priors from \textit{Planck}: by $\sim3\sigma$ for $\Omega_m$, $h$ and $n_s$, and $\sim10\sigma$ for $\ln(10^{10}A_s)$. For \textit{Euclid} the largest shift is $\sim1.2\sigma$, in $\ln(10^{10}A_s)$. Consistently including light-cone effects removes this bias; however, it introduces additional nuisance parameters in the form of the relativistic biases, $n_{\rm evol}$ and $\mathcal{Q}$. Before looking at the resulting constraints on $\fNL$ we first consider how well these bias parameters can be determined.

\subsection{Information on the relativistic biases}\label{sec:beQ_constraints}

Once relativistic effects are included in the model, $n_{\rm evol}$ and $\mathcal{Q}$ become nuisance parameters in each redshift bin, and the uncertainty on these biases propagates into any parameter that is degenerate with the relativistic signal -- such as local-type $\fNL$ (\cref{sec:png_constraints}) and the amplitude of the local relativistic effects themselves (\cref{sec:results2GR}). Throughout, we bracket this with the `fixed' and `free' cases of \cref{sec:nuisance}. Where a real analysis lies between the two depends on how well the biases can be determined independently of the clustering data, so here we consider both what the clustering data themselves say about $n_{\rm evol}$ and $\mathcal{Q}$ and how the biases can be determined independently of them.

In the even multipoles the relativistic biases enter only at $\mathcal{O}(\mathcal{H}/k)^2$, where they are degenerate with each other and share the $k^{-2}$ scaling of local-type PNG, while $\mathcal{Q}$ additionally enters through the lensing contribution. The leading-order $\mathcal{O}(\mathcal{H}/k)$ relativistic terms, which are sensitive to $b_e$ (or $n_{\rm evol}$) and $\mathcal{Q}$ (through \cref{eq:be_nevol}), instead enter into the odd-parity multipoles: the bispectrum dipole (and octupole) for a single tracer, and the odd cross-power spectrum multipoles for a multi-tracer analysis. The odd multipoles are therefore the natural way to constrain these biases from the clustering data itself; in particular $b_e$ ($n_{\rm evol}$) as the magnification bias is already partially constrained by the lensing signal\footnote{In the peak-background split, $b_e$ and $b_{\Psi}$ are closely related: if one assumes the tracer number density evolves only through $\sigma_8(z)$ then $b_e \approx f\, b_{\Psi}/2$, which for a universal mass function gives $b_e = \delta_c f (b_1 - 1)$ \citep{Jeong_2012}. This relationship holds beyond the universal mass function, but not for evolution unrelated to $\sigma_8(z)$ \citep{Sullivan_2025,Nascimento_2026}. We discuss this relation, its connection to $n_{\rm evol}$, and what it implies for constraining $b_{\Psi}$ directly from relativistic effects in \cref{sec:be_bpsi}.}.

\begin{figure*}
\centering
\includegraphics[width=\linewidth]{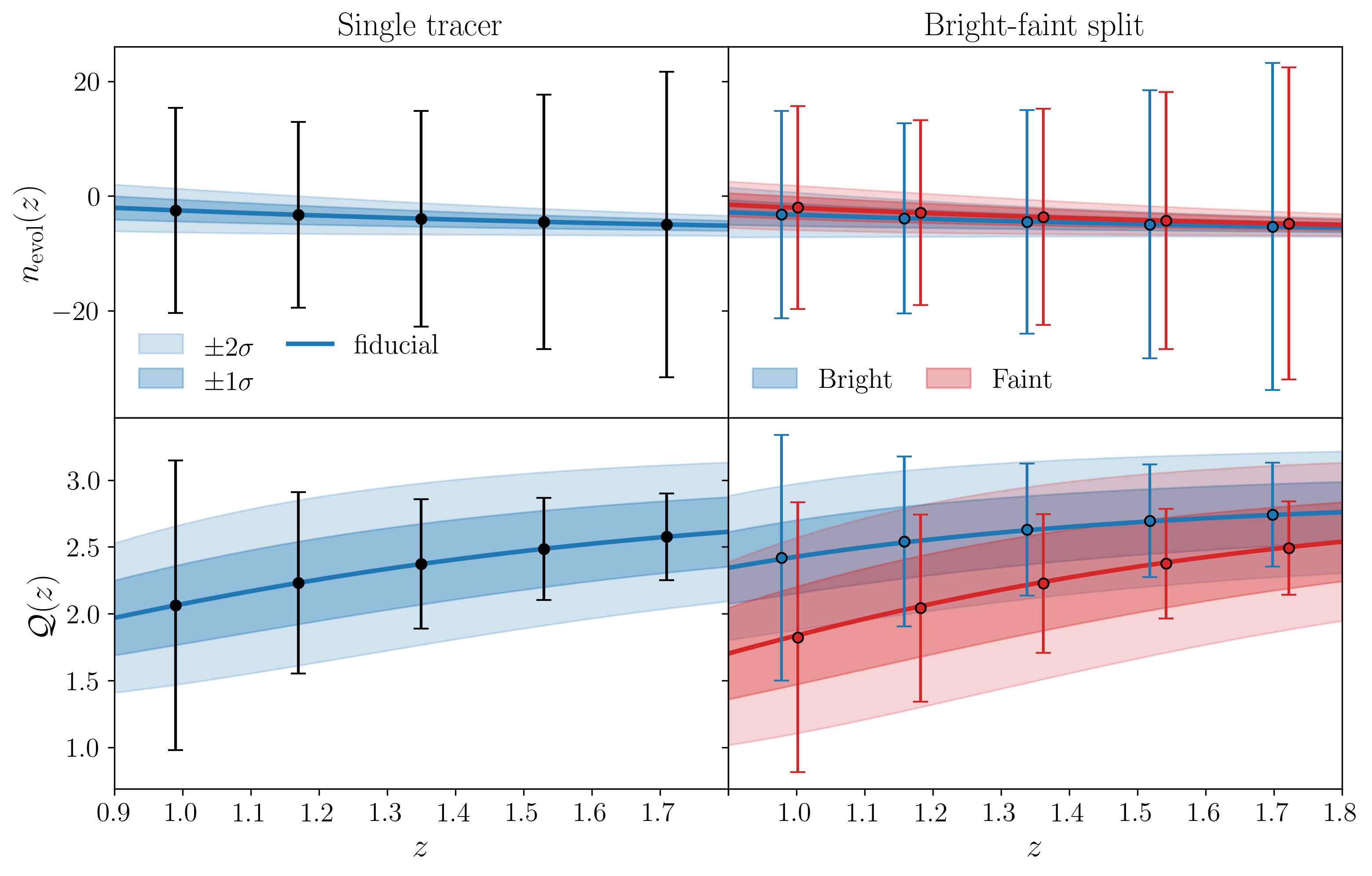}
\caption{Constraints on the count slope $n_{\rm evol}$ (\cref{eq:be_nevol}; top) and the magnification bias (bottom) for a \textit{Euclid}-like H$\alpha$ survey, for a single-tracer analysis (left) and a bright-faint split at a fixed flux, for which each sample has its own $n_{\rm evol}$ (right). The shaded bands show the $1$ and $2\sigma$ priors forward-modelled from the uncertainty in the `Model $3$' luminosity function (\cref{sec:lumfunc_prior}). The error bars show the marginalised $1\sigma$ Fisher constraints from a joint power spectrum and bispectrum analysis without luminosity function priors, including the bispectrum dipole (and, for the bright-faint split, the odd cross-power spectrum multipoles), centred on the fiducial value in each redshift bin. For the bright-faint split, the bright and faint error bars are offset slightly in redshift for clarity.}\label{fig:beQ_st}
\end{figure*}

\Cref{fig:beQ_st} shows the marginalised constraints on $n_{\rm evol}$ and $\mathcal{Q}$ from a joint power spectrum and bispectrum analysis of the \textit{Euclid}-like survey, including the odd multipoles, together with the luminosity function priors discussed below. The clustering data barely constrain the count slope: for a single tracer, $\sigma(n_{\rm evol})\approx16$--$26$ in each redshift bin, compared with fiducial values $|n_{\rm evol}|\approx2.5$--$5$. The magnification bias is much better determined, with $\sigma(\mathcal{Q})$ falling from $\approx1$ at $z\approx1$ to $\approx0.3$ at $z\approx1.7$, as the lensing signal grows with redshift. As we show in \cref{sec:png_constraints}, the weak constraint on $n_{\rm evol}$ matters little for local-type $\fNL$ once the bispectrum dipole is included, since the remaining degeneracy is then with $\mathcal{Q}$.

The picture is similar for the bright-faint split (right-hand panels of \cref{fig:beQ_st}). The single-bin posteriors in \cref{fig:beQ_singlebin_BF} (in the appendix) show that the clustering data mainly constrain combinations of the bright and faint biases, with strong positive correlations between $n^B_{\rm evol}$ and $n^F_{\rm evol}$ and between $\mathcal{Q}^B$ and $\mathcal{Q}^F$. This is consistent with the odd-parity cross-power being sourced by the difference between these biases.

Both biases can, however, be potentially determined independently of the clustering. Unlike the evolution bias, $n_{\rm evol}$ and $\mathcal{Q}$ are properties of the observed sample at the survey flux cut (\cref{sec:magnification_evolution}), so they can be measured from the catalogue itself, without assuming a form for the luminosity function. The count slope is the logarithmic slope of the redshift distribution of the sample, and can be measured by binning the catalogue in redshift. Its precision is set less by the number of galaxies than by how well the redshift dependence of the completeness, purity and redshift success rate of the sample is known, since any unmodelled trend in these enters the measured slope directly. In addition, \cref{eq:be_nevol} assumes that the sample is defined by a single flux cut: any further selection adds terms to it, so that $n_{\rm evol}$ and $\mathcal{Q}$ no longer determine $b_e$ on their own \citep{Sullivan_2025,Nascimento_2026}. For the BOSS luminous red galaxies, for example, \citet{Sullivan_2025} find that the other criteria used to select the samples affect the redshift evolution of their number density more than the magnification correction does. The magnification bias is the response of the number counts to a small change in the flux (or magnitude) cut, and can be measured by perturbing the fluxes in the catalogue and re-applying the full sample selection \citep{Zhou_2023,Elvin-Poole_2023,Wenzl_2024}, or, where the selection depends on more than the flux, by injecting synthetic sources into the survey images \citep{Everett_2022,Elvin-Poole_2023}. The latter may be better suited to the H$\alpha$ sample, for which the efficiency of the redshift measurement depends on the signal-to-noise ratio of the emission line rather than being set by a sharp flux cut \citep[e.g.][]{Gabarra_2023}. For current samples, finite-difference estimates of $\mathcal{Q}$ reach per-cent-level statistical precision \citep{Zhou_2023,Wenzl_2024}.

To consider sample-motivated priors on the relativistic biases in our forecasts, we use existing measurements of the luminosity function of the target population, propagated into priors on $n_{\rm evol}$ and $\mathcal{Q}$, following \citet{Wang_2020}. We use the \textit{Euclid}-like sample as an example of this approach, with priors forward-modelled from the uncertainty in the `Model $3$' H$\alpha$ luminosity function of \citet{Pozzetti_2016} (\cref{sec:lumfunc_prior}), shown by the shaded bands in \cref{fig:beQ_st}. These priors, $\sigma(n_{\rm evol})\approx1$--$2$ and $\sigma(\mathcal{Q})\approx0.3$ in each redshift bin, are tighter than the clustering constraints on $n_{\rm evol}$ by an order of magnitude or more, and comparable to those on $\mathcal{Q}$ at $z\gtrsim1.5$. However, they include only the statistical uncertainty on the parameters of a single model. In \cref{sec:lumfunc_prior} we show that different luminosity function models that give the same number density agree closely on $n_{\rm evol}$, since matching the number density largely fixes the redshift distribution, but differ in $\mathcal{Q}$ by several times its statistical uncertainty. The model dependence of such priors therefore enters mainly through $\mathcal{Q}$ and thus the actual results will be dependent on the true magnification bias of the sample.

We quantify how well the relativistic biases must be known to approach the `fixed' constraints in \cref{sec:prior_req} for local-type $\fNL$ and in \cref{sec:results2GR} for the amplitude of the local relativistic effects.

\subsection{Constraints}\label{sec:png_constraints}

\begin{figure*}
\centering
\includegraphics[width=\linewidth]{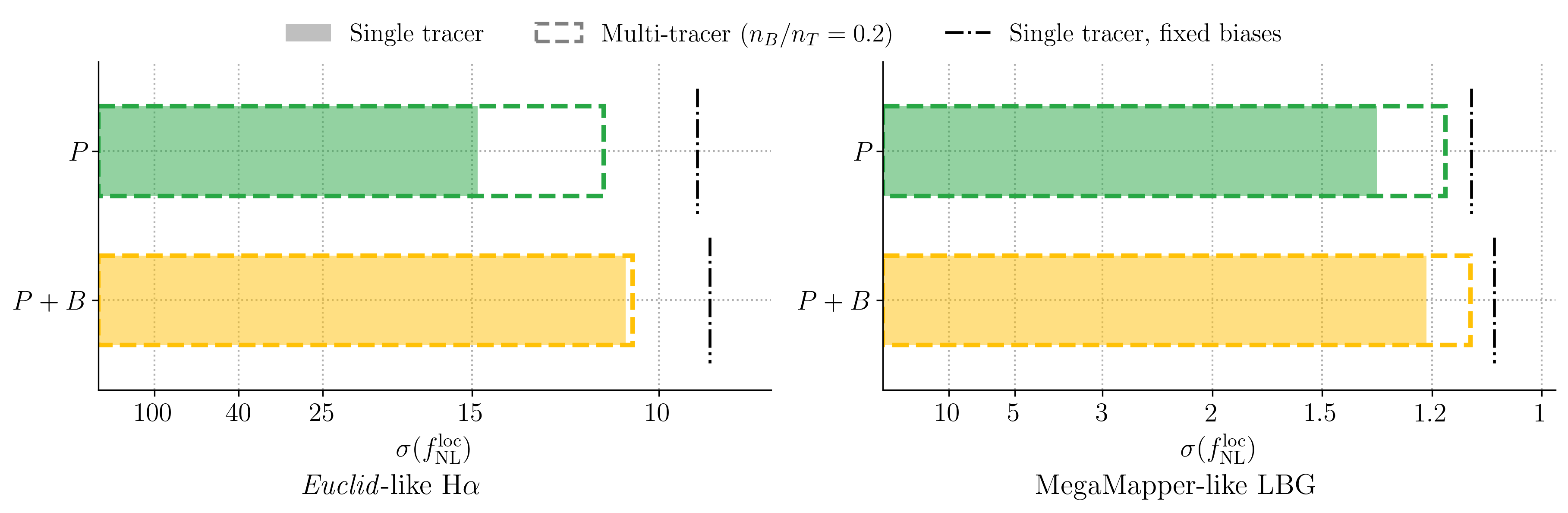}
\caption{Constraints on local-type $\fNL$ for a \textit{Euclid}-like H$\alpha$ survey (left) and a MegaMapper-like LBG survey (right), for a power spectrum ($P$) and a joint power spectrum and bispectrum ($P+B$) analysis. The filled bars show the single-tracer constraints and the dashed unfilled bars a bright-faint split with a fixed bright fraction, $n_B/n_T=0.2$, in every redshift bin (so that both samples share the same $n_{\rm evol}$). In both cases $n_{\rm evol}$ and $\mathcal{Q}$ are marginalised over in each redshift bin with no luminosity function prior, while the black dash-dotted lines show the single-tracer constraint with them held at their fiducial values. All multipoles are included: $\ell=0,2,4$ for the power spectrum (plus the odd cross-multipoles $\ell=1,3$ for the bright-faint split) and $\ell=0,1,2$ for the bispectrum. The horizontal axes are linear in $1/\sigma$.}\label{fig:fnl_con_surveys}
\end{figure*}

\begin{table*}
\centering

\caption{Forecast $1\sigma$ constraints on $\fNL$ for each shape, with $n_{\rm evol}$ and $\mathcal{Q}$ free (no prior), with the luminosity function (LF) prior of \cref{sec:lumfunc_prior}, and held fixed at their fiducial values. ST denotes a single tracer and MT a bright-faint split with a fixed bright fraction $n_B/n_T=0.2$. All multipoles are included, as in \cref{fig:fnl_con_surveys}. We do not forward model LF priors for the MegaMapper-like sample. Each shape is constrained independently, with the other two held at zero.}\label{tab:fnl_summary}

\begin{tabular}{lll ccc ccc ccc}
\toprule
 & & & \multicolumn{3}{c}{$\sigma(\fNL^{\rm loc})$} & \multicolumn{3}{c}{$\sigma(\fNL^{\rm eq})$} & \multicolumn{3}{c}{$\sigma(\fNL^{\rm orth})$} \\
\cmidrule(lr){4-6} \cmidrule(lr){7-9} \cmidrule(lr){10-12}
Survey & Data & Tracer & free & LF & fixed & free & LF & fixed & free & LF & fixed \\
\midrule
\textit{Euclid}-like H$\alpha$ & $P$ & ST & 14.8 & 9.76 & 9.36 & 41.3 & 39.7 & 39.6 & 71.4 & 68.6 & 68.4 \\
 &  & MT & 11.1 & 9.64 & 9.29 & 39.5 & 38.9 & 38.8 & 68.3 & 67.2 & 67.0 \\
 & $P+B$ & ST & 10.6 & 9.58 & 9.17 & 34.8 & 34.4 & 34.3 & 40.6 & 40.3 & 40.2 \\
 &  & MT & 10.5 & 9.47 & 9.10 & 34.2 & 34.0 & 33.9 & 40.5 & 40.2 & 40.2 \\
\midrule
MegaMapper-like LBG & $P$ & ST & 1.33 & -- & 1.12 & 5.93 & -- & 5.83 & 10.2 & -- & 10.1 \\
 &  & MT & 1.17 & -- & 1.07 & 2.89 & -- & 2.88 & 5.00 & -- & 4.97 \\
 & $P+B$ & ST & 1.21 & -- & 1.08 & 5.83 & -- & 5.79 & 9.28 & -- & 9.23 \\
 &  & MT & 1.12 & -- & 1.01 & 2.87 & -- & 2.86 & 4.86 & -- & 4.84 \\
\bottomrule
\end{tabular}
\end{table*}

The forecast constraints on $\fNL$ for each shape are summarised in \cref{tab:fnl_summary}. We compare the `fixed' and `free' treatments of the relativistic biases (\cref{sec:nuisance}) with, for the \textit{Euclid}-like survey, the luminosity function priors of \cref{sec:lumfunc_prior}. The `fixed' constraint approximates the error of an analysis that neglects relativistic corrections. The headline result is that marginalising over the relativistic biases increases $\sigma(\fNL^{\rm loc})$ by up to $\sim60\%$ (for the \textit{Euclid}-like power spectrum), so that an analysis that neglects relativistic corrections would underestimate its error on local-type $\fNL$ by the same factor and hence overstate the significance of any detection. Below we first consider the information from the bispectrum and the bright-faint split with the biases fixed (\cref{sec:png_info}), then the cost of marginalising over them and how much of it the clustering data themselves recover (\cref{sec:png_degeneracy}), and finally the impact of external priors (\cref{sec:prior_req}).

\subsubsection{Information from the bispectrum and the bright-faint split}\label{sec:png_info}

With $n_{\rm evol}$ and $\mathcal{Q}$ fixed, adding the bispectrum to a single-tracer analysis of the \textit{Euclid}-like survey improves the constraint on local-type $\fNL$ by only $\sim2\%$ ($9.36\to9.17$). For the MegaMapper-like survey the gain is similarly small, $\sim4\%$ ($1.12\to1.08$). The local-type signal lies predominantly in squeezed configurations, where triangles sharing a long mode are strongly correlated with each other and with the power spectrum \citep{Biagetti_2022,Salvalaggio_2024}; these triangles largely measure the same scale-dependent bias as the power spectrum\footnote{With a Gaussian bispectrum covariance the gain is instead $\sim12\%$, as it lacks the off-diagonal $PB$ cross-covariance and underestimates the variance of squeezed triangles.}. This is consistent with the modest gain from the bispectrum in BOSS \citep{Cabass_2022b}; raising the bispectrum $k_{\rm max}$ to $0.15\,\aMpch$ increases the gain only to $\sim7\%$.

With the relativistic biases fixed, the gain from a bright-faint split on local-type $\fNL$ comes from the cancellation of sample variance between the two samples \citep{Seljak_2009}, which depends on how much their ratios of PNG to linear bias, $b_\Psi/b_1$, differ. For the peak-background split biases (\cref{sec:png_bias_params}) this ratio tends to $2\delta_c$ for highly biased tracers, so the bright and faint samples differ little: by $\lesssim10\%$ for the \textit{Euclid}-like sample, and from $\sim20\%$ at $z\approx2.3$ to $\sim2\%$ at $z\approx4.9$ for the MegaMapper-like sample. The split therefore changes the local-type constraint by $\lesssim1\%$ for the \textit{Euclid}-like survey and by $\sim5\%$ for the MegaMapper-like survey (\cref{tab:fnl_summary}).

\begin{figure*}
\centering
\includegraphics[width=\linewidth]{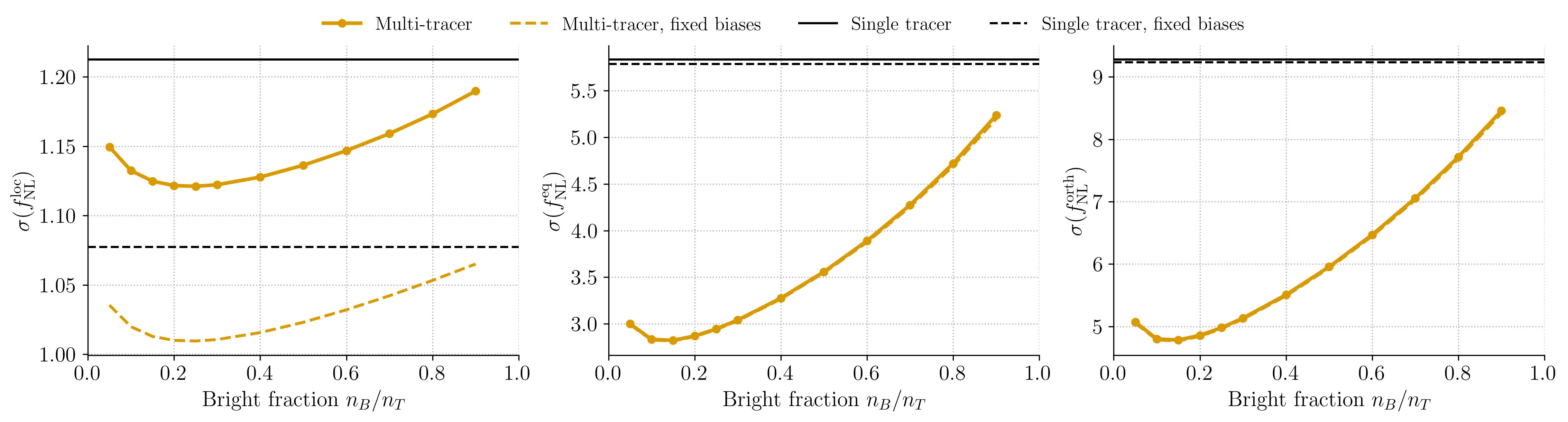}
\caption{Dependence of the bright-faint split constraints on the bright fraction, $n_B/n_T$, held fixed at every redshift, for local (left), equilateral (centre) and orthogonal (right) type $\fNL$, for a joint power spectrum and bispectrum analysis of a MegaMapper-like LBG survey. The coloured lines show the bright-faint split and the black horizontal lines the single-tracer constraints. Solid lines marginalise over $n_{\rm evol}$ and $\mathcal{Q}$ in each redshift bin with an uninformative prior (with both samples of the split sharing $n_{\rm evol}$), while dashed lines hold them at their fiducial values. All multipoles are included, as in \cref{fig:fnl_con_surveys}.}\label{fig:fnl_frac_MM}
\end{figure*}

 The dependence of the joint bright-faint split constraints on the bright fraction is shown for the MegaMapper-like sample in \cref{fig:fnl_frac_MM}, where all three shapes favour small bright samples. The local-type constraint depends only weakly on the split, loosening slowly as the bright sample grows, from $\sigma(\fNL^{\rm loc})\approx1.12$ for $n_B/n_T\approx0.15$--$0.3$ to $1.19$ at $n_B/n_T=0.9$, and is below the single-tracer constraint for every fraction we consider. The equilateral and orthogonal constraints are tightest at $n_B/n_T\approx0.1$--$0.15$ and degrade by $\sim85\%$ and $\sim75\%$, respectively, by $n_B/n_T=0.9$. We therefore use $n_B/n_T=0.2$ unless otherwise stated, which is within $2\%$ of the best constraint for all shapes.

\subsubsection{Degeneracy with the relativistic biases}\label{sec:png_degeneracy}

\begin{figure*}
\centering
\includegraphics[width=0.85\linewidth]{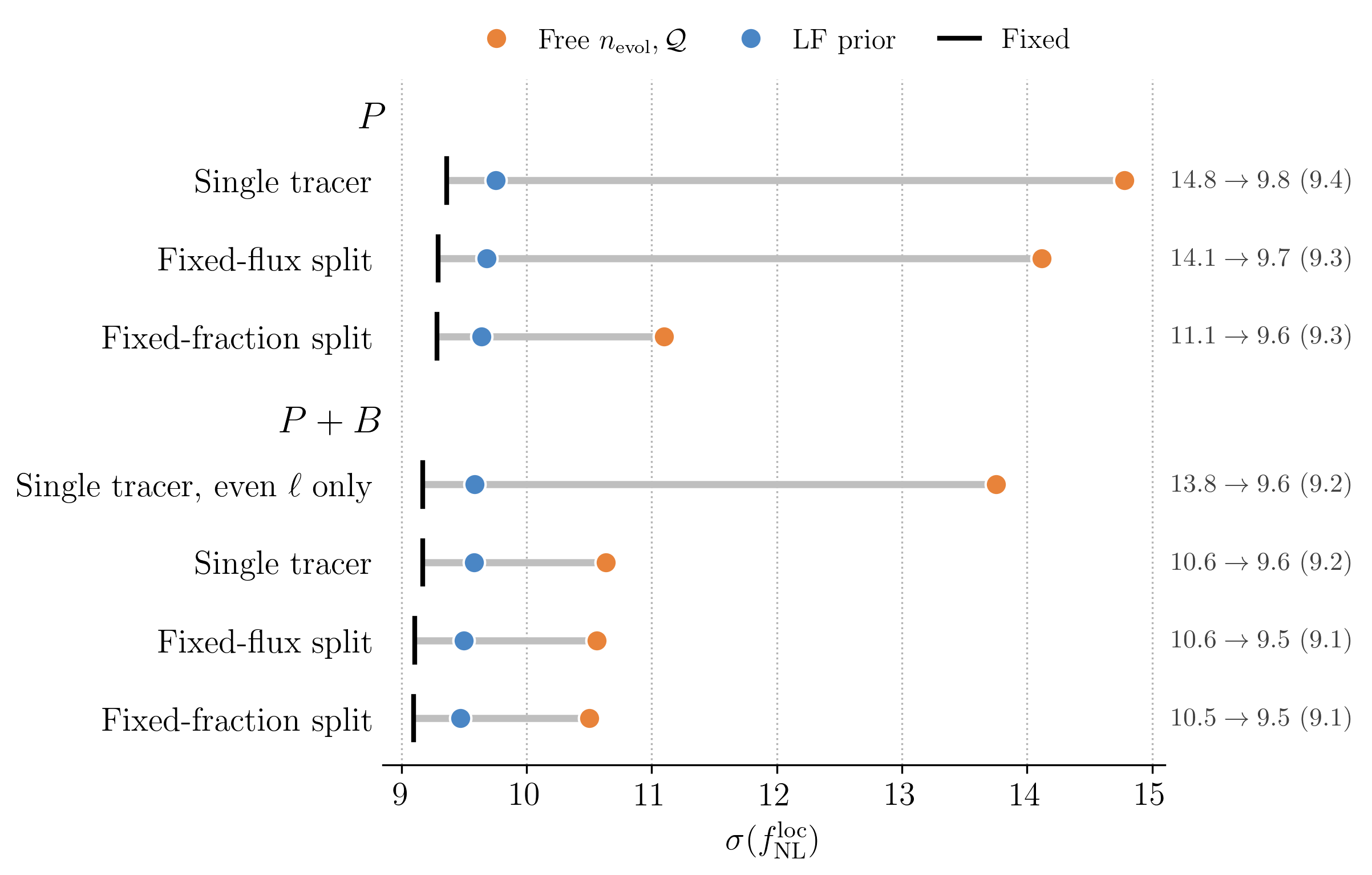}

\caption{Forecast $1\sigma$ constraints on local-type $\fNL$ from each analysis, for a \textit{Euclid}-like H$\alpha$ survey. The orange points marginalise over $n_{\rm evol}$ and $\mathcal{Q}$ with no prior, the blue points add the priors forward-modelled from the `Model $3$' luminosity function (\cref{sec:lumfunc_prior}), and the black ticks mark the `fixed' case, so that each grey line spans the full cost of marginalising over the relativistic biases. The numbers on the right give $\sigma(\fNL^{\rm loc})$ without $\to$ with the luminosity function prior, with the `fixed' case in brackets. The bright-faint splits use either a fixed split flux, $F_s=3\times10^{-16}\,\mathrm{erg\,cm^{-2}\,s^{-1}}$, with independent $n_{\rm evol}$ for each sample, or a fixed bright fraction $n_B/n_T=0.2$ at every redshift, for which both samples share the same $n_{\rm evol}$. Unless stated otherwise all multipoles, including the odd-parity ones, are used.}\label{fig:fnl_con}
\end{figure*}

The cost of marginalising over the relativistic biases is broken down by analysis for the \textit{Euclid}-like survey in \cref{fig:fnl_con}. It is largest for the single-tracer power spectrum, for which $\sigma(\fNL^{\rm loc})$ increases by $\sim60\%$ ($9.4\to14.8$), as in the even multipoles $n_{\rm evol}$ and $\mathcal{Q}$ enter at $\mathcal{O}(\mathcal{H}/k)^2$ with the same $k^{-2}$ scaling as local-type PNG (\cref{sec:beQ_constraints}). The odd-parity multipoles add no PNG signal\footnote{Subdominant local-type PNG signals enter the odd-parity multipoles with $k^{-3}$ dependence through the relativistic effects \cite[e.g.][]{Maartens_2021} or through wide-separation type corrections to the PNG terms themselves, but these are ignored in our analysis.}, but they are sensitive to the leading-order $\mathcal{H}/k$ relativistic terms and so help to break the degeneracy between $\fNL$ and the relativistic biases.

The even bispectrum multipoles partially weaken the degeneracy ($14.8\to13.8$), and the non-zero bispectrum dipole generated by local relativistic projection effects tightens the constraint by a further $\sim23\%$ (to $10.6$), leaving it $\sim16\%$ above the `fixed' case. The dipole does this by breaking the degeneracy between $n_{\rm evol}$ and $\fNL$ (\cref{fig:fnl_con_pk_bk}), even though $n_{\rm evol}$ itself remains poorly constrained by the clustering ($\sigma(n_{\rm evol})\approx16$--$26$; \cref{fig:beQ_st}), so that the remaining loss is almost entirely due to the magnification bias (\cref{fig:fnl_prior_req}), which is still constrained mainly through the lensing contribution.

A bright-faint split doubles the per-bin linear and magnification biases, and the count slope too if the split is at a fixed flux, so its additional information must outweigh these extra nuisance parameters. For a split at a fixed flux, for which each sample has its own $n_{\rm evol}$, the power spectrum constraint barely improves on the single tracer ($14.8\to14.1$), whereas for a fixed bright fraction, for which both samples share a single $n_{\rm evol}$ (\cref{sec:bf_split}), it improves by $\sim25\%$ (to $11.1$). Once the bispectrum dipole is included, however, either split adds little ($\sim1\%$), as the dipole has already broken the degeneracy with $n_{\rm evol}$.

\subsubsection{Priors on the relativistic biases}\label{sec:prior_req}

Applying the luminosity function priors for the \textit{Euclid}-like survey (blue points in \cref{fig:fnl_con}) improves the constraints on local-type $\fNL$ almost to the `fixed' limit. Since these priors are tighter than the clustering constraints on the relativistic biases (\cref{fig:beQ_st}), they leave little for the odd-parity multipoles to add. However, these priors depend on the luminosity function model (\cref{sec:beQ_constraints}), so the blue points should potentially be regarded as optimistic.

\begin{figure*}
\centering
\includegraphics[width=0.85\linewidth]{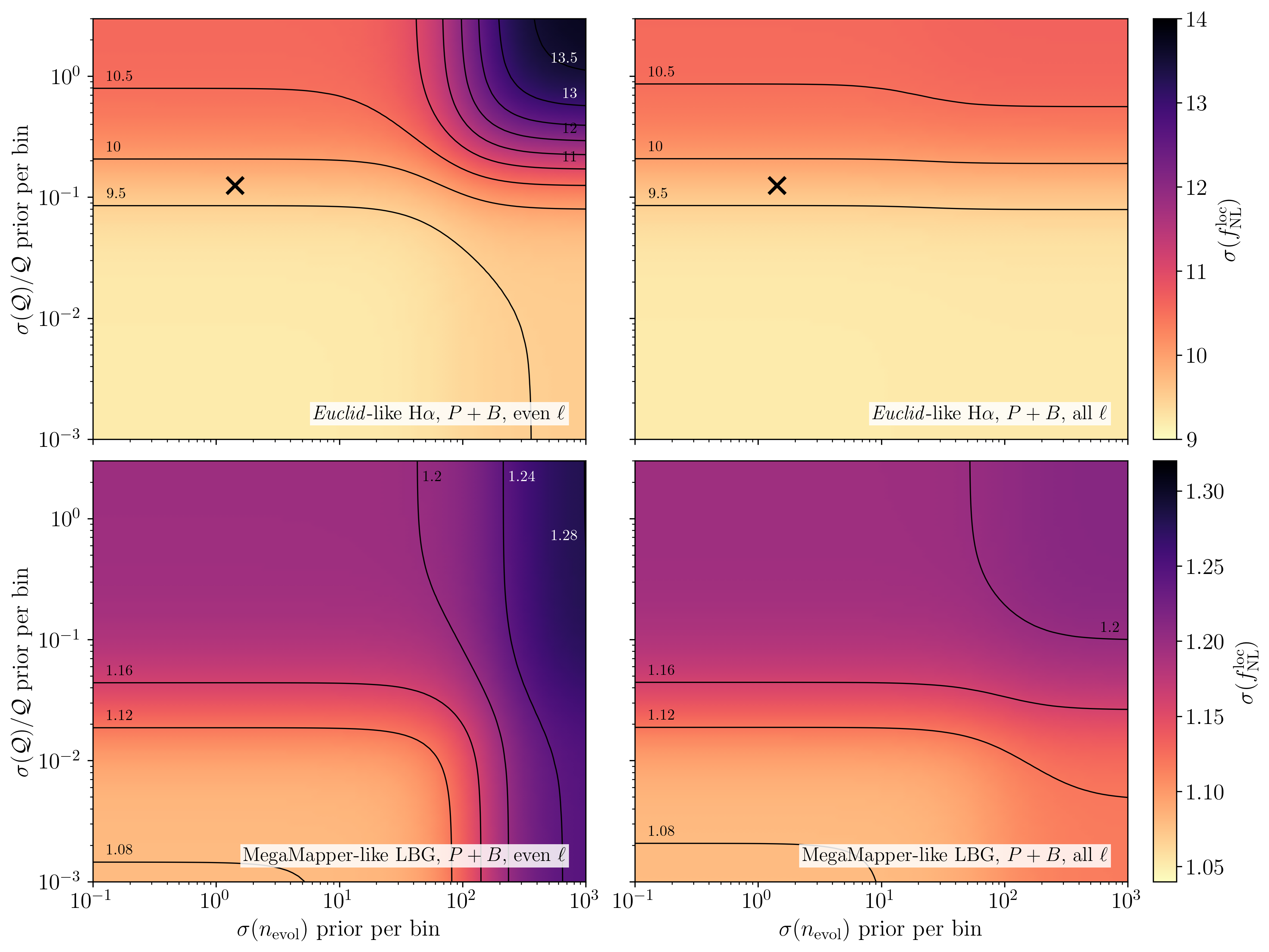}

\caption{Forecast $1\sigma$ constraint on local-type $\fNL$ as a function of the widths of Gaussian priors on $n_{\rm evol}$ and on the fractional magnification bias, $\sigma(\mathcal{Q})/\mathcal{Q}$, for a single-tracer \textit{Euclid}-like H$\alpha$ survey (top) and MegaMapper-like LBG survey (bottom), from a joint power spectrum ($\ell=0,2,4$) and bispectrum analysis with the even bispectrum multipoles only ($\ell=0,2$; left) and with the dipole included ($\ell=0,1,2$; right). Each survey has its own colour scale. With $n_{\rm evol}$ and $\mathcal{Q}$ held at their fiducial values the constraints are $\sigma(\fNL^{\rm loc})=9.2$ (\textit{Euclid}) and $1.08$ (MegaMapper), compared with $13.8$ and $10.6$ (\textit{Euclid}, left and right) and $1.28$ and $1.21$ (MegaMapper) with both free. The same prior width is applied in every redshift bin, with $\mathcal{Q}$ evaluated at the centre of each bin, and added to the Fisher matrix, which includes the \textit{Planck} prior and marginalises over $b_1$, $b_2$ and $b_{\Gamma_2}$ in each bin, on an $80\times80$ logarithmically spaced grid. The crosses mark the priors forward-modelled from the `Model $3$' luminosity function (\cref{sec:lumfunc_prior}), placed at their mean width over the redshift bins. L-shaped contours indicate that a prior on either bias alone breaks the degeneracy with $\fNL$, while horizontal contours indicate that only $\mathcal{Q}$ matters; including the bispectrum dipole makes the contours close to horizontal (see text).}\label{fig:fnl_prior_req}
\end{figure*}

\Cref{fig:fnl_prior_req} shows how precisely the relativistic biases must be known to recover the `fixed' constraint on local-type $\fNL$. For the \textit{Euclid}-like power spectrum (and similarly for a joint analysis with the even multipoles) the two biases are strongly degenerate: with both free, $\sigma(\fNL^{\rm loc})$ is $58\%$ larger than with both fixed, but knowing either one removes most of this loss ($15\%$ with $n_{\rm evol}$ fixed, $3\%$ with $\mathcal{Q}$ fixed). However, including the bispectrum dipole makes the constraint almost insensitive to $n_{\rm evol}$: with the dipole the loss with both free drops to $16\%$, almost all of it due to $\mathcal{Q}$, and fixing $\mathcal{Q}$ alone brings the constraint to within $0.2\%$ of the fixed case. The luminosity function priors constrain $n_{\rm evol}$ well enough to break the degeneracy in every case, leaving a $\sim4$--$5\%$ degradation set by $\sigma(\mathcal{Q})/\mathcal{Q}\simeq10$--$14\%$; reaching $1\%$ would require $\sigma(\mathcal{Q})/\mathcal{Q}\lesssim4\%$ in each redshift bin, which is of the same order as the statistical precision of current finite-difference estimates of $\mathcal{Q}$ from the catalogue (\cref{sec:beQ_constraints}; \citealt{Zhou_2023,Wenzl_2024}).

For the MegaMapper-like sample the power spectrum constraint degrades by $19\%$. The joint constraint degrades by $19\%$ ($12\%$ with the dipole) and, with $n_{\rm evol}$ fixed, requires $\sigma(\mathcal{Q})/\mathcal{Q}\lesssim0.8\%$ to come within $1\%$. The large lensing contribution at these redshifts constrains $\mathcal{Q}$ from the clustering itself, to $\approx13\%$ in the lowest redshift bin and $\approx2\%$ above $z\approx3.4$. A prior of $\sigma(\mathcal{Q})/\mathcal{Q}\approx10\%$, comparable to the luminosity function priors for the \textit{Euclid}-like sample, therefore tightens the joint constraint by only $\lesssim1\%$, and a prior on $\mathcal{Q}$ helps appreciably only once it reaches the per-cent level, where it begins to compete with the clustering constraint.

Finally, since the odd multipoles constrain $n_{\rm evol}$ and $\mathcal{Q}$ directly from the clustering (\cref{sec:beQ_constraints}), they also provide a consistency test of any external determination of the biases.

\subsubsection{Non-local PNG}\label{sec:nonlocal_png}

Equilateral and orthogonal PNG are far less degenerate with relativistic effects than the local shape: the scale-dependent bias in \cref{eq:first_order} scales as $k^{\alpha-2}/T(k)$, i.e.\ as $k^{0}$ for both shapes, compared with the $k^{-2}$ scaling shared by local-type PNG and the leading even-parity $(\mathcal{H}/k)^2$ relativistic terms. The conclusions of the local case therefore carry over but are far less pronounced. For the single-tracer joint forecasts\footnote{Power spectrum information alone poorly constrains the equilateral and orthogonal shapes: for the single-tracer \textit{Euclid}-like survey we forecast $\sigma(\fNL^{\rm eq})\approx41$ and $\sigma(\fNL^{\rm orth})\approx71$ with $n_{\rm evol}$ and $\mathcal{Q}$ free, $\sim4\%$ larger than with them fixed (\cref{tab:fnl_summary}). Since the two shapes have the same scale-dependent bias, up to the ratio of their PNG biases, the power spectrum alone cannot separate them, and its orthogonal constraint is a fixed multiple of the equilateral one.}, neglecting relativistic effects leads one to underestimate the errors on $\fNL^{\rm eq}$ and $\fNL^{\rm orth}$ by only $\lesssim2\%$, and, in contrast to the local case, priors on the relativistic biases make little difference ($\lesssim1\%$) for either the single-tracer or bright-faint split analysis in a \textit{Euclid}-like survey. This remains true for the MegaMapper-like sample, for which marginalising over them loosens the joint constraints on both shapes by $\lesssim1\%$ (\cref{tab:fnl_summary}), even though neglecting relativistic effects biases the orthogonal shape by $\sim3\sigma$ (\cref{sec:png_bias}).

The bright-faint split gives a large gain for the MegaMapper-like sample, improving $\sigma(\fNL^{\rm eq})$ from $5.8$ to $2.9$ and $\sigma(\fNL^{\rm orth})$ from $9.3$ to $4.9$ (\cref{tab:fnl_summary}), but for a \textit{Euclid}-like survey it changes the joint constraints by $\lesssim2\%$, giving $\sigma(\fNL^{\rm eq}) \approx 34$ and $\sigma(\fNL^{\rm orth}) \approx 40$ with or without the split. For reference, the current \textit{Planck} constraints are $\sigma(\fNL^{\rm eq}) = 47$ and $\sigma(\fNL^{\rm orth}) = 24$ \citep{Akrami_2020}: the \textit{Euclid}-like survey would improve on the equilateral one by a factor of $\sim1.4$, but its orthogonal constraint is $\sim1.7$ times weaker than \textit{Planck}'s. Fitting the equilateral and orthogonal shapes jointly loosens the single-tracer constraints, to $\sigma(\fNL^{\rm eq})\approx40$ and $\sigma(\fNL^{\rm orth})\approx47$ for the \textit{Euclid}-like survey (correlation coefficient $+0.52$) and to $13.1$ and $20.9$ for the MegaMapper-like survey ($+0.90$), as the two shapes are then separated only by the bispectrum, and their correlation depends on the uncertain ratio of their PNG biases.  We caution that these constraints derive mainly from bispectrum configurations near $k_\mathrm{max}$, and are therefore the results most sensitive to our tree-level modelling.

\begin{figure*}
\centering
\includegraphics[width=\linewidth]{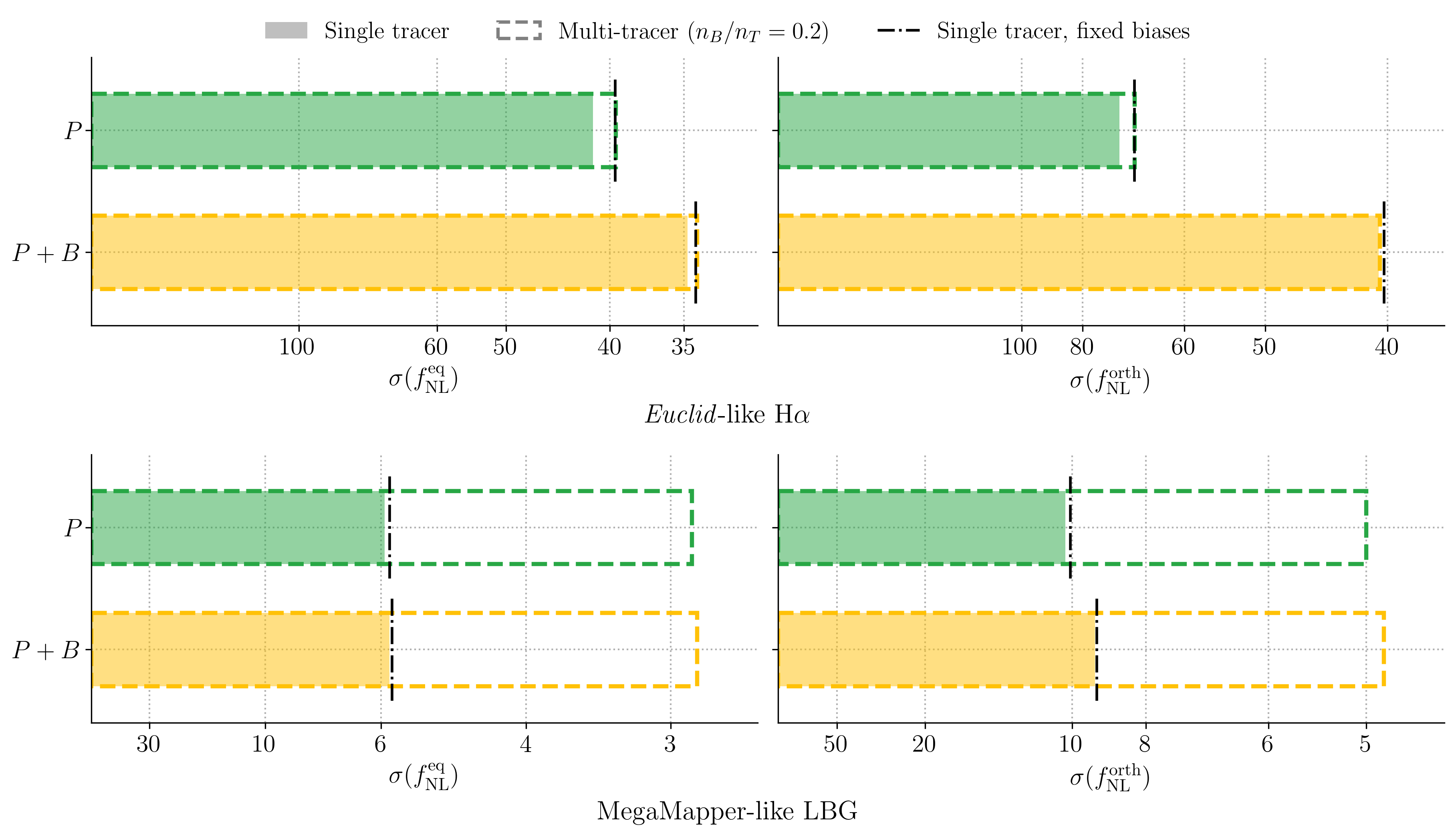}
\caption{\label{fig:fnl_con_eo} Constraints on equilateral (left) and orthogonal (right) type $\fNL$ for a \textit{Euclid}-like H$\alpha$ survey (top) and a MegaMapper-like LBG survey (bottom). The filled bars show the single-tracer constraints and the dashed unfilled bars a bright-faint split with a fixed bright fraction, $n_B/n_T=0.2$, in every redshift bin (so that both samples share the same $n_{\rm evol}$). In both cases $n_{\rm evol}$ and $\mathcal{Q}$ are marginalised over in each redshift bin with no luminosity function prior, while the black dash-dotted lines show the single-tracer constraint with them held at their fiducial values. All multipoles are included: $\ell=0,2,4$ for the power spectrum (plus the odd cross-multipoles $\ell=1,3$ for the bright-faint split) and $\ell=0,1,2$ for the bispectrum. The horizontal axes are linear in $1/\sigma$.}
\end{figure*}

\section{Results II: Local Relativistic Effects}\label{sec:results2GR}

This section focuses on forecasts for the detection of local relativistic projection effects. For forecasts on the detectability of integrated relativistic effects in the power spectrum see Appendix H of \citetalias{Addis_2025}. We group all the local relativistic effects together and quantify their contribution through a global amplitude parameter, $\alpha_\mathrm{GR}$, with a fiducial value of unity. We first consider the impact of wide-separation corrections on this analysis, and then the constraining power of multi-tracer and joint power spectrum and bispectrum analyses, particularly once the uncertainty in the relativistic biases is included. In all of these forecasts we marginalise over the same set of parameters as in the PNG analyses with \textit{Planck} priors (\cref{sec:forecast_details}) and with $\fNL=0$ held fixed for every shape.

The odd multipoles are particularly sensitive to local relativistic effects, as this is where the leading-order $\mathcal{H}/k$ signal enters. One can access this signal in our standard summary statistics either through the bispectrum, which has a non-zero dipole (and octupole) outside the equilateral limit, or through multi-tracer approaches.

We find that single-tracer approaches (including the bispectrum dipole) are insufficient for confident detections of local relativistic effects, and we therefore focus mainly on bright-faint split approaches.

\subsection{Biases}

\begin{figure}
\centering
\includegraphics[width=\linewidth]{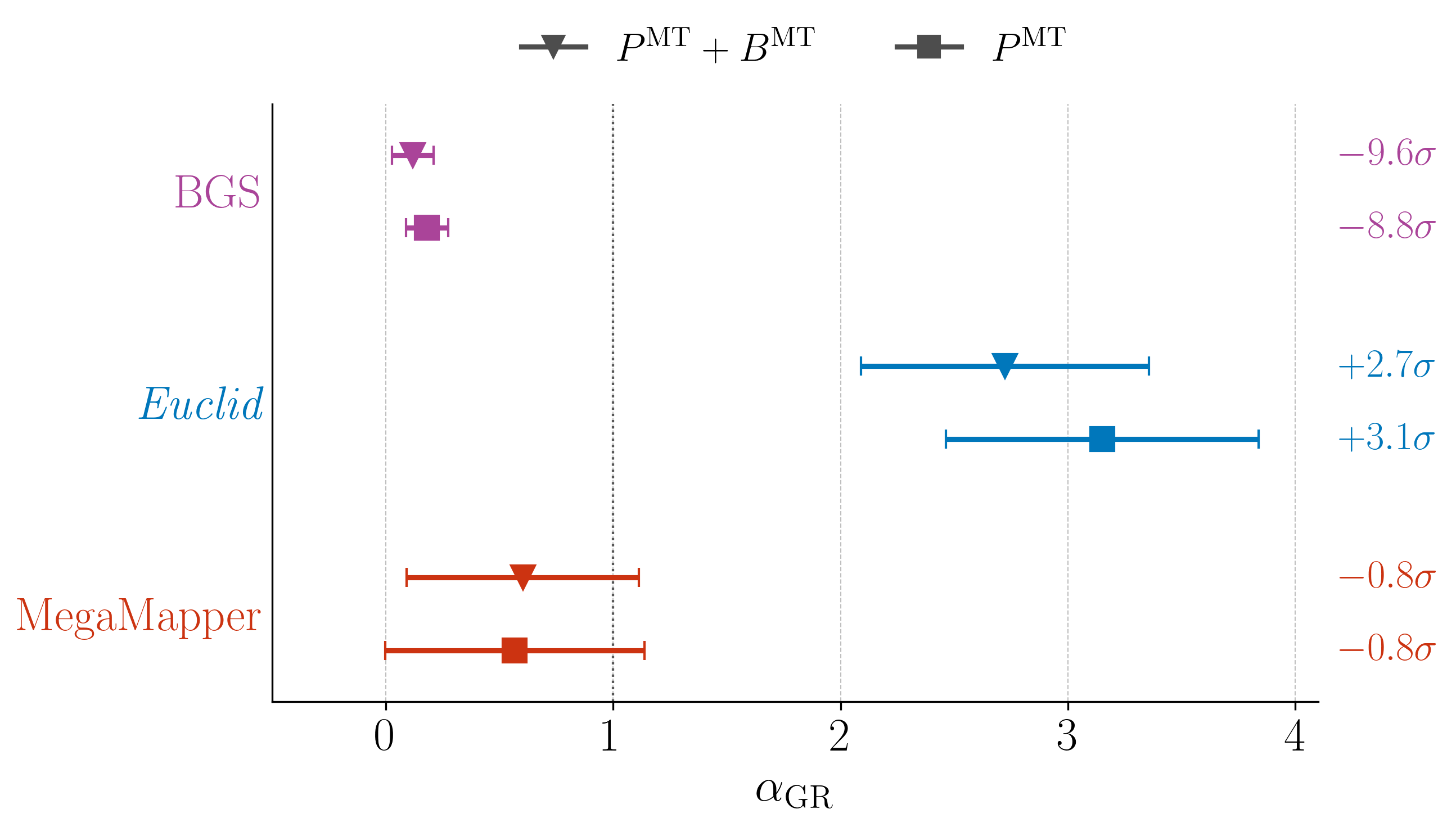}
\caption{Forecasted $1\sigma$ constraints on the amplitude of local relativistic effects, $\alpha_\mathrm{GR}$, for a DESI-like BGS survey (purple), a \textit{Euclid}-like H$\alpha$ survey (blue) and a MegaMapper-like LBG survey (red), if we neglect wide-separation corrections beyond the zeroth order of the wide-separation expansion. The triangles show a multi-tracer (bright-faint) joint power spectrum and bispectrum analysis, and the squares a multi-tracer power spectrum analysis, as in \cref{fig:snr_GR}. The relativistic biases, $n_{\rm evol}$ and $\mathcal{Q}$, are held at their fiducial values, and each sample is split at a fixed bright fraction, $n_B/n_T=0.2$ for the DESI-like BGS and MegaMapper-like samples and $n_B/n_T=0.4$ for the \textit{Euclid}-like sample, as in \cref{fig:snr_GR}. The numbers on the right give the shift of each from the fiducial value, $\alpha_\mathrm{GR}=1$ (dotted line), in units of its $1\sigma$ error.}\label{fig:ws_bias}
\end{figure}

\Cref{fig:ws_bias} shows the constraints on $\alpha_\mathrm{GR}$ if one ignores wide-separation type corrections in the bright-faint split analyses. The impact of ignoring wide-separation effects is twofold: first, the unaccounted-for signal biases the inference on the local relativistic effects, in both the odd and even multipoles, owing to their shared $k$-dependence. This is particularly prominent for the lower-redshift BGS survey, for which the inferred value of $\alpha_\mathrm{GR}$ is shifted by $\approx9\sigma$ in the power spectrum analysis and $\approx10\sigma$ in the joint analysis, due to a larger wide-angle contribution. At higher redshifts the radial evolution is the dominant part of the wide-separation corrections and this biases the inference for the \textit{Euclid}-like survey by $\approx3\sigma$ and the MegaMapper-like survey by less than $1\sigma$. Second, the wide-separation terms themselves contain local relativistic signal; the first-order wide-separation corrections to the odd-parity $\mathcal{H}/k$ local relativistic term generate a relevant $\approx (\mathcal{H}/k)^2$ contribution (see Section 3 of \citetalias{Addis_2024} for a discussion of the scale and redshift dependence of the wide-separation corrections). Including them tightens the constraint on $\alpha_\mathrm{GR}$ by $\sim15\%$ for BGS and MegaMapper and by only $\sim1\%$ for \textit{Euclid}. This contribution is included in the constraints discussed in the next section.

\subsection{Constraints}

\begin{figure}
\centering
\includegraphics[width=\linewidth]{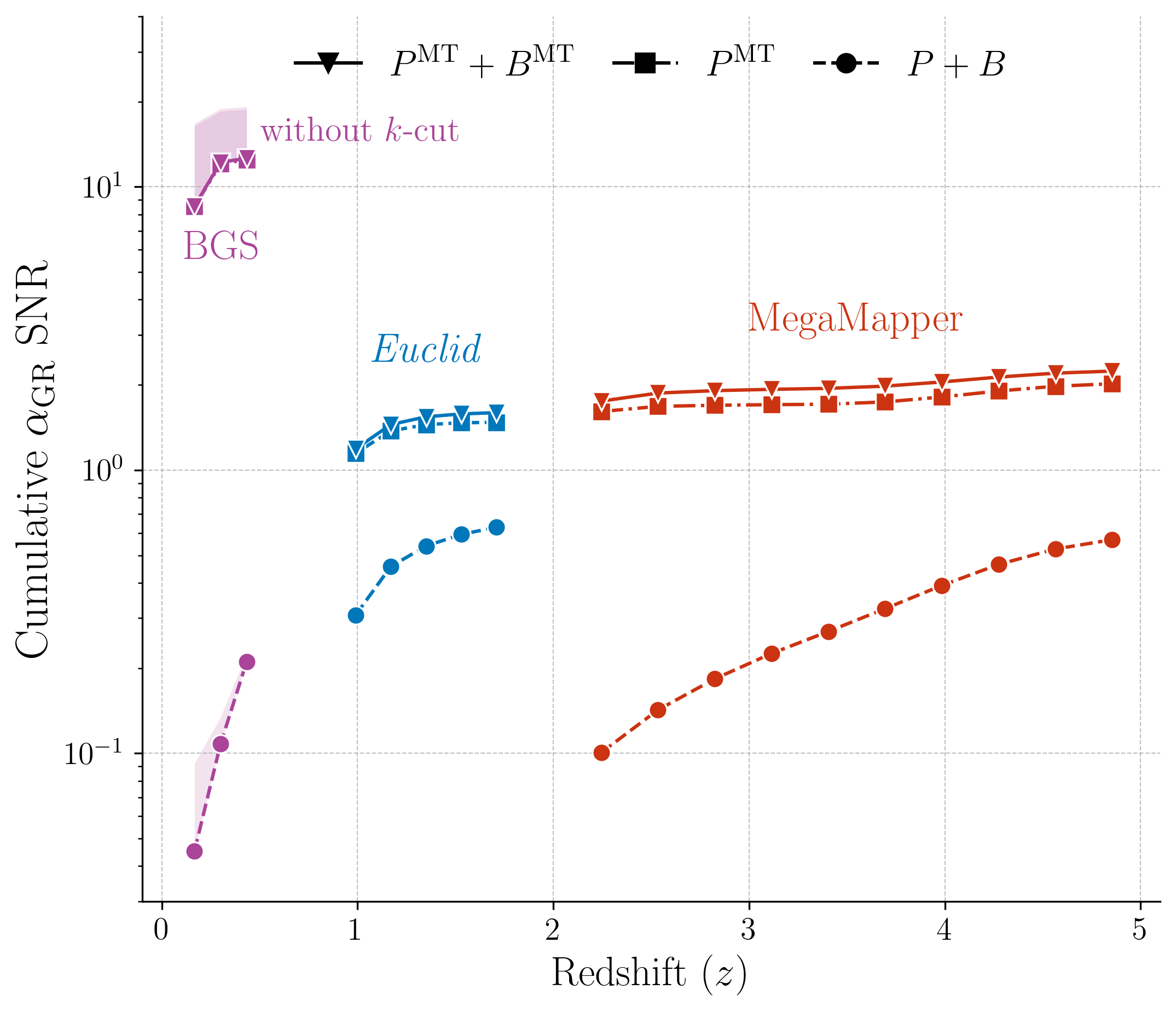}

\caption{Forecasted cumulative marginalised SNR on the amplitude of local relativistic effects, $\alpha_\mathrm{GR}$, for a DESI-like BGS survey (purple), a \textit{Euclid}-like H$\alpha$ survey (blue) and a MegaMapper-like LBG survey (red), over redshift bins, for different data vectors, with $n_{\rm evol}$ and $\mathcal{Q}$ held at their fiducial values. The DESI-like BGS and MegaMapper-like samples are split at a fixed bright fraction, $n_B/n_T=0.2$, and the \textit{Euclid}-like sample at $n_B/n_T=0.4$: the SNR depends only weakly on the bright fraction around these values, which give the highest SNR of the fractions we consider for the \textit{Euclid}-like sample and are within $1\%$ and $2\%$ of it for the DESI-like BGS and MegaMapper-like samples, respectively (see \cref{fig:frac_GR_BGS}). For the DESI-like BGS survey, the shaded bands extend each curve to the SNR without the wide-separation-based minimum $k$-cut, and therefore give some indication of the additional signal on scales not accessible to the 3D Fourier statistics; for the other two surveys the cut removes no $k$-bins.}\label{fig:snr_GR}
\end{figure}

\begin{table}
\centering
\caption{Forecast SNR on the amplitude of local relativistic effects, $\alpha_\mathrm{GR}$, over all redshift bins, with $n_{\rm evol}$ and $\mathcal{Q}$ marginalised over in each redshift bin with no prior (free) and held at their fiducial values (fixed), for the data vectors of \cref{fig:snr_GR}. ST denotes a single tracer ($\ell=0,2,4$ for the power spectrum and $\ell=0,1,2$ for the bispectrum) and MT a bright-faint split at a fixed bright fraction, $n_B/n_T=0.2$ for the DESI-like BGS and MegaMapper-like samples and $0.4$ for the \textit{Euclid}-like sample, which also includes the odd cross-power spectrum multipoles.}\label{tab:gr_summary}
\begin{tabular}{lll cc}
\toprule
 & & & \multicolumn{2}{c}{SNR on $\alpha_\mathrm{GR}$} \\
\cmidrule(lr){4-5}
Survey & Data & Tracer & free & fixed \\
\midrule
DESI-like BGS & $P+B$ & ST & 0.13 & 0.21 \\
 & $P$ & MT & 0.82 & 12.5 \\
 & $P+B$ & MT & 0.94 & 12.6 \\
\midrule
\textit{Euclid}-like H$\alpha$ & $P+B$ & ST & 0.16 & 0.63 \\
 & $P$ & MT & 0.56 & 1.48 \\
 & $P+B$ & MT & 0.98 & 1.60 \\
\midrule
MegaMapper-like LBG & $P+B$ & ST & 0.42 & 0.57 \\
 & $P$ & MT & 0.82 & 2.02 \\
 & $P+B$ & MT & 1.21 & 2.24 \\
\bottomrule
\end{tabular}
\end{table}

The forecasted cumulative marginalised SNR on the amplitude of local relativistic effects, $\alpha_\mathrm{GR}$, is plotted in \cref{fig:snr_GR} and listed in \cref{tab:gr_summary} for the DESI-like BGS, \textit{Euclid}-like and MegaMapper-like surveys\footnote{We do not consider a bright-faint split for the \textit{SPHEREx}-like survey because its samples are selected on photometric-redshift precision rather than on flux, so the magnification biases of the bright and faint subsamples cannot be reliably derived from the luminosity function.}. In \cref{fig:snr_GR}, $n_{\rm evol}$ and $\mathcal{Q}$ are held at their fiducial values.

With $n_{\rm evol}$ and $\mathcal{Q}$ fixed, we forecast that local relativistic effects should be detectable at high significance for a DESI-like BGS survey with a bright-faint split, with a cumulative SNR of $\approx12.5$ for the multi-tracer power spectrum and $\approx12.6$ for the joint analysis. For the \textit{Euclid}-like and MegaMapper-like surveys the multi-tracer power spectrum gives an SNR of $\approx1.5$ and $\approx2$, respectively, rising to $\approx1.6$ and $\approx2.2$ with the multi-tracer bispectrum, so that local relativistic effects are at best marginally detectable.\footnote{Including the $m\neq0$ bispectrum multipoles, which we otherwise neglect (\cref{sec:theory}), increases these SNRs, with $n_{\rm evol}$ and $\mathcal{Q}$ fixed, by $\lesssim3\%$ for the multi-tracer joint analyses and by $\sim3$--$8\%$ for the single-tracer joint analyses.}

The single-tracer joint analysis gives an SNR below $1$ for all three surveys ($\approx0.2$ for BGS and $\approx0.6$ for the other two), and as such we forecast that bright-faint split approaches are far more important than the bispectrum dipole for detecting local relativistic effects.

This result is contrary to much of the existing literature, which forecasts a detectable bispectrum dipole in single-tracer analyses for both \textit{Euclid} and DESI BGS \citep{Maartens_2020,Noorikuhani_2022,Addis_2024,Rossiter_2024}. Part of this difference arises from a sign error in the tree-level bispectrum of \citet{Umeh_2017}: the second-order kernel contracted with the linear fields at $\bm{k}_1$ and $\bm{k}_2$ must be evaluated at $(-\bm{k}_1,-\bm{k}_2)$ (\cref{eq:bk_tree}), i.e.\ complex conjugated (\cref{sec:GR_theory}). This leaves the Newtonian bispectrum unchanged but, rather than simply flipping the sign of the imaginary part, changes the relative sign of the first- and second-order relativistic contributions to both its imaginary and real parts. In our forecasts the correction strengthens cancellations between these terms and so reduces the SNR of the local relativistic contribution; differences in the survey modelling, covariance and nuisance parameters also contribute.

For the multi-tracer analysis, as in the local PNG case, the bispectrum adds little information beyond the power spectrum once the non-Gaussian covariance is included. With the biases fixed, the multi-tracer bispectrum adds $\sim1\%$ to the SNR for BGS, $\sim8\%$ for \textit{Euclid} and $\sim11\%$ for MegaMapper. With the biases free, however, it raises the SNR by $\sim15$--$75\%$ (\cref{tab:gr_summary}), as it helps to break the degeneracy of $\alpha_\mathrm{GR}$ with the relativistic biases.

If $n_{\rm evol}$ and $\mathcal{Q}$ are instead marginalised over in each redshift bin with no prior, the constraints on $\alpha_\mathrm{GR}$ loosen by factors of $\sim15$ and $\sim13$ for the multi-tracer power spectrum and joint analyses of the DESI-like BGS survey, and $\sim2.6$ and $\sim1.6$ for the \textit{Euclid}-like survey. Below we quantify how well the biases must be determined externally in order to detect local relativistic effects.

\subsection{BGS}\label{sec:gr_bgs}

The DESI-like BGS survey gives a larger SNR than the \textit{Euclid}-like and MegaMapper-like samples despite its smaller volume, largely because of its lower redshift coverage. At leading order in $\mathcal{H}/k$, the relativistic signal of a bright-faint split appears mainly in the dipole of the cross-power spectrum (\cref{eq:pk_tree,eq:K1_GR}),
\begin{equation}
\mathrm{Im}\,P^{BF}_1(k)=f\,\frac{\mathcal{H}}{k}\left[\left(b^F_1\gamma^B_1-b^B_1\gamma^F_1\right)+\frac{3}{5}f\left(\gamma^B_1-\gamma^F_1\right)\right]D^2P(k),\label{eq:pbf_dipole}
\end{equation}%
which is set by the difference between the Doppler coefficients of the two samples. Grouping the terms of $\gamma_1$ (\cref{sec:GR_theory}) by their origin,
\begin{equation}
\gamma_1=\underbrace{b_e}_{\rm evolution}-\underbrace{\left(\frac{2}{x\mathcal{H}}+\frac{\mathcal{H}'}{\mathcal{H}^2}\right)}_{\rm volume}+\underbrace{2\mathcal{Q}\left(\frac{1}{x\mathcal{H}}-1\right)}_{\text{Doppler magnification}}.\label{eq:gamma1_be}
\end{equation}
The Doppler shift displaces each galaxy radially, by an amount $\propto v_\parallel/\mathcal{H}$, at a fixed observed redshift. Each term of \cref{eq:gamma1_be} is one consequence of this: the comoving number density is different at the galaxy's true redshift (evolution), the redshift shell has a different volume (volume), and the galaxy's angular diameter distance, $d_A$, and hence its flux at the survey cut, changes (Doppler magnification; \citealt{Bonvin_2008,Bacon_2014}), with $1/(x\mathcal{H})-1=\mathrm{d}\ln d_A/\mathrm{d}\ln(1+z)$.

\begin{figure}
\centering
\includegraphics[width=\linewidth]{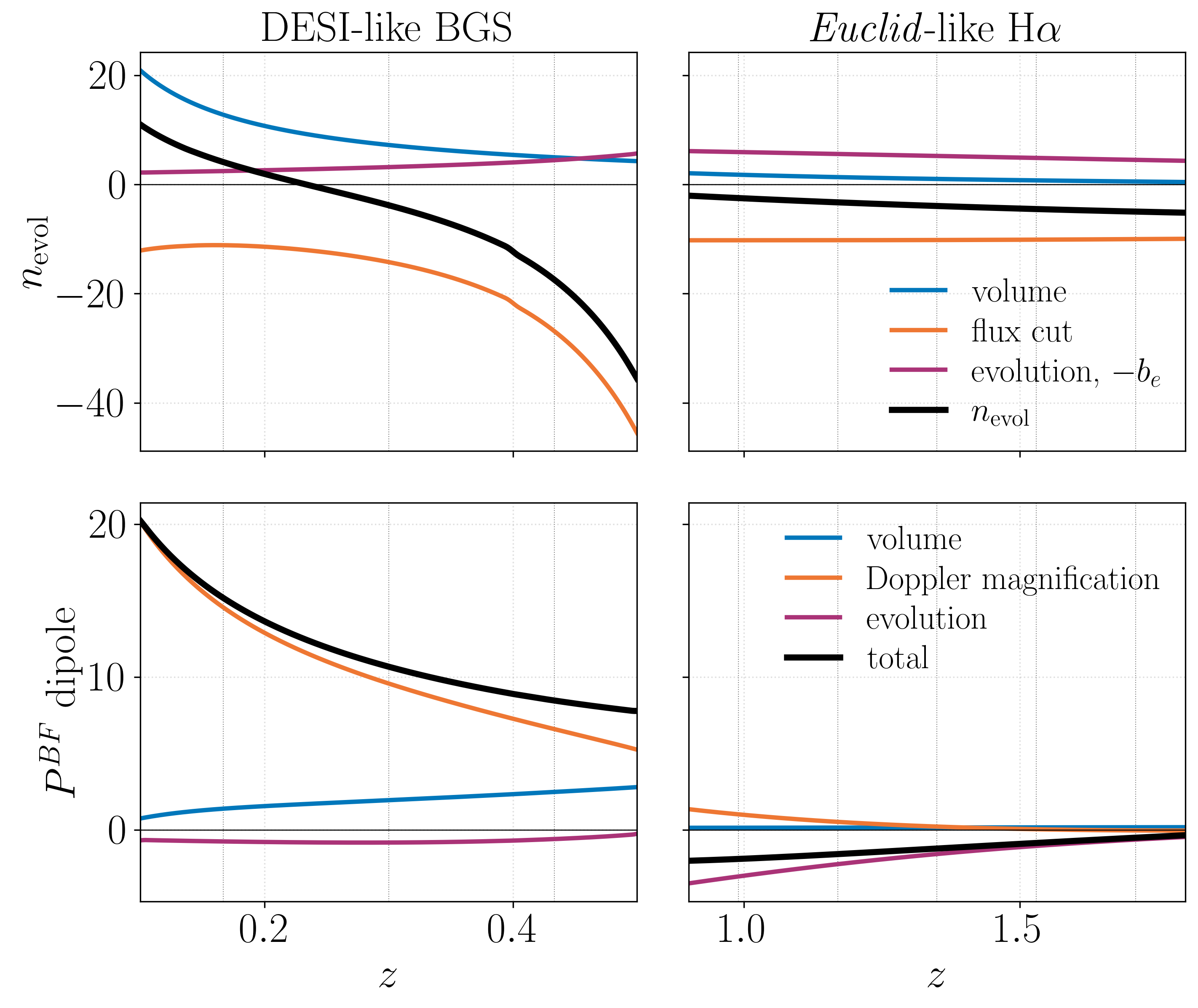}
\caption{Top: contributions to the count slope, $n_{\rm evol}$ (\cref{eq:be_nevol}), of the DESI-like BGS (left) and \textit{Euclid}-like H$\alpha$ (right) samples: the volume, $\mathrm{d}\ln(x^2/H)/\mathrm{d}\ln(1+z)$, the flux cut, $-\mathcal{Q}\,\mathrm{d}\ln L_c/\mathrm{d}\ln(1+z)$, where $L_c$ is the luminosity at the flux cut (including the K-correction for BGS), and the evolution, $-b_e$. The kink in the top-left panel at $z=0.4$ arises because the mean colours of the \citet{Smith_2017} colour model, which set the K-correction, are held constant above this redshift. Bottom: the terms of \cref{eq:gamma1_be} in the dipole of the cross-power spectrum, \cref{eq:pbf_dipole}, divided by $f(\mathcal{H}/k)D^2P$, for the bright-faint splits of \cref{fig:snr_GR}. The vertical dotted lines mark the centres of the redshift bins.}\label{fig:nevol}
\end{figure}

The volume term depends only on redshift, so it is common to the two samples and enters \cref{eq:pbf_dipole} only multiplied by $b^F_1-b^B_1$, whereas Doppler magnification enters through the difference in their magnification biases. Both of its factors, $1/(x\mathcal{H})-1$ and $\mathcal{Q}^B-\mathcal{Q}^F$, are larger for BGS. Since $d_A\propto z$ at low redshift, $1/(x\mathcal{H})-1\approx1/z$, which is $\approx6$ in our lowest BGS redshift bin, whereas it vanishes at the maximum of $d_A$, $z\approx1.6$, within the \textit{Euclid}-like range. The magnification biases differ by $\mathcal{Q}^B-\mathcal{Q}^F\approx0.9$--$1.4$ for BGS, compared with $\approx0.25$--$0.6$ for the \textit{Euclid}-like split. Doppler magnification therefore provides $\approx95\%$ of the BGS dipole in our lowest redshift bin and $\approx80\%$ in the highest, whereas the \textit{Euclid}-like dipole is an order of magnitude smaller and set mainly by the difference in evolution biases (\cref{fig:nevol}). This $1/z$ growth is also why Doppler magnification dominates over lensing only at low redshift \citep{Bonvin_2008,Bacon_2014,Bonvin_2017}, where it could also be detected by correlating BGS counts with galaxy sizes \citep{Ye_2026}.

Removing the $1/(x\mathcal{H})$ terms of $\gamma_1$ and $\gamma_2$ at fixed $b_e$ and $\mathcal{Q}$, together with their counterparts in the second-order kernel, lowers the SNR of the multi-tracer power spectrum on the local relativistic terms alone, without the wide-separation corrections that raise it to $\approx12.5$ (\cref{tab:gr_summary}), from $\approx11$ to $\approx3$. These terms are not explicit in the count-slope form used in our forecasts (\cref{eq:gamma1}), as they are part of $n_{\rm evol}$ (\cref{eq:be_nevol}; \cref{fig:nevol}, top).

\begin{figure}
\centering
\includegraphics[width=\linewidth]{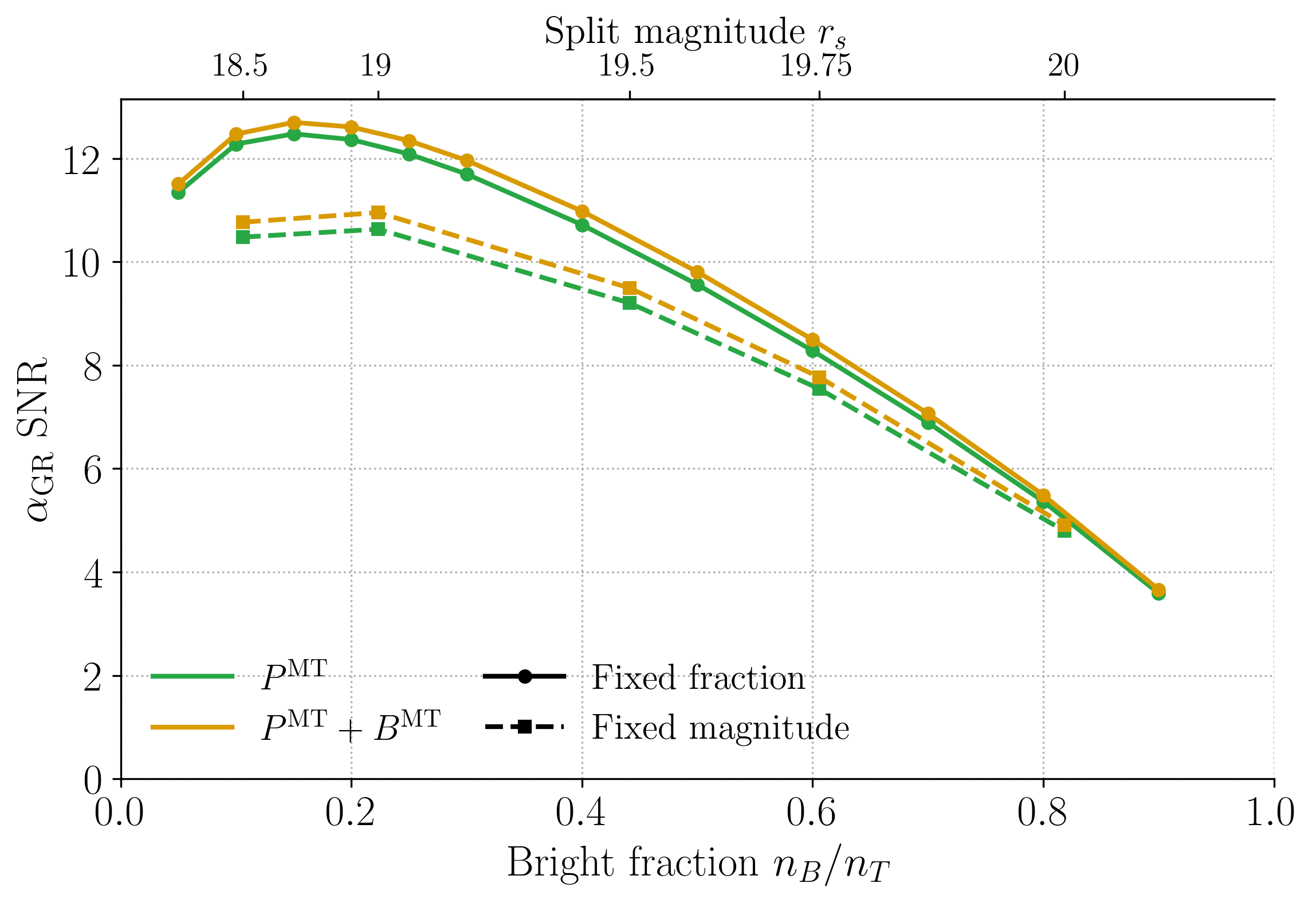}
\caption{SNR on $\alpha_\mathrm{GR}$ as a function of the bright fraction for a bright-faint split of a DESI-like BGS survey, for a multi-tracer power spectrum (green) and a multi-tracer joint power spectrum and bispectrum (gold) analysis, with $n_{\rm evol}$ and $\mathcal{Q}$ held at their fiducial values, as in \cref{fig:ws_bias,fig:snr_GR}. The solid lines split the sample at a fixed bright fraction, $n_B/n_T$, at every redshift, and the dashed lines at a fixed apparent magnitude, $r_s$ (top axis), placed at the overall bright fraction of the survey, $N_B/N_T$, over all redshift bins. Odd multipoles are included.}\label{fig:frac_GR_BGS}
\end{figure}

The dependence of the BGS constraints on the split is shown in \cref{fig:frac_GR_BGS}. For a fixed bright fraction, the SNR on $\alpha_\mathrm{GR}$ depends only weakly on the split for $0.1\lesssim n_B/n_T\lesssim0.3$, peaking at $n_B/n_T\approx0.15$, and falls by a factor of $\sim3.5$ by $n_B/n_T=0.9$. A fixed bright fraction also outperforms every fixed-magnitude split we consider, by at least $\sim15\%$, and we therefore split the BGS sample at $n_B/n_T=0.2$ in \cref{fig:ws_bias,fig:snr_GR}, whose SNR is within $1\%$ of that of the best fraction and about a third higher than that of a split at $r_s=19.5$. 

With the biases free, the SNR at $n_B/n_T=0.2$ falls below $1$ (\cref{tab:gr_summary}), so a detection requires prior knowledge of the relativistic biases.

\begin{figure}
\centering
\includegraphics[width=\linewidth]{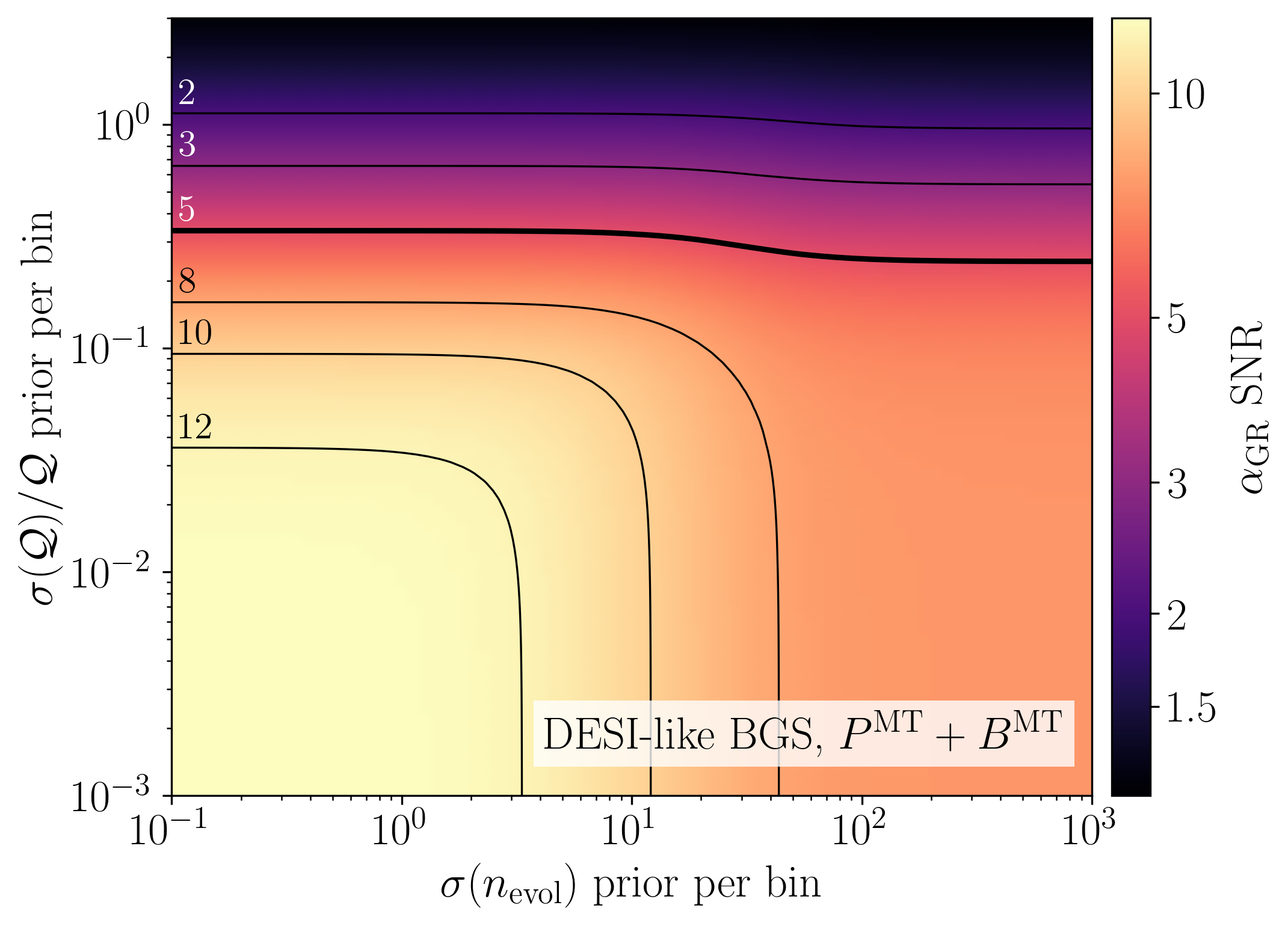}

\caption{Forecast SNR on the amplitude of local relativistic effects, $\alpha_\mathrm{GR}$, as a function of the widths of Gaussian priors on $n_{\rm evol}$ and on the fractional magnification bias, $\sigma(\mathcal{Q})/\mathcal{Q}$, in each redshift bin, for a DESI-like BGS survey split at a fixed bright fraction, $n_B/n_T=0.2$, from a joint multi-tracer power spectrum and bispectrum analysis, as in \cref{fig:snr_GR}. The thick contour marks an SNR of $5$. With $n_{\rm evol}$ and $\mathcal{Q}$ held at their fiducial values the SNR is $12.6$, compared with $0.94$ with both free. The multi-tracer power spectrum alone gives a very similar picture, with SNRs of $12.5$ and $0.82$ with the biases fixed and free, respectively. The two samples share a single $n_{\rm evol}$ in each redshift bin (\cref{sec:bf_split}). The prior on $\mathcal{Q}$ is a fractional error on the magnification biases of the total and bright samples, propagated to the faint sample through \cref{eq:faint_bias}. The prior on $n_{\rm evol}$ is absolute, as $n_{\rm evol}$ crosses zero at the boundary of the two lowest BGS redshift bins, $z\approx0.23$: its fiducial values at the centres of the three BGS bins are $4.1$, $-3.8$ and $-17.4$.}\label{fig:GR_prior_req}
\end{figure}

\Cref{fig:GR_prior_req} shows how well the relativistic biases must be known to recover these constraints, for the BGS sample split at $n_B/n_T=0.2$. As for local-type $\fNL$ with the \textit{Euclid}-like survey (\cref{fig:fnl_prior_req}), the loss is driven mainly by $\mathcal{Q}$, but it is far larger: with $n_{\rm evol}$ fixed but $\mathcal{Q}$ free in each redshift bin, $\sigma(\alpha_\mathrm{GR})$ increases by a factor of $\sim13$ for the joint analysis, whereas with $\mathcal{Q}$ fixed but $n_{\rm evol}$ free it increases by only $\sim70\%$. This reflects the structure of the multi-tracer signal: for a fixed bright fraction the two samples share $n_{\rm evol}$, which therefore enters the leading-order dipole of the cross-power spectrum only multiplied by $b^B_1-b^F_1\approx0.1$--$0.4$, whereas $\mathcal{Q}^B$ and $\mathcal{Q}^F$ enter it directly. Coming within $5\%$ of the fixed case requires $\sigma(\mathcal{Q})/\mathcal{Q}\lesssim4\%$ with $n_{\rm evol}$ fixed, or $\sigma(n_{\rm evol})\lesssim3$ with $\mathcal{Q}$ fixed, in each redshift bin; given that $|n_{\rm evol}|\approx4$ and $\approx17$ in the upper two redshift bins, this is a much weaker requirement on $n_{\rm evol}$ than on $\mathcal{Q}$. The requirement on $\mathcal{Q}$ is comparable to the per-cent-level statistical precision of current finite-difference estimates (\cref{sec:beQ_constraints}), so whether it can be met depends mainly on how well their systematic errors are controlled. These requirements are for recovering the fixed constraint, however, and a detection is far less demanding: an SNR of $5$ requires the magnification bias to be known only to $\approx25\%$ in each redshift bin, even with $n_{\rm evol}$ free.

Finally, the shaded bands for the BGS survey in \cref{fig:snr_GR} show the SNR if we do not impose the wide-separation-based minimum $k$-cut of \cref{sec:data_vector}, i.e.\ if we retain modes with $k < 2\pi/x(z_\mathrm{min})$. On these scales we keep the same signal model, including the wide-separation corrections, although the wide-separation expansion no longer converges there, so that the top of each band differs from the curve below it only in the $k$-bins it includes. These bands therefore give a rough indication of the signal that is inaccessible to the standard 3D Fourier statistics, but which could in principle be recovered in a basis that retains the full wide-separation information, such as the spherical Fourier--Bessel basis discussed in \cref{sec:intro}. The cut is relevant mainly for BGS (and \textit{SPHEREx}), owing to their low-redshift coverage. For BGS it removes the first $k$-bin of the lowest redshift bin, yet this lowers the cumulative SNR by $\sim 34\%$ for both the multi-tracer power spectrum and the joint multi-tracer analysis (and by $\sim 7\%$ for the single-tracer joint analysis), as the relativistic corrections are largest on the largest scales. We note, however, that a more aggressive cut would likely be required in practice, so this should be regarded as a lower limit on the signal lost to the 3D Fourier basis.

\section{Summary and Conclusions}

Building on the formalism of \citetalias{Addis_2024} and \citetalias{Addis_2025} for relativistic and wide-separation effects in the bispectrum and power spectrum, implemented in \textsc{CosmoWAP}, this work has presented joint multi-tracer power spectrum and bispectrum forecasts of local, equilateral and orthogonal PNG and of the local relativistic effects themselves, for Stage IV (DESI-like BGS, \textit{Euclid}-like H$\alpha$ and \textit{SPHEREx}-like) and Stage V (MegaMapper-like LBG) surveys. To this end we have derived the multi-tracer bispectrum multipole covariance, including the odd-parity multipoles, the leading non-Gaussian contributions and the cross-covariance with the power spectrum; the non-Gaussian terms matter mainly for flattened, squeezed triangles and largely cancel in the odd multipoles (\cref{sec:cov}). We have also modelled the second-order biases and the PNG biases of all three shapes with a peak-background split. We parametrise the relativistic biases by the count slope, $n_{\rm evol}$, and the magnification bias, $\mathcal{Q}$, which, unlike the evolution bias, are observable properties of the sample.

Before we reiterate our key results, we stress that our forecasts are limited in scope and note the numerous shortcomings -- many of which are discussed above, and which we briefly list below for clarity:
\begin{itemize}
    \item We restrict our analysis to tree-level perturbation theory, omitting higher-order non-linear physics and Fingers-of-God damping, and instead apply a minimal $k$-cut. While our forecasts generally focus on large-scale effects, this is less true for non-local PNG, which lacks the steep $k$-dependence of other effects. Furthermore, the bispectrum intrinsically correlates small and large scales, which is especially crucial in the squeezed limit.
    \item Window convolution effects will be relevant for an analysis as they dampen signal on large scales and shift signal into different multipoles. Specifically, the odd-parity multipoles are particularly sensitive to leakage from the window-convolved even-parity signal. 
    \item We group all local relativistic effects together and do not attempt to detect individual terms but just use the combined contribution as an indicator for the detectability of these effects.
    \item All of our forecast results depend on the bias modelling choices we have made, in particular the linear bias, luminosity function and HOD.
    \item We do not compute integrated contributions to the bispectrum -- we could expect these to be relevant, as they are in the power spectrum (see \citetalias{Addis_2025}), but we leave their computation to future work as they are non-trivial to include. The size of these contributions is unclear; neglecting them would likely lead to a bias in the inference on large scales, but how their inclusion affects the constraining power on a particular parameter depends on their shape, scale and redshift dependence and how that is degenerate with the parameter in question.
    \item Luminosity function models that give the same number density agree closely on $n_{\rm evol}$ but differ in $\mathcal{Q}$ by several times its statistical uncertainty. Therefore, if one is to take prior information on these terms from a model of the luminosity function then this must be done carefully, with particular caution taken for the magnification bias.
    \item We ignore the cosmology dependence of our HODs \citep[e.g.][]{Smith_2024} and luminosity functions.
    \item As detailed in \cref{sec:forecast_details}, we omit uncertainties for the PNG bias parameters, $b_{\Psi}$ and $b_{\Psi \delta}$. We do not attempt to break their degeneracy with $\fNL$; instead, we hold them fixed to the universal mass function estimates. Much of the information, including for non-local shapes, comes from these biases, so the constraints should be regarded as optimistic.
    \item Our non-Gaussian covariance is approximate (\cref{sec:cov_ng}): the $PT$ term uses a separable squeezed-limit form for all configurations, and the non-Gaussian terms are evaluated in the plane-parallel constant-redshift limit, without wide-separation contributions. We neglect integrated contributions to all of the covariance terms, since without a radial window function the lensing contribution to the power spectrum is sharply peaked at $\mu=0$, and the covariance strongly overweights this peak. We also neglect the connected 6-point function, non-Gaussian contributions to the power spectrum covariance, $\textsf{C}_{PP}$, and super-sample covariance.
\end{itemize}
Thus, the ratio of the constraints with and without a proper treatment of these light-cone effects is often more relevant to the reader than their absolute size. With that in mind, our main results are as follows.

\subsection{Biases from neglecting light-cone effects}

We forecast that neglecting light-cone effects would bias the inference on local-type $\fNL$ for a joint power spectrum and bispectrum analysis for both a \textit{Euclid}-like H$\alpha$ and a MegaMapper-like LBG survey, at $\sim2\sigma$ and $\sim25\sigma$ respectively, which could lead to a false detection of local-type PNG (\cref{fig:fnl_bias}). This shift is driven by lensing, which pushes $\fNL^{\rm loc}$ to positive values; neglecting only the local relativistic and wide-separation terms shifts it by $\lesssim0.4\sigma$. This result is perhaps straightforward for those used to angular analyses, but is less intuitively understood by the community when considering 3D Fourier statistics. The precise size of the shift depends on the survey window and covariance modelling: a radial window function roughly halves the MegaMapper-like shift, but it remains large and positive in every set-up we have tested. By contrast, the shift is negligible for our \textit{SPHEREx}-like sample, whose lower redshift coverage and magnification bias suppress the lensing contribution. Since integrated contributions to the bispectrum are not included, the lensing contribution enters these forecasts through the power spectrum alone. For the MegaMapper-like survey, neglecting light-cone effects also biases the orthogonal shape, by $\sim3\sigma$, and the cosmological parameters despite the \textit{Planck} priors, by up to $\sim10\sigma$ for $A_s$, while for the \textit{Euclid}-like survey the shifts in the non-local shapes remain within the forecast errors. Lensing must therefore be modelled for precision PNG analyses, while the local relativistic terms also matter for the uncertainty on $\fNL$, through the relativistic biases they introduce (\cref{sec:conc_png}).

Neglecting wide-separation corrections also biases the inferred amplitude of local relativistic effects, $\alpha_\mathrm{GR}$. For the bright-faint analyses with $n_{\rm evol}$ and $\mathcal{Q}$ fixed, the shift is $\approx9$--$10\sigma$ for the DESI-like BGS survey, mainly through the wide-angle terms, $\approx3\sigma$ for the \textit{Euclid}-like survey and less than $1\sigma$ for the MegaMapper-like survey (\cref{fig:ws_bias}). The wide-separation corrections also carry relativistic signal of their own, and including them tightens the constraint on $\alpha_\mathrm{GR}$ by $\sim15\%$ for BGS and the MegaMapper-like survey.

\subsection{Local-type PNG and the relativistic biases}\label{sec:conc_png}

Modelling light-cone effects introduces $n_{\rm evol}$ and $\mathcal{Q}$ as nuisance parameters in each redshift bin. Uncertainty in these biases must therefore be propagated into the PNG constraints, which are otherwise overly optimistic. In the even multipoles both biases enter at $\mathcal{O}(\mathcal{H}/k)^2$, with the same $k^{-2}$ scaling as local-type PNG, so for the \textit{Euclid}-like survey marginalising over them with no prior increases $\sigma(\fNL^{\rm loc})$ by $\sim60\%$ for the power spectrum and by $\sim50\%$ for a joint power spectrum and bispectrum analysis of the even multipoles (\cref{fig:fnl_con}).

The odd-parity multipoles, which carry the leading-order $\mathcal{H}/k$ relativistic terms, recover most of this loss: including the bispectrum dipole reduces it to $\sim15\%$ by breaking the degeneracy between $n_{\rm evol}$ and $\fNL^{\rm loc}$, without requiring a precise measurement of $n_{\rm evol}$ itself. For the power spectrum alone, a bright-faint split at a fixed bright fraction, whose two samples share one $n_{\rm evol}$, recovers part of the loss, whereas a fixed-flux split, with a count slope for each sample, barely helps; neither adds much once the dipole is included.

The remaining loss is driven by $\mathcal{Q}$, which is therefore the limiting nuisance parameter (\cref{fig:fnl_prior_req}). For the \textit{Euclid}-like survey, coming within $1\%$ of the `fixed' constraint requires $\sigma(\mathcal{Q})/\mathcal{Q}\lesssim4\%$ in each redshift bin, comparable to the statistical precision of current catalogue estimates. Priors forward-modelled from the luminosity function come within $\sim5\%$, but depend on its functional form, as models with similar number densities disagree on $\mathcal{Q}$. For the MegaMapper-like survey, the stronger lensing signal constrains $\mathcal{Q}$ from the clustering itself, so the joint constraint degrades by only $\sim12\%$, and an external prior on $\mathcal{Q}$ helps appreciably only once it reaches the per-cent level.

\subsection{What the bispectrum and the bright-faint split add for PNG}

Both the bispectrum and the bright-faint split bring in further nuisance parameters ($b_2$ and $b_{\Gamma_2}$, and a second set of sample biases), so their additional information must outweigh the cost of marginalising over them. With the relativistic biases fixed, neither gains much on local-type $\fNL$. The bispectrum improves the constraint by only a few per cent, as the squeezed triangles that carry the signal are strongly correlated with each other and with the power spectrum (a Gaussian covariance would give $\sim12\%$). The split gains $\lesssim6\%$, as the bright and faint parts of a single population have similar $b_\Psi/b_1$, which tends to $2\delta_c$ for highly biased tracers; larger gains would need distinct tracer populations \citep{Seljak_2009}. For local-type PNG, the main benefit of the bispectrum is instead its dipole, which breaks the degeneracy with the relativistic biases (\cref{sec:conc_png}). The bispectrum may also help to break the degeneracy between $b_\Psi$ and $\fNL$ (\cref{sec:bias_priors}).

The bispectrum is essential for competitive constraints on equilateral and orthogonal PNG: the \textit{Euclid}-like survey gives $\sigma(\fNL^{\rm eq})\approx34$ and $\sigma(\fNL^{\rm orth})\approx40$, improving on \textit{Planck} by a factor of $\sim1.4$ for the equilateral shape but not for the orthogonal one. These shapes are much less degenerate with the relativistic terms: marginalising over $n_{\rm evol}$ and $\mathcal{Q}$ loosens their joint power spectrum and bispectrum constraints by $\lesssim2\%$ for both the \textit{Euclid}-like and MegaMapper-like surveys. For the MegaMapper-like survey, the split improves the joint power spectrum and bispectrum constraints on both shapes by a factor of $\sim2$, to $\sigma(\fNL^{\rm eq})\approx2.9$ and $\sigma(\fNL^{\rm orth})\approx4.9$ (\cref{tab:fnl_summary}). The non-local constraints derive mainly from configurations near $k_{\rm max}$ and are therefore particularly sensitive to our tree-level modelling.

\subsection{Detecting local relativistic effects}

For the surveys and scales considered, the single-tracer joint analyses give an SNR on $\alpha_\mathrm{GR}$ below $1$, even with $n_{\rm evol}$ and $\mathcal{Q}$ fixed. With these biases fixed, a bright-faint split gives an SNR of $\approx13$ for the DESI-like BGS survey, driven mainly by the cross-power spectrum, compared with $\approx1.5$--$2.2$ for the \textit{Euclid}-like and MegaMapper-like surveys (\cref{tab:gr_summary}). BGS does better despite its smaller volume mainly because of Doppler magnification, which grows as $1/z$ at low redshift, and the larger difference between the magnification biases of its bright and faint samples (\cref{sec:gr_bgs}). The adopted BGS split at a fixed bright fraction, $n_B/n_T=0.2$, is near-optimal with the biases fixed and outperforms every fixed-magnitude split we consider (\cref{fig:frac_GR_BGS}).

Earlier forecasts, including \citetalias{Addis_2024}, predicted a detectable single-tracer bispectrum dipole. The discrepancy can be linked to a sign error in the construction of the tree-level bispectrum in \citet{Umeh_2017}: the second-order kernel must be evaluated at $-\bm{k}_1$ and $-\bm{k}_2$, though the kernel itself is unchanged. Correcting this strengthens cancellations between terms and reduces the dipole; the full non-Gaussian covariance and marginalising over nuisance parameters also contribute (\cref{sec:results2GR}).

Marginalising over $n_{\rm evol}$ and $\mathcal{Q}$ with uninformative priors lowers the BGS joint SNR to $\approx0.9$, so a detection requires independent information on these biases (\cref{fig:GR_prior_req}). This requirement falls mainly on $\mathcal{Q}$. For a fixed-fraction cut the two samples share $n_{\rm evol}$, which therefore enters the leading dipole only through the small difference in their linear biases: with $\mathcal{Q}$ fixed, freeing $n_{\rm evol}$ lowers the SNR only to $\approx7$, whereas with $\mathcal{Q}$ free, fixing $n_{\rm evol}$ makes almost no difference. An SNR of $5$ requires $\sigma(\mathcal{Q})/\mathcal{Q}\approx25\%$ in each redshift bin (even with $n_{\rm evol}$ free), while recovering the fixed-bias SNR to within $5\%$ requires $\sigma(\mathcal{Q})/\mathcal{Q}\lesssim4\%$ with $n_{\rm evol}$ fixed. The multi-tracer bispectrum helps to break the degeneracies between $\alpha_\mathrm{GR}$ and the relativistic biases, raising the SNR without priors by $\sim15$--$75\%$, compared with $\sim1$--$11\%$ with the biases fixed (\cref{tab:gr_summary}), but not enough for a detection.

Finally, for the largest scales in the lower redshift bins of the BGS survey one would expect the perturbative wide-separation expansion to break down. Removing these scales lowers the multi-tracer SNR by about a third even with a minimal $k$-cut, as the relativistic corrections are largest on the largest scales. Therefore to capture the full information one should use alternative bases to the standard 3D Fourier summary statistics.

\subsection{Outlook}

Of the limitations listed above, the window function is the most pressing, both for the odd multipoles and because it must be treated before the integrated terms can be included in the covariance; we also leave the integrated contributions to the bispectrum to future work. Additionally, an alternative basis that retains the full wide-separation information, such as the spherical Fourier--Bessel basis, could in principle recover the full large-scale information where our perturbative wide-separation expansion breaks down.

Our results identify that determining the magnification bias of the sample is a priority for both local-type PNG inference and the detection of local relativistic effects. As properties of the observed sample, $n_{\rm evol}$ and $\mathcal{Q}$ can be measured from the catalogue, $n_{\rm evol}$ from the slope of the redshift distribution and $\mathcal{Q}$ by perturbing the flux cut and re-applying the sample selection, while the odd multipoles provide a complementary consistency test. For BGS, moderately informative constraints on $\mathcal{Q}$ can enable a significant detection; approaching the full forecast sensitivity requires measurements at the few-per-cent level, together with sufficient knowledge of $n_{\rm evol}$. Establishing these measurements with controlled selection and completeness errors is therefore an important next step.

\section*{Acknowledgements}
CA is supported by a studentship from the UK Science and Technology Facilities Council (STFC). CC is supported by STFC (UK) grant ST/X000931/1. DK is supported by the Dutch Research Council (NWO) (grant no. \href{https://www.nwo.nl/projecten/iolfn50333}{OCENW.M.22.307}). LLMs were used to improve the formatting and readability of the manuscript as well as debugging, optimisation and testing in \textsc{CosmoWAP}. The authors accept full responsibility for the accuracy and interpretation of the results, and the information presented in the text.

\section*{Code Availability}

\textsc{CosmoWAP} is publicly available at \url{https://github.com/craddis1/CosmoWAP} and the companion repository \textsc{MathWAP} is available at \url{https://github.com/craddis1/MathWAP}. The results in this paper were produced with \textsc{CosmoWAP} v0.10.0 (\url{https://github.com/craddis1/CosmoWAP/tree/v0.10.0}), which is archived on Zenodo (\href{https://doi.org/10.5281/zenodo.24373137}{doi:10.5281/zenodo.24373137}) and is also available on PyPI. A full description of \textsc{CosmoWAP} is given in \citet{cosmowap_paper}. \textsc{CosmoWAP} utilises the public code \href{https://github.com/lesgourg/class_public}{\textsc{class}} \citep{Lesgourgues2011,blas2011}, and the following \textit{Python} packages and libraries: \href{https://numpy.org/}{\textsc{numpy}} \citep{harris2020}, \href{https://scipy.org/}{\textsc{scipy}} \citep{2020SciPy}, \href{https://matplotlib.org/}{\textsc{matplotlib}} \citep{hunter2007}, \href{https://numba.pydata.org/}{\textsc{NUMBA}} \citep{NUMBA_2015}, \href{https://samreay.github.io/ChainConsumer/ChainConsumer}{\textsc{ChainConsumer}} \citep{Hinton_2016}, \href{https://cobaya.readthedocs.io/en/latest/}{\textsc{cobaya}} \citep{Torrado_2021} and \href{https://alessiospuriomancini.github.io/cosmopower/}{\textsc{CosmoPower}} \citep{Mancini_2022}.

\section*{Data Availability}
The results in this article are all derived using the public code \textsc{CosmoWAP}.

%%%%%%%%%%%%%%%%%%%% REFERENCES %%%%%%%%%%%%%%%%%%

\bibliographystyle{mnras}
\bibliography{main}

@INPROCEEDINGS{Wirth_2026,
       author = {{Wirth}, Gregory D. and {Wang}, Yun and {Robberto}, Massimo and {Dickinson}, Mark and {Greene}, Thomas and {Kruk}, Jeffrey and {Percival}, Will and {Schlieder}, Joshua E. and {Faisst}, Andreas and {Hillenbrand}, Lynne and et al.},
        title = "{The SIRMOS mission concept: a space-based near-IR spectroscopic survey of 100 million galaxies in the Euclid Wide Survey footprint}",
    booktitle = {Space Telescopes and Instrumentation 2026: Optical, Infrared, and Millimeter Wave},
       editor = {{Coyle}, Laura E. and {Perrin}, Marshall D. and {Matsuura}, Shuji},
         year = 2026,
        month = aug,
       series = {Society of Photo-Optical Instrumentation Engineers (SPIE) Conference Series},
       volume = {14145},
          eid = {141451D},
        pages = {141451D},
          doi = {10.1117/12.3104870},
archivePrefix = {arXiv},
       eprint = {2610.35945},
 primaryClass = {astro-ph.IM},
       adsurl = {https://ui.adsabs.harvard.edu/abs/2026arXiv261035945W}
}

@ARTICLE{Nascimento_2026,
       author = {{Nascimento}, Caio B. de S. and {Dalal}, Neal},
        title = "{Breaking the $f_{\text{NL}}$--$b_{\Phi}$ degeneracy with the time evolution of tracer number counts}",
      journal = {arXiv e-prints},
         year = 2026,
        month = jul,
          eid = {arXiv:2607.18229},
        pages = {arXiv:2607.18229},
          doi = {10.48550/arXiv.2607.18229},
archivePrefix = {arXiv},
       eprint = {2607.18229},
 primaryClass = {astro-ph.CO},
       adsurl = {https://ui.adsabs.harvard.edu/abs/2026arXiv260718229N}
}

@INPROCEEDINGS{NUMBA_2015,
       author = {{Lam}, Siu Kwan and {Pitrou}, Antoine and {Seibert}, Stanley},
        title = "{Numba: A LLVM-based Python JIT Compiler}",
    booktitle = {Proc. Second Workshop on the LLVM Compiler Infrastructure in HPC},
         year = 2015,
        month = nov,
        pages = {1-6},
          doi = {10.1145/2833157.2833162},
       adsurl = {https://ui.adsabs.harvard.edu/abs/2015llvm.confE...1L}
}

@ARTICLE{Wang_2020a,
       author = {{Wang}, M.~S. and {Avila}, S. and {Bianchi}, D. and {Crittenden}, R. and {Percival}, W.~J.},
        title = "{Hybrid-basis inference for large-scale galaxy clustering: combining spherical and Cartesian Fourier analyses}",
      journal = {\jcap},
         year = 2020,
        month = oct,
       volume = {2020},
       number = {10},
          eid = {022},
        pages = {022},
          doi = {10.1088/1475-7516/2020/10/022},
archivePrefix = {arXiv},
       eprint = {2007.14962},
 primaryClass = {astro-ph.CO},
       adsurl = {https://ui.adsabs.harvard.edu/abs/2020JCAP...10..022W}
}

@ARTICLE{Wen_2026,
       author = {{Wen}, Robin Y. and {Grasshorn Gebhardt}, Henry S. and {Dor{\'e}}, Olivier},
        title = "{Impact and measurability of linear relativistic effects in galaxy surveys}",
      journal = {arXiv e-prints},
         year = 2026,
        month = jul,
          eid = {arXiv:2607.06697},
        pages = {arXiv:2607.06697},
          doi = {10.48550/arXiv.2607.06697},
archivePrefix = {arXiv},
       eprint = {2607.06697},
 primaryClass = {astro-ph.CO},
       adsurl = {https://ui.adsabs.harvard.edu/abs/2026arXiv260706697W}
}

@ARTICLE{Wen_2025,
       author = {{Wen}, Robin Y. and {Gebhardt}, Henry S. Grasshorn and {Heinrich}, Chen and {Dor{\'e}}, Olivier},
        title = "{Separating angular and radial modes with spherical-Fourier Bessel power spectrum on all scales and implications for systematics mitigation}",
      journal = {\prd},
         year = 2025,
        month = sep,
       volume = {112},
       number = {6},
          eid = {063518},
        pages = {063518},
          doi = {10.1103/hqc3-tbxc},
archivePrefix = {arXiv},
       eprint = {2506.06902},
 primaryClass = {astro-ph.CO},
       adsurl = {https://ui.adsabs.harvard.edu/abs/2025PhRvD.112f3518W}
}

@ARTICLE{Yoo_2013,
       author = {{Yoo}, Jaiyul and {Desjacques}, Vincent},
        title = "{All-sky analysis of the general relativistic galaxy power spectrum}",
      journal = {\prd},
         year = 2013,
        month = jul,
       volume = {88},
       number = {2},
          eid = {023502},
        pages = {023502},
          doi = {10.1103/PhysRevD.88.023502},
archivePrefix = {arXiv},
       eprint = {1301.4501},
 primaryClass = {astro-ph.CO},
       adsurl = {https://ui.adsabs.harvard.edu/abs/2013PhRvD..88b3502Y}
}

@ARTICLE{Heavens_1995,
       author = {{Heavens}, A.~F. and {Taylor}, A.~N.},
        title = "{A spherical harmonic analysis of redshift space}",
      journal = {\mnras},
         year = 1995,
        month = jul,
       volume = {275},
       number = {2},
        pages = {483-497},
          doi = {10.1093/mnras/275.2.483},
archivePrefix = {arXiv},
       eprint = {astro-ph/9409027},
 primaryClass = {astro-ph},
       adsurl = {https://ui.adsabs.harvard.edu/abs/1995MNRAS.275..483H}
}

@ARTICLE{Fisher_1995a,
       author = {{Fisher}, K.~B. and {Lahav}, O. and {Hoffman}, Y. and {Lynden-Bell}, D. and {Zaroubi}, S.},
        title = "{Wiener reconstruction of density, velocity and potential fields from all-sky galaxy redshift surveys}",
      journal = {\mnras},
         year = 1995,
        month = feb,
       volume = {272},
       number = {4},
        pages = {885-908},
          doi = {10.1093/mnras/272.4.885},
archivePrefix = {arXiv},
       eprint = {astro-ph/9406009},
 primaryClass = {astro-ph},
       adsurl = {https://ui.adsabs.harvard.edu/abs/1995MNRAS.272..885F}
}

@ARTICLE{Dai_2012,
       author = {{Dai}, Liang and {Kamionkowski}, Marc and {Jeong}, Donghui},
        title = "{Total angular momentum waves for scalar, vector, and tensor fields}",
      journal = {\prd},
         year = 2012,
        month = dec,
       volume = {86},
       number = {12},
          eid = {125013},
        pages = {125013},
          doi = {10.1103/PhysRevD.86.125013},
archivePrefix = {arXiv},
       eprint = {1209.0761},
 primaryClass = {astro-ph.CO},
       adsurl = {https://ui.adsabs.harvard.edu/abs/2012PhRvD..86l5013D}
}

@ARTICLE{Lukens_2026,
       author = {{Lukens}, Gregory and {Jeong}, Donghui},
        title = "{PowerFull: fast and accurate computation of the relativistic angular galaxy power spectrum including local primordial non-Gaussianity}",
      journal = {\jcap},
         year = 2026,
        month = nov,
       volume = {2026},
       number = {11},
          eid = {051},
        pages = {051},
          doi = {10.1088/1475-7516/2026/11/051},
archivePrefix = {arXiv},
       eprint = {2608.18334},
 primaryClass = {astro-ph.CO},
       adsurl = {https://ui.adsabs.harvard.edu/abs/2026arXiv260818334L}
}

@ARTICLE{Lake_2018,
       author = {{Lake}, S.~E. and {Wright}, E.~L. and {Assef}, R.~J. and {Jarrett}, T.~H. and {Petty}, S. and {Stanford}, S.~A. and {Stern}, D. and {Tsai}, C.-W.},
        title = "{The 2.4 {\ensuremath{\mu}}m Galaxy Luminosity Function as Measured Using WISE. III. Measurement Results}",
      journal = {\apj},
         year = 2018,
        month = oct,
       volume = {866},
       number = {1},
          eid = {45},
        pages = {45},
          doi = {10.3847/1538-4357/aadd47},
archivePrefix = {arXiv},
       eprint = {1702.07829},
 primaryClass = {astro-ph.GA},
       adsurl = {https://ui.adsabs.harvard.edu/abs/2018ApJ...866...45L}
}

@ARTICLE{Camera_2026,
       author = {{Camera}, S. and {Addis}, C. and {Guedezounme}, S.~L. and {Montano}, F. and {Montandon}, T. and {Novara}, M. and {Rossiter}, S.~J. and {Asorey}, J. and {Castello}, S. and {Clarkson}, C. and {Fonseca}, J. and {Frusciante}, N. and {Gomes}, C. and {Grimm}, N. and {Nasirudin}, A. and {Oliveira}, L.~G.~T. and {Pace}, F. and {Sakr}, Z. and {Berti}, M. and {Bonvin}, C. and {Bull}, P. and {Di Dio}, E. and {Cress}, C. and {Heneka}, C. and {Kloeckner}, H.-R. and {Maartens}, R. and {Martins}, C.~J.~A.~P. and {Parkinson}, D. and {Saraf}, C.~S.},
        title = "{Beyond {\ensuremath{\Lambda}}CDM with the SKA Observatory -- I: Probing Gravity on Cosmological Scales}",
      journal = {arXiv e-prints},
         year = 2026,
        month = jul,
          eid = {arXiv:2607.11917},
        pages = {arXiv:2607.11917},
archivePrefix = {arXiv},
       eprint = {2607.11917},
 primaryClass = {gr-qc},
       adsurl = {https://ui.adsabs.harvard.edu/abs/2026arXiv260711917C}
}

@ARTICLE{Guandalin_2023,
       author = {{Guandalin}, Caroline and {Piat}, Jade and {Clarkson}, Chris and {Maartens}, Roy},
        title = "{Theoretical Systematics in Testing the Cosmological Principle with the Kinematic Quasar Dipole}",
      journal = {\apj},
         year = 2023,
        month = aug,
       volume = {953},
       number = {2},
          eid = {144},
        pages = {144},
          doi = {10.3847/1538-4357/acdf46},
archivePrefix = {arXiv},
       eprint = {2212.04925},
 primaryClass = {astro-ph.CO},
       adsurl = {https://ui.adsabs.harvard.edu/abs/2023ApJ...953..144G}
}

@ARTICLE{Neal_2005,
       author = {{Neal}, Radford M.},
        title = "{Taking Bigger Metropolis Steps by Dragging Fast Variables}",
      journal = {arXiv Mathematics e-prints},
         year = 2005,
        month = feb,
          eid = {math/0502099},
        pages = {math/0502099},
          doi = {10.48550/arXiv.math/0502099},
archivePrefix = {arXiv},
       eprint = {math/0502099},
 primaryClass = {math.ST},
       adsurl = {https://ui.adsabs.harvard.edu/abs/2005math......2099N}
}

@ARTICLE{Dube_2026,
       author = {{Dube}, Yolanda and {Maartens}, Roy and {Dinda}, Bikash R. and {Guedezounme}, S{\^e}cloka L. and {Jolicoeur}, Sheean},
        title = "{Impact of lensing magnification on the power spectrum turnover}",
      journal = {arXiv e-prints},
         year = 2026,
        month = jun,
          eid = {arXiv:2606.18908},
        pages = {arXiv:2606.18908},
archivePrefix = {arXiv},
       eprint = {2606.18908},
 primaryClass = {astro-ph.CO},
       adsurl = {https://ui.adsabs.harvard.edu/abs/2026arXiv260618908D}
}

@ARTICLE{Cooray_2002,
       author = {{Cooray}, Asantha and {Sheth}, Ravi},
        title = "{Halo models of large scale structure}",
      journal = {\physrep},
         year = 2002,
        month = dec,
       volume = {372},
       number = {1},
        pages = {1-129},
          doi = {10.1016/S0370-1573(02)00276-4},
archivePrefix = {arXiv},
       eprint = {astro-ph/0206508},
 primaryClass = {astro-ph},
       adsurl = {https://ui.adsabs.harvard.edu/abs/2002PhR...372....1C}
}

@ARTICLE{Schmidt_2010,
       author = {{Schmidt}, Fabian and {Kamionkowski}, Marc},
        title = "{Halo clustering with nonlocal non-Gaussianity}",
      journal = {\prd},
         year = 2010,
        month = nov,
       volume = {82},
       number = {10},
          eid = {103002},
        pages = {103002},
          doi = {10.1103/PhysRevD.82.103002},
archivePrefix = {arXiv},
       eprint = {1008.0638},
 primaryClass = {astro-ph.CO},
       adsurl = {https://ui.adsabs.harvard.edu/abs/2010PhRvD..82j3002S}
}

@ARTICLE{Karagiannis_2018,
       author = {{Karagiannis}, Dionysios and {Lazanu}, Andrei and {Liguori}, Michele and {Raccanelli}, Alvise and {Bartolo}, Nicola and {Verde}, Licia},
        title = "{Constraining primordial non-Gaussianity with bispectrum and power spectrum from upcoming optical and radio surveys}",
      journal = {\mnras},
         year = 2018,
        month = jul,
       volume = {478},
       number = {1},
        pages = {1341-1376},
          doi = {10.1093/mnras/sty1029},
archivePrefix = {arXiv},
       eprint = {1801.09280},
 primaryClass = {astro-ph.CO},
       adsurl = {https://ui.adsabs.harvard.edu/abs/2018MNRAS.478.1341K}
}

@ARTICLE{Mo_1997,
       author = {{Mo}, H.~J. and {Jing}, Y.~P. and {White}, S.~D.~M.},
        title = "{High-order correlations of peaks and haloes: a step towards understanding galaxy biasing}",
      journal = {\mnras},
         year = 1997,
        month = jan,
       volume = {284},
       number = {1},
        pages = {189-201},
          doi = {10.1093/mnras/284.1.189},
archivePrefix = {arXiv},
       eprint = {astro-ph/9603039},
 primaryClass = {astro-ph},
       adsurl = {https://ui.adsabs.harvard.edu/abs/1997MNRAS.284..189M}
}

@ARTICLE{Schmidt_2013,
       author = {{Schmidt}, Fabian and {Jeong}, Donghui and {Desjacques}, Vincent},
        title = "{Peak-background split, renormalization, and galaxy clustering}",
      journal = {\prd},
         year = 2013,
        month = jul,
       volume = {88},
       number = {2},
          eid = {023515},
        pages = {023515},
          doi = {10.1103/PhysRevD.88.023515},
archivePrefix = {arXiv},
       eprint = {1212.0868},
 primaryClass = {astro-ph.CO},
       adsurl = {https://ui.adsabs.harvard.edu/abs/2013PhRvD..88b3515S}
}

@ARTICLE{Barreira_2022a,
       author = {{Barreira}, Alexandre},
        title = "{Can we actually constrain f$_{NL}$ using the scale-dependent bias effect? An illustration of the impact of galaxy bias uncertainties using the BOSS DR12 galaxy power spectrum}",
      journal = {\jcap},
         year = 2022,
        month = nov,
       volume = {2022},
       number = {11},
          eid = {013},
        pages = {013},
          doi = {10.1088/1475-7516/2022/11/013},
archivePrefix = {arXiv},
       eprint = {2205.05673},
 primaryClass = {astro-ph.CO},
       adsurl = {https://ui.adsabs.harvard.edu/abs/2022JCAP...11..013B}
}

@ARTICLE{Linde_euclid_2026,
       author = {{Euclid Collaboration} and {Linde}, D. and {Moradinezhad Dizgah}, A. and {Parimbelli}, G. and {Pardede}, K. and {Sefusatti}, E. and {Cagliari}, M.~S. and {D'Amico}, G. and {Desjacques}, V. and {Eggemeier}, A. and {Biagetti}, M. and {Veropalumbo}, A. and {Camacho Quevedo}, B. and {Chudaykin}, A. and {Crocce}, M. and {Castiblanco}, L. and {Castorina}, E. and {Farina}, A. and {Guidi}, M. and {Karcher}, M. and {Pezzotta}, A. and {Pugno}, A. and {Altieri}, B. and {Andreon}, S. and {Auricchio}, N. and {Baccigalupi}, C. and {Baldi}, M. and {Bardelli}, S. and {Battaglia}, P. and {Biviano}, A. and {Branchini}, E. and {Brescia}, M. and {Camera}, S. and {Canas-Herrera}, G. and {Capobianco}, V. and {Carbone}, C. and {Carretero}, J. and {Casas}, S. and {Castellano}, M. and {Castignani}, G. and {Cavuoti}, S. and {Chambers}, K.~C. and {Cimatti}, A. and {Colodro-Conde}, C. and {Congedo}, G. and {Conversi}, L. and {Copin}, Y. and {Courbin}, F. and {Courtois}, H.~M. and {Degaudenzi}, H. and {de la Torre}, S. and {De Lucia}, G. and {Dole}, H. and {Douspis}, M. and {Dubath}, F. and {Dupac}, X. and {Escoffier}, S. and {Farina}, M. and {Farinelli}, R. and {Ferriol}, S. and {Finelli}, F. and {Fosalba}, P. and {Fotopoulou}, S. and {Frailis}, M. and {Fumana}, M. and {Galeotta}, S. and {George}, K. and {Gillis}, B. and {Giocoli}, C. and {Gracia-Carpio}, J. and {Grazian}, A. and {Grupp}, F. and {Haugan}, S.~V.~H. and {Holmes}, W. and {Hormuth}, F. and {Hornstrup}, A. and {Jahnke}, K. and {Joachimi}, B. and {Kermiche}, S. and {Kiessling}, A. and {Kubik}, B. and {Kunz}, M. and {Kurki-Suonio}, H. and {Le Brun}, A.~M.~C. and {Ligori}, S. and {Lilje}, P.~B. and {Lindholm}, V. and {Lloro}, I. and {Mainetti}, G. and {Mansutti}, O. and {Marggraf}, O. and {Martinelli}, M. and {Martinet}, N. and {Marulli}, F. and {Massey}, R.~J. and {Medinaceli}, E. and {Mei}, S. and {Meneghetti}, M. and {Merlin}, E. and {Meylan}, G. and {Mora}, A. and {Moresco}, M. and {Moscardini}, L. and {Neissner}, C. and {Niemi}, S.-M. and {Nightingale}, J.~W. and {Padilla}, C. and {Paltani}, S. and {Pasian}, F. and {Pedersen}, K. and {Percival}, W.~J. and {Pettorino}, V. and {Pires}, S. and {Polenta}, G. and {Poncet}, M. and {Popa}, L.~A. and {Raison}, F. and {Renzi}, A. and {Rhodes}, J. and {Riccio}, G. and {Romelli}, E. and {Roncarelli}, M. and {Saglia}, R. and {Sakr}, Z. and {Sanchez}, A.~G. and {Sapone}, D. and {Sartoris}, B. and {Secroun}, A. and {Seidel}, G. and {Sihvola}, E. and {Simon}, P. and {Sirignano}, C. and {Sirri}, G. and {Spurio Mancini}, A. and {Stanco}, L. and {Tallada-Crespi}, P. and {Taylor}, A.~N. and {Tereno}, I. and {Tessore}, N. and {Toft}, S. and {Toledo-Moreo}, R. and {Torradeflot}, F. and {Tutusaus}, I. and {Valenziano}, L. and {Valiviita}, J. and {Vassallo}, T. and {Verdoes Kleijn}, G. and {Wang}, Y. and {Weller}, J. and {Zamorani}, G. and {Zerbi}, F.~M. and {Zucca}, E. and {Ballardini}, M. and {Bozzo}, E. and {Burigana}, C. and {Cabanac}, R. and {Calabrese}, M. and {Castro}, T. and {Escartin Vigo}, J.~A. and {Garcia-Bellido}, J. and {Macias-Perez}, J. and {Maoli}, R. and {Martin-Fleitas}, J. and {Mauri}, N. and {Metcalf}, R.~B. and {Monaco}, P. and {Pontinen}, M. and {Risso}, I. and {Scottez}, V. and {Sereno}, M. and {Tenti}, M. and {Tucci}, M. and {Viel}, M. and {Wiesmann}, M. and {Akrami}, Y. and {Andika}, I.~T. and {Angora}, G. and {Archidiacono}, M. and {Atrio-Barandela}, F. and {Avila}, S. and {Bazzanini}, L. and {Bel}, J. and {Bertacca}, D. and {Bethermin}, M. and {Beutler}, F. and {Blanchard}, A. and {Blot}, L. and {Bohringer}, H. and {Bonici}, M. and {Borgani}, S. and {Brown}, M.~L. and {Bruton}, S. and {Calabro}, A. and {Caro}, F. and {Carvalho}, C.~S. and {Cogato}, F. and {Cooray}, A.~R. and {Davini}, S. and {Desprez}, G. and {Diaz-Sanchez}, A.},
        title = "{Euclid preparation: Testing multi-field inflation with galaxy power spectrum and bispectrum}",
      journal = {arXiv e-prints},
         year = 2026,
        month = may,
          eid = {arXiv:2605.21436},
        pages = {arXiv:2605.21436},
          doi = {10.48550/arXiv.2605.21436},
archivePrefix = {arXiv},
       eprint = {2605.21436},
 primaryClass = {astro-ph.CO},
       adsurl = {https://ui.adsabs.harvard.edu/abs/2026arXiv260521436E}
}

@ARTICLE{Fondi_2024,
       author = {{Fondi}, Emanuele and {Verde}, Licia and {Villaescusa-Navarro}, Francisco and {Baldi}, Marco and {Coulton}, William R. and {Jung}, Gabriel and {Karagiannis}, Dionysios and {Liguori}, Michele and {Ravenni}, Andrea and {Wandelt}, Benjamin D.},
        title = "{Taming assembly bias for primordial non-Gaussianity}",
      journal = {\jcap},
         year = 2024,
        month = feb,
       volume = {2024},
       number = {2},
          eid = {048},
        pages = {048},
          doi = {10.1088/1475-7516/2024/02/048},
archivePrefix = {arXiv},
       eprint = {2311.10088},
 primaryClass = {astro-ph.CO},
       adsurl = {https://ui.adsabs.harvard.edu/abs/2024JCAP...02..048F}
}

@ARTICLE{PNGUNIT_2024,
       author = {{Guti{\'e}rrez Adame}, Adri{\'a}n and {Avila}, Santiago and {Gonzalez-Perez}, Violeta and {Yepes}, Gustavo and {Pellejero}, Marcos and {Wang}, Mike S. and {Chuang}, Chia-Hsun and {Feng}, Yu and {Garcia-Bellido}, Juan and {Knebe}, Alexander},
        title = "{PNG-UNITsims: Halo clustering response to primordial non-Gaussianities as a function of mass}",
      journal = {\aap},
         year = 2024,
        month = sep,
       volume = {689},
          eid = {A69},
        pages = {A69},
          doi = {10.1051/0004-6361/202349037},
archivePrefix = {arXiv},
       eprint = {2312.12405},
 primaryClass = {astro-ph.CO},
       adsurl = {https://ui.adsabs.harvard.edu/abs/2024A&A...689A..69G}
}

@ARTICLE{Grossi_2009,
       author = {{Grossi}, M. and {Verde}, L. and {Carbone}, C. and {Dolag}, K. and {Branchini}, E. and {Iannuzzi}, F. and {Matarrese}, S. and {Moscardini}, L.},
        title = "{Large-scale non-Gaussian mass function and halo bias: tests on N-body simulations}",
      journal = {\mnras},
         year = 2009,
        month = sep,
       volume = {398},
       number = {1},
        pages = {321-332},
          doi = {10.1111/j.1365-2966.2009.15150.x},
archivePrefix = {arXiv},
       eprint = {0902.2013},
 primaryClass = {astro-ph.CO},
       adsurl = {https://ui.adsabs.harvard.edu/abs/2009MNRAS.398..321G}
}

@ARTICLE{Barreira_2020,
       author = {{Barreira}, Alexandre and {Cabass}, Giovanni and {Schmidt}, Fabian and {Pillepich}, Annalisa and {Nelson}, Dylan},
        title = "{Galaxy bias and primordial non-Gaussianity: insights from galaxy formation simulations with IllustrisTNG}",
      journal = {\jcap},
         year = 2020,
        month = dec,
       volume = {2020},
       number = {12},
          eid = {013},
        pages = {013},
          doi = {10.1088/1475-7516/2020/12/013},
archivePrefix = {arXiv},
       eprint = {2006.09368},
 primaryClass = {astro-ph.CO},
       adsurl = {https://ui.adsabs.harvard.edu/abs/2020JCAP...12..013B}
}

@ARTICLE{Mancini_2022,
       author = {{Spurio Mancini}, Alessio and {Piras}, Davide and {Alsing}, Justin and {Joachimi}, Benjamin and {Hobson}, Michael P.},
        title = "{COSMOPOWER: emulating cosmological power spectra for accelerated Bayesian inference from next-generation surveys}",
      journal = {\mnras},
         year = 2022,
        month = apr,
       volume = {511},
       number = {2},
        pages = {1771-1788},
          doi = {10.1093/mnras/stac064},
archivePrefix = {arXiv},
       eprint = {2106.03846},
 primaryClass = {astro-ph.CO},
       adsurl = {https://ui.adsabs.harvard.edu/abs/2022MNRAS.511.1771S}
}

@ARTICLE{Torrado_2021,
       author = {{Torrado}, Jes{\'u}s and {Lewis}, Antony},
        title = "{Cobaya: code for Bayesian analysis of hierarchical physical models}",
      journal = {\jcap},
         year = 2021,
        month = may,
       volume = {2021},
       number = {5},
          eid = {057},
        pages = {057},
          doi = {10.1088/1475-7516/2021/05/057},
archivePrefix = {arXiv},
       eprint = {2005.05290},
 primaryClass = {astro-ph.IM},
       adsurl = {https://ui.adsabs.harvard.edu/abs/2021JCAP...05..057T}
}

@ARTICLE{Lewis_2013,
       author = {{Lewis}, Antony},
        title = "{Efficient sampling of fast and slow cosmological parameters}",
      journal = {\prd},
         year = 2013,
        month = may,
       volume = {87},
       number = {10},
          eid = {103529},
        pages = {103529},
          doi = {10.1103/PhysRevD.87.103529},
archivePrefix = {arXiv},
       eprint = {1304.4473},
 primaryClass = {astro-ph.CO},
       adsurl = {https://ui.adsabs.harvard.edu/abs/2013PhRvD..87j3529L}
}

@ARTICLE{Lewis_2002,
       author = {{Lewis}, Antony and {Bridle}, Sarah},
        title = "{Cosmological parameters from CMB and other data: A Monte Carlo approach}",
      journal = {\prd},
         year = 2002,
        month = nov,
       volume = {66},
       number = {10},
          eid = {103511},
        pages = {103511},
          doi = {10.1103/PhysRevD.66.103511},
archivePrefix = {arXiv},
       eprint = {astro-ph/0205436},
 primaryClass = {astro-ph},
       adsurl = {https://ui.adsabs.harvard.edu/abs/2002PhRvD..66j3511L}
}

@ARTICLE{Hinton_2016,
       author = {{Hinton}, Samuel R.},
        title = "{ChainConsumer}",
      journal = {The Journal of Open Source Software},
         year = 2016,
        month = aug,
       volume = {1},
       number = {4},
          eid = {00045},
        pages = {00045},
          doi = {10.21105/joss.00045},
       adsurl = {https://ui.adsabs.harvard.edu/abs/2016JOSS....1...45H}
}

@ARTICLE{Giannantonio_2010,
       author = {{Giannantonio}, Tommaso and {Porciani}, Cristiano},
        title = "{Structure formation from non-Gaussian initial conditions: Multivariate biasing, statistics, and comparison with N-body simulations}",
      journal = {\prd},
         year = 2010,
        month = mar,
       volume = {81},
       number = {6},
          eid = {063530},
        pages = {063530},
          doi = {10.1103/PhysRevD.81.063530},
archivePrefix = {arXiv},
       eprint = {0911.0017},
 primaryClass = {astro-ph.CO},
       adsurl = {https://ui.adsabs.harvard.edu/abs/2010PhRvD..81f3530G}
}

@ARTICLE{Pozzetti_2016,
       author = {{Pozzetti}, L. and {Hirata}, C.~M. and {Geach}, J.~E. and {Cimatti}, A. and {Baugh}, C. and {Cucciati}, O. and {Merson}, A. and {Norberg}, P. and {Shi}, D.},
        title = "{Modelling the number density of H{\ensuremath{\alpha}} emitters for future spectroscopic near-IR space missions}",
      journal = {\aap},
         year = 2016,
        month = may,
       volume = {590},
          eid = {A3},
        pages = {A3},
          doi = {10.1051/0004-6361/201527081},
archivePrefix = {arXiv},
       eprint = {1603.01453},
 primaryClass = {astro-ph.GA},
       adsurl = {https://ui.adsabs.harvard.edu/abs/2016A&A...590A...3P}
}

@ARTICLE{Euclid_2020,
       author = {{Euclid Collaboration} and {Blanchard}, A. and {Camera}, S. and {Carbone}, C. and {Cardone}, V.~F. and {Casas}, S. and {Clesse}, S. and {Ili{\'c}}, S. and {Kilbinger}, M. and {Kitching}, T. and {Kunz}, M. and {Lacasa}, F. and {Linder}, E. and {Majerotto}, E. and {Markovi{\v{c}}}, K. and {Martinelli}, M. and {Pettorino}, V. and {Pourtsidou}, A. and {Sakr}, Z. and {S{\'a}nchez}, A.~G. and {Sapone}, D. and {Tutusaus}, I. and {Yahia-Cherif}, S. and {Yankelevich}, V. and {Andreon}, S. and {Aussel}, H. and {Balaguera-Antol{\'\i}nez}, A. and {Baldi}, M. and {Bardelli}, S. and {Bender}, R. and {Biviano}, A. and {Bonino}, D. and {Boucaud}, A. and {Bozzo}, E. and {Branchini}, E. and {Brau-Nogue}, S. and {Brescia}, M. and {Brinchmann}, J. and {Burigana}, C. and {Cabanac}, R. and {Capobianco}, V. and {Cappi}, A. and {Carretero}, J. and {Carvalho}, C.~S. and {Casas}, R. and {Castander}, F.~J. and {Castellano}, M. and {Cavuoti}, S. and {Cimatti}, A. and {Cledassou}, R. and {Colodro-Conde}, C. and {Congedo}, G. and {Conselice}, C.~J. and {Conversi}, L. and {Copin}, Y. and {Corcione}, L. and {Coupon}, J. and {Courtois}, H.~M. and {Cropper}, M. and {Da Silva}, A. and {de la Torre}, S. and {Di Ferdinando}, D. and {Dubath}, F. and {Ducret}, F. and {Duncan}, C.~A.~J. and {Dupac}, X. and {Dusini}, S. and {Fabbian}, G. and {Fabricius}, M. and {Farrens}, S. and {Fosalba}, P. and {Fotopoulou}, S. and {Fourmanoit}, N. and {Frailis}, M. and {Franceschi}, E. and {Franzetti}, P. and {Fumana}, M. and {Galeotta}, S. and {Gillard}, W. and {Gillis}, B. and {Giocoli}, C. and {G{\'o}mez-Alvarez}, P. and {Graci{\'a}-Carpio}, J. and {Grupp}, F. and {Guzzo}, L. and {Hoekstra}, H. and {Hormuth}, F. and {Israel}, H. and {Jahnke}, K. and {Keihanen}, E. and {Kermiche}, S. and {Kirkpatrick}, C.~C. and {Kohley}, R. and {Kubik}, B. and {Kurki-Suonio}, H. and {Ligori}, S. and {Lilje}, P.~B. and {Lloro}, I. and {Maino}, D. and {Maiorano}, E. and {Marggraf}, O. and {Martinet}, N. and {Marulli}, F. and {Massey}, R. and {Medinaceli}, E. and {Mei}, S. and {Mellier}, Y. and {Metcalf}, B. and {Metge}, J.~J. and {Meylan}, G. and {Moresco}, M. and {Moscardini}, L. and {Munari}, E. and {Nichol}, R.~C. and {Niemi}, S. and {Nucita}, A.~A. and {Padilla}, C. and {Paltani}, S. and {Pasian}, F. and {Percival}, W.~J. and {Pires}, S. and {Polenta}, G. and {Poncet}, M. and {Pozzetti}, L. and {Racca}, G.~D. and {Raison}, F. and {Renzi}, A. and {Rhodes}, J. and {Romelli}, E. and {Roncarelli}, M. and {Rossetti}, E. and {Saglia}, R. and {Schneider}, P. and {Scottez}, V. and {Secroun}, A. and {Sirri}, G. and {Stanco}, L. and {Starck}, J. -L. and {Sureau}, F. and {Tallada-Cresp{\'\i}}, P. and {Tavagnacco}, D. and {Taylor}, A.~N. and {Tenti}, M. and {Tereno}, I. and {Toledo-Moreo}, R. and {Torradeflot}, F. and {Valenziano}, L. and {Vassallo}, T. and {Verdoes Kleijn}, G.~A. and {Viel}, M. and {Wang}, Y. and {Zacchei}, A. and {Zoubian}, J. and {Zucca}, E.},
        title = "{Euclid preparation. VII. Forecast validation for Euclid cosmological probes}",
      journal = {\aap},
         year = 2020,
        month = oct,
       volume = {642},
          eid = {A191},
        pages = {A191},
          doi = {10.1051/0004-6361/202038071},
archivePrefix = {arXiv},
       eprint = {1910.09273},
 primaryClass = {astro-ph.CO},
       adsurl = {https://ui.adsabs.harvard.edu/abs/2020A&A...642A.191E}
}

@ARTICLE{Byun_2017,
       author = {{Byun}, Joyce and {Eggemeier}, Alexander and {Regan}, Donough and {Seery}, David and {Smith}, Robert E.},
        title = "{Towards optimal cosmological parameter recovery from compressed bispectrum statistics}",
      journal = {\mnras},
         year = 2017,
        month = oct,
       volume = {471},
       number = {2},
        pages = {1581-1618},
          doi = {10.1093/mnras/stx1681},
archivePrefix = {arXiv},
       eprint = {1705.04392},
 primaryClass = {astro-ph.CO},
       adsurl = {https://ui.adsabs.harvard.edu/abs/2017MNRAS.471.1581B}
}

@ARTICLE{Lazeryas_2023,
       author = {{Lazeyras}, Titouan and {Barreira}, Alexandre and {Schmidt}, Fabian and {Desjacques}, Vincent},
        title = "{Assembly bias in the local PNG halo bias and its implication for f $_{NL}$ constraints}",
      journal = {\jcap},
         year = 2023,
        month = jan,
       volume = {2023},
       number = {1},
          eid = {023},
        pages = {023},
          doi = {10.1088/1475-7516/2023/01/023},
archivePrefix = {arXiv},
       eprint = {2209.07251},
 primaryClass = {astro-ph.CO},
       adsurl = {https://ui.adsabs.harvard.edu/abs/2023JCAP...01..023L}
}

@ARTICLE{Dizgah_2021,
       author = {{Moradinezhad Dizgah}, Azadeh and {Biagetti}, Matteo and {Sefusatti}, Emiliano and {Desjacques}, Vincent and {Nore{\~n}a}, Jorge},
        title = "{Primordial non-Gaussianity from biased tracers: likelihood analysis of real-space power spectrum and bispectrum}",
      journal = {\jcap},
         year = 2021,
        month = may,
       volume = {2021},
       number = {5},
          eid = {015},
        pages = {015},
          doi = {10.1088/1475-7516/2021/05/015},
archivePrefix = {arXiv},
       eprint = {2010.14523},
 primaryClass = {astro-ph.CO},
       adsurl = {https://ui.adsabs.harvard.edu/abs/2021JCAP...05..015M}
}

@ARTICLE{Karagiannis_2021,
       author = {{Karagiannis}, Dionysios and {Fonseca}, Jos{\'e} and {Maartens}, Roy and {Camera}, Stefano},
        title = "{Probing primordial non-Gaussianity with the power spectrum and bispectrum of future 21 cm intensity maps}",
      journal = {Physics of the Dark Universe},
         year = 2021,
        month = may,
       volume = {32},
          eid = {100821},
        pages = {100821},
          doi = {10.1016/j.dark.2021.100821},
archivePrefix = {arXiv},
       eprint = {2010.07034},
 primaryClass = {astro-ph.CO},
       adsurl = {https://ui.adsabs.harvard.edu/abs/2021PDU....3200821K}
}

@ARTICLE{Afshordi_2008,
       author = {{Afshordi}, Niayesh and {Tolley}, Andrew J.},
        title = "{Primordial non-Gaussianity, statistics of collapsed objects, and the integrated Sachs-Wolfe effect}",
      journal = {\prd},
         year = 2008,
        month = dec,
       volume = {78},
       number = {12},
          eid = {123507},
        pages = {123507},
          doi = {10.1103/PhysRevD.78.123507},
archivePrefix = {arXiv},
       eprint = {0806.1046},
 primaryClass = {astro-ph},
       adsurl = {https://ui.adsabs.harvard.edu/abs/2008PhRvD..78l3507A}
}

@ARTICLE{Slosar_2008,
       author = {{Slosar}, An{\v{z}}e and {Hirata}, Christopher and {Seljak}, Uro{\v{s}} and {Ho}, Shirley and {Padmanabhan}, Nikhil},
        title = "{Constraints on local primordial non-Gaussianity from large scale structure}",
      journal = {\jcap},
         year = 2008,
        month = aug,
       volume = {2008},
       number = {8},
          eid = {031},
        pages = {031},
          doi = {10.1088/1475-7516/2008/08/031},
archivePrefix = {arXiv},
       eprint = {0805.3580},
 primaryClass = {astro-ph},
       adsurl = {https://ui.adsabs.harvard.edu/abs/2008JCAP...08..031S}
}

@ARTICLE{Matarrese_2008,
       author = {{Matarrese}, Sabino and {Verde}, Licia},
        title = "{The Effect of Primordial Non-Gaussianity on Halo Bias}",
      journal = {\apjl},
         year = 2008,
        month = apr,
       volume = {677},
       number = {2},
        pages = {L77},
          doi = {10.1086/587840},
archivePrefix = {arXiv},
       eprint = {0801.4826},
 primaryClass = {astro-ph},
       adsurl = {https://ui.adsabs.harvard.edu/abs/2008ApJ...677L..77M}
}

@ARTICLE{Komatsu_2001,
       author = {{Komatsu}, Eiichiro and {Spergel}, David N.},
        title = "{Acoustic signatures in the primary microwave background bispectrum}",
      journal = {\prd},
         year = 2001,
        month = mar,
       volume = {63},
       number = {6},
          eid = {063002},
        pages = {063002},
          doi = {10.1103/PhysRevD.63.063002},
archivePrefix = {arXiv},
       eprint = {astro-ph/0005036},
 primaryClass = {astro-ph},
       adsurl = {https://ui.adsabs.harvard.edu/abs/2001PhRvD..63f3002K}
}

@ARTICLE{Salopek_1990,
       author = {{Salopek}, D.~S. and {Bond}, J.~R.},
        title = "{Nonlinear evolution of long-wavelength metric fluctuations in inflationary models}",
      journal = {\prd},
         year = 1990,
        month = dec,
       volume = {42},
       number = {12},
        pages = {3936-3962},
          doi = {10.1103/PhysRevD.42.3936},
       adsurl = {https://ui.adsabs.harvard.edu/abs/1990PhRvD..42.3936S}
}

@ARTICLE{Creminelli_2006,
       author = {{Creminelli}, Paolo and {Nicolis}, Alberto and {Senatore}, Leonardo and {Tegmark}, Max and {Zaldarriaga}, Matias},
        title = "{Limits on non-Gaussianities from WMAP data}",
      journal = {\jcap},
         year = 2006,
        month = may,
       volume = {2006},
       number = {5},
          eid = {004},
        pages = {004},
          doi = {10.1088/1475-7516/2006/05/004},
archivePrefix = {arXiv},
       eprint = {astro-ph/0509029},
 primaryClass = {astro-ph},
       adsurl = {https://ui.adsabs.harvard.edu/abs/2006JCAP...05..004C}
}

@ARTICLE{Senatore_2010,
       author = {{Senatore}, Leonardo and {Smith}, Kendrick M. and {Zaldarriaga}, Matias},
        title = "{Non-Gaussianities in single field inflation and their optimal limits from the WMAP 5-year data}",
      journal = {\jcap},
         year = 2010,
        month = jan,
       volume = {2010},
       number = {1},
          eid = {028},
        pages = {028},
          doi = {10.1088/1475-7516/2010/01/028},
archivePrefix = {arXiv},
       eprint = {0905.3746},
 primaryClass = {astro-ph.CO},
       adsurl = {https://ui.adsabs.harvard.edu/abs/2010JCAP...01..028S}
}

@ARTICLE{Mcdonald_2009,
       author = {{McDonald}, Patrick and {Roy}, Arabindo},
        title = "{Clustering of dark matter tracers: generalizing bias for the coming era of precision LSS}",
      journal = {\jcap},
         year = 2009,
        month = aug,
       volume = {2009},
       number = {8},
          eid = {020},
        pages = {020},
          doi = {10.1088/1475-7516/2009/08/020},
archivePrefix = {arXiv},
       eprint = {0902.0991},
 primaryClass = {astro-ph.CO},
       adsurl = {https://ui.adsabs.harvard.edu/abs/2009JCAP...08..020M}
}

@ARTICLE{Baldauf_2011,
       author = {{Baldauf}, Tobias and {Seljak}, Uro{\v{s}} and {Senatore}, Leonardo},
        title = "{Primordial non-Gaussianity in the bispectrum of the halo density field}",
      journal = {\jcap},
         year = 2011,
        month = apr,
       volume = {2011},
       number = {4},
          eid = {006},
        pages = {006},
          doi = {10.1088/1475-7516/2011/04/006},
archivePrefix = {arXiv},
       eprint = {1011.1513},
 primaryClass = {astro-ph.CO},
       adsurl = {https://ui.adsabs.harvard.edu/abs/2011JCAP...04..006B}
}

@ARTICLE{Barreira_2023,
       author = {{Barreira}, Alexandre and {Krause}, Elisabeth},
        title = "{Towards optimal and robust f\_NL constraints with multi-tracer analyses}",
      journal = {\jcap},
         year = 2023,
        month = oct,
       volume = {2023},
       number = {10},
          eid = {044},
        pages = {044},
          doi = {10.1088/1475-7516/2023/10/044},
archivePrefix = {arXiv},
       eprint = {2302.09066},
 primaryClass = {astro-ph.CO},
       adsurl = {https://ui.adsabs.harvard.edu/abs/2023JCAP...10..044B}
}

@ARTICLE{Seljak_2009,
       author = {{Seljak}, Uro{\v{s}}},
        title = "{Extracting Primordial Non-Gaussianity without Cosmic Variance}",
      journal = {\prl},
         year = 2009,
        month = jan,
       volume = {102},
       number = {2},
          eid = {021302},
        pages = {021302},
          doi = {10.1103/PhysRevLett.102.021302},
archivePrefix = {arXiv},
       eprint = {0807.1770},
 primaryClass = {astro-ph},
       adsurl = {https://ui.adsabs.harvard.edu/abs/2009PhRvL.102b1302S}
}

@ARTICLE{Bonvin_2014_b,
       author = {{Bonvin}, Camille},
        title = "{Isolating relativistic effects in large-scale structure}",
      journal = {Classical and Quantum Gravity},
         year = 2014,
        month = dec,
       volume = {31},
       number = {23},
          eid = {234002},
        pages = {234002},
          doi = {10.1088/0264-9381/31/23/234002},
archivePrefix = {arXiv},
       eprint = {1409.2224},
 primaryClass = {astro-ph.CO},
       adsurl = {https://ui.adsabs.harvard.edu/abs/2014CQGra..31w4002B}
}

@ARTICLE{Blanco_2024,
       author = {{Blanco}, Daniel Sobral and {Bonvin}, Camille and {Clarkson}, Chris and {Maartens}, Roy},
        title = "{Using relativistic effects in large-scale structure to constrain astrophysical properties of galaxy populations}",
      journal = {\jcap},
         year = 2024,
        month = dec,
       volume = {2024},
       number = {12},
          eid = {029},
        pages = {029},
          doi = {10.1088/1475-7516/2024/12/029},
archivePrefix = {arXiv},
       eprint = {2406.19908},
 primaryClass = {astro-ph.CO},
       adsurl = {https://ui.adsabs.harvard.edu/abs/2024JCAP...12..029B}
}

@misc{mathwap,
       author = {{Addis}, Chris},
        title = "{MathWAP: Compute power spectra bispectra contributions from peturbation theory}",
 howpublished = {Astrophysics Source Code Library, record ascl:2507.019},
         year = 2025,
        month = jul,
          eid = {ascl:2507.019},
archivePrefix = {ascl},
       eprint = {2507.019},
       adsurl = {https://ui.adsabs.harvard.edu/abs/2025ascl.soft07019A}
}

@misc{cosmowap,
       author = {{Addis}, Chris},
        title = "{CosmoWAP: Power spectra and bispectra analyzer}",
 howpublished = {Astrophysics Source Code Library, record ascl:2507.020},
         year = 2025,
        month = jul,
          eid = {ascl:2507.020},
archivePrefix = {ascl},
       eprint = {2507.020},
       adsurl = {https://ui.adsabs.harvard.edu/abs/2025ascl.soft07020A}
}

@unpublished{cosmowap_paper,
       author = {{Addis}, Chris},
        title = "{\textsc{CosmoWAP}: A Forecasting Package for Large-Scale Structure on a Light-cone}",
         note = {in prep.},
         year = 2026
}

@ARTICLE{Guedezounme_2025,
       author = {{Guedezounme}, S{\^e}cloka L. and {Jolicoeur}, Sheean and {Maartens}, Roy},
        title = "{Primordial non-Gaussianity {\textemdash} the effects of relativistic and wide-angle corrections to the power spectrum}",
      journal = {\jcap},
         year = 2025,
        month = jul,
       volume = {2025},
       number = {7},
          eid = {063},
        pages = {063},
          doi = {10.1088/1475-7516/2025/07/063},
archivePrefix = {arXiv},
       eprint = {2412.06553},
 primaryClass = {astro-ph.CO},
       adsurl = {https://ui.adsabs.harvard.edu/abs/2025JCAP...07..063G}
}

@ARTICLE{Wang_2020,
       author = {{Wang}, Mike (Shengbo) and {Beutler}, Florian and {Bacon}, David},
        title = "{Impact of relativistic effects on the primordial non-Gaussianity signature in the large-scale clustering of quasars}",
      journal = {\mnras},
         year = 2020,
        month = dec,
       volume = {499},
       number = {2},
        pages = {2598-2607},
          doi = {10.1093/mnras/staa2998},
archivePrefix = {arXiv},
       eprint = {2007.01802},
 primaryClass = {astro-ph.CO},
       adsurl = {https://ui.adsabs.harvard.edu/abs/2020MNRAS.499.2598W}
}

@ARTICLE{Dio_2017,
       author = {{Di Dio}, E. and {Perrier}, H. and {Durrer}, R. and {Marozzi}, G. and {Moradinezhad Dizgah}, A. and {Nore{\~n}a}, J. and {Riotto}, A.},
        title = "{Non-Gaussianities due to relativistic corrections to the observed galaxy bispectrum}",
      journal = {\jcap},
         year = 2017,
        month = mar,
       volume = {2017},
       number = {3},
          eid = {006},
        pages = {006},
          doi = {10.1088/1475-7516/2017/03/006},
archivePrefix = {arXiv},
       eprint = {1611.03720},
 primaryClass = {astro-ph.CO},
       adsurl = {https://ui.adsabs.harvard.edu/abs/2017JCAP...03..006D}
}

@ARTICLE{Jeong_2012,
       author = {{Jeong}, Donghui and {Schmidt}, Fabian and {Hirata}, Christopher M.},
        title = "{Large-scale clustering of galaxies in general relativity}",
      journal = {\prd},
         year = 2012,
        month = jan,
       volume = {85},
       number = {2},
          eid = {023504},
        pages = {023504},
          doi = {10.1103/PhysRevD.85.023504},
archivePrefix = {arXiv},
       eprint = {1107.5427},
 primaryClass = {astro-ph.CO},
       adsurl = {https://ui.adsabs.harvard.edu/abs/2012PhRvD..85b3504J}
}

@ARTICLE{Bruni_2012,
       author = {{Bruni}, Marco and {Crittenden}, Robert and {Koyama}, Kazuya and {Maartens}, Roy and {Pitrou}, Cyril and {Wands}, David},
        title = "{Disentangling non-Gaussianity, bias, and general relativistic effects in the galaxy distribution}",
      journal = {\prd},
         year = 2012,
        month = feb,
       volume = {85},
       number = {4},
          eid = {041301},
        pages = {041301},
          doi = {10.1103/PhysRevD.85.041301},
archivePrefix = {arXiv},
       eprint = {1106.3999},
 primaryClass = {astro-ph.CO},
       adsurl = {https://ui.adsabs.harvard.edu/abs/2012PhRvD..85d1301B}
}

@ARTICLE{Camera_2015,
       author = {{Camera}, S. and {Maartens}, R. and {Santos}, M.~G.},
        title = "{Einstein's legacy in galaxy surveys.}",
      journal = {\mnras},
         year = 2015,
        month = jul,
       volume = {451},
        pages = {L80-L84},
          doi = {10.1093/mnrasl/slv069},
archivePrefix = {arXiv},
       eprint = {1412.4781},
 primaryClass = {astro-ph.CO},
       adsurl = {https://ui.adsabs.harvard.edu/abs/2015MNRAS.451L..80C}
}

@ARTICLE{Bonvin_2023,
       author = {{Bonvin}, Camille and {Lepori}, Francesca and {Schulz}, Sebastian and {Tutusaus}, Isaac and {Adamek}, Julian and {Fosalba}, Pablo},
        title = "{A case study for measuring the relativistic dipole of a galaxy cross-correlation with the Dark Energy Spectroscopic Instrument}",
      journal = {\mnras},
         year = 2023,
        month = nov,
       volume = {525},
       number = {3},
        pages = {4611-4627},
          doi = {10.1093/mnras/stad2567},
archivePrefix = {arXiv},
       eprint = {2306.04213},
 primaryClass = {astro-ph.CO},
       adsurl = {https://ui.adsabs.harvard.edu/abs/2023MNRAS.525.4611B}
}

@ARTICLE{Ferramacho_2014,
       author = {{Ferramacho}, L.~D. and {Santos}, M.~G. and {Jarvis}, M.~J. and {Camera}, S.},
        title = "{Radio galaxy populations and the multitracer technique: pushing the limits on primordial non-Gaussianity}",
      journal = {\mnras},
         year = 2014,
        month = aug,
       volume = {442},
       number = {3},
        pages = {2511-2518},
          doi = {10.1093/mnras/stu1015},
archivePrefix = {arXiv},
       eprint = {1402.2290},
 primaryClass = {astro-ph.CO},
       adsurl = {https://ui.adsabs.harvard.edu/abs/2014MNRAS.442.2511F}
}

@ARTICLE{Alonso_2015,
       author = {{Alonso}, David and {Bull}, Philip and {Ferreira}, Pedro G. and {Maartens}, Roy and {Santos}, M{\'a}rio G.},
        title = "{Ultra-large-scale Cosmology in Next-generation Experiments with Single Tracers}",
      journal = {\apj},
         year = 2015,
        month = dec,
       volume = {814},
       number = {2},
          eid = {145},
        pages = {145},
          doi = {10.1088/0004-637X/814/2/145},
archivePrefix = {arXiv},
       eprint = {1505.07596},
 primaryClass = {astro-ph.CO},
       adsurl = {https://ui.adsabs.harvard.edu/abs/2015ApJ...814..145A}
}

@ARTICLE{Mueller_2019,
       author = {{Mueller}, Eva-Maria and {Percival}, Will J. and {Ruggeri}, Rossana},
        title = "{Optimizing primordial non-Gaussianity measurements from galaxy surveys}",
      journal = {\mnras},
         year = 2019,
        month = may,
       volume = {485},
       number = {3},
        pages = {4160-4166},
          doi = {10.1093/mnras/sty3150},
archivePrefix = {arXiv},
       eprint = {1702.05088},
 primaryClass = {astro-ph.CO},
       adsurl = {https://ui.adsabs.harvard.edu/abs/2019MNRAS.485.4160M}
}

@ARTICLE{Amendola_2018,
       author = {{Amendola}, Luca and {Appleby}, Stephen and {Avgoustidis}, Anastasios and {Bacon}, David and {Baker}, Tessa and {Baldi}, Marco and {Bartolo}, Nicola and {Blanchard}, Alain and {Bonvin}, Camille and {Borgani}, Stefano and {Branchini}, Enzo and {Burrage}, Clare and {Camera}, Stefano and {Carbone}, Carmelita and {Casarini}, Luciano and {Cropper}, Mark and {de Rham}, Claudia and {Dietrich}, J{\"o}rg P. and {Di Porto}, Cinzia and {Durrer}, Ruth and {Ealet}, Anne and {Ferreira}, Pedro G. and {Finelli}, Fabio and {Garc{\'\i}a-Bellido}, Juan and {Giannantonio}, Tommaso and {Guzzo}, Luigi and {Heavens}, Alan and {Heisenberg}, Lavinia and {Heymans}, Catherine and {Hoekstra}, Henk and {Hollenstein}, Lukas and {Holmes}, Rory and {Hwang}, Zhiqi and {Jahnke}, Knud and {Kitching}, Thomas D. and {Koivisto}, Tomi and {Kunz}, Martin and {La Vacca}, Giuseppe and {Linder}, Eric and {March}, Marisa and {Marra}, Valerio and {Martins}, Carlos and {Majerotto}, Elisabetta and {Markovic}, Dida and {Marsh}, David and {Marulli}, Federico and {Massey}, Richard and {Mellier}, Yannick and {Montanari}, Francesco and {Mota}, David F. and {Nunes}, Nelson J. and {Percival}, Will and {Pettorino}, Valeria and {Porciani}, Cristiano and {Quercellini}, Claudia and {Read}, Justin and {Rinaldi}, Massimiliano and {Sapone}, Domenico and {Sawicki}, Ignacy and {Scaramella}, Roberto and {Skordis}, Constantinos and {Simpson}, Fergus and {Taylor}, Andy and {Thomas}, Shaun and {Trotta}, Roberto and {Verde}, Licia and {Vernizzi}, Filippo and {Vollmer}, Adrian and {Wang}, Yun and {Weller}, Jochen and {Zlosnik}, Tom and {The Euclid Theory Working Group}},
        title = "{Cosmology and fundamental physics with the Euclid satellite}",
      journal = {Living Reviews in Relativity},
         year = 2018,
        month = dec,
       volume = {21},
       number = {1},
          eid = {2},
        pages = {2},
          doi = {10.1007/s41114-017-0010-3},
archivePrefix = {arXiv},
       eprint = {1606.00180},
 primaryClass = {astro-ph.CO},
       adsurl = {https://ui.adsabs.harvard.edu/abs/2018LRR....21....2A}
}

@ARTICLE{FontRibera_2014,
       author = {{Font-Ribera}, Andreu and {McDonald}, Patrick and {Mostek}, Nick and {Reid}, Beth A. and {Seo}, Hee-Jong and {Slosar}, An{\v{z}}e},
        title = "{DESI and other Dark Energy experiments in the era of neutrino mass measurements}",
      journal = {\jcap},
         year = 2014,
        month = may,
       volume = {2014},
       number = {5},
          eid = {023},
        pages = {023},
          doi = {10.1088/1475-7516/2014/05/023},
archivePrefix = {arXiv},
       eprint = {1308.4164},
 primaryClass = {astro-ph.CO},
       adsurl = {https://ui.adsabs.harvard.edu/abs/2014JCAP...05..023F}
}

@ARTICLE{Giannantonio_2012,
       author = {{Giannantonio}, Tommaso and {Porciani}, Cristiano and {Carron}, Julien and {Amara}, Adam and {Pillepich}, Annalisa},
        title = "{Constraining primordial non-Gaussianity with future galaxy surveys}",
      journal = {\mnras},
         year = 2012,
        month = jun,
       volume = {422},
       number = {4},
        pages = {2854-2877},
          doi = {10.1111/j.1365-2966.2012.20604.x},
archivePrefix = {arXiv},
       eprint = {1109.0958},
 primaryClass = {astro-ph.CO},
       adsurl = {https://ui.adsabs.harvard.edu/abs/2012MNRAS.422.2854G}
}

@ARTICLE{Abacus_2021,
       author = {{Maksimova}, Nina A. and {Garrison}, Lehman H. and {Eisenstein}, Daniel J. and {Hadzhiyska}, Boryana and {Bose}, Sownak and {Satterthwaite}, Thomas P.},
        title = "{ABACUSSUMMIT: a massive set of high-accuracy, high-resolution N-body simulations}",
      journal = {\mnras},
         year = 2021,
        month = dec,
       volume = {508},
       number = {3},
        pages = {4017-4037},
          doi = {10.1093/mnras/stab2484},
archivePrefix = {arXiv},
       eprint = {2110.11398},
 primaryClass = {astro-ph.CO},
       adsurl = {https://ui.adsabs.harvard.edu/abs/2021MNRAS.508.4017M}
}

@ARTICLE{Akrami_2020,
       author = {{Planck Collaboration} and {Akrami}, Y. and {Arroja}, F. and {Ashdown}, M. and {Aumont}, J. and {Baccigalupi}, C. and {Ballardini}, M. and {Banday}, A.~J. and {Barreiro}, R.~B. and {Bartolo}, N. and {Basak}, S. and {Benabed}, K. and {Bernard}, J. -P. and {Bersanelli}, M. and {Bielewicz}, P. and {Bond}, J.~R. and {Borrill}, J. and {Bouchet}, F.~R. and {Bucher}, M. and {Burigana}, C. and {Butler}, R.~C. and {Calabrese}, E. and {Cardoso}, J. -F. and {Casaponsa}, B. and {Challinor}, A. and {Chiang}, H.~C. and {Colombo}, L.~P.~L. and {Combet}, C. and {Crill}, B.~P. and {Cuttaia}, F. and {de Bernardis}, P. and {de Rosa}, A. and {de Zotti}, G. and {Delabrouille}, J. and {Delouis}, J. -M. and {Di Valentino}, E. and {Diego}, J.~M. and {Dor{\'e}}, O. and {Douspis}, M. and {Ducout}, A. and {Dupac}, X. and {Dusini}, S. and {Efstathiou}, G. and {Elsner}, F. and {En{\ss}lin}, T.~A. and {Eriksen}, H.~K. and {Fantaye}, Y. and {Fergusson}, J. and {Fernandez-Cobos}, R. and {Finelli}, F. and {Frailis}, M. and {Fraisse}, A.~A. and {Franceschi}, E. and {Frolov}, A. and {Galeotta}, S. and {Galli}, S. and {Ganga}, K. and {G{\'e}nova-Santos}, R.~T. and {Gerbino}, M. and {Gonz{\'a}lez-Nuevo}, J. and {G{\'o}rski}, K.~M. and {Gratton}, S. and {Gruppuso}, A. and {Gudmundsson}, J.~E. and {Hamann}, J. and {Handley}, W. and {Hansen}, F.~K. and {Herranz}, D. and {Hivon}, E. and {Huang}, Z. and {Jaffe}, A.~H. and {Jones}, W.~C. and {Jung}, G. and {Keih{\"a}nen}, E. and {Keskitalo}, R. and {Kiiveri}, K. and {Kim}, J. and {Krachmalnicoff}, N. and {Kunz}, M. and {Kurki-Suonio}, H. and {Lamarre}, J. -M. and {Lasenby}, A. and {Lattanzi}, M. and {Lawrence}, C.~R. and {Le Jeune}, M. and {Levrier}, F. and {Lewis}, A. and {Liguori}, M. and {Lilje}, P.~B. and {Lindholm}, V. and {L{\'o}pez-Caniego}, M. and {Ma}, Y. -Z. and {Mac{\'\i}as-P{\'e}rez}, J.~F. and {Maggio}, G. and {Maino}, D. and {Mandolesi}, N. and {Marcos-Caballero}, A. and {Maris}, M. and {Martin}, P.~G. and {Mart{\'\i}nez-Gonz{\'a}lez}, E. and {Matarrese}, S. and {Mauri}, N. and {McEwen}, J.~D. and {Meerburg}, P.~D. and {Meinhold}, P.~R. and {Melchiorri}, A. and {Mennella}, A. and {Migliaccio}, M. and {Miville-Desch{\^e}nes}, M. -A. and {Molinari}, D. and {Moneti}, A. and {Montier}, L. and {Morgante}, G. and {Moss}, A. and {M{\"u}nchmeyer}, M. and {Natoli}, P. and {Oppizzi}, F. and {Pagano}, L. and {Paoletti}, D. and {Partridge}, B. and {Patanchon}, G. and {Perrotta}, F. and {Pettorino}, V. and {Piacentini}, F. and {Polenta}, G. and {Puget}, J. -L. and {Rachen}, J.~P. and {Racine}, B. and {Reinecke}, M. and {Remazeilles}, M. and {Renzi}, A. and {Rocha}, G. and {Rubi{\~n}o-Mart{\'\i}n}, J.~A. and {Ruiz-Granados}, B. and {Salvati}, L. and {Savelainen}, M. and {Scott}, D. and {Shellard}, E.~P.~S. and {Shiraishi}, M. and {Sirignano}, C. and {Sirri}, G. and {Smith}, K. and {Spencer}, L.~D. and {Stanco}, L. and {Sunyaev}, R. and {Suur-Uski}, A. -S. and {Tauber}, J.~A. and {Tavagnacco}, D. and {Tenti}, M. and {Toffolatti}, L. and {Tomasi}, M. and {Trombetti}, T. and {Valiviita}, J. and {Van Tent}, B. and {Vielva}, P. and {Villa}, F. and {Vittorio}, N. and {Wandelt}, B.~D. and {Wehus}, I.~K. and {Zacchei}, A. and {Zonca}, A.},
        title = "{Planck 2018 results. IX. Constraints on primordial non-Gaussianity}",
      journal = {\aap},
         year = 2020,
        month = sep,
       volume = {641},
          eid = {A9},
        pages = {A9},
          doi = {10.1051/0004-6361/201935891},
archivePrefix = {arXiv},
       eprint = {1905.05697},
 primaryClass = {astro-ph.CO},
       adsurl = {https://ui.adsabs.harvard.edu/abs/2020A&A...641A...9P}
}

@ARTICLE{2016VillaRampf,
       author = {{Villa}, Eleonora and {Rampf}, Cornelius},
        title = "{Relativistic perturbations in {\ensuremath{\Lambda}}CDM: Eulerian \& Lagrangian approaches}",
      journal = {\jcap},
         year = 2016,
        month = jan,
       volume = {2016},
       number = {1},
        pages = {030-030},
          doi = {10.1088/1475-7516/2016/01/030},
archivePrefix = {arXiv},
       eprint = {1505.04782},
 primaryClass = {gr-qc},
       adsurl = {https://ui.adsabs.harvard.edu/abs/2016JCAP...01..030V}
}

@ARTICLE{Castello_2022,
       author = {{Castello}, Sveva and {Grimm}, Nastassia and {Bonvin}, Camille},
        title = "{Rescuing constraints on modified gravity using gravitational redshift in large-scale structure}",
      journal = {\prd},
         year = 2022,
        month = oct,
       volume = {106},
       number = {8},
          eid = {083511},
        pages = {083511},
          doi = {10.1103/PhysRevD.106.083511},
archivePrefix = {arXiv},
       eprint = {2204.11507},
 primaryClass = {astro-ph.CO},
       adsurl = {https://ui.adsabs.harvard.edu/abs/2022PhRvD.106h3511C}
}

@ARTICLE{Blanco_2023,
       author = {{Sobral Blanco}, Daniel and {Bonvin}, Camille},
        title = "{Measuring the distortion of time with relativistic effects in large-scale structure}",
      journal = {\mnras},
         year = 2023,
        month = feb,
       volume = {519},
       number = {1},
        pages = {L39-L44},
          doi = {10.1093/mnrasl/slac124},
archivePrefix = {arXiv},
       eprint = {2205.02567},
 primaryClass = {astro-ph.CO},
       adsurl = {https://ui.adsabs.harvard.edu/abs/2023MNRAS.519L..39S}
}

@ARTICLE{Castello_2025,
       author = {{Castello}, Sveva and {Zheng}, Ziyang and {Bonvin}, Camille and {Amendola}, Luca},
        title = "{Testing the equivalence principle across the Universe: A model-independent approach with galaxy multitracing}",
      journal = {\prd},
         year = 2025,
        month = jun,
       volume = {111},
       number = {12},
          eid = {123559},
        pages = {123559},
          doi = {10.1103/1my7-zklj},
archivePrefix = {arXiv},
       eprint = {2412.08627},
 primaryClass = {astro-ph.CO},
       adsurl = {https://ui.adsabs.harvard.edu/abs/2025PhRvD.111l3559C}
}

@ARTICLE{Bonvin_2020,
       author = {{Bonvin}, Camille and {Oliveira Franco}, Felipe and {Fleury}, Pierre},
        title = "{A null test of the equivalence principle using relativistic effects in galaxy surveys}",
      journal = {\jcap},
         year = 2020,
        month = aug,
       volume = {2020},
       number = {8},
          eid = {004},
        pages = {004},
          doi = {10.1088/1475-7516/2020/08/004},
archivePrefix = {arXiv},
       eprint = {2004.06457},
 primaryClass = {astro-ph.CO},
       adsurl = {https://ui.adsabs.harvard.edu/abs/2020JCAP...08..004B}
}

@ARTICLE{Lombriser_2013,
       author = {{Lombriser}, Lucas and {Yoo}, Jaiyul and {Koyama}, Kazuya},
        title = "{Relativistic effects in galaxy clustering in a parametrized post-Friedmann universe}",
      journal = {\prd},
         year = 2013,
        month = may,
       volume = {87},
       number = {10},
          eid = {104019},
        pages = {104019},
          doi = {10.1103/PhysRevD.87.104019},
archivePrefix = {arXiv},
       eprint = {1301.3132},
 primaryClass = {astro-ph.CO},
       adsurl = {https://ui.adsabs.harvard.edu/abs/2013PhRvD..87j4019L}
}

@ARTICLE{Bonvin_2018,
       author = {{Bonvin}, Camille and {Fleury}, Pierre},
        title = "{Testing the equivalence principle on cosmological scales}",
      journal = {\jcap},
         year = 2018,
        month = may,
       volume = {2018},
       number = {5},
          eid = {061},
        pages = {061},
          doi = {10.1088/1475-7516/2018/05/061},
archivePrefix = {arXiv},
       eprint = {1803.02771},
 primaryClass = {astro-ph.CO},
       adsurl = {https://ui.adsabs.harvard.edu/abs/2018JCAP...05..061B}
}

@ARTICLE{Novara_2025,
   title={Detecting local and integrated relativistic effects by multi-tracing a single galaxy population in harmonic space},
   volume={50},
   ISSN={2212-6864},
   url={http://dx.doi.org/10.1016/j.dark.2025.102160},
   DOI={10.1016/j.dark.2025.102160},
   journal={Physics of the Dark Universe},
   publisher={Elsevier BV},
   author={Novara, Marco and Montano, Federico and Camera, Stefano},
   year={2025},
   month=dec, pages={102160} 
}

@ARTICLE{Gaztanaga_2017,
       author = {{Gaztanaga}, Enrique and {Bonvin}, Camille and {Hui}, Lam},
        title = "{Measurement of the dipole in the cross-correlation function of galaxies}",
      journal = {\jcap},
         year = 2017,
        month = jan,
       volume = {2017},
       number = {1},
          eid = {032},
        pages = {032},
          doi = {10.1088/1475-7516/2017/01/032},
archivePrefix = {arXiv},
       eprint = {1512.03918},
 primaryClass = {astro-ph.CO},
       adsurl = {https://ui.adsabs.harvard.edu/abs/2017JCAP...01..032G}
}

@ARTICLE{Bonvin_2016,
       author = {{Bonvin}, Camille and {Hui}, Lam and {Gaztanaga}, Enrique},
        title = "{Optimising the measurement of relativistic distortions in large-scale structure}",
      journal = {\jcap},
         year = 2016,
        month = aug,
       volume = {2016},
       number = {8},
          eid = {021},
        pages = {021},
          doi = {10.1088/1475-7516/2016/08/021},
archivePrefix = {arXiv},
       eprint = {1512.03566},
 primaryClass = {astro-ph.CO},
       adsurl = {https://ui.adsabs.harvard.edu/abs/2016JCAP...08..021B}
}

@ARTICLE{Montano_2024b,
       author = {{Montano}, Federico and {Camera}, Stefano},
        title = "{Detecting relativistic Doppler by multi-tracing a single galaxy population}",
      journal = {Physics of the Dark Universe},
         year = 2024,
        month = dec,
       volume = {46},
          eid = {101634},
        pages = {101634},
          doi = {10.1016/j.dark.2024.101634},
archivePrefix = {arXiv},
       eprint = {2407.06284},
 primaryClass = {astro-ph.CO},
       adsurl = {https://ui.adsabs.harvard.edu/abs/2024PDU....4601634M}
}

@ARTICLE{Montano_2024a,
       author = {{Montano}, Federico and {Camera}, Stefano},
        title = "{Detecting relativistic Doppler in galaxy clustering with tailored galaxy samples}",
      journal = {Physics of the Dark Universe},
         year = 2024,
        month = dec,
       volume = {46},
          eid = {101570},
        pages = {101570},
          doi = {10.1016/j.dark.2024.101570},
archivePrefix = {arXiv},
       eprint = {2309.12400},
 primaryClass = {astro-ph.CO},
       adsurl = {https://ui.adsabs.harvard.edu/abs/2024PDU....4601570M}
}

@ARTICLE{Wotjak_2011,
       author = {{Wojtak}, Rados{\l}aw and {Hansen}, Steen H. and {Hjorth}, Jens},
        title = "{Gravitational redshift of galaxies in clusters as predicted by general relativity}",
      journal = {\nat},
         year = 2011,
        month = sep,
       volume = {477},
       number = {7366},
        pages = {567-569},
          doi = {10.1038/nature10445},
archivePrefix = {arXiv},
       eprint = {1109.6571},
 primaryClass = {astro-ph.CO},
       adsurl = {https://ui.adsabs.harvard.edu/abs/2011Natur.477..567W}
}

@ARTICLE{Iftach_2015,
       author = {{Sadeh}, Iftach and {Feng}, Low Lerh and {Lahav}, Ofer},
        title = "{Gravitational Redshift of Galaxies in Clusters from the Sloan Digital Sky Survey and the Baryon Oscillation Spectroscopic Survey}",
      journal = {\prl},
         year = 2015,
        month = feb,
       volume = {114},
       number = {7},
          eid = {071103},
        pages = {071103},
          doi = {10.1103/PhysRevLett.114.071103},
archivePrefix = {arXiv},
       eprint = {1410.5262},
 primaryClass = {astro-ph.CO},
       adsurl = {https://ui.adsabs.harvard.edu/abs/2015PhRvL.114g1103S}
}

@ARTICLE{euclid_2,
       author = {{Euclid Collaboration} and {Mellier}, Y. and {Abdurro'uf} and {Acevedo Barroso}, J.~A. and {Ach{\'u}carro}, A. and {Adamek}, J. and {Adam}, R. and {Addison}, G.~E. and {Aghanim}, N. and {Aguena}, M. and {Ajani}, V. and {Akrami}, Y. and {Al-Bahlawan}, A. and {Alavi}, A. and {Albuquerque}, I.~S. and {Alestas}, G. and {Alguero}, G. and {Allaoui}, A. and {Allen}, S.~W. and {Allevato}, V. and {Alonso-Tetilla}, A.~V. and {Altieri}, B. and {Alvarez-Candal}, A. and {Alvi}, S. and {Amara}, A. and {Amendola}, L. and {Amiaux}, J. and {Andika}, I.~T. and {Andreon}, S. and {Andrews}, A. and {Angora}, G. and {Angulo}, R.~E. and {Annibali}, F. and {Anselmi}, A. and {Anselmi}, S. and {Arcari}, S. and {Archidiacono}, M. and {Aric{\`o}}, G. and {Arnaud}, M. and {Arnouts}, S. and {Asgari}, M. and {Asorey}, J. and {Atayde}, L. and {Atek}, H. and {Atrio-Barandela}, F. and {Aubert}, M. and {Aubourg}, E. and {Auphan}, T. and {Auricchio}, N. and {Aussel}, B. and {Aussel}, H. and {Avelino}, P.~P. and {Avgoustidis}, A. and {Avila}, S. and {Awan}, S. and {Azzollini}, R. and {Baccigalupi}, C. and {Bachelet}, E. and {Bacon}, D. and {Baes}, M. and {Bagley}, M.~B. and {Bahr-Kalus}, B. and {Balaguera-Antolinez}, A. and {Balbinot}, E. and {Balcells}, M. and {Baldi}, M. and {Baldry}, I. and {Balestra}, A. and {Ballardini}, M. and {Ballester}, O. and {Balogh}, M. and {Ba{\~n}ados}, E. and {Barbier}, R. and {Bardelli}, S. and {Baron}, M. and {Barreiro}, T. and {Barrena}, R. and {Barriere}, J. -C. and {Barros}, B.~J. and {Barthelemy}, A. and {Bartolo}, N. and {Basset}, A. and {Battaglia}, P. and {Battisti}, A.~J. and {Baugh}, C.~M. and {Baumont}, L. and {Bazzanini}, L. and {Beaulieu}, J. -P. and {Beckmann}, V. and {Belikov}, A.~N. and {Bel}, J. and {Bellagamba}, F. and {Bella}, M. and {Bellini}, E. and {Benabed}, K. and {Bender}, R. and {Benevento}, G. and {Bennett}, C.~L. and {Benson}, K. and {Bergamini}, P. and {Bermejo-Climent}, J.~R. and {Bernardeau}, F. and {Bertacca}, D. and {Berthe}, M. and {Berthier}, J. and {Bethermin}, M. and {Beutler}, F. and {Bevillon}, C. and {Bhargava}, S. and {Bhatawdekar}, R. and {Bianchi}, D. and {Bisigello}, L. and {Biviano}, A. and {Blake}, R.~P. and {Blanchard}, A. and {Blazek}, J. and {Blot}, L. and {Bosco}, A. and {Bodendorf}, C. and {Boenke}, T. and {B{\"o}hringer}, H. and {Boldrini}, P. and {Bolzonella}, M. and {Bonchi}, A. and {Bonici}, M. and {Bonino}, D. and {Bonino}, L. and {Bonvin}, C. and {Bon}, W. and {Booth}, J.~T. and {Borgani}, S. and {Borlaff}, A.~S. and {Borsato}, E. and {Bose}, B. and {Botticella}, M.~T. and {Boucaud}, A. and {Bouche}, F. and {Boucher}, J.~S. and {Boutigny}, D. and {Bouvard}, T. and {Bouwens}, R. and {Bouy}, H. and {Bowler}, R.~A.~A. and {Bozza}, V. and {Bozzo}, E. and {Branchini}, E. and {Brando}, G. and {Brau-Nogue}, S. and {Brekke}, P. and {Bremer}, M.~N. and {Brescia}, M. and {Breton}, M. -A. and {Brinchmann}, J. and {Brinckmann}, T. and {Brockley-Blatt}, C. and {Brodwin}, M. and {Brouard}, L. and {Brown}, M.~L. and {Bruton}, S. and {Bucko}, J. and {Buddelmeijer}, H. and {Buenadicha}, G. and {Buitrago}, F. and {Burger}, P. and {Burigana}, C. and {Busillo}, V. and {Busonero}, D. and {Cabanac}, R. and {Cabayol-Garcia}, L. and {Cagliari}, M.~S. and {Caillat}, A. and {Caillat}, L. and {Calabrese}, M. and {Calabro}, A. and {Calderone}, G. and {Calura}, F. and {Camacho Quevedo}, B. and {Camera}, S. and {Campos}, L. and {Ca{\~n}as-Herrera}, G. and {Candini}, G.~P. and {Cantiello}, M. and {Capobianco}, V. and {Cappellaro}, E. and {Cappelluti}, N. and {Cappi}, A. and {Caputi}, K.~I. and {Cara}, C. and {Carbone}, C. and {Cardone}, V.~F. and {Carella}, E. and {Carlberg}, R.~G. and {Carle}, M. and {Carminati}, L. and {Caro}, F. and {Carrasco}, J.~M. and {Carretero}, J. and {Carrilho}, P. and {Carron Duque}, J. and {Carry}, B.},
        title = "{Euclid: I. Overview of the Euclid mission}",
      journal = {\aap},
         year = 2025,
        month = may,
       volume = {697},
          eid = {A1},
        pages = {A1},
          doi = {10.1051/0004-6361/202450810},
archivePrefix = {arXiv},
       eprint = {2405.13491},
 primaryClass = {astro-ph.CO},
       adsurl = {https://ui.adsabs.harvard.edu/abs/2025A&A...697A...1E}
}

@ARTICLE{Bull_2024,
       author = {{Bull}, Philip and {Guandalin}, Caroline and {Addis}, Chris},
        title = "{Modes of the Dark Ages 21 cm field accessible to a lunar radio interferometer}",
      journal = {Philosophical Transactions of the Royal Society of London Series A},
         year = 2024,
        month = may,
       volume = {382},
       number = {2271},
          eid = {20230072},
        pages = {20230072},
          doi = {10.1098/rsta.2023.0072},
archivePrefix = {arXiv},
       eprint = {2403.16955},
 primaryClass = {astro-ph.CO},
       adsurl = {https://ui.adsabs.harvard.edu/abs/2024RSPTA.38230072B}
}

@ARTICLE{Montano_2026,
       author = {{Montano}, Federico and {Rossiter}, Samantha J. and {Addis}, Chris and {Hammond}, Jessie and {Camera}, Stefano and {Clarkson}, Chris and {Yousry Elkhashab}, Mohamed and {Guidi}, Massimo and {Lahav}, Ofer and {Aric{\`o}}, Giovanni and {Contarini}, Sofia and {Dayal}, Pratika and {Degni}, Giulia and {Farina}, Antonio and {Ir{\v{s}}i{\v{c}}}, Vid and {Marulli}, Federico and {Sarpa}, Elena and {Sartori}, Simone and {Sefusatti}, Emiliano and {Verdiani}, Francesco and {Verza}, Giovanni},
        title = "{Probing General Relativity on Cosmological Scales in the 2040s}",
      journal = {arXiv e-prints},
         year = 2026,
        month = jan,
          eid = {arXiv:2601.05320},
        pages = {arXiv:2601.05320},
          doi = {10.48550/arXiv.2601.05320},
archivePrefix = {arXiv},
       eprint = {2601.05320},
 primaryClass = {astro-ph.CO},
       adsurl = {https://ui.adsabs.harvard.edu/abs/2026arXiv260105320M}
}

@ARTICLE{Rossiter_2026,
       author = {{Rossiter}, Samantha Josephine and {Camera}, Stefano and {Montano}, Federico and {Clarkson}, Chris and {Karagiannis}, Dionysios and {Maartens}, Roy},
        title = "{Halving and Doubling: Boosting the Detection of Relativistic Effects in the Galaxy Bispectrum with Optimal Subsample Selection}",
      journal = {arXiv e-prints},
         year = 2026,
        month = feb,
          eid = {arXiv:2603.00244},
        pages = {arXiv:2603.00244},
          doi = {10.48550/arXiv.2603.00244},
archivePrefix = {arXiv},
       eprint = {2603.00244},
 primaryClass = {astro-ph.CO},
       adsurl = {https://ui.adsabs.harvard.edu/abs/2026arXiv260300244R}
}

@ARTICLE{WST_2024,
       author = {{Mainieri}, Vincenzo and {Anderson}, Richard I. and {Brinchmann}, Jarle and {Cimatti}, Andrea and {Ellis}, Richard S. and {Hill}, Vanessa and {Kneib}, Jean-Paul and {McLeod}, Anna F. and {Opitom}, Cyrielle and {Roth}, Martin M. and {Sanchez-Saez}, Paula and {Smiljanic}, Rodolfo and {Tolstoy}, Eline and {Bacon}, Roland and {Randich}, Sofia and {Adamo}, Angela and {Annibali}, Francesca and {Arevalo}, Patricia and {Audard}, Marc and {Barsanti}, Stefania and {Battaglia}, Giuseppina and {Bayo Aran}, Amelia M. and {Belfiore}, Francesco and {Bellazzini}, Michele and {Bellini}, Emilio and {Beltran}, Maria Teresa and {Berni}, Leda and {Bianchi}, Simone and {Biazzo}, Katia and {Bisero}, Sofia and {Bisogni}, Susanna and {Bland-Hawthorn}, Joss and {Blondin}, Stephane and {Bodensteiner}, Julia and {Boffin}, Henri M.~J. and {Bonito}, Rosaria and {Bono}, Giuseppe and {Bouche}, Nicolas F. and {Bowman}, Dominic and {Braga}, Vittorio F. and {Bragaglia}, Angela and {Branchesi}, Marica and {Brucalassi}, Anna and {Bryant}, Julia J. and {Bryson}, Ian and {Busa}, Innocenza and {Camera}, Stefano and {Carbone}, Carmelita and {Casali}, Giada and {Casali}, Mark and {Casasola}, Viviana and {Castro}, Norberto and {Catelan}, Marcio and {Cavallo}, Lorenzo and {Chiappini}, Cristina and {Cioni}, Maria-Rosa and {Colless}, Matthew and {Colzi}, Laura and {Contarini}, Sofia and {Couch}, Warrick and {D'Ammando}, Filippo and {d'Assignies D.}, William and {D'Orazi}, Valentina and {da Silva}, Ronaldo and {Dainotti}, Maria Giovanna and {Damiani}, Francesco and {Danielski}, Camilla and {De Cia}, Annalisa and {de Jong}, Roelof S. and {Dhawan}, Suhail and {Dierickx}, Philippe and {Driver}, Simon P. and {Dupletsa}, Ulyana and {Escoffier}, Stephanie and {Escorza}, Ana and {Fabrizio}, Michele and {Fiorentino}, Giuliana and {Fontana}, Adriano and {Fontani}, Francesco and {Forero Sanchez}, Daniel and {Franois}, Patrick and {Galindo-Guil}, Francisco Jose and {Gallazzi}, Anna Rita and {Galli}, Daniele and {Garcia}, Miriam and {Garcia-Rojas}, Jorge and {Garilli}, Bianca and {Grand}, Robert and {Guarcello}, Mario Giuseppe and {Hazra}, Nandini and {Helmi}, Amina and {Herrero}, Artemio and {Iglesias}, Daniela and {Ilic}, Dragana and {Irsic}, Vid and {Ivanov}, Valentin D. and {Izzo}, Luca and {Jablonka}, Pascale and {Joachimi}, Benjamin and {Kakkad}, Darshan and {Kamann}, Sebastian and {Koposov}, Sergey and {Kordopatis}, Georges and {Kovacevic}, Andjelka B. and {Kraljic}, Katarina and {Kuncarayakti}, Hanindyo and {Kwon}, Yuna and {La Forgia}, Fiorangela and {Lahav}, Ofer and {Laigle}, Clotilde and {Lazzarin}, Monica and {Leaman}, Ryan and {Leclercq}, Floriane and {Lee}, Khee-Gan and {Lee}, David and {Lehnert}, Matt D. and {Lira}, Paulina and {Loffredo}, Eleonora and {Lucatello}, Sara and {Magrini}, Laura and {Maguire}, Kate and {Mahler}, Guillaume and {Zahra Majidi}, Fatemeh and {Malavasi}, Nicola and {Mannucci}, Filippo and {Marconi}, Marcella and {Martin}, Nicolas and {Marulli}, Federico and {Massari}, Davide and {Matsuno}, Tadafumi and {Mattheee}, Jorryt and {McGee}, Sean and {Merc}, Jaroslav and {Merle}, Thibault and {Miglio}, Andrea and {Migliorini}, Alessandra and {Minchev}, Ivan and {Minniti}, Dante and {Miret-Roig}, Nuria and {Monreal Ibero}, Ana and {Montano}, Federico and {Montet}, Ben T. and {Moresco}, Michele and {Moretti}, Chiara and {Moscardini}, Lauro and {Moya}, Andres and {Mueller}, Oliver and {Nanayakkara}, Themiya and {Nicholl}, Matt and {Nordlander}, Thomas and {Onori}, Francesca and {Padovani}, Marco and {Pala}, Anna Francesca and {Panda}, Swayamtrupta and {Pandey-Pommier}, Mamta and {Pasquini}, Luca and {Pawlak}, Michal and {Pessi}, Priscila J. and {Pisani}, Alice and {Popovic}, Lukav C. and {Prisinzano}, Loredana and {Raddi}, Roberto and {Rainer}, Monica and {Rebassa-Mansergas}, Alberto and {Richard}, Johan and {Rigault}, Mickael and {Rocher}, Antoine and {Romano}, Donatella and {Rosati}, Piero and {Sacco}, Germano and {Sanchez-Janssen}, Ruben and {Sander}, Andreas A.~C. and {Sanders}, Jason L. and {Sargent}, Mark and {Sarpa}, Elena and {Schimd}, Carlo and {Schipani}, Pietro and {Sefusatti}, Emiliano and {Smith}, Graham P. and {Spina}, Lorenzo and {Steinmetz}, Matthias and {Tacchella}, Sandro and {Tautvaisiene}, Grazina and {Theissen}, Christopher and {Thomas}, Guillaume and {Ting}, Yuan-Sen and {Travouillon}, Tony and {Tresse}, Laurence and {Trivedi}, Oem and {Tsantaki}, Maria and {Tsedrik}, Maria and {Urrutia}, Tanya and {Valenti}, Elena and {Van der Swaelmen}, Mathieu and {Van Eck}, Sophie and {Verdiani}, Francesco and {Verdier}, Aurelien and {Vergani}, Susanna Diana and {Verhamme}, Anne and {Vernet}, Joel},
        title = "{The Wide-field Spectroscopic Telescope (WST) Science White Paper}",
      journal = {arXiv e-prints},
         year = 2024,
        month = mar,
          eid = {arXiv:2403.05398},
        pages = {arXiv:2403.05398},
          doi = {10.48550/arXiv.2403.05398},
archivePrefix = {arXiv},
       eprint = {2403.05398},
 primaryClass = {astro-ph.IM},
       adsurl = {https://ui.adsabs.harvard.edu/abs/2024arXiv240305398M}
}

@ARTICLE{inflation_2022,
       author = {{Ach{\'u}carro}, Ana and {Biagetti}, Matteo and {Braglia}, Matteo and {Cabass}, Giovanni and {Caldwell}, Robert and {Castorina}, Emanuele and {Chen}, Xingang and {Coulton}, William and {Flauger}, Raphael and {Fumagalli}, Jacopo and {Ivanov}, Mikhail M. and {Lee}, Hayden and {Maleknejad}, Azadeh and {Meerburg}, P. Daniel and {Moradinezhad Dizgah}, Azadeh and {Palma}, Gonzalo A. and {Pimentel}, Guilherme L. and {Renaux-Petel}, S{\'e}bastien and {Wallisch}, Benjamin and {Wandelt}, Benjamin D. and {Witkowski}, Lukas T. and {Kimmy Wu}, W.~L.},
        title = "{Inflation: Theory and Observations}",
      journal = {arXiv e-prints},
         year = 2022,
        month = mar,
          eid = {arXiv:2203.08128},
        pages = {arXiv:2203.08128},
          doi = {10.48550/arXiv.2203.08128},
archivePrefix = {arXiv},
       eprint = {2203.08128},
 primaryClass = {astro-ph.CO},
       adsurl = {https://ui.adsabs.harvard.edu/abs/2022arXiv220308128A}
}

@ARTICLE{Feldman_1994,
       author = {{Feldman}, Hume A. and {Kaiser}, Nick and {Peacock}, John A.},
        title = "{Power-Spectrum Analysis of Three-dimensional Redshift Surveys}",
      journal = {\apj},
         year = 1994,
        month = may,
       volume = {426},
        pages = {23},
          doi = {10.1086/174036},
archivePrefix = {arXiv},
       eprint = {astro-ph/9304022},
 primaryClass = {astro-ph},
       adsurl = {https://ui.adsabs.harvard.edu/abs/1994ApJ...426...23F}
}

@ARTICLE{Beutler_2020,
       author = {{Beutler}, Florian and {Di Dio}, Enea},
        title = "{Modeling relativistic contributions to the halo power spectrum dipole}",
      journal = {\jcap},
         year = 2020,
        month = jul,
       volume = {2020},
       number = {7},
          eid = {048},
        pages = {048},
          doi = {10.1088/1475-7516/2020/07/048},
archivePrefix = {arXiv},
       eprint = {2004.08014},
 primaryClass = {astro-ph.CO},
       adsurl = {https://ui.adsabs.harvard.edu/abs/2020JCAP...07..048B}
}

@ARTICLE{Yoo_2015,
       author = {{Yoo}, Jaiyul and {Seljak}, Uro{\v{s}}},
        title = "{Wide-angle effects in future galaxy surveys}",
      journal = {\mnras},
         year = 2015,
        month = feb,
       volume = {447},
       number = {2},
        pages = {1789-1805},
          doi = {10.1093/mnras/stu2491},
archivePrefix = {arXiv},
       eprint = {1308.1093},
 primaryClass = {astro-ph.CO},
       adsurl = {https://ui.adsabs.harvard.edu/abs/2015MNRAS.447.1789Y}
}

@article{Reimberg_2016,
        author = {{Reimberg}, Paulo and {Bernardeau}, Francis and {Pitrou}, Cyril},
        title = "{Redshift-space distortions with wide angular separations}",
      journal = {\jcap},
         year = 2016,
        month = jan,
       volume = {2016},
       number = {1},
        pages = {048-048},
          doi = {10.1088/1475-7516/2016/01/048},
archivePrefix = {arXiv},
       eprint = {1506.06596},
 primaryClass = {astro-ph.CO},
       adsurl = {https://ui.adsabs.harvard.edu/abs/2016JCAP...01..048R}
}

@ARTICLE{Challinor_2011,
       author = {{Challinor}, Anthony and {Lewis}, Antony},
        title = "{Linear power spectrum of observed source number counts}",
      journal = {\prd},
         year = 2011,
        month = aug,
       volume = {84},
       number = {4},
          eid = {043516},
        pages = {043516},
          doi = {10.1103/PhysRevD.84.043516},
archivePrefix = {arXiv},
       eprint = {1105.5292},
 primaryClass = {astro-ph.CO},
       adsurl = {https://ui.adsabs.harvard.edu/abs/2011PhRvD..84d3516C}
}

@ARTICLE{Macdonald_2009,
       author = {{McDonald}, Patrick},
        title = "{Gravitational redshift and other redshift-space distortions of the imaginary part of the power spectrum}",
      journal = {\jcap},
         year = 2009,
        month = nov,
       volume = {2009},
       number = {11},
          eid = {026},
        pages = {026},
          doi = {10.1088/1475-7516/2009/11/026},
archivePrefix = {arXiv},
       eprint = {0907.5220},
 primaryClass = {astro-ph.CO},
       adsurl = {https://ui.adsabs.harvard.edu/abs/2009JCAP...11..026M}
}

@ARTICLE{Zheng_2005,
       author = {{Zheng}, Zheng and {Berlind}, Andreas A. and {Weinberg}, David H. and {Benson}, Andrew J. and {Baugh}, Carlton M. and {Cole}, Shaun and {Dav{\'e}}, Romeel and {Frenk}, Carlos S. and {Katz}, Neal and {Lacey}, Cedric G.},
        title = "{Theoretical Models of the Halo Occupation Distribution: Separating Central and Satellite Galaxies}",
      journal = {\apj},
         year = 2005,
        month = nov,
       volume = {633},
       number = {2},
        pages = {791-809},
          doi = {10.1086/466510},
archivePrefix = {arXiv},
       eprint = {astro-ph/0408564},
 primaryClass = {astro-ph},
       adsurl = {https://ui.adsabs.harvard.edu/abs/2005ApJ...633..791Z}
}

@ARTICLE{Zehavi_2011,
       author = {{Zehavi}, Idit and {Zheng}, Zheng and {Weinberg}, David H. and {Blanton}, Michael R. and {Bahcall}, Neta A. and {Berlind}, Andreas A. and {Brinkmann}, Jon and {Frieman}, Joshua A. and {Gunn}, James E. and {Lupton}, Robert H. and {Nichol}, Robert C. and {Percival}, Will J. and {Schneider}, Donald P. and {Skibba}, Ramin A. and {Strauss}, Michael A. and {Tegmark}, Max and {York}, Donald G.},
        title = "{Galaxy Clustering in the Completed SDSS Redshift Survey: The Dependence on Color and Luminosity}",
      journal = {\apj},
         year = 2011,
        month = jul,
       volume = {736},
       number = {1},
          eid = {59},
        pages = {59},
          doi = {10.1088/0004-637X/736/1/59},
archivePrefix = {arXiv},
       eprint = {1005.2413},
 primaryClass = {astro-ph.CO},
       adsurl = {https://ui.adsabs.harvard.edu/abs/2011ApJ...736...59Z}
}

@ARTICLE{Smith_2017,
       author = {{Smith}, Alex and {Cole}, Shaun and {Baugh}, Carlton and {Zheng}, Zheng and {Angulo}, Ra{\'u}l and {Norberg}, Peder and {Zehavi}, Idit},
        title = "{A lightcone catalogue from the Millennium-XXL simulation}",
      journal = {\mnras},
         year = 2017,
        month = oct,
       volume = {470},
       number = {4},
        pages = {4646-4661},
          doi = {10.1093/mnras/stx1432},
archivePrefix = {arXiv},
       eprint = {1701.06581},
 primaryClass = {astro-ph.CO},
       adsurl = {https://ui.adsabs.harvard.edu/abs/2017MNRAS.470.4646S}
}

@ARTICLE{Loveday_2012,
       author = {{Loveday}, J. and {Norberg}, P. and {Baldry}, I.~K. and others},
        title = "{Galaxy and Mass Assembly (GAMA): ugriz galaxy luminosity functions}",
      journal = {\mnras},
         year = 2012,
       volume = {420},
        pages = {1239-1262},
          doi = {10.1111/j.1365-2966.2011.20111.x},
archivePrefix = {arXiv},
       eprint = {1111.0166},
 primaryClass = {astro-ph.CO}
}

@ARTICLE{Smith_2024,
       author = {{Smith}, A. and {Grove}, C. and {Cole}, S. and {Norberg}, P. and {Zarrouk}, P. and {Yuan}, S. and {Aguilar}, J. and {Ahlen}, S. and {Brooks}, D. and {Claybaugh}, T. and {de la Macorra}, A. and {Doel}, P. and {Forero-Romero}, J.~E. and {Gazta{\~n}aga}, E. and {Gontcho}, S. Gontcho A. and {Hahn}, C. and {Kehoe}, R. and {Kremin}, A. and {Levi}, M.~E. and {Manera}, M. and {Meisner}, A. and {Miquel}, R. and {Moustakas}, J. and {Nie}, J. and {Percival}, W.~J. and {Rezaie}, M. and {Rossi}, G. and {Sanchez}, E. and {Seo}, H. and {Tarl{\'e}}, G. and {Zhou}, Z.},
        title = "{Generating mock galaxy catalogues for flux-limited samples like the DESI Bright Galaxy Survey}",
      journal = {\mnras},
         year = 2024,
        month = jul,
       volume = {532},
       number = {1},
        pages = {903-919},
          doi = {10.1093/mnras/stae1503},
archivePrefix = {arXiv},
       eprint = {2312.08792},
 primaryClass = {astro-ph.CO},
       adsurl = {https://ui.adsabs.harvard.edu/abs/2024MNRAS.532..903S}
}

@ARTICLE{Smith_2009,
       author = {{Smith}, Robert E.},
        title = "{Covariance of cross-correlations: towards efficient measures for large-scale structure}",
      journal = {\mnras},
         year = 2009,
        month = dec,
       volume = {400},
       number = {2},
        pages = {851-865},
          doi = {10.1111/j.1365-2966.2009.15490.x},
archivePrefix = {arXiv},
       eprint = {0810.1960},
 primaryClass = {astro-ph},
       adsurl = {https://ui.adsabs.harvard.edu/abs/2009MNRAS.400..851S}
}

@ARTICLE{Addis_2025,
       author = {{Addis}, Chris and {Guedezounme}, S{\^e}cloka L. and {Hammond}, Jessie and {Clarkson}, Chris and {Montano}, Federico and {Camera}, Stefano and {Jolicoeur}, Sheean and {Maartens}, Roy},
        title = "{Unbiased analysis of primordial non-Gaussianity: the multipoles of the full relativistic power spectrum}",
      journal = {\jcap},
         year = 2026,
        month = jun,
       volume = {2026},
       number = {6},
          eid = {039},
        pages = {039},
          doi = {10.1088/1475-7516/2026/06/039},
archivePrefix = {arXiv},
       eprint = {2511.09466},
 primaryClass = {astro-ph.CO},
       adsurl = {https://ui.adsabs.harvard.edu/abs/2026JCAP...06..039A}
}

@ARTICLE{Oddo_2020,
       author = {{Oddo}, Andrea and {Sefusatti}, Emiliano and {Porciani}, Cristiano and {Monaco}, Pierluigi and {S{\'a}nchez}, Ariel G.},
        title = "{Toward a robust inference method for the galaxy bispectrum: likelihood function and model selection}",
      journal = {\jcap},
         year = 2020,
        month = mar,
       volume = {2020},
       number = {3},
          eid = {056},
        pages = {056},
          doi = {10.1088/1475-7516/2020/03/056},
archivePrefix = {arXiv},
       eprint = {1908.01774},
 primaryClass = {astro-ph.CO}
}

@ARTICLE{Rizzo_2023,
       author = {{Rizzo}, Federico and {Moretti}, Chiara and {Pardede}, Kevin and {Eggemeier}, Alexander and {Oddo}, Andrea and {Sefusatti}, Emiliano and {Porciani}, Cristiano and {Monaco}, Pierluigi},
        title = "{The halo bispectrum multipoles in redshift space}",
      journal = {\jcap},
         year = 2023,
        month = jan,
       volume = {2023},
       number = {1},
          eid = {031},
        pages = {031},
          doi = {10.1088/1475-7516/2023/01/031},
archivePrefix = {arXiv},
       eprint = {2204.13628},
 primaryClass = {astro-ph.CO},
       adsurl = {https://ui.adsabs.harvard.edu/abs/2023JCAP...01..031R}
}

@ARTICLE{Ivanov_2023,
       author = {{Ivanov}, Mikhail M. and {Philcox}, Oliver H.~E. and {Cabass}, Giovanni and {Nishimichi}, Takahiro and {Simonovi{\'c}}, Marko and {Zaldarriaga}, Matias},
        title = "{Cosmology with the galaxy bispectrum multipoles: Optimal estimation and application to BOSS data}",
      journal = {\prd},
         year = 2023,
        month = apr,
       volume = {107},
       number = {8},
          eid = {083515},
        pages = {083515},
          doi = {10.1103/PhysRevD.107.083515},
archivePrefix = {arXiv},
       eprint = {2302.04414},
 primaryClass = {astro-ph.CO},
       adsurl = {https://ui.adsabs.harvard.edu/abs/2023PhRvD.107h3515I}
}

@ARTICLE{Ono_2018,
       author = {{Ono}, Yoshiaki and {Ouchi}, Masami and {Harikane}, Yuichi and {Toshikawa}, Jun and {Rauch}, Michael and {Yuma}, Suraphong and {Sawicki}, Marcin and {Shibuya}, Takatoshi and {Shimasaku}, Kazuhiro and {Oguri}, Masamune and {Willott}, Chris and {Akhlaghi}, Mohammad and {Akiyama}, Masayuki and {Coupon}, Jean and {Kashikawa}, Nobunari and {Komiyama}, Yutaka and {Konno}, Akira and {Lin}, Lihwai and {Matsuoka}, Yoshiki and {Miyazaki}, Satoshi and {Nagao}, Tohru and {Nakajima}, Kimihiko and {Silverman}, John and {Tanaka}, Masayuki and {Taniguchi}, Yoshiaki and {Wang}, Shiang-Yu},
        title = "{Great Optically Luminous Dropout Research Using Subaru HSC (GOLDRUSH). I. UV luminosity functions at z {\ensuremath{\sim}} 4-7 derived with the half-million dropouts on the 100 deg$^{2}$ sky}",
      journal = {\pasj},
         year = 2018,
        month = jan,
       volume = {70},
          eid = {S10},
        pages = {S10},
          doi = {10.1093/pasj/psx103},
archivePrefix = {arXiv},
       eprint = {1704.06004},
 primaryClass = {astro-ph.GA},
       adsurl = {https://ui.adsabs.harvard.edu/abs/2018PASJ...70S..10O}
}

@ARTICLE{Schechter_1976,
       author = {{Schechter}, P.},
        title = "{An analytic expression for the luminosity function for galaxies.}",
      journal = {\apj},
         year = 1976,
        month = jan,
       volume = {203},
        pages = {297-306},
          doi = {10.1086/154079},
       adsurl = {https://ui.adsabs.harvard.edu/abs/1976ApJ...203..297S}
}

@ARTICLE{Cole_2000,
       author = {{Cole}, Shaun and {Lacey}, Cedric G. and {Baugh}, Carlton M. and {Frenk}, Carlos S.},
        title = "{Hierarchical galaxy formation}",
      journal = {\mnras},
         year = 2000,
        month = nov,
       volume = {319},
       number = {1},
        pages = {168-204},
          doi = {10.1046/j.1365-8711.2000.03879.x},
archivePrefix = {arXiv},
       eprint = {astro-ph/0007281},
 primaryClass = {astro-ph},
       adsurl = {https://ui.adsabs.harvard.edu/abs/2000MNRAS.319..168C}
}

@ARTICLE{Pan_2020,
       author = {{Pan}, Hengxing and {Obreschkow}, Danail and {Howlett}, Cullan and {Lagos}, Claudia del P. and {Elahi}, Pascal J. and {Baugh}, Carlton and {Gonzalez-Perez}, Violeta},
        title = "{Multiwavelength consensus of large-scale linear bias}",
      journal = {\mnras},
         year = 2020,
        month = mar,
       volume = {493},
       number = {1},
        pages = {747-764},
          doi = {10.1093/mnras/staa222},
archivePrefix = {arXiv},
       eprint = {1909.12069},
 primaryClass = {astro-ph.CO},
       adsurl = {https://ui.adsabs.harvard.edu/abs/2020MNRAS.493..747P}
}

@ARTICLE{Wilson_2019,
       author = {{Wilson}, M.~J. and {White}, Martin},
        title = "{Cosmology with dropout selection: straw-man surveys \& CMB lensing}",
      journal = {\jcap},
         year = 2019,
        month = oct,
       volume = {2019},
       number = {10},
          eid = {015},
        pages = {015},
          doi = {10.1088/1475-7516/2019/10/015},
archivePrefix = {arXiv},
       eprint = {1904.13378},
 primaryClass = {astro-ph.CO},
       adsurl = {https://ui.adsabs.harvard.edu/abs/2019JCAP...10..015W}
}

@article{wfirst,
      author = {{Spergel}, D. and {Gehrels}, N. and {Breckinridge}, J. and {Donahue}, M. and {Dressler}, A. and {Gaudi}, B.~S. and {Greene}, T. and {Guyon}, O. and {Hirata}, C. and {Kalirai}, J. and {Kasdin}, N.~J. and {Moos}, W. and {Perlmutter}, S. and {Postman}, M. and {Rauscher}, B. and {Rhodes}, J. and {Wang}, Y. and {Weinberg}, D. and {Centrella}, J. and {Traub}, W. and {Baltay}, C. and {Colbert}, J. and {Bennett}, D. and {Kiessling}, A. and {Macintosh}, B. and {Merten}, J. and {Mortonson}, M. and {Penny}, M. and {Rozo}, E. and {Savransky}, D. and {Stapelfeldt}, K. and {Zu}, Y. and {Baker}, C. and {Cheng}, E. and {Content}, D. and {Dooley}, J. and {Foote}, M. and {Goullioud}, R. and {Grady}, K. and {Jackson}, C. and {Kruk}, J. and {Levine}, M. and {Melton}, M. and {Peddie}, C. and {Ruffa}, J. and {Shaklan}, S.},
        title = "{Wide-Field InfraRed Survey Telescope-Astrophysics Focused Telescope Assets WFIRST-AFTA Final Report}",
      journal = {arXiv e-prints},
         year = 2013,
        month = may,
          eid = {arXiv:1305.5422},
        pages = {arXiv:1305.5422},
          doi = {10.48550/arXiv.1305.5422},
archivePrefix = {arXiv},
       eprint = {1305.5422},
 primaryClass = {astro-ph.IM},
       adsurl = {https://ui.adsabs.harvard.edu/abs/2013arXiv1305.5422S}
}

@article{Noorikuhani_2022,
       author = {{Noorikuhani}, Milad and {Scoccimarro}, Roman},
        title = "{Wide-angle and relativistic effects in Fourier-space clustering statistics}",
      journal = {\prd},
         year = 2023,
        month = apr,
       volume = {107},
       number = {8},
          eid = {083528},
        pages = {083528},
          doi = {10.1103/PhysRevD.107.083528},
archivePrefix = {arXiv},
       eprint = {2207.12383},
 primaryClass = {astro-ph.CO},
       adsurl = {https://ui.adsabs.harvard.edu/abs/2023PhRvD.107h3528N}
}

@article{castorina2018,
        author = {{Castorina}, Emanuele and {White}, Martin},
        title = "{Beyond the plane-parallel approximation for redshift surveys}",
      journal = {\mnras},
         year = 2018,
        month = jun,
       volume = {476},
       number = {4},
        pages = {4403-4417},
          doi = {10.1093/mnras/sty410},
archivePrefix = {arXiv},
       eprint = {1709.09730},
 primaryClass = {astro-ph.CO},
       adsurl = {https://ui.adsabs.harvard.edu/abs/2018MNRAS.476.4403C}
}

@article{beutler2014,
  title={The clustering of galaxies in the {SDSS-III} {B}aryon {O}scillation {S}pectroscopic {S}urvey: testing gravity with redshift space distortions using the power spectrum multipoles},
  author={Beutler, Florian and Saito, Shun and Seo, Hee-Jong and Brinkmann, Jon and Dawson, Kyle S and Eisenstein, Daniel J and Font-Ribera, Andreu and Ho, Shirley and McBride, Cameron K and Montesano, Francesco and others},
  journal={Monthly Notices of the Royal Astronomical Society},
  volume={443},
  number={2},
  pages={1065--1089},
  year={2014},
  publisher={Oxford University Press}
}

@ARTICLE{Beutler_2021,
       author = {{Beutler}, Florian and {McDonald}, Patrick},
        title = "{Unified galaxy power spectrum measurements from 6dFGS, BOSS, and eBOSS}",
      journal = {\jcap},
         year = 2021,
        month = nov,
       volume = {2021},
       number = {11},
          eid = {031},
        pages = {031},
          doi = {10.1088/1475-7516/2021/11/031},
archivePrefix = {arXiv},
       eprint = {2106.06324},
 primaryClass = {astro-ph.CO}
}

@article{beutler2019,
       author = {{Beutler}, Florian and {Castorina}, Emanuele and {Zhang}, Pierre},
        title = "{Interpreting measurements of the anisotropic galaxy power spectrum}",
      journal = {\jcap},
         year = 2019,
        month = mar,
       volume = {2019},
       number = {3},
          eid = {040},
        pages = {040},
          doi = {10.1088/1475-7516/2019/03/040},
archivePrefix = {arXiv},
       eprint = {1810.05051},
 primaryClass = {astro-ph.CO},
       adsurl = {https://ui.adsabs.harvard.edu/abs/2019JCAP...03..040B}
}

@article{tellarini2016,
	author = {{Tellarini}, Matteo and {Ross}, Ashley J. and {Tasinato}, Gianmassimo and {Wands}, David},
        title = "{Galaxy bispectrum, primordial non-Gaussianity and redshift space distortions}",
      journal = {\jcap},
         year = 2016,
        month = jun,
       volume = {2016},
       number = {6},
          eid = {014},
        pages = {014},
          doi = {10.1088/1475-7516/2016/06/014},
archivePrefix = {arXiv},
       eprint = {1603.06814},
 primaryClass = {astro-ph.CO},
       adsurl = {https://ui.adsabs.harvard.edu/abs/2016JCAP...06..014T}
}

@article{scoccimarro1999,
	author = {{Scoccimarro}, Rom{\'a}n and {Couchman}, H.~M.~P. and {Frieman}, Joshua A.},
        title = "{The Bispectrum as a Signature of Gravitational Instability in Redshift Space}",
      journal = {\apj},
         year = 1999,
        month = jun,
       volume = {517},
       number = {2},
        pages = {531-540},
          doi = {10.1086/307220},
archivePrefix = {arXiv},
       eprint = {astro-ph/9808305},
 primaryClass = {astro-ph},
       adsurl = {https://ui.adsabs.harvard.edu/abs/1999ApJ...517..531S}
}

@INPROCEEDINGS{hamilton1998,
        author = {{Hamilton}, A.~J.~S.},
        title = "{Linear Redshift Distortions: a Review}",
    booktitle = {The Evolving Universe},
         year = 1998,
       editor = {{Hamilton}, Donald},
       series = {Astrophysics and Space Science Library},
       volume = {231},
        month = jan,
        journal={The Evolving Universe: Selected Topics on Large-Scale Structure and on the Properties of Galaxies},
        pages = {185},
          doi = {10.1007/978-94-011-4960-0_17},
archivePrefix = {arXiv},
       eprint = {astro-ph/9708102},
 primaryClass = {astro-ph},
       adsurl = {https://ui.adsabs.harvard.edu/abs/1998ASSL..231..185H}
}

@article{Scoccimarro_2015,
       author = {{Scoccimarro}, Rom{\'a}n},
        title = "{Fast estimators for redshift-space clustering}",
      journal = {\prd},
         year = 2015,
        month = oct,
       volume = {92},
       number = {8},
          eid = {083532},
        pages = {083532},
          doi = {10.1103/PhysRevD.92.083532},
archivePrefix = {arXiv},
       eprint = {1506.02729},
 primaryClass = {astro-ph.CO},
       adsurl = {https://ui.adsabs.harvard.edu/abs/2015PhRvD..92h3532S}
}

@article{Zaroubi_1993,
       author = {{Zaroubi}, Saleem and {Hoffman}, Yehuda},
        title = "{Clustering in Redshift Space: Linear Theory}",
      journal = {\apj},
         year = 1996,
        month = may,
       volume = {462},
        pages = {25},
          doi = {10.1086/177124},
archivePrefix = {arXiv},
       eprint = {astro-ph/9311013},
 primaryClass = {astro-ph},
       adsurl = {https://ui.adsabs.harvard.edu/abs/1993astro.ph.11013Z}
}

@ARTICLE{Pardede_2023,
       author = {{Pardede}, Kevin and {Di Dio}, Enea and {Castorina}, Emanuele},
        title = "{Wide-angle effects in the galaxy bispectrum}",
      journal = {\jcap},
         year = 2023,
        month = sep,
       volume = {2023},
       number = {9},
          eid = {030},
        pages = {030},
          doi = {10.1088/1475-7516/2023/09/030},
archivePrefix = {arXiv},
       eprint = {2302.12789},
 primaryClass = {astro-ph.CO},
       adsurl = {https://ui.adsabs.harvard.edu/abs/2023JCAP...09..030P}
}

@article{Sugiyama_2018,
       author = {{Sugiyama}, Naonori S. and {Saito}, Shun and {Beutler}, Florian and {Seo}, Hee-Jong},
        title = "{A complete FFT-based decomposition formalism for the redshift-space bispectrum}",
      journal = {\mnras},
         year = 2019,
        month = mar,
       volume = {484},
       number = {1},
        pages = {364-384},
          doi = {10.1093/mnras/sty3249},
archivePrefix = {arXiv},
       eprint = {1803.02132},
 primaryClass = {astro-ph.CO},
       adsurl = {https://ui.adsabs.harvard.edu/abs/2019MNRAS.484..364S}
}

@article{Bianchi_2015,
       author = {{Bianchi}, Davide and {Gil-Mar{\'\i}n}, H{\'e}ctor and {Ruggeri}, Rossana and {Percival}, Will J.},
        title = "{Measuring line-of-sight-dependent Fourier-space clustering using FFTs}",
      journal = {\mnras},
         year = 2015,
        month = oct,
       volume = {453},
       number = {1},
        pages = {L11-L15},
          doi = {10.1093/mnrasl/slv090},
archivePrefix = {arXiv},
       eprint = {1505.05341},
 primaryClass = {astro-ph.CO},
       adsurl = {https://ui.adsabs.harvard.edu/abs/2015MNRAS.453L..11B}
}

@article{Bernardeau_2001,
       author = {{Bernardeau}, F. and {Colombi}, S. and {Gazta{\~n}aga}, E. and {Scoccimarro}, R.},
        title = "{Large-scale structure of the Universe and cosmological perturbation theory}",
      journal = {\physrep},
         year = 2002,
        month = sep,
       volume = {367},
       number = {1-3},
        pages = {1-248},
          doi = {10.1016/S0370-1573(02)00135-7},
archivePrefix = {arXiv},
       eprint = {astro-ph/0112551},
 primaryClass = {astro-ph},
       adsurl = {https://ui.adsabs.harvard.edu/abs/2002PhR...367....1B}
}

@article{euclid,
      author = {{Laureijs}, R. and {Amiaux}, J. and {Arduini}, S. and {Augu{\`e}res}, J. -L. and {Brinchmann}, J. and {Cole}, R. and {Cropper}, M. and {Dabin}, C. and {Duvet}, L. and {Ealet}, A. and {Garilli}, B. and {Gondoin}, P. and {Guzzo}, L. and {Hoar}, J. and {Hoekstra}, H. and {Holmes}, R. and {Kitching}, T. and {Maciaszek}, T. and {Mellier}, Y. and {Pasian}, F. and {Percival}, W. and {Rhodes}, J. and {Saavedra Criado}, G. and {Sauvage}, M. and {Scaramella}, R. and {Valenziano}, L. and {Warren}, S. and {Bender}, R. and {Castander}, F. and {Cimatti}, A. and {Le F{\`e}vre}, O. and {Kurki-Suonio}, H. and {Levi}, M. and {Lilje}, P. and {Meylan}, G. and {Nichol}, R. and {Pedersen}, K. and {Popa}, V. and {Rebolo Lopez}, R. and {Rix}, H. -W. and {Rottgering}, H. and {Zeilinger}, W. and {Grupp}, F. and {Hudelot}, P. and {Massey}, R. and {Meneghetti}, M. and {Miller}, L. and {Paltani}, S. and {Paulin-Henriksson}, S. and {Pires}, S. and {Saxton}, C. and {Schrabback}, T. and {Seidel}, G. and {Walsh}, J. and {Aghanim}, N. and {Amendola}, L. and {Bartlett}, J. and {Baccigalupi}, C. and {Beaulieu}, J. -P. and {Benabed}, K. and {Cuby}, J. -G. and {Elbaz}, D. and {Fosalba}, P. and {Gavazzi}, G. and {Helmi}, A. and {Hook}, I. and {Irwin}, M. and {Kneib}, J. -P. and {Kunz}, M. and {Mannucci}, F. and {Moscardini}, L. and {Tao}, C. and {Teyssier}, R. and {Weller}, J. and {Zamorani}, G. and {Zapatero Osorio}, M.~R. and {Boulade}, O. and {Foumond}, J.~J. and {Di Giorgio}, A. and {Guttridge}, P. and {James}, A. and {Kemp}, M. and {Martignac}, J. and {Spencer}, A. and {Walton}, D. and {Bl{\"u}mchen}, T. and {Bonoli}, C. and {Bortoletto}, F. and {Cerna}, C. and {Corcione}, L. and {Fabron}, C. and {Jahnke}, K. and {Ligori}, S. and {Madrid}, F. and {Martin}, L. and {Morgante}, G. and {Pamplona}, T. and {Prieto}, E. and {Riva}, M. and {Toledo}, R. and {Trifoglio}, M. and {Zerbi}, F. and {Abdalla}, F. and {Douspis}, M. and {Grenet}, C. and {Borgani}, S. and {Bouwens}, R. and {Courbin}, F. and {Delouis}, J. -M. and {Dubath}, P. and {Fontana}, A. and {Frailis}, M. and {Grazian}, A. and {Koppenh{\"o}fer}, J. and {Mansutti}, O. and {Melchior}, M. and {Mignoli}, M. and {Mohr}, J. and {Neissner}, C. and {Noddle}, K. and {Poncet}, M. and {Scodeggio}, M. and {Serrano}, S. and {Shane}, N. and {Starck}, J. -L. and {Surace}, C. and {Taylor}, A. and {Verdoes-Kleijn}, G. and {Vuerli}, C. and {Williams}, O.~R. and {Zacchei}, A. and {Altieri}, B. and {Escudero Sanz}, I. and {Kohley}, R. and {Oosterbroek}, T. and {Astier}, P. and {Bacon}, D. and {Bardelli}, S. and {Baugh}, C. and {Bellagamba}, F. and {Benoist}, C. and {Bianchi}, D. and {Biviano}, A. and {Branchini}, E. and {Carbone}, C. and {Cardone}, V. and {Clements}, D. and {Colombi}, S. and {Conselice}, C. and {Cresci}, G. and {Deacon}, N. and {Dunlop}, J. and {Fedeli}, C. and {Fontanot}, F. and {Franzetti}, P. and {Giocoli}, C. and {Garcia-Bellido}, J. and {Gow}, J. and {Heavens}, A. and {Hewett}, P. and {Heymans}, C. and {Holland}, A. and {Huang}, Z. and {Ilbert}, O. and {Joachimi}, B. and {Jennins}, E. and {Kerins}, E. and {Kiessling}, A. and {Kirk}, D. and {Kotak}, R. and {Krause}, O. and {Lahav}, O. and {van Leeuwen}, F. and {Lesgourgues}, J. and {Lombardi}, M. and {Magliocchetti}, M. and {Maguire}, K. and {Majerotto}, E. and {Maoli}, R. and {Marulli}, F. and {Maurogordato}, S. and {McCracken}, H. and {McLure}, R. and {Melchiorri}, A. and {Merson}, A. and {Moresco}, M. and {Nonino}, M. and {Norberg}, P. and {Peacock}, J. and {Pello}, R. and {Penny}, M. and {Pettorino}, V. and {Di Porto}, C. and {Pozzetti}, L. and {Quercellini}, C. and {Radovich}, M. and {Rassat}, A. and {Roche}, N. and {Ronayette}, S. and {Rossetti}, E. and {Sartoris}, B. and {Schneider}, P. and {Semboloni}, E. and {Serjeant}, S. and {Simpson}, F. and {Skordis}, C. and {Smadja}, G. and {Smartt}, S. and {Spano}, P. and {Spiro}, S. and {Sullivan}, M. and {Tilquin}, A. and {Trotta}, R. and {Verde}, L. and {Wang}, Y. and {Williger}, G. and {Zhao}, G. and {Zoubian}, J. and {Zucca}, E.},
        title = "{Euclid Definition Study Report}",
      journal = {arXiv e-prints},
         year = 2011,
        month = oct,
          eid = {arXiv:1110.3193},
        pages = {arXiv:1110.3193},
          doi = {10.48550/arXiv.1110.3193},
archivePrefix = {arXiv},
       eprint = {1110.3193},
 primaryClass = {astro-ph.CO},
       adsurl = {https://ui.adsabs.harvard.edu/abs/2011arXiv1110.3193L}
}

@article{desi,
  author = {{DESI Collaboration} and {Aghamousa}, Amir and {Aguilar}, Jessica and {Ahlen}, Steve and {Alam}, Shadab and {Allen}, Lori E. and {Allende Prieto}, Carlos and {Annis}, James and {Bailey}, Stephen and {Balland}, Christophe and {Ballester}, Otger and {Baltay}, Charles and {Beaufore}, Lucas and {Bebek}, Chris and {Beers}, Timothy C. and {Bell}, Eric F. and {Bernal}, Jos{\'e} Luis and {Besuner}, Robert and {Beutler}, Florian and {Blake}, Chris and {Bleuler}, Hannes and {Blomqvist}, Michael and {Blum}, Robert and {Bolton}, Adam S. and {Briceno}, Cesar and {Brooks}, David and {Brownstein}, Joel R. and {Buckley-Geer}, Elizabeth and {Burden}, Angela and {Burtin}, Etienne and {Busca}, Nicolas G. and {Cahn}, Robert N. and {Cai}, Yan-Chuan and {Cardiel-Sas}, Laia and {Carlberg}, Raymond G. and {Carton}, Pierre-Henri and {Casas}, Ricard and {Castander}, Francisco J. and {Cervantes-Cota}, Jorge L. and {Claybaugh}, Todd M. and {Close}, Madeline and {Coker}, Carl T. and {Cole}, Shaun and {Comparat}, Johan and {Cooper}, Andrew P. and {Cousinou}, M. -C. and {Crocce}, Martin and {Cuby}, Jean-Gabriel and {Cunningham}, Daniel P. and {Davis}, Tamara M. and {Dawson}, Kyle S. and {de la Macorra}, Axel and {De Vicente}, Juan and {Delubac}, Timoth{\'e}e and {Derwent}, Mark and {Dey}, Arjun and {Dhungana}, Govinda and {Ding}, Zhejie and {Doel}, Peter and {Duan}, Yutong T. and {Ealet}, Anne and {Edelstein}, Jerry and {Eftekharzadeh}, Sarah and {Eisenstein}, Daniel J. and {Elliott}, Ann and {Escoffier}, St{\'e}phanie and {Evatt}, Matthew and {Fagrelius}, Parker and {Fan}, Xiaohui and {Fanning}, Kevin and {Farahi}, Arya and {Farihi}, Jay and {Favole}, Ginevra and {Feng}, Yu and {Fernandez}, Enrique and {Findlay}, Joseph R. and {Finkbeiner}, Douglas P. and {Fitzpatrick}, Michael J. and {Flaugher}, Brenna and {Flender}, Samuel and {Font-Ribera}, Andreu and {Forero-Romero}, Jaime E. and {Fosalba}, Pablo and {Frenk}, Carlos S. and {Fumagalli}, Michele and {Gaensicke}, Boris T. and {Gallo}, Giuseppe and {Garcia-Bellido}, Juan and {Gaztanaga}, Enrique and {Pietro Gentile Fusillo}, Nicola and {Gerard}, Terry and {Gershkovich}, Irena and {Giannantonio}, Tommaso and {Gillet}, Denis and {Gonzalez-de-Rivera}, Guillermo and {Gonzalez-Perez}, Violeta and {Gott}, Shelby and {Graur}, Or and {Gutierrez}, Gaston and {Guy}, Julien and {Habib}, Salman and {Heetderks}, Henry and {Heetderks}, Ian and {Heitmann}, Katrin and {Hellwing}, Wojciech A. and {Herrera}, David A. and {Ho}, Shirley and {Holland}, Stephen and {Honscheid}, Klaus and {Huff}, Eric and {Hutchinson}, Timothy A. and {Huterer}, Dragan and {Hwang}, Ho Seong and {Illa Laguna}, Joseph Maria and {Ishikawa}, Yuzo and {Jacobs}, Dianna and {Jeffrey}, Niall and {Jelinsky}, Patrick and {Jennings}, Elise and {Jiang}, Linhua and {Jimenez}, Jorge and {Johnson}, Jennifer and {Joyce}, Richard and {Jullo}, Eric and {Juneau}, St{\'e}phanie and {Kama}, Sami and {Karcher}, Armin and {Karkar}, Sonia and {Kehoe}, Robert and {Kennamer}, Noble and {Kent}, Stephen and {Kilbinger}, Martin and {Kim}, Alex G. and {Kirkby}, David and {Kisner}, Theodore and {Kitanidis}, Ellie and {Kneib}, Jean-Paul and {Koposov}, Sergey and {Kovacs}, Eve and {Koyama}, Kazuya and {Kremin}, Anthony and {Kron}, Richard and {Kronig}, Luzius and {Kueter-Young}, Andrea and {Lacey}, Cedric G. and {Lafever}, Robin and {Lahav}, Ofer and {Lambert}, Andrew and {Lampton}, Michael and {Landriau}, Martin and {Lang}, Dustin and {Lauer}, Tod R. and {Le Goff}, Jean-Marc and {Le Guillou}, Laurent and {Le Van Suu}, Auguste and {Lee}, Jae Hyeon and {Lee}, Su-Jeong and {Leitner}, Daniela and {Lesser}, Michael and {Levi}, Michael E. and {L'Huillier}, Benjamin and {Li}, Baojiu and {Liang}, Ming and {Lin}, Huan and {Linder}, Eric and {Loebman}, Sarah R. and {Luki{\'c}}, Zarija and {Ma}, Jun and {MacCrann}, Niall and {Magneville}, Christophe and {Makarem}, Laleh and {Manera}, Marc and {Manser}, Christopher J. and {Marshall}, Robert and {Martini}, Paul and {Massey}, Richard and {Matheson}, Thomas and {McCauley}, Jeremy and {McDonald}, Patrick and {McGreer}, Ian D. and {Meisner}, Aaron and {Metcalfe}, Nigel and {Miller}, Timothy N. and {Miquel}, Ramon and {Moustakas}, John and {Myers}, Adam and {Naik}, Milind and {Newman}, Jeffrey A. and {Nichol}, Robert C. and {Nicola}, Andrina and {Nicolati da Costa}, Luiz and {Nie}, Jundan and {Niz}, Gustavo and {Norberg}, Peder and {Nord}, Brian and {Norman}, Dara and {Nugent}, Peter and {O'Brien}, Thomas and {Oh}, Minji and {Olsen}, Knut A.~G. and {Padilla}, Cristobal and {Padmanabhan}, Hamsa and {Padmanabhan}, Nikhil and {Palanque-Delabrouille}, Nathalie and {Palmese}, Antonella and {Pappalardo}, Daniel and {P{\^a}ris}, Isabelle and {Park}, Changbom and {Patej}, Anna and {Peacock}, John A. and {Peiris}, Hiranya V. and {Peng}, Xiyan and {Percival}, Will J. and {Perruchot}, Sandrine and {Pieri}, Matthew M. and {Pogge}, Richard and {Pollack}, Jennifer E. and {Poppett}, Claire and {Prada}, Francisco and {Prakash}, Abhishek and {Probst}, Ronald G. and {Rabinowitz}, David and {Raichoor}, Anand and {Ree}, Chang Hee and {Refregier}, Alexandre and {Regal}, Xavier and {Reid}, Beth and {Reil}, Kevin and {Rezaie}, Mehdi and {Rockosi}, Constance M. and {Roe}, Natalie and {Ronayette}, Samuel and {Roodman}, Aaron and {Ross}, Ashley J. and {Ross}, Nicholas P. and {Rossi}, Graziano and {Rozo}, Eduardo and {Ruhlmann-Kleider}, Vanina and {Rykoff}, Eli S. and {Sabiu}, Cristiano and {Samushia}, Lado and {Sanchez}, Eusebio and {Sanchez}, Javier and {Schlegel}, David J. and {Schneider}, Michael and {Schubnell}, Michael and {Secroun}, Aur{\'e}lia and {Seljak}, Uros and {Seo}, Hee-Jong and {Serrano}, Santiago and {Shafieloo}, Arman and {Shan}, Huanyuan and {Sharples}, Ray and {Sholl}, Michael J. and {Shourt}, William V. and {Silber}, Joseph H. and {Silva}, David R. and {Sirk}, Martin M. and {Slosar}, Anze and {Smith}, Alex and {Smoot}, George F. and {Som}, Debopam and {Song}, Yong-Seon and {Sprayberry}, David and {Staten}, Ryan and {Stefanik}, Andy and {Tarle}, Gregory and {Sien Tie}, Suk and {Tinker}, Jeremy L. and {Tojeiro}, Rita and {Valdes}, Francisco and {Valenzuela}, Octavio and {Valluri}, Monica and {Vargas-Magana}, Mariana and {Verde}, Licia and {Walker}, Alistair R. and {Wang}, Jiali and {Wang}, Yuting and {Weaver}, Benjamin A. and {Weaverdyck}, Curtis and {Wechsler}, Risa H. and {Weinberg}, David H. and {White}, Martin and {Yang}, Qian and {Yeche}, Christophe and {Zhang}, Tianmeng and {Zhao}, Gong-Bo and {Zheng}, Yi and {Zhou}, Xu and {Zhou}, Zhimin and {Zhu}, Yaling and {Zou}, Hu and {Zu}, Ying},
        title = "{The DESI Experiment Part I: Science,Targeting, and Survey Design}",
      journal = {arXiv e-prints},
         year = 2016,
        month = oct,
          eid = {arXiv:1611.00036},
        pages = {arXiv:1611.00036},
          doi = {10.48550/arXiv.1611.00036},
archivePrefix = {arXiv},
       eprint = {1611.00036},
 primaryClass = {astro-ph.IM},
       adsurl = {https://ui.adsabs.harvard.edu/abs/2016arXiv161100036D}
}

@article{Dalal_2008,
        author = {{Dalal}, Neal and {Dor{\'e}}, Olivier and {Huterer}, Dragan and {Shirokov}, Alexander},
        title = "{Imprints of primordial non-Gaussianities on large-scale structure: Scale-dependent bias and abundance of virialized objects}",
      journal = {\prd},
         year = 2008,
        month = jun,
       volume = {77},
       number = {12},
          eid = {123514},
        pages = {123514},
          doi = {10.1103/PhysRevD.77.123514},
archivePrefix = {arXiv},
       eprint = {0710.4560},
 primaryClass = {astro-ph},
       adsurl = {https://ui.adsabs.harvard.edu/abs/2008PhRvD..77l3514D}
}

@ARTICLE{Wilson_2017,
       author = {{Wilson}, M.~J. and {Peacock}, J.~A. and {Taylor}, A.~N. and {de la Torre}, S.},
        title = "{Rapid modelling of the redshift-space power spectrum multipoles for a masked density field}",
      journal = {\mnras},
         year = 2017,
        month = jan,
       volume = {464},
       number = {3},
        pages = {3121-3130},
          doi = {10.1093/mnras/stw2576},
archivePrefix = {arXiv},
       eprint = {1511.07799},
 primaryClass = {astro-ph.CO},
       adsurl = {https://ui.adsabs.harvard.edu/abs/2017MNRAS.464.3121W}
}

@ARTICLE{Jolicoeur_2018,
    author = {{Jolicoeur}, Sheean and {Umeh}, Obinna and {Maartens}, Roy and {Clarkson}, Chris},
        title = "{Imprints of local lightcone projection effects on the galaxy bispectrum. Part III. Relativistic corrections from nonlinear dynamical evolution on large-scales}",
      journal = {\jcap},
         year = 2018,
        month = mar,
       volume = {2018},
       number = {3},
          eid = {036},
        pages = {036},
          doi = {10.1088/1475-7516/2018/03/036},
archivePrefix = {arXiv},
       eprint = {1711.01812},
 primaryClass = {astro-ph.CO},
       adsurl = {https://ui.adsabs.harvard.edu/abs/2018JCAP...03..036J}
}

@article{Tram_2016,
   author = {{Tram}, Thomas and {Fidler}, Christian and {Crittenden}, Robert and {Koyama}, Kazuya and {Pettinari}, Guido W. and {Wands}, David},
        title = "{The intrinsic matter bispectrum in {\ensuremath{\Lambda}}CDM}",
      journal = {\jcap},
         year = 2016,
        month = may,
       volume = {2016},
       number = {5},
          eid = {058},
        pages = {058},
          doi = {10.1088/1475-7516/2016/05/058},
archivePrefix = {arXiv},
       eprint = {1602.05933},
 primaryClass = {astro-ph.CO},
       adsurl = {https://ui.adsabs.harvard.edu/abs/2016JCAP...05..058T}
}

@ARTICLE{Desjacques_2018,
       author = {{Desjacques}, Vincent and {Jeong}, Donghui and {Schmidt}, Fabian},
        title = "{Large-scale galaxy bias}",
      journal = {\physrep},
         year = 2018,
        month = feb,
       volume = {733},
        pages = {1-193},
          doi = {10.1016/j.physrep.2017.12.002},
archivePrefix = {arXiv},
       eprint = {1611.09787},
 primaryClass = {astro-ph.CO},
       adsurl = {https://ui.adsabs.harvard.edu/abs/2018PhR...733....1D}
}

@article{harris2020,
 author = {{Harris}, Charles R. and {Millman}, K. Jarrod and {van der Walt}, St{\'e}fan J. and {Gommers}, Ralf and {Virtanen}, Pauli and {Cournapeau}, David and {Wieser}, Eric and {Taylor}, Julian and {Berg}, Sebastian and {Smith}, Nathaniel J. and {Kern}, Robert and {Picus}, Matti and {Hoyer}, Stephan and {van Kerkwijk}, Marten H. and {Brett}, Matthew and {Haldane}, Allan and {del R{\'\i}o}, Jaime Fern{\'a}ndez and {Wiebe}, Mark and {Peterson}, Pearu and {G{\'e}rard-Marchant}, Pierre and {Sheppard}, Kevin and {Reddy}, Tyler and {Weckesser}, Warren and {Abbasi}, Hameer and {Gohlke}, Christoph and {Oliphant}, Travis E.},
        title = "{Array programming with NumPy}",
      journal = {\nat},
         year = 2020,
        month = sep,
       volume = {585},
       number = {7825},
        pages = {357-362},
          doi = {10.1038/s41586-020-2649-2},
archivePrefix = {arXiv},
       eprint = {2006.10256},
 primaryClass = {cs.MS},
       adsurl = {https://ui.adsabs.harvard.edu/abs/2020Natur.585..357H}
}

@ARTICLE{2020SciPy,
  author = {{Virtanen}, Pauli and {Gommers}, Ralf and {Oliphant}, Travis E. and {Haberland}, Matt and {Reddy}, Tyler and {Cournapeau}, David and {Burovski}, Evgeni and {Peterson}, Pearu and {Weckesser}, Warren and {Bright}, Jonathan and {van der Walt}, St{\'e}fan J. and {Brett}, Matthew and {Wilson}, Joshua and {Millman}, K. Jarrod and {Mayorov}, Nikolay and {Nelson}, Andrew R.~J. and {Jones}, Eric and {Kern}, Robert and {Larson}, Eric and {Carey}, C.~J. and {Polat}, {\.I}lhan and {Feng}, Yu and {Moore}, Eric W. and {VanderPlas}, Jake and {Laxalde}, Denis and {Perktold}, Josef and {Cimrman}, Robert and {Henriksen}, Ian and {Quintero}, E.~A. and {Harris}, Charles R. and {Archibald}, Anne M. and {Ribeiro}, Ant{\^o}nio H. and {Pedregosa}, Fabian and {van Mulbregt}, Paul and {SciPy 1. 0 Contributors}},
        title = "{SciPy 1.0: fundamental algorithms for scientific computing in Python}",
      journal = {Nature Methods},
         year = 2020,
        month = feb,
       volume = {17},
        pages = {261-272},
          doi = {10.1038/s41592-019-0686-2},
archivePrefix = {arXiv},
       eprint = {1907.10121},
 primaryClass = {cs.MS},
       adsurl = {https://ui.adsabs.harvard.edu/abs/2020NatMe..17..261V}
}

@article{hunter2007,
  author = {{Hunter}, John D.},
        title = "{Matplotlib: A 2D Graphics Environment}",
      journal = {Computing in Science and Engineering},
         year = 2007,
        month = may,
       volume = {9},
       number = {3},
        pages = {90-95},
          doi = {10.1109/MCSE.2007.55},
       adsurl = {https://ui.adsabs.harvard.edu/abs/2007CSE.....9...90H}
}

@article{blas2011,
   title={The Cosmic Linear Anisotropy Solving System (CLASS). Part II: Approximation schemes},
   volume={2011},
   ISSN={1475-7516},
   url={http://dx.doi.org/10.1088/1475-7516/2011/07/034},
   DOI={10.1088/1475-7516/2011/07/034},
   number={07},
   journal={Journal of Cosmology and Astroparticle Physics},
   publisher={IOP Publishing},
   author={Blas, Diego and Lesgourgues, Julien and Tram, Thomas},
   year={2011},
   month={Jul},
   pages={034–034}
}

@article{Lesgourgues2011,
       author = {{Lesgourgues}, Julien},
        title = "{The Cosmic Linear Anisotropy Solving System (CLASS) I: Overview}",
      journal = {arXiv e-prints},
         year = 2011,
        month = apr,
          eid = {arXiv:1104.2932},
        pages = {arXiv:1104.2932},
          doi = {10.48550/arXiv.1104.2932},
archivePrefix = {arXiv},
       eprint = {1104.2932},
 primaryClass = {astro-ph.IM},
       adsurl = {https://ui.adsabs.harvard.edu/abs/2011arXiv1104.2932L}
}

@ARTICLE{Maartens_2022,
       author = {{Maartens}, Roy and {Fonseca}, Jos{\'e} and {Camera}, Stefano and {Jolicoeur}, Sheean and {Viljoen}, Jan-Albert and {Clarkson}, Chris},
        title = "{Magnification and evolution biases in large-scale structure surveys}",
      journal = {\jcap},
         year = 2021,
        month = dec,
       volume = {2021},
       number = {12},
          eid = {009},
        pages = {009},
          doi = {10.1088/1475-7516/2021/12/009},
archivePrefix = {arXiv},
       eprint = {2107.13401},
 primaryClass = {astro-ph.CO},
       adsurl = {https://ui.adsabs.harvard.edu/abs/2021JCAP...12..009M}
}

@ARTICLE{Maartens_2020,
       author = {{Maartens}, Roy and {Jolicoeur}, Sheean and {Umeh}, Obinna and {De Weerd}, Eline M. and {Clarkson}, Chris and {Camera}, Stefano},
        title = "{Detecting the relativistic galaxy bispectrum}",
      journal = {\jcap},
         year = 2020,
        month = mar,
       volume = {2020},
       number = {3},
          eid = {065},
        pages = {065},
          doi = {10.1088/1475-7516/2020/03/065},
archivePrefix = {arXiv},
       eprint = {1911.02398},
 primaryClass = {astro-ph.CO},
       adsurl = {https://ui.adsabs.harvard.edu/abs/2020JCAP...03..065M}
}

@ARTICLE{Planck_2018,
       author = {{Planck Collaboration} and {Aghanim}, N. and {Akrami}, Y. and {Ashdown}, M. and {Aumont}, J. and {Baccigalupi}, C. and {Ballardini}, M. and {Banday}, A.~J. and {Barreiro}, R.~B. and {Bartolo}, N. and {Basak}, S. and {Battye}, R. and {Benabed}, K. and {Bernard}, J. -P. and {Bersanelli}, M. and {Bielewicz}, P. and {Bock}, J.~J. and {Bond}, J.~R. and {Borrill}, J. and {Bouchet}, F.~R. and {Boulanger}, F. and {Bucher}, M. and {Burigana}, C. and {Butler}, R.~C. and {Calabrese}, E. and {Cardoso}, J. -F. and {Carron}, J. and {Challinor}, A. and {Chiang}, H.~C. and {Chluba}, J. and {Colombo}, L.~P.~L. and {Combet}, C. and {Contreras}, D. and {Crill}, B.~P. and {Cuttaia}, F. and {de Bernardis}, P. and {de Zotti}, G. and {Delabrouille}, J. and {Delouis}, J. -M. and {Di Valentino}, E. and {Diego}, J.~M. and {Dor{\'e}}, O. and {Douspis}, M. and {Ducout}, A. and {Dupac}, X. and {Dusini}, S. and {Efstathiou}, G. and {Elsner}, F. and {En{\ss}lin}, T.~A. and {Eriksen}, H.~K. and {Fantaye}, Y. and {Farhang}, M. and {Fergusson}, J. and {Fernandez-Cobos}, R. and {Finelli}, F. and {Forastieri}, F. and {Frailis}, M. and {Fraisse}, A.~A. and {Franceschi}, E. and {Frolov}, A. and {Galeotta}, S. and {Galli}, S. and {Ganga}, K. and {G{\'e}nova-Santos}, R.~T. and {Gerbino}, M. and {Ghosh}, T. and {Gonz{\'a}lez-Nuevo}, J. and {G{\'o}rski}, K.~M. and {Gratton}, S. and {Gruppuso}, A. and {Gudmundsson}, J.~E. and {Hamann}, J. and {Handley}, W. and {Hansen}, F.~K. and {Herranz}, D. and {Hildebrandt}, S.~R. and {Hivon}, E. and {Huang}, Z. and {Jaffe}, A.~H. and {Jones}, W.~C. and {Karakci}, A. and {Keih{\"a}nen}, E. and {Keskitalo}, R. and {Kiiveri}, K. and {Kim}, J. and {Kisner}, T.~S. and {Knox}, L. and {Krachmalnicoff}, N. and {Kunz}, M. and {Kurki-Suonio}, H. and {Lagache}, G. and {Lamarre}, J. -M. and {Lasenby}, A. and {Lattanzi}, M. and {Lawrence}, C.~R. and {Le Jeune}, M. and {Lemos}, P. and {Lesgourgues}, J. and {Levrier}, F. and {Lewis}, A. and {Liguori}, M. and {Lilje}, P.~B. and {Lilley}, M. and {Lindholm}, V. and {L{\'o}pez-Caniego}, M. and {Lubin}, P.~M. and {Ma}, Y. -Z. and {Mac{\'\i}as-P{\'e}rez}, J.~F. and {Maggio}, G. and {Maino}, D. and {Mandolesi}, N. and {Mangilli}, A. and {Marcos-Caballero}, A. and {Maris}, M. and {Martin}, P.~G. and {Martinelli}, M. and {Mart{\'\i}nez-Gonz{\'a}lez}, E. and {Matarrese}, S. and {Mauri}, N. and {McEwen}, J.~D. and {Meinhold}, P.~R. and {Melchiorri}, A. and {Mennella}, A. and {Migliaccio}, M. and {Millea}, M. and {Mitra}, S. and {Miville-Desch{\^e}nes}, M. -A. and {Molinari}, D. and {Montier}, L. and {Morgante}, G. and {Moss}, A. and {Natoli}, P. and {N{\o}rgaard-Nielsen}, H.~U. and {Pagano}, L. and {Paoletti}, D. and {Partridge}, B. and {Patanchon}, G. and {Peiris}, H.~V. and {Perrotta}, F. and {Pettorino}, V. and {Piacentini}, F. and {Polastri}, L. and {Polenta}, G. and {Puget}, J. -L. and {Rachen}, J.~P. and {Reinecke}, M. and {Remazeilles}, M. and {Renzi}, A. and {Rocha}, G. and {Rosset}, C. and {Roudier}, G. and {Rubi{\~n}o-Mart{\'\i}n}, J.~A. and {Ruiz-Granados}, B. and {Salvati}, L. and {Sandri}, M. and {Savelainen}, M. and {Scott}, D. and {Shellard}, E.~P.~S. and {Sirignano}, C. and {Sirri}, G. and {Spencer}, L.~D. and {Sunyaev}, R. and {Suur-Uski}, A. -S. and {Tauber}, J.~A. and {Tavagnacco}, D. and {Tenti}, M. and {Toffolatti}, L. and {Tomasi}, M. and {Trombetti}, T. and {Valenziano}, L. and {Valiviita}, J. and {Van Tent}, B. and {Vibert}, L. and {Vielva}, P. and {Villa}, F. and {Vittorio}, N. and {Wandelt}, B.~D. and {Wehus}, I.~K. and {White}, M. and {White}, S.~D.~M. and {Zacchei}, A. and {Zonca}, A.},
        title = "{Planck 2018 results. VI. Cosmological parameters}",
      journal = {\aap},
         year = 2020,
        month = sep,
       volume = {641},
          eid = {A6},
        pages = {A6},
          doi = {10.1051/0004-6361/201833910},
archivePrefix = {arXiv},
       eprint = {1807.06209},
 primaryClass = {astro-ph.CO},
       adsurl = {https://ui.adsabs.harvard.edu/abs/2020A&A...641A...6P}
}

@ARTICLE{Maartens_2021,
       author = {{Maartens}, Roy and {Jolicoeur}, Sheean and {Umeh}, Obinna and {De Weerd}, Eline M. and {Clarkson}, Chris},
        title = "{Local primordial non-Gaussianity in the relativistic galaxy bispectrum}",
      journal = {\jcap},
         year = 2021,
        month = apr,
       volume = {2021},
       number = {4},
          eid = {013},
        pages = {013},
          doi = {10.1088/1475-7516/2021/04/013},
archivePrefix = {arXiv},
       eprint = {2011.13660},
 primaryClass = {astro-ph.CO},
       adsurl = {https://ui.adsabs.harvard.edu/abs/2021JCAP...04..013M}
}

@ARTICLE{Chan_2017,
        author = {{Chan}, Kwan Chuen and {Blot}, Linda},
        title = "{Assessment of the information content of the power spectrum and bispectrum}",
      journal = {\prd},
         year = 2017,
        month = jul,
       volume = {96},
       number = {2},
          eid = {023528},
        pages = {023528},
          doi = {10.1103/PhysRevD.96.023528},
archivePrefix = {arXiv},
       eprint = {1610.06585},
 primaryClass = {astro-ph.CO},
       adsurl = {https://ui.adsabs.harvard.edu/abs/2017PhRvD..96b3528C}
}

@ARTICLE{Biagetti_2022,
       author = {{Biagetti}, Matteo and {Castiblanco}, Lina and {Nore{\~n}a}, Jorge and {Sefusatti}, Emiliano},
        title = "{The covariance of squeezed bispectrum configurations}",
      journal = {\jcap},
         year = 2022,
        month = sep,
       volume = {2022},
       number = {9},
          eid = {009},
        pages = {009},
          doi = {10.1088/1475-7516/2022/09/009},
archivePrefix = {arXiv},
       eprint = {2111.05887},
 primaryClass = {astro-ph.CO},
       adsurl = {https://ui.adsabs.harvard.edu/abs/2022JCAP...09..009B}
}

@ARTICLE{Barreira_2019,
       author = {{Barreira}, Alexandre},
        title = "{The squeezed matter bispectrum covariance with responses}",
      journal = {\jcap},
         year = 2019,
        month = mar,
       volume = {2019},
       number = {3},
          eid = {008},
        pages = {008},
          doi = {10.1088/1475-7516/2019/03/008},
archivePrefix = {arXiv},
       eprint = {1901.01243},
 primaryClass = {astro-ph.CO},
       adsurl = {https://ui.adsabs.harvard.edu/abs/2019JCAP...03..008B}
}

@ARTICLE{Salvalaggio_2024,
       author = {{Salvalaggio}, Jacopo and {Castiblanco}, Lina and {Nore{\~n}a}, Jorge and {Sefusatti}, Emiliano and {Monaco}, Pierluigi},
        title = "{Bispectrum non-Gaussian Covariance in Redshift Space}",
      journal = {\jcap},
         year = 2024,
        month = aug,
       volume = {2024},
       number = {8},
          eid = {046},
        pages = {046},
          doi = {10.1088/1475-7516/2024/08/046},
archivePrefix = {arXiv},
       eprint = {2403.08634},
 primaryClass = {astro-ph.CO},
       adsurl = {https://ui.adsabs.harvard.edu/abs/2024JCAP...08..046S}
}

@ARTICLE{Dore_SPHEREx,
       author = {{Dor{\'e}}, Olivier and {Bock}, Jamie and {Ashby}, Matthew and {Capak}, Peter and {Cooray}, Asantha and {de Putter}, Roland and {Eifler}, Tim and {Flagey}, Nicolas and {Gong}, Yan and {Habib}, Salman and {Heitmann}, Katrin and {Hirata}, Chris and {Jeong}, Woong-Seob and {Katti}, Raj and {Korngut}, Phil and {Krause}, Elisabeth and {Lee}, Dae-Hee and {Masters}, Daniel and {Mauskopf}, Phil and {Melnick}, Gary and {Mennesson}, Bertrand and {Nguyen}, Hien and {{\"O}berg}, Karin and {Pullen}, Anthony and {Raccanelli}, Alvise and {Smith}, Roger and {Song}, Yong-Seon and {Tolls}, Volker and {Unwin}, Steve and {Venumadhav}, Tejaswi and {Viero}, Marco and {Werner}, Mike and {Zemcov}, Mike},
        title = "{Cosmology with the SPHEREX All-Sky Spectral Survey}",
      journal = {arXiv e-prints},
         year = 2014,
        month = dec,
          eid = {arXiv:1412.4872},
        pages = {arXiv:1412.4872},
          doi = {10.48550/arXiv.1412.4872},
archivePrefix = {arXiv},
       eprint = {1412.4872},
 primaryClass = {astro-ph.CO},
       adsurl = {https://ui.adsabs.harvard.edu/abs/2014arXiv1412.4872D}
}

@ARTICLE{Heinrich_2024,
       author = {{Heinrich}, Chen and {Dor{\'e}}, Olivier and {Krause}, Elisabeth},
        title = "{Measuring $f_{\rm NL}$ with the SPHEREx multitracer redshift space bispectrum}",
      journal = {\prd},
         year = 2024,
       volume = {109},
       number = {12},
          eid = {123511},
        pages = {123511},
          doi = {10.1103/PhysRevD.109.123511},
archivePrefix = {arXiv},
       eprint = {2311.13082},
 primaryClass = {astro-ph.CO},
}

@ARTICLE{Bonvin_2011,
       author = {{Bonvin}, Camille and {Durrer}, Ruth},
        title = "{What galaxy surveys really measure}",
      journal = {\prd},
         year = 2011,
        month = sep,
       volume = {84},
       number = {6},
          eid = {063505},
        pages = {063505},
          doi = {10.1103/PhysRevD.84.063505},
archivePrefix = {arXiv},
       eprint = {1105.5280},
 primaryClass = {astro-ph.CO},
       adsurl = {https://ui.adsabs.harvard.edu/abs/2011PhRvD..84f3505B}
}

@ARTICLE{Yoo_2010,
       author = {{Yoo}, Jaiyul},
        title = "{General relativistic description of the observed galaxy power spectrum: Do we understand what we measure?}",
      journal = {\prd},
         year = 2010,
        month = oct,
       volume = {82},
       number = {8},
          eid = {083508},
        pages = {083508},
          doi = {10.1103/PhysRevD.82.083508},
archivePrefix = {arXiv},
       eprint = {1009.3021},
 primaryClass = {astro-ph.CO},
       adsurl = {https://ui.adsabs.harvard.edu/abs/2010PhRvD..82h3508Y}
}

@ARTICLE{Umeh_2017,
       author = {{Umeh}, Obinna and {Jolicoeur}, Sheean and {Maartens}, Roy and {Clarkson}, Chris},
        title = "{A general relativistic signature in the galaxy bispectrum: the local effects of observing on the lightcone}",
      journal = {\jcap},
         year = 2017,
        month = mar,
       volume = {2017},
       number = {3},
          eid = {034},
        pages = {034},
          doi = {10.1088/1475-7516/2017/03/034},
archivePrefix = {arXiv},
       eprint = {1610.03351},
 primaryClass = {astro-ph.CO},
       adsurl = {https://ui.adsabs.harvard.edu/abs/2017JCAP...03..034U}
}

@ARTICLE{Megamapper,
       author = {{Schlegel}, David J. and {Kollmeier}, Juna A. and {Aldering}, Greg and {Bailey}, Stephen and {Baltay}, Charles and {Bebek}, Christopher and {BenZvi}, Segev and {Besuner}, Robert and {Blanc}, Guillermo and {Bolton}, Adam S. and {Bonaca}, Ana and {Bouri}, Mohamed and {Brooks}, David and {Buckley-Geer}, Elizabeth and {Cai}, Zheng and {Crane}, Jeffrey and {Demina}, Regina and {DeRose}, Joseph and {Dey}, Arjun and {Doel}, Peter and {Fan}, Xiaohui and {Ferraro}, Simone and {Finkbeiner}, Douglas and {Font-Ribera}, Andreu and {Gontcho}, Satya Gontcho A and {Green}, Daniel and {Gutierrez}, Gaston and {Guy}, Julien and {Heetderks}, Henry and {Huterer}, Dragan and {Infante}, Leopoldo and {Jelinsky}, Patrick and {Karagiannis}, Dionysios and {Kent}, Stephen M. and {Kim}, Alex G. and {Kneib}, Jean-Paul and {Kremin}, Anthony and {Kronig}, Luzius and {Konidaris}, Nick and {Lahav}, Ofer and {Lampton}, Michael L. and {Landriau}, Martin and {Lang}, Dustin and {Leauthaud}, Alexie and {Levi}, Michael E. and {Liguori}, Michele and {Linder}, Eric V. and {Magneville}, Christophe and {Martini}, Paul and {Mateo}, Mario and {McDonald}, Patrick and {Miller}, Christopher J. and {Moustakas}, John and {Myers}, Adam D. and {Mulchaey}, John and {Newman}, Jeffrey A. and {Nugent}, Peter E. and {Padmanabhan}, Nikhil and {Palanque-Delabrouille}, Nathalie and {Piro}, Antonella Palmese Anthony L. and {Poppett}, Claire and {Prochaska}, Jason X. and {Pullen}, Anthony R. and {Rabinowitz}, David and {Raichoor}, Anand and {Ramirez}, Solange and {Rix}, Hans-Walter and {Ross}, Ashley J. and {Samushia}, Lado and {Schaan}, Emmanuel and {Schubnell}, Michael and {Seljak}, Uros and {Seo}, Hee-Jong and {Shectman}, Stephen A. and {Schlafly}, Edward F. and {Silber}, Joseph and {Simon}, Joshua D. and {Slepian}, Zachary and {Slosar}, An{\v{z}}e and {Soares-Santos}, Marcelle and {Tarl{\'e}}, Greg and {Thompson}, Ian and {Valluri}, Monica and {Wechsler}, Risa H. and {White}, Martin and {Wilson}, Michael J. and {Y{\`e}che}, Christophe and {Zaritsky}, Dennis and {Zhou}, Rongpu},
        title = "{The MegaMapper: A Stage-5 Spectroscopic Instrument Concept for the Study of Inflation and Dark Energy}",
      journal = {arXiv e-prints},
         year = 2022,
        month = sep,
          eid = {arXiv:2209.04322},
        pages = {arXiv:2209.04322},
          doi = {10.48550/arXiv.2209.04322},
archivePrefix = {arXiv},
       eprint = {2209.04322},
 primaryClass = {astro-ph.IM},
       adsurl = {https://ui.adsabs.harvard.edu/abs/2022arXiv220904322S}
}

@ARTICLE{Karagiannis_2024,
       author = {{Karagiannis}, Dionysios and {Maartens}, Roy and {Fonseca}, Jos{\'e} and {Camera}, Stefano and {Clarkson}, Chris},
        title = "{Multi-tracer power spectra and bispectra: formalism}",
      journal = {\jcap},
         year = 2024,
        month = mar,
       volume = {2024},
       number = {3},
          eid = {034},
        pages = {034},
          doi = {10.1088/1475-7516/2024/03/034},
archivePrefix = {arXiv},
       eprint = {2305.04028},
 primaryClass = {astro-ph.CO},
       adsurl = {https://ui.adsabs.harvard.edu/abs/2024JCAP...03..034K}
}

@ARTICLE{Cabass_2022b,
       author = {{Cabass}, Giovanni and {Ivanov}, Mikhail M. and {Philcox}, Oliver H.~E. and {Simonovi{\'c}}, Marko and {Zaldarriaga}, Matias},
        title = "{Constraints on multifield inflation from the BOSS galaxy survey}",
      journal = {\prd},
         year = 2022,
        month = aug,
       volume = {106},
       number = {4},
          eid = {043506},
        pages = {043506},
          doi = {10.1103/PhysRevD.106.043506},
archivePrefix = {arXiv},
       eprint = {2204.01781},
 primaryClass = {astro-ph.CO},
       adsurl = {https://ui.adsabs.harvard.edu/abs/2022PhRvD.106d3506C}
}

@ARTICLE{Cabass_2022a,
       author = {{Cabass}, Giovanni and {Ivanov}, Mikhail M. and {Philcox}, Oliver H.~E. and {Simonovi{\'c}}, Marko and {Zaldarriaga}, Matias},
        title = "{Constraints on Single-Field Inflation from the BOSS Galaxy Survey}",
      journal = {\prl},
         year = 2022,
        month = jul,
       volume = {129},
       number = {2},
          eid = {021301},
        pages = {021301},
          doi = {10.1103/PhysRevLett.129.021301},
archivePrefix = {arXiv},
       eprint = {2201.07238},
 primaryClass = {astro-ph.CO},
       adsurl = {https://ui.adsabs.harvard.edu/abs/2022PhRvL.129b1301C}
}

@ARTICLE{Benabou_2024,
       author = {{Benabou}, Joshua N. and {Testa}, Adriano and {Heinrich}, Chen and {Gebhardt}, Henry S. Grasshorn and {Dor{\'e}}, Olivier},
        title = "{Galaxy bispectrum in the spherical Fourier-Bessel basis}",
      journal = {\prd},
         year = 2024,
        month = may,
       volume = {109},
       number = {10},
          eid = {103507},
        pages = {103507},
          doi = {10.1103/PhysRevD.109.103507},
archivePrefix = {arXiv},
       eprint = {2312.15992},
 primaryClass = {astro-ph.CO},
       adsurl = {https://ui.adsabs.harvard.edu/abs/2024PhRvD.109j3507B}
}

@ARTICLE{Rossiter_2024,
       author = {{Rossiter}, Samantha and {Camera}, Stefano and {Clarkson}, Chris and {Maartens}, Roy},
        title = "{Decoupling local primordial non-Gaussianity from relativistic effects in the galaxy bispectrum}",
      journal = {\jcap},
         year = 2025,
        month = jul,
       volume = {2025},
       number = {7},
          eid = {055},
        pages = {055},
          doi = {10.1088/1475-7516/2025/07/055},
archivePrefix = {arXiv},
       eprint = {2407.06301},
 primaryClass = {astro-ph.CO},
       adsurl = {https://ui.adsabs.harvard.edu/abs/2024arXiv240706301R}
}

@ARTICLE{Wang_2024,
       author = {{Wang}, Mike Shengbo and {Beutler}, Florian and {Aguilar}, J. and {Ahlen}, S. and {Bianchi}, D. and {Brooks}, D. and {Claybaugh}, T. and {de la Macorra}, A. and {Doel}, P. and {Font-Ribera}, A. and {Gazta{\~n}aga}, E. and {Gutierrez}, G. and {Honscheid}, K. and {Howlett}, C. and {Kirkby}, D. and {Lambert}, A. and {Landriau}, M. and {Miquel}, R. and {Niz}, G. and {Prada}, F. and {P{\'e}rez-R{\`a}fols}, I. and {Rossi}, G. and {Sanchez}, E. and {Schlegel}, D. and {Schubnell}, M. and {Sprayberry}, D. and {Tarl{\'e}}, G. and {Weaver}, B.~A.},
        title = "{Window convolution of the galaxy clustering bispectrum}",
      journal = {\jcap},
         year = 2025,
        month = jun,
       volume = {2025},
       number = {6},
          eid = {031},
        pages = {031},
          doi = {10.1088/1475-7516/2025/06/031},
archivePrefix = {arXiv},
       eprint = {2411.14947},
 primaryClass = {astro-ph.CO},
       adsurl = {https://ui.adsabs.harvard.edu/abs/2024arXiv241114947W}
}

@ARTICLE{Karagiannis_2020,
       author = {{Karagiannis}, Dionysios and {Slosar}, An{\v{z}}e and {Liguori}, Michele},
        title = "{Forecasts on primordial non-Gaussianity from 21 cm intensity mapping experiments}",
      journal = {\jcap},
         year = 2020,
        month = nov,
       volume = {2020},
       number = {11},
          eid = {052},
        pages = {052},
          doi = {10.1088/1475-7516/2020/11/052},
archivePrefix = {arXiv},
       eprint = {1911.03964},
 primaryClass = {astro-ph.CO},
       adsurl = {https://ui.adsabs.harvard.edu/abs/2020JCAP...11..052K}
}

@ARTICLE{Tinker_2010,
       author = {{Tinker}, Jeremy L. and {Robertson}, Brant E. and {Kravtsov}, Andrey V. and {Klypin}, Anatoly and {Warren}, Michael S. and {Yepes}, Gustavo and {Gottl{\"o}ber}, Stefan},
        title = "{The Large-scale Bias of Dark Matter Halos: Numerical Calibration and Model Tests}",
      journal = {\apj},
         year = 2010,
        month = dec,
       volume = {724},
       number = {2},
        pages = {878-886},
          doi = {10.1088/0004-637X/724/2/878},
archivePrefix = {arXiv},
       eprint = {1001.3162},
 primaryClass = {astro-ph.CO},
       adsurl = {https://ui.adsabs.harvard.edu/abs/2010ApJ...724..878T}
}

@ARTICLE{Addis_2024,
       author = {{Addis}, Chris and {Guandalin}, Caroline and {Clarkson}, Chris},
        title = "{Multipoles of the galaxy bispectrum on a light cone: wide-separation and relativistic corrections}",
      journal = {\jcap},
         year = 2025,
        month = apr,
       volume = {2025},
       number = {4},
          eid = {080},
        pages = {080},
          doi = {10.1088/1475-7516/2025/04/080},
archivePrefix = {arXiv},
       eprint = {2407.00168},
 primaryClass = {astro-ph.CO},
       adsurl = {https://ui.adsabs.harvard.edu/abs/2024arXiv240700168A}
}

@ARTICLE{Yankelevich_2018,
       author = {{Yankelevich}, Victoria and {Porciani}, Cristiano},
        title = "{Cosmological information in the redshift-space bispectrum}",
      journal = {\mnras},
         year = 2019,
        month = feb,
       volume = {483},
       number = {2},
        pages = {2078-2099},
          doi = {10.1093/mnras/sty3143},
archivePrefix = {arXiv},
       eprint = {1807.07076},
 primaryClass = {astro-ph.CO},
       adsurl = {https://ui.adsabs.harvard.edu/abs/2019MNRAS.483.2078Y}
}

@ARTICLE{Geach_2012,
       author = {{Geach}, J.~E. and {Sobral}, D. and {Hickox}, R.~C. and {Wake}, D.~A. and {Smail}, Ian and {Best}, P.~N. and {Baugh}, C.~M. and {Stott}, J.~P.},
        title = "{The clustering of H{\ensuremath{\alpha}} emitters at z=2.23 from HiZELS}",
      journal = {\mnras},
         year = 2012,
        month = oct,
       volume = {426},
       number = {1},
        pages = {679-689},
          doi = {10.1111/j.1365-2966.2012.21725.x},
archivePrefix = {arXiv},
       eprint = {1206.4052},
 primaryClass = {astro-ph.CO},
       adsurl = {https://ui.adsabs.harvard.edu/abs/2012MNRAS.426..679G}
}

@ARTICLE{Gonzalez_2018,
       author = {{Gonzalez-Perez}, V. and {Comparat}, J. and {Norberg}, P. and {Baugh}, C.~M. and {Contreras}, S. and {Lacey}, C. and {McCullagh}, N. and {Orsi}, A. and {Helly}, J. and {Humphries}, J.},
        title = "{The host dark matter haloes of [O II] emitters at 0.5 < z < 1.5}",
      journal = {\mnras},
         year = 2018,
        month = mar,
       volume = {474},
       number = {3},
        pages = {4024-4038},
          doi = {10.1093/mnras/stx2807},
archivePrefix = {arXiv},
       eprint = {1708.07628},
 primaryClass = {astro-ph.GA},
       adsurl = {https://ui.adsabs.harvard.edu/abs/2018MNRAS.474.4024G}
}

@ARTICLE{Sullivan_2025,
       author = {{Sullivan}, James M. and {Seljak}, Uro{\v{s}}},
        title = "{Local primordial non-Gaussian bias from time evolution}",
      journal = {\prd},
         year = 2025,
        month = oct,
       volume = {112},
       number = {8},
          eid = {083522},
        pages = {083522},
          doi = {10.1103/8jvj-66zc},
archivePrefix = {arXiv},
       eprint = {2503.21736},
 primaryClass = {astro-ph.CO},
       adsurl = {https://ui.adsabs.harvard.edu/abs/2025PhRvD.112h3522S}
}

@ARTICLE{Sugiyama_2020,
       author = {{Sugiyama}, Naonori S. and {Saito}, Shun and {Beutler}, Florian and {Seo}, Hee-Jong},
        title = "{Perturbation theory approach to predict the covariance matrices of the galaxy power spectrum and bispectrum in redshift space}",
      journal = {\mnras},
         year = 2020,
       volume = {497},
       number = {2},
        pages = {1684--1711},
          doi = {10.1093/mnras/staa1940},
archivePrefix = {arXiv},
       eprint = {1908.06234},
 primaryClass = {astro-ph.CO}
}

@ARTICLE{Montano_2026_cov,
       author = {{Montano}, Federico and {Camera}, Stefano and {Sefusatti}, Emiliano and {Yousry Elkhashab}, Mohamed},
        title = "{Complex yet Hermitian: Gaussian covariance of cross-correlation and multi-tracer power spectra}",
      journal = {arXiv e-prints},
         year = 2026,
        month = jun,
          eid = {arXiv:2606.12551},
        pages = {arXiv:2606.12551},
          doi = {10.48550/arXiv.2606.12551},
archivePrefix = {arXiv},
       eprint = {2606.12551},
 primaryClass = {astro-ph.CO}
}

@ARTICLE{Chan_2018,
       author = {{Chan}, Kwan Chuen and {Moradinezhad Dizgah}, Azadeh and {Nore{\~n}a}, Jorge},
        title = "{Bispectrum Supersample Covariance}",
      journal = {\prd},
         year = 2018,
        month = feb,
       volume = {97},
       number = {4},
          eid = {043532},
        pages = {043532},
          doi = {10.1103/PhysRevD.97.043532},
archivePrefix = {arXiv},
       eprint = {1709.02473},
 primaryClass = {astro-ph.CO}
}

@ARTICLE{Elkhashab_2022,
       author = {{Elkhashab}, Mohamed Yousry and {Porciani}, Cristiano and {Bertacca}, Daniele},
        title = "{The large-scale monopole of the power spectrum in a Euclid-like survey: wide-angle effects, lensing, and the 'finger of the observer'}",
      journal = {\mnras},
         year = 2022,
        month = jan,
       volume = {509},
       number = {2},
        pages = {1626-1645},
          doi = {10.1093/mnras/stab3010},
archivePrefix = {arXiv},
       eprint = {2108.13424},
 primaryClass = {astro-ph.CO},
       adsurl = {https://ui.adsabs.harvard.edu/abs/2022MNRAS.509.1626E}
}

@ARTICLE{Wenzl_2024,
       author = {{Wenzl}, Lukas and {Chen}, Shi-Fan and {Bean}, Rachel},
        title = "{Magnification bias estimators for realistic surveys: an application to the BOSS survey}",
      journal = {\mnras},
         year = 2024,
       volume = {527},
       number = {2},
        pages = {1760-1773},
          doi = {10.1093/mnras/stad3314},
archivePrefix = {arXiv},
       eprint = {2308.05892},
 primaryClass = {astro-ph.CO}
}

@ARTICLE{Zhou_2023,
       author = {{Zhou}, Rongpu and {Ferraro}, Simone and {White}, Martin and others},
        title = "{DESI luminous red galaxy samples for cross-correlations}",
      journal = {\jcap},
         year = 2023,
       volume = {2023},
       number = {11},
        pages = {097},
          doi = {10.1088/1475-7516/2023/11/097},
archivePrefix = {arXiv},
       eprint = {2309.06443},
 primaryClass = {astro-ph.CO}
}

@ARTICLE{Elvin-Poole_2023,
       author = {{Elvin-Poole}, J. and {MacCrann}, N. and {Everett}, S. and others},
        title = "{Dark Energy Survey Year 3 results: magnification modelling and impact on cosmological constraints from galaxy clustering and galaxy-galaxy lensing}",
      journal = {\mnras},
         year = 2023,
       volume = {523},
       number = {3},
        pages = {3649-3670},
          doi = {10.1093/mnras/stad1594},
archivePrefix = {arXiv},
       eprint = {2209.09782},
 primaryClass = {astro-ph.CO}
}

@ARTICLE{Everett_2022,
       author = {{Everett}, S. and {Yanny}, B. and {Kuropatkin}, N. and others},
        title = "{Dark Energy Survey Year 3 Results: Measuring the Survey Transfer Function with Balrog}",
      journal = {\apjs},
         year = 2022,
       volume = {258},
       number = {1},
          eid = {15},
        pages = {15},
          doi = {10.3847/1538-4365/ac26c1},
archivePrefix = {arXiv},
       eprint = {2012.12825},
 primaryClass = {astro-ph.CO}
}

@ARTICLE{Gabarra_2023,
       author = {{Euclid Collaboration} and {Gabarra}, L. and others},
        title = "{Euclid preparation. XXX. Performance assessment of the NISP Red-Grism through spectroscopic simulations for the Wide and Deep surveys}",
      journal = {\aap},
         year = 2023,
       volume = {676},
          eid = {A34},
        pages = {A34},
          doi = {10.1051/0004-6361/202346177},
archivePrefix = {arXiv},
       eprint = {2302.09372},
 primaryClass = {astro-ph.GA}
}

@ARTICLE{Bonvin_2008,
       author = {{Bonvin}, Camille},
        title = "{Effect of peculiar motion in weak lensing}",
      journal = {\prd},
         year = 2008,
       volume = {78},
       number = {12},
          eid = {123530},
        pages = {123530},
          doi = {10.1103/PhysRevD.78.123530},
archivePrefix = {arXiv},
       eprint = {0810.0180},
 primaryClass = {astro-ph}
}

@ARTICLE{Bacon_2014,
       author = {{Bacon}, D.~J. and {Andrianomena}, S. and {Clarkson}, C. and {Bolejko}, K. and {Maartens}, R.},
        title = "{Cosmology with Doppler lensing}",
      journal = {\mnras},
         year = 2014,
       volume = {443},
       number = {3},
        pages = {1900-1915},
          doi = {10.1093/mnras/stu1270},
archivePrefix = {arXiv},
       eprint = {1401.3694},
 primaryClass = {astro-ph.CO}
}

@ARTICLE{Bonvin_2017,
       author = {{Bonvin}, Camille and {Andrianomena}, Sambatra and {Bacon}, David and {Clarkson}, Chris and {Maartens}, Roy and {Moloi}, Teboho and {Bull}, Philip},
        title = "{Dipolar modulation in the size of galaxies: the effect of Doppler magnification}",
      journal = {\mnras},
         year = 2017,
       volume = {472},
       number = {4},
        pages = {3936},
          doi = {10.1093/mnras/stx2049},
archivePrefix = {arXiv},
       eprint = {1610.05946},
 primaryClass = {astro-ph.CO}
}

@ARTICLE{Ye_2026,
       author = {{Ye}, Isabelle and {Bull}, Philip and {Guandalin}, Caroline and {Clarkson}, Chris and {Nasirudin}, Ainulnabilah},
        title = "{Prospects for measuring the Doppler magnification dipole with LSST and DESI}",
      journal = {\mnras},
         year = 2026,
       volume = {547},
       number = {4},
          doi = {10.1093/mnras/stag506},
archivePrefix = {arXiv},
       eprint = {2511.17858},
 primaryClass = {astro-ph.CO}
}

%%%%%%%%%%%%%%%%% APPENDICES %%%%%%%%%%%%%%%%%%%%%

\appendix
\crefalias{section}{appendix}
\crefalias{subsection}{appendix}

\section{Bias Modelling}

In this section, we outline some details of our bias modelling machinery. First, we provide an overview of the peak-background split (PBS) approach, and then we discuss our luminosity-function-dependent biases and the computation of these biases for our bright and faint samples.

The PBS approach splits the density field into a long-wavelength mode and noisy, short-wavelength fluctuations that sit atop it. In this picture, an overdense, long `background' mode effectively lowers the local threshold for gravitational collapse. The halo clustering bias parameters can be derived from the derivatives of the halo mass function (HMF) with respect to the background density—see \citet{Desjacques_2018} and Appendix C of \citet{Karagiannis_2018} for a detailed review. Thus, given a cosmology and a theoretical HMF model\footnote{We use the HMF model presented in \citet{Tinker_2010} using the fitted parameter values given for $\Delta=200$ from their Table 4.}, we can compute the halo bias parameters. These can then be related to the galaxy bias parameters, as detailed in \cref{sec:galaxy_bias}.

\subsection{Magnification and evolution bias}\label{sec:magnification_evolution}

The magnification bias, $\mathcal{Q}$, relates to the galaxies lost or gained by a sample at a certain flux or magnitude cut due to perturbations of their observed flux, such as lensing magnification and Doppler effects, while the evolution bias, $b_e$, describes the physical change in number density due to formation and mergers. These biases can be expressed as derivatives of the number density and therefore are closely related to the survey luminosity function.

For a flux- (or magnitude-)limited survey, the evolution, $b_e$, and magnification, $\mathcal{Q}$, biases are defined with partial derivatives at the luminosity cut $L_c$ \citep{Challinor_2011,Alonso_2015}
\begin{equation}\label{eq:Q_be}
b_e(z, L_c) = -\frac{\partial \, \mathrm{ln} \,n_g}{\partial \, \mathrm{ln}(1+z)}\bigg|_{L_{\mathrm{c}}}, \, \, \mathcal{Q}(z, L_c)= - \frac{\partial \, \mathrm{ln} \, n_g}{\partial \, \mathrm{ln} \, L}\bigg|_z,
\end{equation}
where $n_g$ is the comoving number density and $|_x$ refers to what is being held fixed; see \citet{Maartens_2022} for a more detailed overview.

While $\mathcal{Q}$ can be measured directly from the slope of the source counts at the flux cut, $b_e$ cannot, as it is defined at fixed luminosity rather than fixed flux. The corresponding observable is the slope of the redshift distribution of the sample, $n_{\rm evol} \equiv \mathrm{d}\ln N/\mathrm{d}\ln(1+z)$ \citep{Blanco_2024}, where $N = x^2 n_g/H$ is the number of sources above the flux cut per unit redshift and solid angle, and $H$ is the Hubble rate. Accounting for the evolution of the luminosity cut and of the volume factor, the two are related by \citep{Alonso_2015,Maartens_2022}
\begin{equation}\label{eq:be_nevol}
b_e = -n_{\rm evol} - \frac{\mathrm{d} \ln H}{\mathrm{d} \ln (1+z)} + \frac{2}{x\mathcal{H}} - 2\left(1 + \frac{1}{x\mathcal{H}}\right)\mathcal{Q},
\end{equation}
where, for surveys that measure fluxes in fixed wavelength bands, the last term also contains a K-correction \citep{Maartens_2022}. We therefore take $n_{\rm evol}$ and $\mathcal{Q}$ as the parameters describing each sample, with $b_e$ following from \cref{eq:be_nevol}, such that it varies with cosmology through the remaining terms.

\subsection{Evolution bias and the PNG bias}\label{sec:be_bpsi}

The PNG bias is the response of the number density to the amplitude of the small-scale fluctuations, $b_\Psi = 2\,\partial\ln n_g/\partial\ln\sigma_8$ (\cref{sec:png_bias_params}). At fixed cosmology $\sigma_R\propto D(z)$, the linear growth factor, so the redshift evolution of the number density carries the same response \citep{Sullivan_2025,Nascimento_2026}. Since $b_e$ is defined at fixed luminosity (\cref{eq:Q_be}), with $\mathrm{d}\ln D = -f\,\mathrm{d}\ln(1+z)$ it is the sum of this response and the evolution of the population at fixed $\sigma_R$ \citep[cf.][equation 9]{Nascimento_2026}, so that
\begin{equation}\label{eq:be_bpsi}
b_\Psi = \frac{2\,b_e}{f} - 2\,\frac{\partial \ln n_g}{\partial \ln D}\bigg|_{\sigma_R,\,L_c}.
\end{equation}
If the last term vanishes, this gives $b_e = f\,b_\Psi/2$, and so $b_e = \delta_c f(b_1-1)$ for a universal mass function \citep{Jeong_2012,Sullivan_2025}.

Written in terms of $n_{\rm evol}$ and $\mathcal{Q}$ through \cref{eq:be_nevol}, \cref{eq:be_bpsi} is the time-evolution estimator of \citet{Sullivan_2025} and \citet{Nascimento_2026} applied to the count slope of a flux-limited sample. The $\mathcal{Q}$ term in \cref{eq:be_nevol}, together with the K-correction, accounts for the drift of the luminosity cut with redshift at fixed flux, which the evolution bias excludes by definition. \citet{Sullivan_2025} give this correction in terms of $\mathcal{Q}$ without the K-correction (their equation 22), while \citet{Nascimento_2026} evaluate it, including the K-correction, by shifting the spectral energy distributions of the galaxies (their equation 13). In our parametrisation it is set by the same $\mathcal{Q}$ that enters the relativistic clustering signal.

The last term in \cref{eq:be_bpsi} is the evolution of the population at fixed luminosity that is not driven by the amplitude of the fluctuations, such as the passive fading of stellar populations, feedback and mergers \citep{Nascimento_2026}, and, for emission-line samples such as the \textit{Euclid}-like H$\alpha$ sample, the decline of the cosmic star formation rate. It must be modelled, for example with simulations, for any time-evolution estimate of $b_\Psi$, and we do not attempt this here.

Applied to the catalogue, the time-evolution estimator shares the limitation of the catalogue measurement of $n_{\rm evol}$ (\cref{sec:beQ_constraints}): any unmodelled redshift dependence of the completeness of the sample biases it. The leading-order relativistic terms in the odd multipoles instead depend on $b_e$ and $\mathcal{Q}$ directly, and so offer a route to $b_\Psi$ through \cref{eq:be_bpsi} that does not rely on the redshift distribution. For the \textit{Euclid}-like survey, however, the clustering constraint on $n_{\rm evol}$ is weak (\cref{sec:beQ_constraints}), and since $\mathcal{Q}$ is much better determined, the constraint on $b_e$ is similarly weak, so this route would at most provide a consistency check. Throughout our forecasts $b_\Psi$ is held fixed (\cref{sec:forecast_details}) and \cref{eq:be_bpsi} is not imposed but \textsc{CosmoWAP} has been extended to allow for forecasts using it.

\subsection{Bright-faint split}\label{sec:bf_split}

The treatment of the biases for the bright sample is reasonably straightforward: they can be treated as a single-tracer survey with a more conservative luminosity threshold. The biases for the faint sample are less simple, as one now has to consider both the flux cut and the flux split. For most biases, including the magnification bias, the faint sample bias can be derived from the biases of the total and bright samples \cite[e.g.][]{Ferramacho_2014, Bonvin_2023}
\begin{equation}\label{eq:faint_bias}
    b^F = \frac{n^T_g b^T - n^B_g b^B}{n^T_g - n^B_g},
\end{equation}
which, as the faint counts are simply $N^F = N^T - N^B$, also holds for $n_{\rm evol}$. However, relating the evolution bias of the faint sample to $n^F_{\rm evol}$ is slightly more involved, as the faint sample is bounded by two luminosity cuts, and we detail this below for a flux-limited survey.

The evolution bias of the faint tracer population is the partial derivative of the faint number density at fixed luminosity cuts
\begin{equation}
    b^F_e = -\left.\frac{\partial \ln \left(n^T_g- n^B_g\right)}{\partial \ln (1 + z)}\right|_{L_c,L_s}.
\end{equation}
As for \cref{eq:be_nevol}, we relate this to the total derivative of the observed faint counts, $N^F = N^T - N^B$, which includes the volume factor and the redshift-dependent luminosity cuts. With upper and lower luminosity cuts, this can be written as
\begin{equation}
\begin{aligned}
    n^F_{\rm evol} = -b^F_e - \frac{\mathrm{d} \ln H}{\mathrm{d} \ln (1+z)} + \frac{2}{x\mathcal{H}} &+ \left. \frac{\partial \ln n^F_g}{\partial \ln L_c} \right|_{z, L_s} \frac{\mathrm{d} \ln L_c}{\mathrm{d} \ln (1 + z)}\\ & + \left. \frac{\partial \ln n^F_g}{\partial \ln L_s} \right|_{z, L_c} \frac{\mathrm{d} \ln L_s}{\mathrm{d} \ln (1 + z)}.
\end{aligned}
\end{equation}
Using the fact that $n_g^T$ and $n_g^B$ are independent of $L_s$ and $L_c$ respectively, we can rewrite the partial derivatives
\begin{equation}
\begin{aligned}
     \left. \frac{\partial \ln n^F_g}{\partial \ln L_c} \right|_{z, L_s} &= \frac{n_g^T}{n_g^F} \left. \frac{\partial \ln n^T_g}{\partial \ln L_c} \right|_{z, L_s}, \\
     \left. \frac{\partial \ln n^F_g}{\partial \ln L_s} \right|_{z, L_c} &= -\frac{n_g^B}{n_g^F} \left. \frac{\partial \ln n^B_g}{\partial \ln L_s} \right|_{z, L_c}.
\end{aligned}
\end{equation}
For a split at fixed flux, both luminosity cuts evolve at the same rate, $\mathrm{d}\ln L_c/\mathrm{d}\ln(1+z) = \mathrm{d}\ln L_s/\mathrm{d}\ln(1+z) = 2[1 + 1/(x\mathcal{H})]$. Thus, using the definition of the magnification bias in \cref{eq:Q_be}, the evolution bias of the faint sample is given by \cref{eq:be_nevol} with the count slope and magnification bias of the faint sample, $n^F_{\rm evol}$ and $\mathcal{Q}^F$. As before, for galaxy surveys, like DESI BGS, that measure fluxes in fixed wavelength bands, the magnification bias term also contains a K-correction. If the split is instead chosen to keep a fixed fraction of bright galaxies at each redshift, the split flux varies with redshift, the two cuts no longer evolve at the same rate, and both the bright and faint evolution biases pick up an additional term proportional to $\mathcal{Q}^B\,\mathrm{d}\ln F_s/\mathrm{d}\ln(1+z)$ \citep{Blanco_2024}. For such a split the bright and faint counts are fixed fractions of the total, $N^B\propto N^F\propto N^T$, so all three samples share the same count slope, $n_{\rm evol}$, and their evolution biases differ only through $\mathcal{Q}^B$, $\mathcal{Q}^F$ and this additional term. In our forecasts with a fixed bright fraction we therefore marginalise over a single $n_{\rm evol}$ per redshift bin, shared by the two samples.

\subsection{Luminosity function priors}\label{sec:lumfunc_prior}

The count slope and magnification bias are the logarithmic slopes of the number of sources above the survey flux cut with respect to redshift and luminosity, respectively (\cref{eq:Q_be,eq:be_nevol}), and are therefore fixed by the survey luminosity function, so existing constraints on the luminosity function can be turned into prior information on $n_{\rm evol}$ and $\mathcal{Q}$. To propagate the luminosity function uncertainty into these biases, we adopt a Monte Carlo forward model, following \citet{Wang_2020}: we draw realisations of the luminosity function parameters from their fitted distributions, compute $n_{\rm evol}$ and $\mathcal{Q}$ at the survey flux cut for each draw, and use the resulting mean and covariance as a Gaussian prior in each redshift bin. For the \textit{Euclid}-like H$\alpha$ survey we use the `Model $3$' luminosity function of \citet{Pozzetti_2016}, with the best-fit parameters and uncertainties given in their Table 2. Since each draw is a realisation of the full luminosity function, this approach automatically captures the correlation between $n_{\rm evol}$ and $\mathcal{Q}$ and, for a bright-faint split, between all of the relativistic biases of the two samples, $\{n^B_{\rm evol},\mathcal{Q}^B,n^F_{\rm evol},\mathcal{Q}^F\}$ (or $\{n_{\rm evol},\mathcal{Q}^B,\mathcal{Q}^F\}$ for a fixed bright fraction, for which the two samples share the same count slope; \cref{sec:bf_split}), which all derive from the same underlying luminosity function. For a single tracer, the resulting prior has $\sigma(n_{\rm evol})$ falling from $\approx1.9$ to $\approx1.0$ across our redshift bins and $\sigma(\mathcal{Q})\approx0.3$ ($10$--$14\%$ of its fiducial value), with a correlation of $\approx-0.8$ between the two in each bin.

These priors are only as reliable as the luminosity function model from which they are derived, and their widths include only the statistical uncertainty on the fitted parameters of a single model. With \textit{Euclid}'s far greater area and depth, the statistical uncertainty on the luminosity function should shrink considerably, leaving systematics\footnote{Such as [N\,\textsc{ii}] deblending, aperture corrections and survey completeness.} and the choice of functional form as the dominant contributions to the error budget. The latter is especially important here: $n_{\rm evol}$ and $\mathcal{Q}$ are derivatives of the luminosity function, so mismodelling of its shape is amplified rather than averaged over. Indeed, \citet{Guandalin_2023} showed that quasar luminosity functions that are all consistent with the same data can yield significantly different relativistic biases.

\begin{figure*}
\centering
\includegraphics[width=\linewidth]{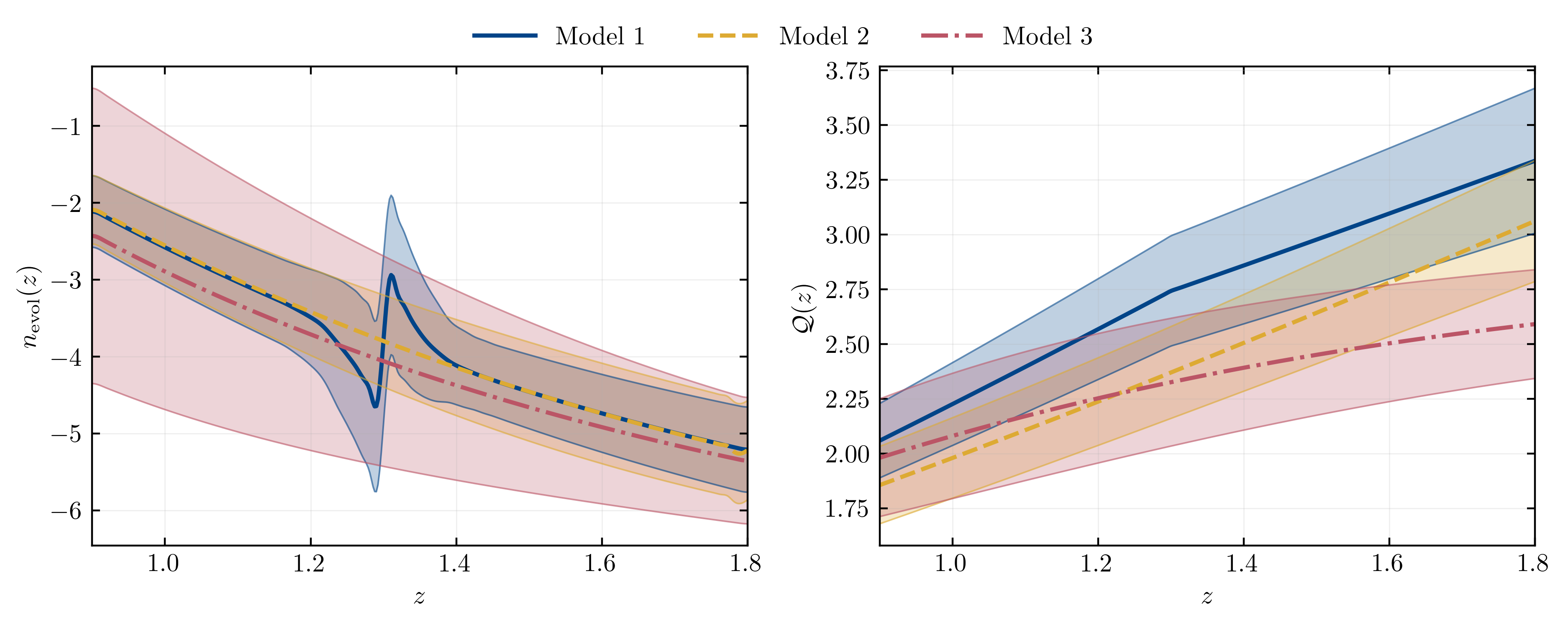}
\caption{Count slope $n_{\rm evol}$ (left) and magnification bias (right) derived from the three H$\alpha$ luminosity function models of \citet{Pozzetti_2016}, with the bands showing the $1\sigma$ uncertainty forward-modelled from the uncertainties on each model's fitted parameters. The flux cuts for Models $1$ and $2$ are chosen as a function of redshift such that all three models give the same comoving number density. The feature in $n_{\rm evol}$ for Model $1$ at $z\approx1.3$ arises from its broken power law form.}\label{fig:3_models}
\end{figure*}

To illustrate this for our H$\alpha$ sample, we forward model $n_{\rm evol}$ and $\mathcal{Q}$ from all three luminosity function models of \citet{Pozzetti_2016}. To isolate the effect of the shape of the luminosity function, we require that all three models share the same number density at every redshift, which we achieve through a redshift-dependent flux cut for Models $1$ and $2$ (so that $\mathcal{Q}$ is necessarily evaluated at a different flux cut for each model). The results are shown in \cref{fig:3_models}. Since matching the number density largely fixes the redshift distribution, the three models agree closely on $n_{\rm evol}$, to within $\approx0.3$ and well inside the `Model $3$' prior, apart from a sharp feature near $z\approx1.3$ that arises from the break in the broken power law of Model $1$. Their magnification biases, however, are mutually inconsistent, particularly at higher redshift: by $z\approx1.8$ Models $1$ and $3$ differ by $\Delta\mathcal{Q}\approx0.75$, several times the statistical width of the `Model $3$' prior, which through \cref{eq:be_nevol} corresponds to $\Delta b_e\approx3$.

The model-to-model scatter in $\mathcal{Q}$ therefore exceeds the statistical uncertainty on which our priors are based, and the forecasts with these priors should be read as optimistic. \citet{Elkhashab_2022} similarly found that, for a \textit{Euclid}-like H$\alpha$ survey, the detectability of lensing in the power spectrum monopole depends on the faint-end slope of the luminosity function. We leave a systematic treatment of luminosity function modelling uncertainty to future work.

\section{Survey details}\label{app:survey_details}

In this appendix we describe the modelling and assumptions for each survey. For MegaMapper and \textit{Euclid} these set-ups are similar to those described in \citetalias{Addis_2025}; however, we also extend the modelling to the second-order and PNG biases using a peak-background split approach. We outline the HODs we use and then briefly describe the remaining assumptions for each survey.

\subsection{Halo occupation distributions}

In our forecasts we use two different HOD models which are briefly described below:

\subsubsection{Two-parameter HOD}
The forecasts for the \textit{Euclid}-like H$\alpha$, MegaMapper-like and \textit{SPHEREx}-like surveys use an HOD, $\langle N(M,z)\rangle$, following \citet{Yankelevich_2018}, based on earlier work from \citet{Geach_2012,Gonzalez_2018}, with two free parameters, $M_0$ and $N_{HO}$. These are fitted to the linear bias and number density models at each redshift, so that both are functions of redshift; $M_0$ can be fitted independently, as the linear bias does not depend on $N_{HO}$. See \textsc{CosmoWAP} for implementation details and best-fit parameters for different cosmologies and luminosity cuts.

\subsubsection{DESI-like BGS HOD}
The DESI-like BGS survey uses a five-parameter HOD model based on \citet{Zheng_2005,Zehavi_2011} and implemented by \citet{Smith_2017}. We adopt the best-fitting HOD parameters of \citet{Smith_2024}, which are smooth functions of the absolute magnitude threshold in ${}^{0.1}M_r$, the $r$ band K-corrected to $z=0.1$. These were fitted on the $z=0.2$ snapshot of the AbacusSummit simulations \citep{Abacus_2021} to reproduce both the clustering and the number densities of magnitude-threshold samples in the mock catalogue of \citet{Smith_2017}, which was itself tuned on SDSS and GAMA data. We extend the fits to other redshifts with the method of \citet{Smith_2017}, following the public \textsc{hodpy} code \href{https://github.com/amjsmith/hodpy}{\faGithub}; at each redshift a target luminosity function, the evolving Schechter fit to GAMA of \citet{Loveday_2012}, gives the number density of galaxies brighter than the absolute magnitude threshold. The threshold is mapped to the one at $z=0.2$ with the same target number density and the HOD parameters are taken there, so that the shape of the HOD is fixed at fixed number density; its three mass parameters are then scaled by a common factor until the HOD reproduces the target number density on our halo mass function. The number density, $\mathcal{Q}$ and $n_{\rm evol}$ are therefore those of the target luminosity function, while the HOD sets the clustering biases. Unlike \textsc{hodpy}, we use the GAMA luminosity function at all redshifts rather than blending to an SDSS one below $z\approx0.15$, as that sharp switch produces a spurious feature in $n_{\rm evol}$.

\subsection{Survey specifications}

\subsubsection{DESI BGS} To forecast for a DESI-like BGS survey we use a three-bin, 15,000 $\rm{deg}^2$ survey over the redshift range $(0.1<z<0.5)$. We fix the $r$-band apparent magnitude cut-off to $m_c=20.175$ and use the colour-dependent GAMA K-corrections of \citet{Smith_2017}, averaged over the colours of their mock galaxies at the magnitude limit, so that a single K-correction is applied at each redshift, to the bright and faint samples alike. For $r<19.8$ the resulting redshift distribution agrees with the GAMA fit of \citet{Smith_2017} to within $3\%$ for $0.1<z<0.4$. K-correcting each galaxy by its own colour, as the mocks do, would instead give a $\mathcal{Q}$ that is $\sim5$–$10\%$ larger for $z\gtrsim0.3$, so our relativistic biases are more approximate at these redshifts.

\subsubsection{Euclid} We consider a five-bin, 15,000 $\rm{deg}^2$ H$\alpha$ \textit{Euclid}-like survey $(0.9<z<1.8)$. Along with the two-parameter HOD, we adopt the `Model $3$' H$\alpha$ luminosity function \citep{Pozzetti_2016,Euclid_2020} and use a flux cut of $F_c = 2\times10^{-16}\,[\mathrm{erg\,cm^{-2}\,s^{-1}}]$. We also adopt a phenomenological fit for the linear bias,
\begin{equation}
    b_1(x,z) = a + b(1+z)^e (1+\mathrm{exp}((x-c)\,d))
\end{equation}
as a function of redshift, $z$, and luminosity, $x = \log_{10}(L/[\mathrm{erg\,s}^{-1}])$. The fitted parameters are taken from \citet[Table 1]{Pan_2020}, their fit to the differential bias of galaxies at luminosity $x$. These fits derive from the semi-analytic galaxy formation model GALFORM \citep{Cole_2000,Gonzalez_2018}. The cumulative linear galaxy bias above the luminosity cut, $L_c$, is then given by \citep[eq.~15]{Pan_2020}
\begin{equation}\label{eq:biasluminosity}
b_1(\ge L_c)= \frac{\int_{L_c}^{+\infty} \dd L \, b_1(L,z) \phi(L,z)}{\int_{L_c}^{+\infty} \dd L \, \phi(L,z)}
\end{equation}
where $\phi(L,z)$ is the H$\alpha$ luminosity function. By incorporating $b_1$ and the comoving number density, $n_g$, derived from the luminosity function, we can constrain the two-parameter HOD to model the remaining bias parameters. See \cref{sec:lumfunc_prior} for how we forward model the uncertainty in the H$\alpha$ luminosity function into priors on $n_{\rm evol}$ and $\mathcal{Q}$.

\subsubsection{MegaMapper}

For the MegaMapper-like forecasts we consider a ten-bin, Stage V 20,000 $\rm{deg}^2$ idealised Lyman-break galaxy (LBG) survey over $(2<z<5)$. To do this, we follow \citet{Wilson_2019} and use a UV Schechter luminosity function \citep{Schechter_1976,Ono_2018} with an apparent magnitude cut, $m_c =24.5$. For the linear clustering bias, we adopt the redshift and apparent magnitude dependent function from \citet{Wilson_2019}, of the form,
\begin{equation}\label{eq:megamapperbias}
b(z,m) = A(m)(1+z) + B(m)(1+z)^2
\end{equation}
where $A(m) = -0.98(m-25) + 0.11$ and $B(m) = 0.12(m-25) + 0.17$. As this is a fit to the measured bias of magnitude-limited samples, we evaluate it directly at the survey cut, $b_1(z) = b(z,m_c)$. As in the \textit{Euclid}-like H$\alpha$ survey, we constrain the HOD using this linear bias model and a number density derived from the luminosity function for a given apparent magnitude cut. Absolute magnitudes are obtained with a K-correction, $K(z) = -2.5\log_{10}(1+z)$, i.e.\ the bandwidth term alone with no colour correction, as the observed band tracks the rest-frame UV.

\subsubsection{SPHEREx}\label{app:spherex} For the \textit{SPHEREx}-like forecasts we consider a ten-bin, all-sky survey over $(0.1<z<4.3)$, retaining $75\%$ of the sky ($\sim 31,000 \, \rm{deg}^2$) \citep{Dore_SPHEREx}. We use only the highest-accuracy redshift sample, $\sigma_z/(1+z) <0.003$. This is partly why our constraints on local-type $\fNL$ are weaker than previous \textit{SPHEREx} forecasts, $\sigma(\fNL^{\rm loc})\approx0.9$ from the power spectrum \citep{Dore_SPHEREx} and $\approx0.7$ from the bispectrum \citep{Heinrich_2024}, which combine all five redshift-accuracy samples as multiple tracers. The power spectrum constraint is driven by the larger samples with less precise redshifts, up to $\sigma_z/(1+z)=0.2$ \citep{Dore_SPHEREx}, and even for the bispectrum, keeping only the two most precise samples weakens the constraint by $\sim90\%$ \citep{Heinrich_2024}. The comoving number density and the abundance-matched linear bias of each subsample are taken directly from the public \textit{SPHEREx} GitHub \href{https://github.com/SPHEREx/Public-products/tree/master}{\faGithub}, and we fit the latter with
\begin{equation}\label{eq:spherexbias}
  b_1(z) = A(1+\beta z)^{\gamma}.
\end{equation}
As for the \textit{Euclid}-like and MegaMapper-like surveys, we constrain the two-parameter HOD with this linear bias and number density to model the remaining bias parameters. Following \citet{Lukens_2026}, we adopt the $2.4 \, \mu$m WISE continuum luminosity function of \citet{Lake_2018}, taking their canonical `High $z$ Prior' fit together with an effective flux cut of $\nu F_\nu = 2\times10^{-16}\,[\mathrm{erg\,cm^{-2}\,s^{-1}}]$ at $2.4\,\mu$m. Since this is a continuum rather than a line luminosity, the limiting luminosity also carries a K-correction; we adopt the same K-correction as in the MegaMapper case and neglect the colour term. We note that this luminosity function is calibrated on $z \lesssim 1$ data, so its use over the full \textit{SPHEREx} range is an extrapolation.

\section{Additional Plots}

\begin{figure*}
\centering
\includegraphics[width=\linewidth]{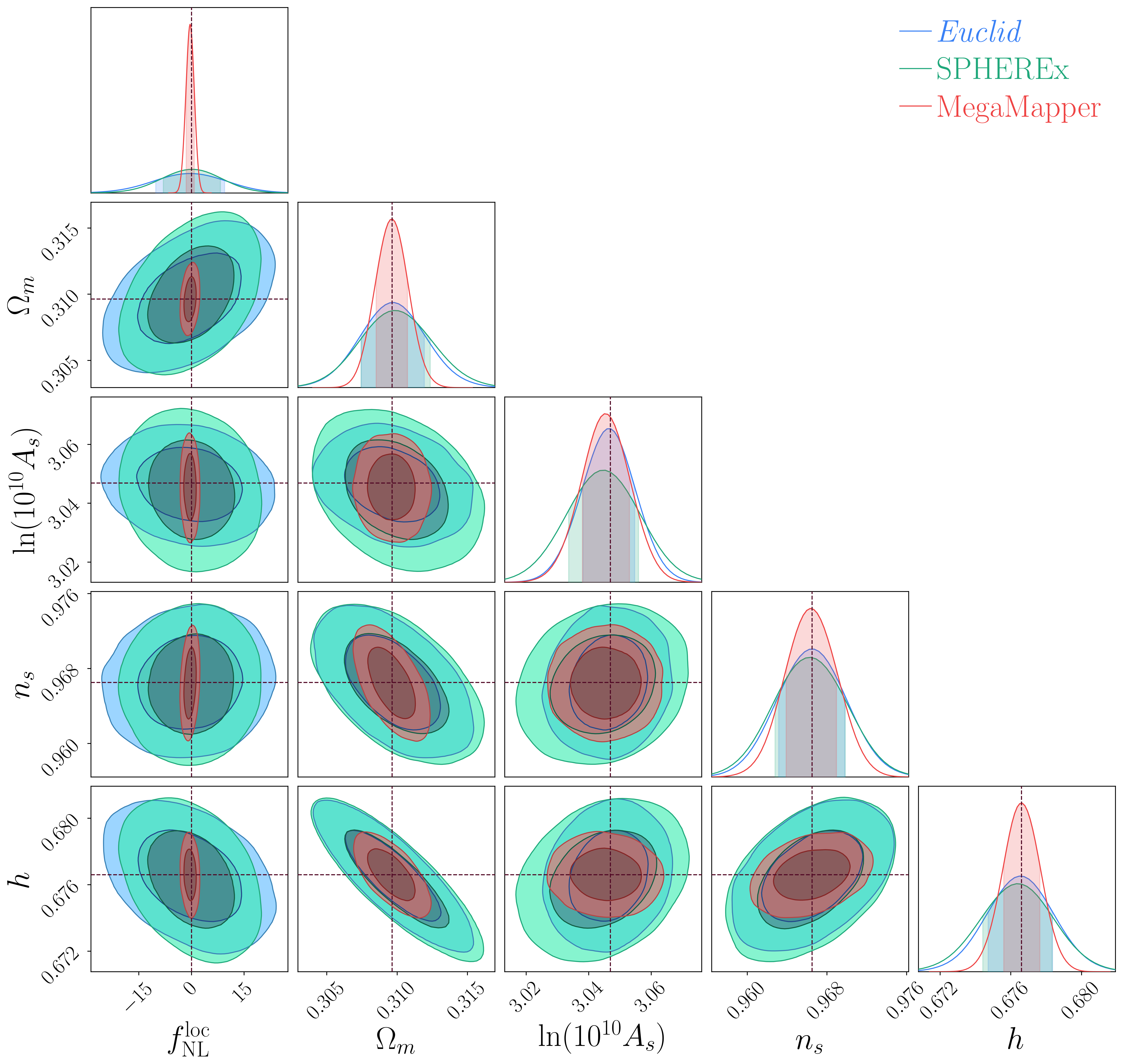}
\caption{As \cref{fig:fnl_bias}, but with the integrated (lensing, ISW and time-delay) contributions omitted from the data, so that the shifts arise only from neglecting the local relativistic and wide-separation effects. All shifts are $\lesssim0.4\sigma$, so the biases in \cref{fig:fnl_bias} are driven by the integrated terms.}\label{fig:fnl_bias_noint}
\end{figure*}

\begin{figure*}
\centering

\includegraphics[width=0.73\linewidth]{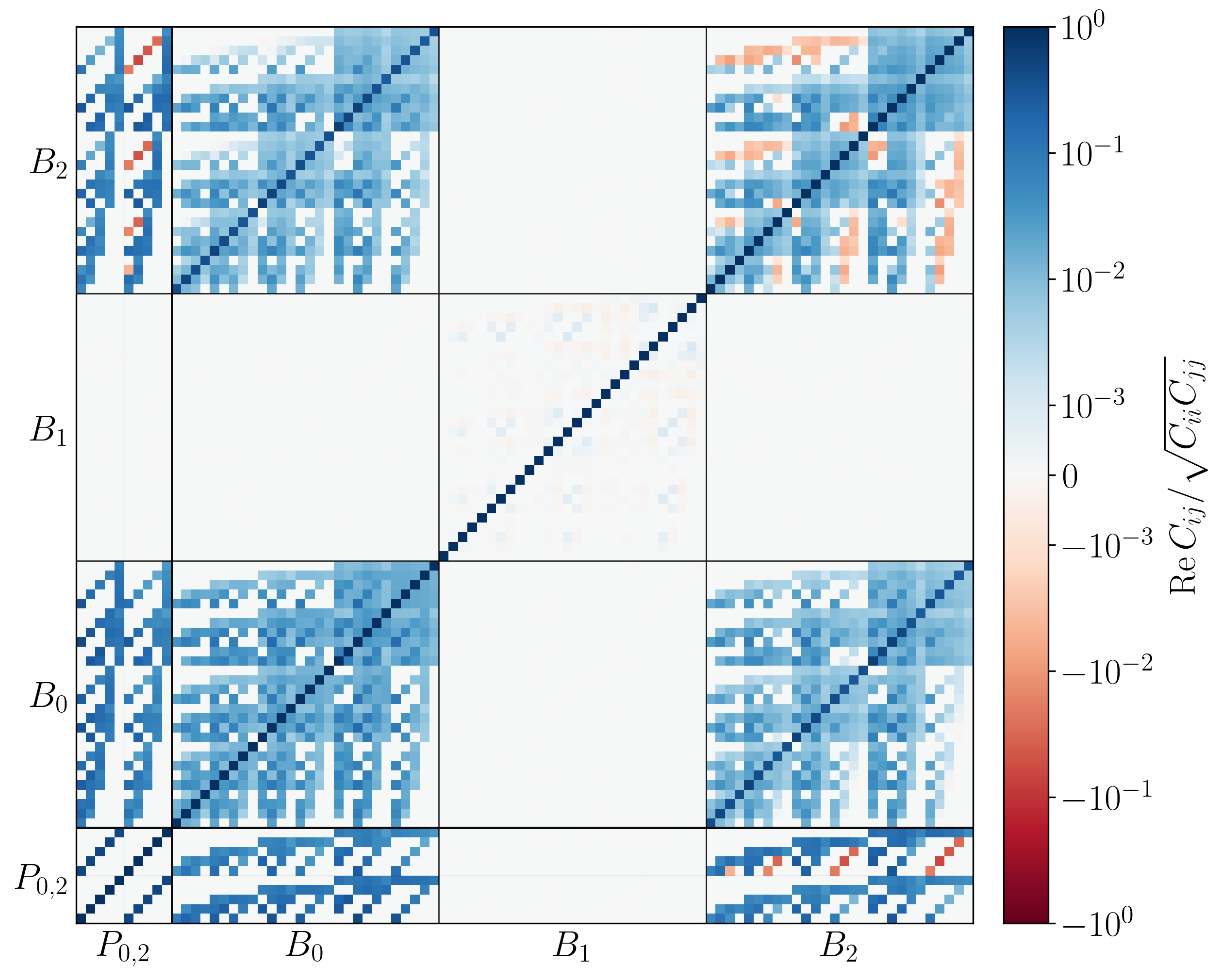}

\includegraphics[width=0.73\linewidth]{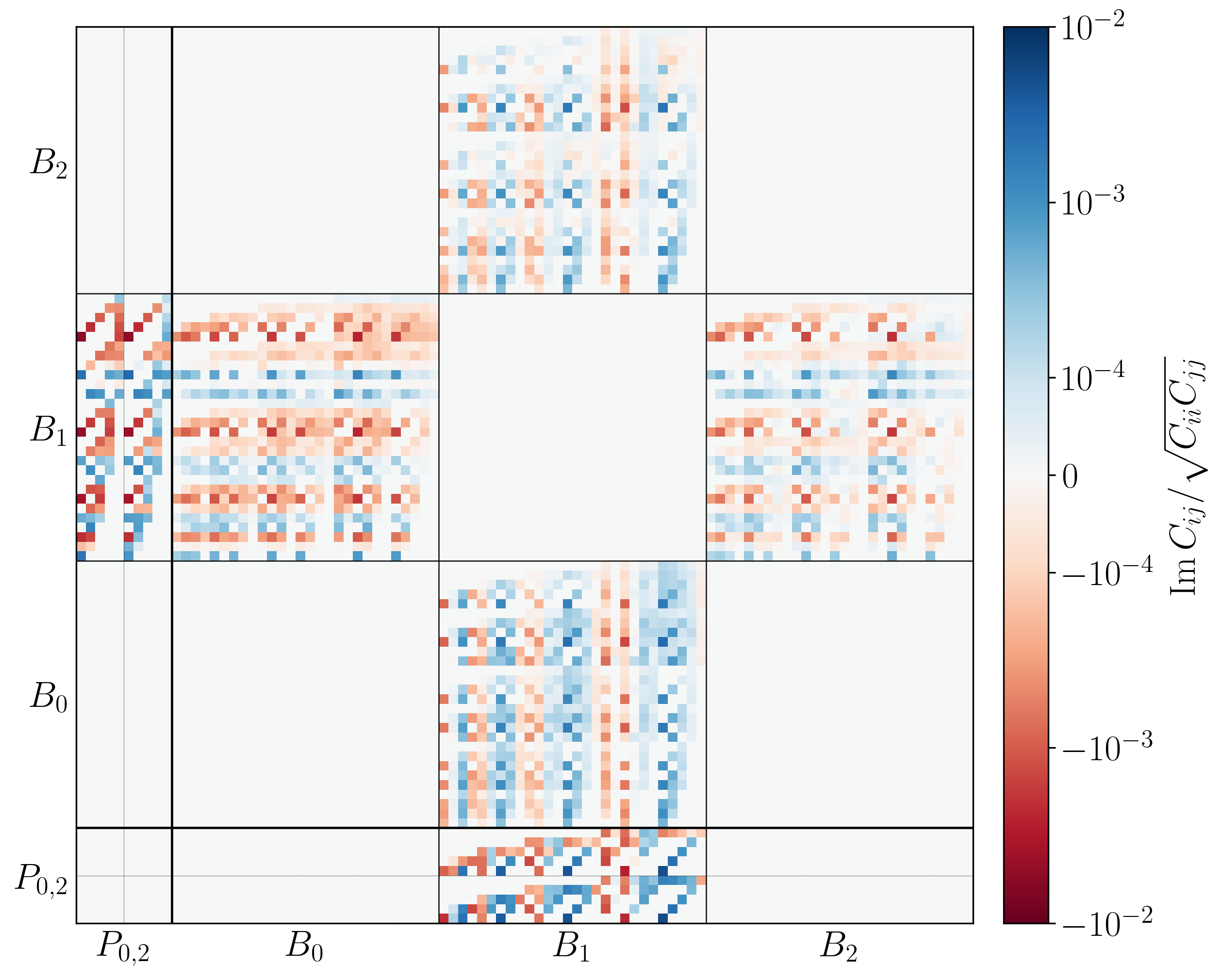}
\caption{Real (top) and imaginary (bottom) parts of the correlation matrix, $C_{ij}/\sqrt{C_{ii}C_{jj}}$, of the joint single-tracer data vector of the power spectrum ($\ell=0,2$) and bispectrum ($\ell=0,1,2$) multipoles, including the non-Gaussian terms of \cref{eq:cov_ng,eq:cov_pb}, for the same survey and redshift bin as \cref{fig:cov_ng}, with $k_{\rm max}=0.1\,[\aMpch]$ for both. Within each block, the rows run over the $k$-bins of the power spectrum, or over the triangle bins in the order of \cref{fig:cov_ng}. The colour scales are logarithmic, and linear below $10^{-3}$ (top) and $10^{-4}$ (bottom), with the imaginary part on a much smaller scale. The imaginary part is antisymmetric and non-zero only between multipoles of opposite parity, so the bispectrum dipole is correlated with the even multipoles and the power spectrum only through it, and only weakly (see \cref{sec:cov_ng}). For visual clarity we use wider bins here, $\Delta k=6k_{\rm f}$.}\label{fig:cov_ng_corr}
\end{figure*}

\begin{figure}
\centering

\includegraphics[width=\linewidth]{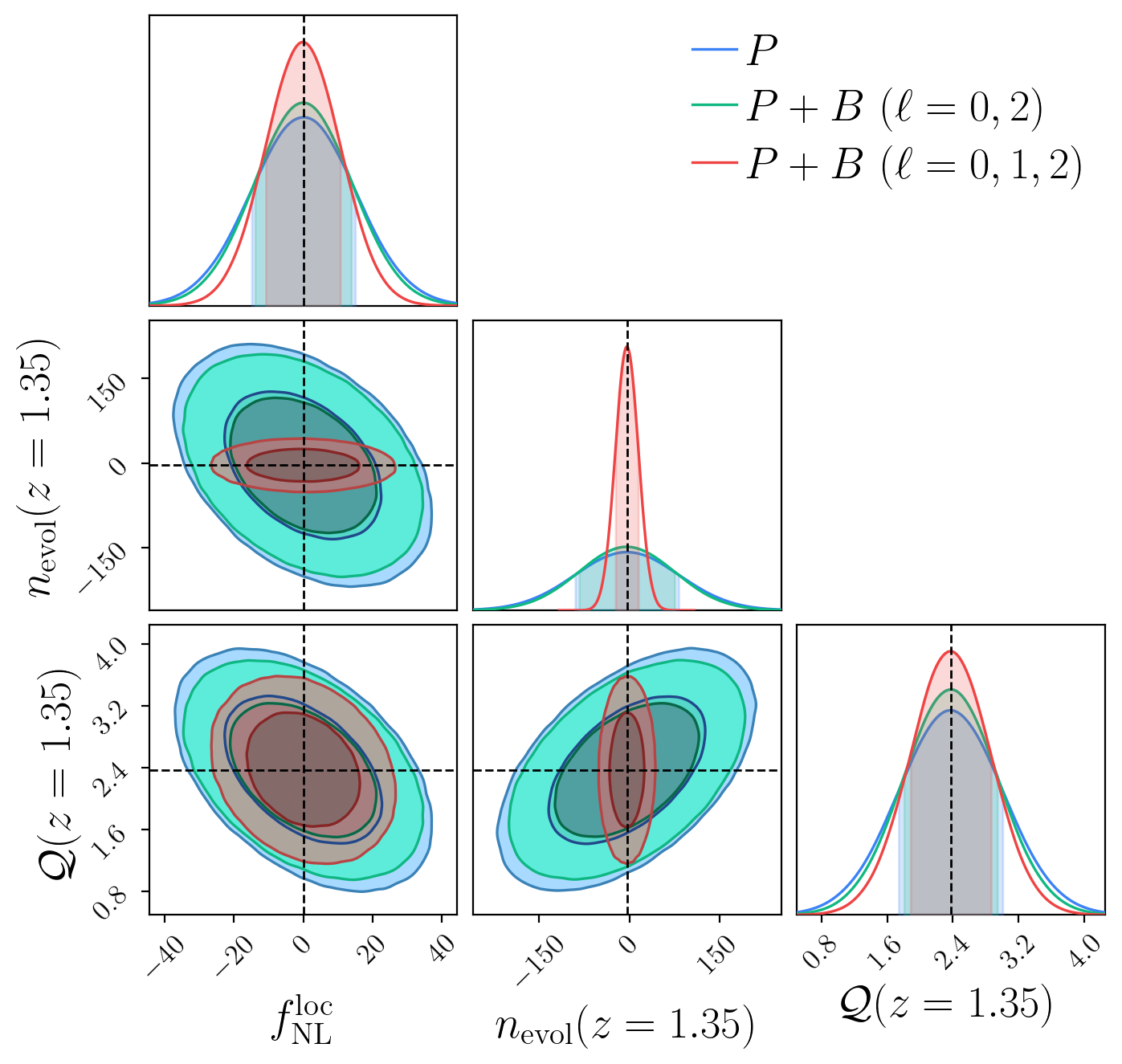}
\caption{Fisher constraints on local-type $\fNL$ and on the relativistic biases, $n_{\rm evol}$ and $\mathcal{Q}$, in the redshift bin centred at $z=1.35$, for a single-tracer \textit{Euclid}-like H$\alpha$ survey, with $1$ and $2\sigma$ contours. The data vectors are the power spectrum ($\ell=0,2,4$; blue) and the joint power spectrum and bispectrum without (green) and with (red) the bispectrum dipole. All other parameters are marginalised over. Without the dipole, $\fNL$ is anticorrelated with both biases. The dipole effectively removes the correlation with $n_{\rm evol}$ and narrows its constraint by a factor of $\sim4$.}\label{fig:fnl_con_pk_bk}
\end{figure}

\begin{figure}
\centering
\includegraphics[width=\linewidth]{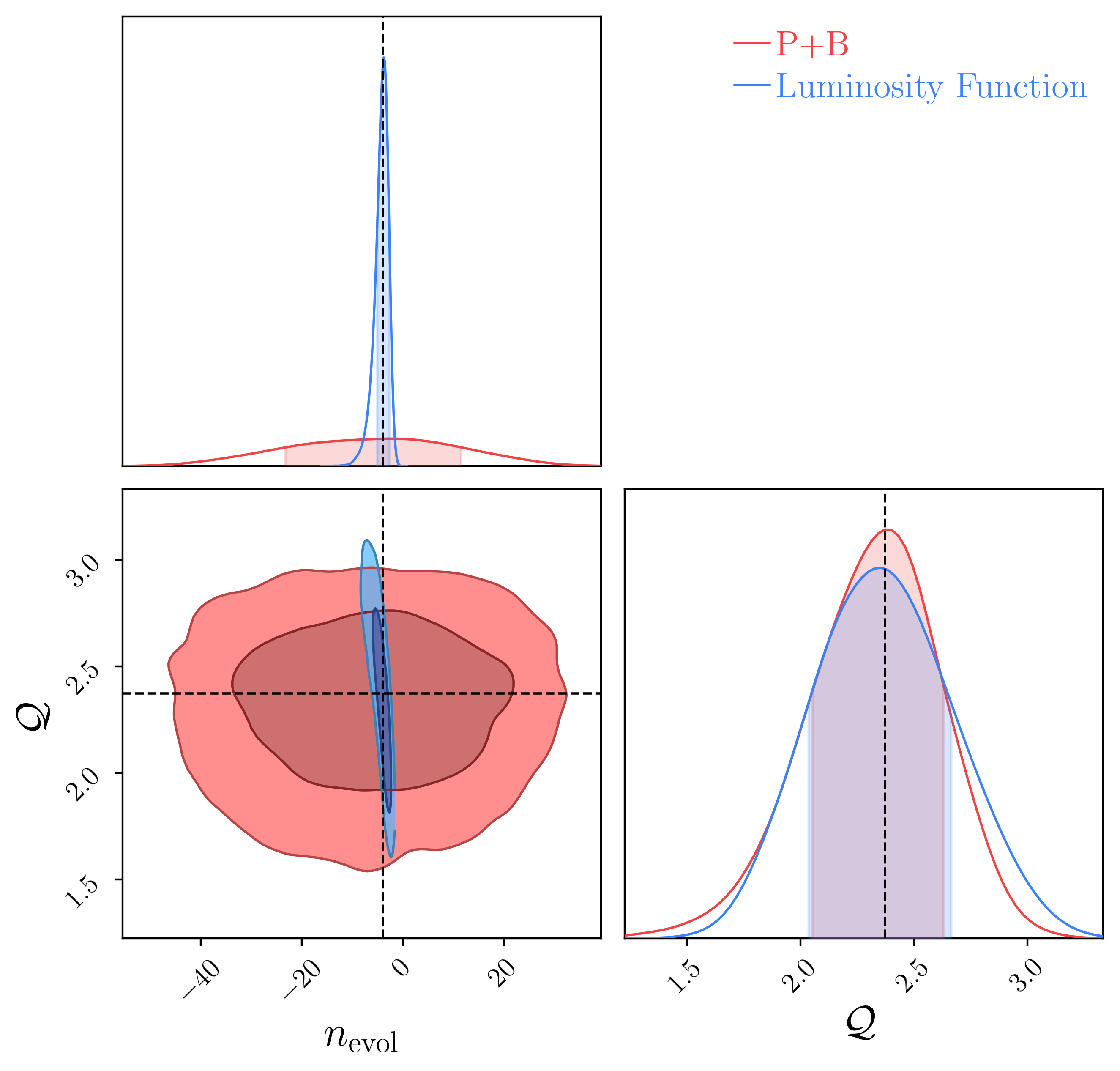}
\caption{Constraints on the relativistic biases for a single-tracer \textit{Euclid}-like H$\alpha$ survey from an MCMC for a single redshift bin of width $\Delta z = 0.18$ centred at $z=1.35$, with $1$ and $2\sigma$ contours. Unlike the Fisher constraints in \cref{fig:beQ_st}, these capture the non-Gaussianity of the posteriors. The joint power spectrum and bispectrum constraints, including the bispectrum dipole, are shown in red and the priors forward-modelled from the `Model $3$' luminosity function in blue.}\label{fig:beQ_singlebin}
\end{figure}

\begin{figure*}
\centering
\includegraphics[width=0.9\linewidth]{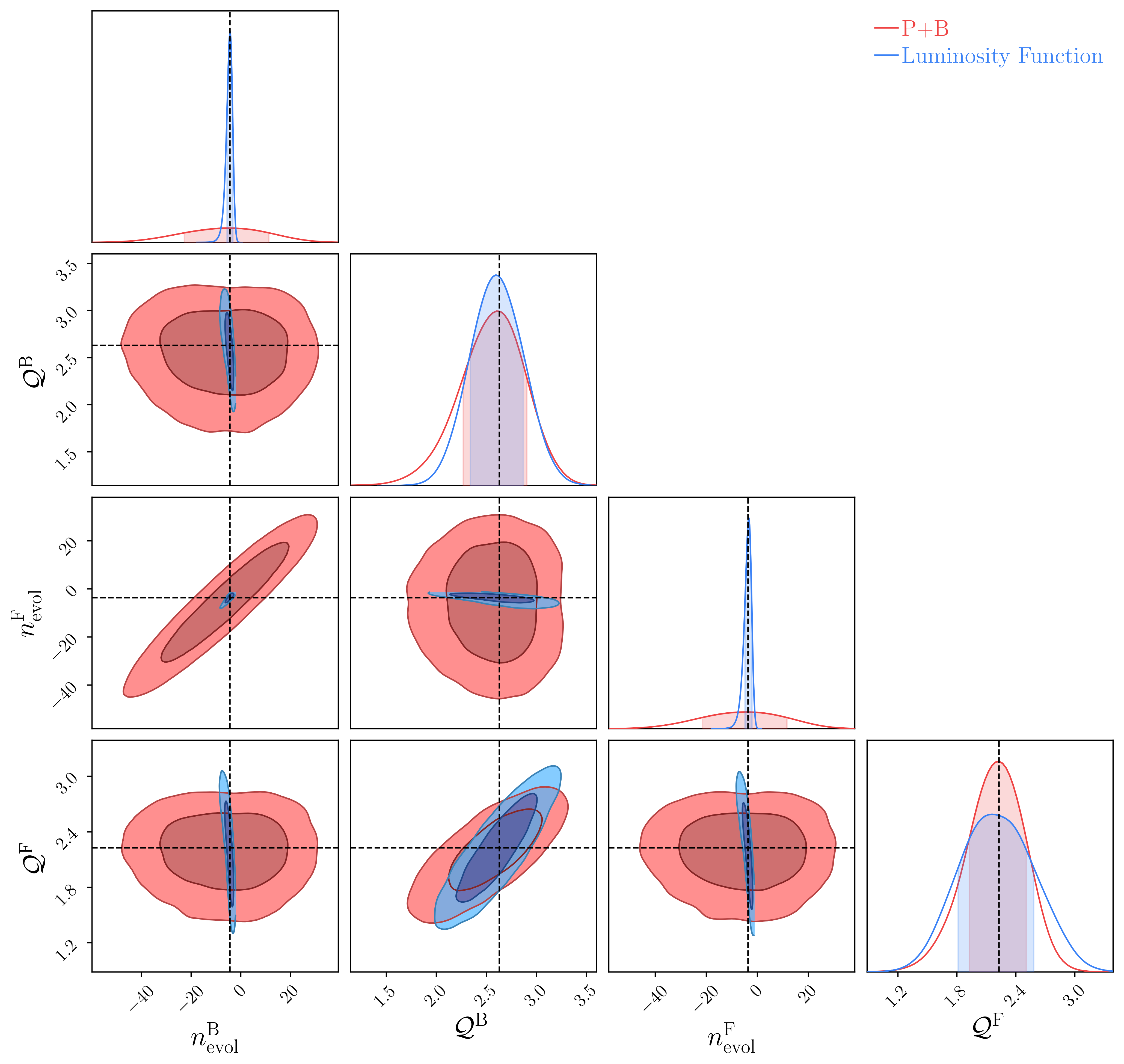}
\caption{As \cref{fig:beQ_singlebin}, but for a bright-faint split at a fixed flux, so that each sample has its own $n_{\rm evol}$. The joint constraints include the bispectrum dipole and the odd cross-power spectrum multipoles.}\label{fig:beQ_singlebin_BF}
\end{figure*}

% Don't change these lines
\bsp	% typesetting comment
\label{lastpage}
\end{document}